\documentclass[prx,showpacs,twocolumn,amsmath,amssymb,floatfix]{revtex4-1}

\usepackage{hyperref}
\usepackage{graphicx}
\usepackage{bm}
\usepackage{color}
\usepackage[applemac]{inputenc}
\usepackage{amssymb}

\newcommand{\red}[1] {\textcolor{black}{#1}}   
\newcommand{\editR}[1] {\textcolor{black}{#1}}
\newcommand{\editE}[1] {\textcolor{black}{#1}}
\newcommand{\editP}[1] {\textcolor{black}{#1}}

\begin{document}  
   
\title{From Andreev to Majorana bound states in hybrid superconductor-semiconductor nanowires}

\author{Elsa Prada$^1$, Pablo San-Jose$^2$, Michiel W. A. de Moor$^3$, Attila Geresdi$^3$, Eduardo J. H. Lee$^1$, Jelena Klinovaja$^4$, Daniel Loss$^4$, Jesper Nyg\aa rd$^5$, Ram\'on Aguado$^2$, Leo P. Kouwenhoven$^{3,6}$}
\affiliation{\\
$^1$Departamento de F\'isica de la Materia Condensada and Condensed Matter Physics Center (IFIMAC), Universidad Aut\'onoma de Madrid, Madrid, Spain\\$^2$Instituto de Ciencia de Materiales de Madrid (ICMM), Consejo Superior de Investigaciones Cient\'{i}ficas (CSIC), Madrid, Spain\\$^3$QuTech and Kavli Institute of Nanoscience, Delft University of Technology, Delft, Netherlands\\$^4$Department of Physics, University of Basel, Basel, Switzerland\\$^5$Center for Quantum Devices, Niels Bohr Institute, University of Copenhagen, Copenhagen, Denmark\\$^6$Microsoft Station Q at Delft University of Technology, Delft, Netherlands.}

\date{\today} 

\begin{abstract}
Electronic excitations above the ground state must overcome an energy gap in superconductors with spatially-homogeneous $s$-wave pairing. In contrast, inhomogeneous superconductors such as those with magnetic impurities or weak links, or heterojunctions containing normal metals \editE{or quantum dots}, can host subgap electronic excitations that are generically known as Andreev bound states (ABSs). With the advent of topological superconductivity, a new kind of ABS with exotic qualities, known as Majorana bound state (MBS), has been discovered. We review the main properties of \editP{ABSs and MBSs}, and the state-of-the-art techniques for their detection. We focus on hybrid superconductor-semiconductor nanowires, possibly coupled to quantum dots, as one of the most flexible and promising experimental platforms. We discuss how the combined effect of spin-orbit coupling and Zeeman field in these wires triggers the transition from ABSs into MBSs. We show theoretical progress beyond minimal models in understanding experiments, including the possibility of \editP{different types of robust zero modes that may emerge without a} band-topological transition. We examine the role of  spatial non-locality, a special property of MBS wavefunctions that, together with non-Abelian braiding, is the key \editP{to realizing topological quantum computation}.
\end{abstract}

\maketitle

\section{Introduction}
\label{sec:introduction}

\begin{figure*}
\includegraphics[width=\textwidth]{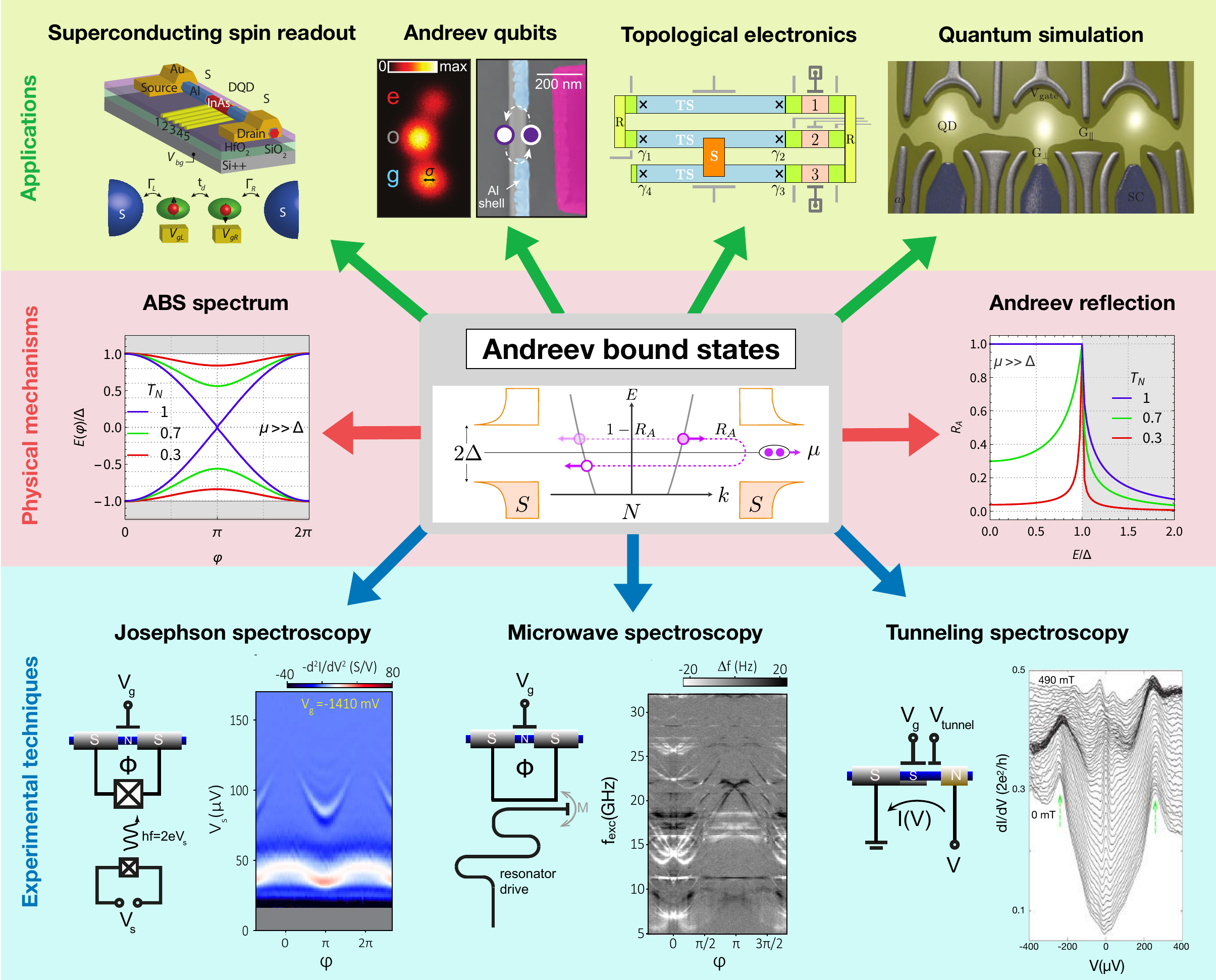}
\caption{\label{Fig1}\textbf{Andreev bound states (ABSs) in semiconducting nanowires}. ABSs are at the heart of several physical mechanisms (central row), experimental techniques (lower row) and applications (upper row) in condensed matter physics. \textbf{Central row:} Andreev reflection at an NS junction, see central panel, is the retro-reflection of an electron into a hole (or viceversa) of opposite spin and \editE{velocity}, with the addition (or removal) of a Cooper pair to the SC condensate. In contrast, normal (specular) reflection leaves the particle and spin quantum numbers unchanged. The probability of each ($R_A$ vs. $1-R_A$ below the gap) depends on normal-state transparency $T_N$ and energy of the incident electron $E$. The central-right panel shows the Andreev reflection probability versus $E$ at an NS junction of $T_N$ in the high density limit (chemical potential $\mu$ much larger than superconducting gap $\Delta$). Multiple coherent Andreev reflections in a short SNS Josephson junction produce an ABS confined to the N region with energy $E(\varphi)$ below the gap. This energy depends on $T_N$ and the phase difference $\varphi$ between the two SCs, see central-left panel.
\textbf{Lower row:} Several experimental techniques have been developed to probe the ABS spectrum, amongst which we highlight Josephson spectroscopy using the AC Josephson effect of a capacitively coupled tunnel junction, microwave spectroscopy through the dispersive shift of a planar resonator, and tunneling spectroscopy using the differential conductance into the nanowire through an opaque barrier. In the bottom-left panel, a nanowire Josephson junction with a gate voltage $V_g$ is embedded in a SQUID loop, which sets a phase bias of $\varphi=2\pi \Phi/\Phi_0$, where $\Phi$ is the applied flux and $\Phi_0=h/2e$ the superconducting flux quantum. The Andreev level excitation frequency $f$ is set by $V_s=hf/2e$, the spectrometer bias voltage.
Nearby we show the measured excitation spectrum in the single channel regime, where the phase-dependent Andreev level (upper line) and the Josephson plasma oscillations (lower line) contribute to the signal. Experimental data are reproduced from Ref. \cite{Woerkom:NP17}. The bottom-central panel showcases experiments using the dispersive shift $\Delta f$ of an inductively coupled planar superconducting microwave resonator. \editE{The data, as a function of the excitation frequency $f_\textrm{exc}$ and phase bias $\varphi$}, are reproduced from Ref. \cite{Tosi:PRX19}. The bottom-right panel shows the setup and an experimental dataset for voltage bias spectroscopy, reproduced from Ref. \cite{Mourik:S12}, where the differential conductance $dI/dV$ is measured as a function of the voltage bias $V$. A tunnel barrier is created by depleting a section of the nanowire by the local gate voltage $V_{\textrm{tunnel}}$.
\textbf{Upper row}: Potential application domains of ABSs in quantum technologies include single spin readout \cite{Estrada-Saldana:PRL18}, Andreev quantum bits \cite{Hays:PRL18}, topological quantum electronics \cite{Plugge:NJP17} and hybrid quantum simulators \cite{Fulga:NJOP13}.}

\end{figure*}

Ever since Kamerlingh Onnes discovered the ``zero resistance state'' of metals at very low temperatures in 1911 \cite{Kamerlingh-Onnes:CPLUL11,vanDelft:PT2010}, the superconducting state of matter \cite{De-Gennes:18,Tinkham:04} has fascinated 
physicists. In the last century, the understanding of superconductivity has evolved extraordinarily and has garnered eight Nobel prizes, turning it into one of the most iconic topics in condensed matter physics \cite{Martin:PT2019}. 
As described by the seminal Bardeen-Cooper-Schrieffer (BCS) theory of superconductivity \cite{Bardeen:PR57}, the characteristic feature of superconductors (SCs) is the macroscopic occupation of bound pairs of electrons, known as Cooper pairs \cite{Cooper:1956}, in the same quantum-coherent ground state. The condensation of Cooper pairs into such ground state is associated with a superconducting complex order parameter $
\bm{\Delta}=\Delta e^{i\varphi}$ \cite{Ginzburg:ZETF50,Cyrot:ROPIP73}, where $\varphi$ is the conjugate of the number of Cooper pairs. In a homogeneous $s$-wave BCS SC, \editP{the spectrum of single-particle  excitations above the ground state develops an energy gap $\Delta$. These gapped excitations} are propagating superpositions of electrons and holes with different energy-dependent weights. However, if the order parameter --also called the pair potential \cite{De-Gennes:18}-- varies in space, $\bm{\Delta}(\bf{r})$, lower energy (`subgap') excitations may develop. Such is the case of states trapped in magnetic flux vortices (so-called Caroli-Matricon-De Gennes states \cite{Caroli:PL64}), at magnetic domains or impurities (Yu-Shiba-Rusinov states \cite{Yu:APS65,Shiba:PTP68,Rusinov:SPJ69}), at weak links between SCs or at normal metal-superconductor (NS)  contacts \cite{Blonder:PRB82}, to name a few. Collectively, these subgap states are dubbed Andreev bound states (ABSs), and are the focus of numerous theoretical and experimental \editP{works}, as well as the basis of promising emerging quantum technologies, see Fig. \ref{Fig1}.

The core physical mechanism behind the formation of ABSs in inhomogeneous systems with $\bm{\Delta}(\bf{r})$ is a remarkable scattering process, predicted by Andreev \cite{Andreev:SPJ64,Andreev:SPJ66}, in which an incoming particle-like excitation can convert into an outgoing hole-like one and viceversa, see central row of Fig. \ref{Fig1}. 
Many of such Andreev scattering events coherently concatenated lead to the formation of subgap ABSs \cite{De-Gennes:PL63,Kulik:SPJ70} that are localized near the region where the pair potential has strong spatial variations (for a recent review see \cite{Sauls:PTRSA18}). 
 
In the last decade, a new twist in the possibilities afforded by the superconducting pairing of electrons has been possible with the advent of topological materials \cite{Hasan:RMP10,Qi:RMP11}. Inspired by notions of topology \cite{Thouless:PRL82}, several authors have predicted the existence
of \editP{new states of matter known collectively as topological superconducting phases, see Refs. \cite{Leijnse:SSAT12,Alicea:RPP12,Beenakker:ARCMP13,Sato:JPSJ16,Aguado:RNC17,Sato:ROPIP17} for reviews. These arise in particular} in so-called $p$-wave SCs, which possess a rare triplet-like pair potential (an exotic form of superconductivity \editP{involving only a single spin band} \cite{Salomaa:PRB88,Volovik:09,Read:PRB00,Kitaev:PU01,Sato:PRB09}). \editP{Topological SC phases are} characterized by the emergence of a rather special type of subgap bound state occurring at topological defects such as vortices, boundaries or domain walls. Importantly, such bound states occur \emph{precisely} at zero energy, and exhibit electron and hole character with exactly equal probability. The second quantization operators describing these states are self-conjugate, $\gamma = \gamma^\dagger$. They are in this sense a condensed matter realisation of the celebrated `particle-equals-antiparticle' states known as Majorana fermions \cite{Majorana:INC137}, and also of so-called Jackiw-Rossi states at field vortices in the Dirac equation \cite{Nishida:PRB10,Jackiw:NPB81}.

As opposed to standard ABSs, which can be pushed out of the gap by continuous deformations of the Hamiltonian, Majorana bound states (MBSs) cannot be removed from zero energy by any local perturbation or local noise that does not close the gap. This \editP{robust zero-energy pinning is a consequence of} the bulk-boundary correspondence principle of band topology \cite{Fukui:JOTPSOJ12}, which predicts that at the boundaries between \editE{materials with different topological indices}, edge states must appear that are protected against perturbations by the topology of the bulk. Quite remarkably, MBSs do not follow fermion statistics, unlike the original particles predicted by Majorana \cite{Majorana:INC137}, but rather possess non-Abelian exchange statistics. Upon exchange of two MBSs (braiding), a non-trivial unitary operation will be performed on them. This property, together with their topological protection against local noise, holds promise for applications in fault-tolerant quantum computing \cite{Kitaev:AOP03,Nayak:RMP08}. 

The interesting connection between Dirac physics, superconductivity and Majorana zero modes was exploited by Fu and Kane in 2008 \cite{Fu:PRL08}, who put forward the conceptual breakthrough of \emph{effectively} creating $p$-wave superconductivity and MBSs out of standard $s$-wave SCs by virtue of the proximity effect acting onto the helical edge states of topological insulators (propagating edge states with spin-momentum locking). 
The possibility of combining different materials to \emph{engineer} the topological superconducting state has spurred immense interest in the physics of Majorana states in hybrid systems. 

Fu and Kane's idea was soon extended to other materials with helical states produced by strong spin-orbit (SO) coupling, but different from topological insulators \cite{Sato:PRB09}. A popular practical proposal was put forward independently by two groups in 2010 (Lutchyn \emph{et al.}~\cite{Lutchyn:PRL10} and Oreg \emph{et al.}~\cite{Oreg:PRL10}), that realizes the conceptual model for one-dimensional (1D) $p$-wave superconductivity proposed by Kitaev in 2001 \cite{Kitaev:PU01}. It was based on 1D low-density semiconducting nanowires under an external magnetic field $B$, which readily allowed its implementation in experiments. The combination of the SO interaction and the Zeeman field $V_Z=g\mu_B B/2$ associated to $B$ generates, for a small chemical potential $\mu$ in the nanowire, a helical phase similar to that of topological insulators but with broken time-reversal symmetry~\cite{Streda:PRL03}. By covering the nanowire with a conventional SC, its spectrum becomes gapped by the proximity effect. In this device, sometimes dubbed a Majorana nanowire, a topological transition in the form of a band inversion was predicted to occur at a critical Zeeman energy $V_Z^c$ of the order of the induced superconducting gap (Box \ref{box:1}).
The material properties necessary to realize this proposal in the lab can be achieved by using e.g. InAs or InSb semiconducting nanowires \cite{Stanescu:JOPCM13,Lutchyn:NRM18}. Hybrid superconducting-semiconducting devices based on such nanowires can be tuned to the topological phase by increasing $B$ and depleting the wires by means of gate voltages. In finite, \editP{but sufficiently long} wires, \editE{zero energy} MBSs emerge in pairs for $V_Z>V_Z^c$, one localized at either end. 
One pair of Majorana states forms a non-local fermion. The occupation of two such fermions defines the elementary qubit in proposals of topological quantum computers \cite{Kitaev:PU01}.

In this work we review the formation and properties of general subgap bound states in  nanowires and nanowire junctions, as they evolve from conventional ABSs in high-density nanowires (Sec. \ref{sec:abs}) to topological Majorana zero modes at low-densities and finite magnetic fields (Sec. \ref{sec:mbs}). We summarize the main experimental approaches currently used for their detection and characterization \cite{Mourik:S12,Woerkom:NP17,Tosi:PRX19}, including ABSs in nanowire quantum dots (QDs) \cite{Lee:PRL12,Lee:NN14,Lee:PRB17,Grove-Rasmussen:NC18,Su:PRL18,Junger:CP19}. Going beyond, we discuss in Sec. \ref{sec:beyond} various physical extensions of the minimal description of Majorana nanowires. These include multimode effects \cite{Potter:PRL10,Potter:PRB11,Lutchyn:PRL11,Lutchyn:PRB11}, renormalized $g$-factors and SO couplings due to strong proximity effect with the parent SC \cite{Cole:PRB15,Reeg:PRB17a,Reeg:PRB18,Vaitiekenas:PRL18,Antipov:PRX18,Dmytruk:PRB18,Pan:PRB19}, effects of the charge density distribution across the wire section and of the electrostatic environment \cite{Vuik:NJP16,Dominguez:NQM17,Mikkelsen:PRX18,Antipov:PRX18,Moor:NJP18,Woods:PRB18,Escribano:BJN18,Winkler:PRB19} or density and pairing inhomogeneities \cite{Prada:PRB12,Kells:PRB12,Stanescu:PRB13,Roy:PRB13,Fleckenstein:PRB18}. Such generalized nanowires have been predicted to sometimes develop 
robust zero modes \cite{Prada:PRB12,Kells:PRB12,Stanescu:PRB13,Roy:PRB13,Stanescu:PRB14,Liu:PRB17a,Fleckenstein:PRB18,Penaranda:PRB18,Reeg:PRB18b,Liu:PRB18,Moore:PRB18a,Moore:PRB18,Vuik:SP19,Awoga:PRL19,Avila:CP19,Stanescu:PRB19,Woods:PRB19} that cannot be classified using the band-topological concepts of uniform Majorana nanowires. They allow nevertheless a classification within the more general context of non-Hermitian topology \cite{Avila:CP19}.
We review the open questions that remain as to their nature (e.g. their location within the wire \cite{Stanescu:PRB13,Penaranda:PRB18,Fleckenstein:PRB18,Moore:PRB18a,Reeg:PRB18b,Vuik:SP19,Stanescu:PRB19,Woods:PRB19}, their degree of fermionic non-locality \cite{Prada:PRB17,Penaranda:PRB18,Deng:PRB18}, their decay into external leads \cite{San-Jose:SR16,Avila:CP19}, their resilience to perturbations \cite{Goldstein:PRB11,Budich:PRB12,Rainis:PRB12,Trif:PRL12,Schmidt:PRB12,Schmidt:PRL13,Scheurer:PRB13,Pedrocchi:PRL15,Dmytruk:PRB15,Sekania:PRB17,Knapp:PRB18,Aseev:PRB18,Lai:PRB18,Aseev:PRB19}) and the conditions for their emergence. Understanding these zero modes without a clear relation to bulk topology is particularly important currently, in view of the many observations of robust zero bias anomalies reported in recent experiments \cite{Deng:S16,Nichele:PRL17,Zhang:NC17}.


\editE{Although we focus on semiconductor nanowires in this review, we note that MBSs are also investigated in other material platforms, including atomic chains \cite{Nadj-Perge:PRB13,Nadj-Perge:S14}, monolayer islands \cite{Menard:NC17,Menard:NC19,Palacio-Morales:SA19}, topological insulators \cite{Fu:PRB09,Wiedenmann:NC16} and planar semiconductor heterostructures \cite{Suominen:PRL17,Nichele:PRL17},
while ABSs were studied in e.g. atomic point contacts \cite{Bretheau:N13,Janvier:S15}, carbon nanotubes \cite{Pillet:NP10,Eichler:PRL07}, graphene \cite{Dirks:NP11}, and nanoparticles \cite{Deacon:PRL10}.}

\section{ABS\lowercase{s} in high-density nanowires and QD\lowercase{s}}
\label{sec:abs}

\subsection{Formation of ABSs}
\label{sec:formabs}

ABSs arise in superconducting systems as the result of an unusual form of quantum confinement caused by so-called Andreev reflection \cite{Andreev:SPJ64,Andreev:SPJ66}. In a metallic system in its normal phase, electrons become specularly reflected at planar interfaces with vacuum or insulating materials. This is known as normal reflection. However, at an NS boundary \cite{Kummel:ZPA69,Blonder:PRB82}, an incoming electron from the N side may transform into an outgoing hole with inverted spin and wave vector. This hole is said to be retro-reflected since both the parallel and normal \editP{velocity} components to the interface change sign, whereas in a normal reflection the parallel component remains the same. This process is known as Andreev reflection, and is accompanied by the injection of a Cooper pair into the SC.  If the interface is highly transparent, below the gap such Andreev process dominates with high probability $R_A\approx 1$, whereas in the opposite limit the electron becomes normal-reflected ($1-R_A\approx 1$), see central panel of Fig. \ref{Fig1}.
\editE{Experimentally, $R_A$ can be characterized based on the finite subgap conductance of the NS interface \cite{Zhang:NC17}. The bias-dependent conductance of short nanowire segments between two SC leads was also used to extract the magnitude of $R_A$, close to one, in various semiconductor nanowires, including SiGe \cite{Xiang:NN06,Ridderbos:PRM19}, InAs \cite{Jespersen:NJP09,Doh:S05,Gunel:JAP12,Goffman:NJP17} and InSb \cite{Nilsson:NL12,Deng:NL12}.}

Consider now an electron in a normal metal between two or more insulating interfaces. When the metallic region is small, multiple coherent normal reflections on the  boundaries leads to the formation of electronic states of quantized energy. A similar process takes place when some or all of the
confining insulators are replaced by SC boundaries \cite{Andreev:SPJ66}. This leads to the
formation of ABSs, which are the superconducting analogue of the above particle-in-a-box states of quantum mechanics. 

The formation of ABSs becomes particularly simple in the common case of high density SCs with negligible SO coupling and zero magnetic field. In such systems the superconducting gap is much smaller than the chemical potential,
$\Delta\ll \mu$, a condition known as the Andreev limit \cite{Beenakker:92}. The NS Andreev
reflection probability $R_A$ then exhibits a simple dependence with normal-state junction transparency $T_N$ and energy $E$ (relative to the SC chemical potential), see Fig.
\ref{Fig1} central-right panel. It reaches $R_A=1$ at $E=\Delta$ from
$R_A\approx T_N^2/(2-T_N)^2$ at $E=0$ \cite{Beenakker:PRB92}. This result assumes a step-like pair potential at the interface, a common approximation known as the rigid boundary-condition \cite{Likharev:RMP79}.

We now combine two such NS interfaces into a 1D SNS junction with normal length $L_N$. A computation in the Andreev limit of the energy $E(\varphi)$ of ABSs below $\Delta$ for fully transparent interfaces, and as a function of SC phase difference $\varphi$ across the junction, yields the following quantization condition \cite{Kulik:SPJ70}
\begin{equation}
\label{ABSquantization}
\varphi - 2\arccos[E(\varphi)/\Delta] - 2\frac{E(\varphi)}{\Delta} \frac{L_N}{\xi} = 2\pi n,
\end{equation}
where $n$ is an integer and $\xi$ is the superconducting coherence length.
A generalization to a multimode junction of finite transparency yields, in the short junction $L_N\ll \xi$ limit~\cite{Kulik:SPJ70,Furusaki:PRB91,Beenakker:PRL91,Beenakker:92,Bagwell:PRB92,Furusaki:SM99}, an explicit $E(\varphi)$ solution for mode $i$ 
\begin{equation}
E_i(\varphi) = \pm \Delta \sqrt{1-T_N^i\sin^2{\frac{\varphi}{2}}},
\label{Eq.1}
\end{equation}
where $T_N^i$ is the normal transmission for each independent scattering-matrix eigenmode $i$ in the normal phase \cite{Landauer:PLA81}.
The presence of such bound states has important consequences for transport, since, as argued by Kulik \cite{Kulik:SPJ70}, it implies that a normal metal can carry a dissipationless supercurrent $I_A(\varphi)$
between two SCs over arbitrarily long lengths, provided that transport is coherent. 
This is the celebrated dc Josephson effect \cite{Josephson:PL62,Josephson:AP65}. At zero temperature and neglecting the contribution to the supercurrent coming from the continuum of states above $\Delta$, $I_A(\varphi)=-(2e/\hbar)\sum_i\partial E_i(\varphi) / \partial \varphi$ (where the factor 2 accounts for spin degeneracy).  

Figure \ref{Fig1} central-left panel illustrates the solution $E(\varphi)$ for different $T_N$ in a single-channel junction. Near $\varphi=0,2\pi$ the ABSs touch the continuum of single quasiparticle states above $\Delta$ while they reach their minimum value at  $\varphi=\pi$, with an $E(\pi)$ that decreases with increasing transparency until reaching an accidental zero-energy crossing as $T_N\to1$ \editP{(assuming $\Delta\ll\mu$)}. It is important to realize that, since $|E(\varphi)|<\Delta$, the ABS wavefunctions are confined to the junction, and exponentially
decay into the bulk of the SC leads on a length scale
\editP{$\xi_\textrm{ABS}=\xi/(\sqrt{T_N}|\sin(\varphi/2)|)$.}

Deviations from the Andreev limit, relevant in low-density nanowires, introduce important corrections to the Andreev reflection $R_A$ and ABS energies $E_i(\varphi)$, and will be discussed in Sec. \ref{sec:mbs}.

\subsection{ABS spectroscopy}
\label{sec:ABSspectroscopy}

Several measurement techniques have been developed to obtain information about ABSs in nanowire Josephson junctions. Here we focus on three broad classes: Josephson spectroscopy, microwave spectroscopy and tunneling spectroscopy, see bottom row of Fig. \ref{Fig1}.

In a Josephson junction, parity-conserving transitions between the ground and excited states with an addition energy of $2E(\varphi)$ [see Eq.~\eqref{Eq.1}] can be created by an incident photon with a frequency of $f=2E(\varphi)/h$, \editP{where $h$ is Planck's constant}. Note that the SC gap $\Delta\approx180\,\mu$eV of Al corresponds to a frequency range of $2\Delta/h\approx 90\,$GHz. A precise treatment of the pair transition leads to an effective microwave impedance $Z(f)$ associated with the transition \cite{Kos:PRB13}. It can be detected via the inelastic Cooper-pair tunneling \cite{Hofheinz:PRL11} in a capacitively coupled auxiliary Josephson junction \cite{Holst:PRL94}, which is sensitive to the environmental impedance seen by this spectrometer junction. The probing frequency $f$ can be set by applying a voltage bias of $V_s=hf/2e$ (Fig. \ref{Fig1} lower-left panel). Measurements of this type confirmed the applicability of the short junction formula Eq.~\eqref{Eq.1} in a wide range of excitation energies in InAs semiconductor channels with epitaxial Al leads and demonstrated that few-channel configurations of high channel transparency can be attained \cite{Woerkom:NP17}.

\editE{The Andreev two-level system [Eq. \eqref{Eq.1}] can also be characterized and manipulated by the well-established toolbox of circuit quantum electrodynamics \cite{Blais:PRA04}, based on the coupling between a resonator with frequency $f_r$ and the junction hosting the Andreev level. In the lowest order, this coupling is described by the Hamiltonian $H_c=M\hat{I}_A \hat{I}_r$, where $M$ is the mutual inductance (Fig.~\ref{Fig1} lower-central panel), and $\hat{I}_A$, $\hat{I}_r$ are the current operators of the Andreev level (see Sec. \ref{sec:formabs}) and the resonator, respectively. It is instructive to note that the supercurrent $I_A$ changes sign between the ground and excited state. Furthermore, the odd parity state with an unpaired quasiparticle yields $I_A=0$. These three states can then be distinguished by the dispersive frequency shift of the coupled resonator, enabling a real time tracking of the junction charge parity \cite{Hays:PRL18}. The characteristic parity lifetimes are measured to be in excess of $100\,\mu$s in InAs nanowire Josephson junctions. In the same experiment, typical relaxation times ranging up to $\sim 10\,\mu$s allowed for the coherent manipulation of the nanowire-based Andreev level quantum bit.}

Direct quasiparticle tunneling into the ABSs can also probe the ABS spectrum (Fig. \ref{Fig1} lower-right panel). These experiments utilize a gate-defined depleted section of the nanowire \cite{Mourik:S12} or an in-situ grown axial tunnel barrier \cite{Car:NL17,Junger:CP19} as the opaque probe junction. This measurement geometry allows for the characterization of energy spectra in proximitized semiconductor segments \cite{Chang:NN15} or quantum dots \cite{Lee:NN14,Grove-Rasmussen:NC18}, and makes non-local correlation experiments possible \cite{Anselmetti:PRB19}. However mesoscopic interference effects in the leads may yield additional features in the differential conductance \cite{Su:PRL18}.

It is worth noting that the ABS spectrum can indirectly be characterized via the measurement of the phase-dependent supercurrent $I_A(\varphi) \sim dE/d\varphi$, which was performed by an inductively coupled SQUID loop \cite{Spanton:NP17,Hart:PRB19}. These experiments yielded strongly skewed current-phase relations, the signature of highly transparent channels in an InAs nanowire with Al superconducting leads. Similarly, the Josephson inductance, $L_J^{-1}\sim  dI_A(\varphi)/d\varphi$ could serve as another probe of the anharmonicity in the current phase relationship \cite{Rifkin:PRB76}. Finally, external tunnel barriers, typically AlO$_x$ of a few atomic layers, attached to a metallic probe also became an established technique to detect ABSs in other systems, such as carbon nanotubes \cite{Pillet:NP10} and graphene flakes
\cite{Dirks:NP11}.

\subsection{ABSs in QDs}

\begin{figure*}
	\includegraphics[width=\textwidth]{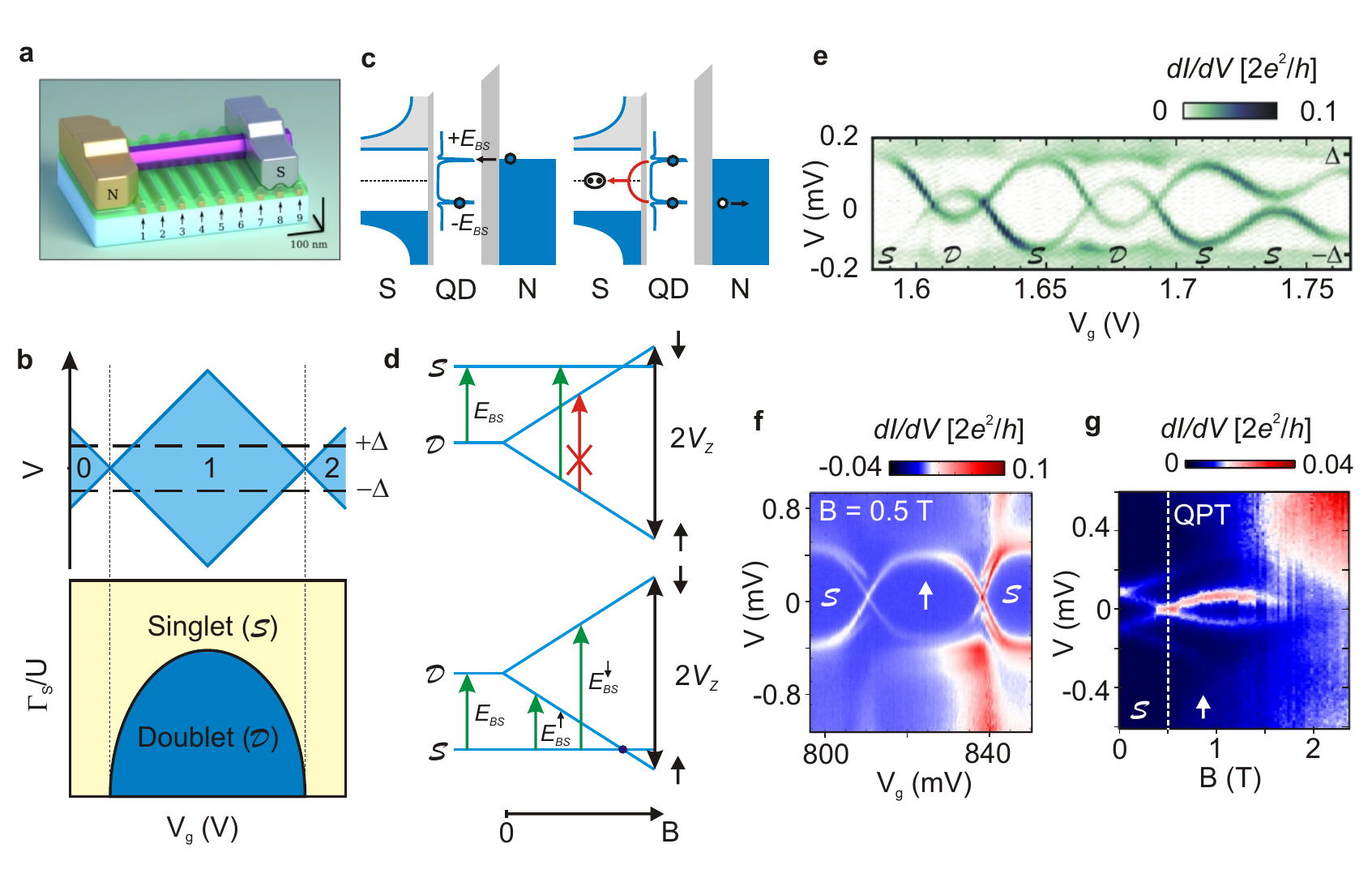}
	\caption{\label{fig:Shiba} \textbf{ABSs in hybrid quantum dots (QDs).} (a) Sketch of a semiconductor nanowire contacted by a normal metal (N) and a superconductor (SC). Local gates can be used to confine a QD, and to tune the dot-electrode tunnel couplings and the dot occupation/parity \cite{Grove-Rasmussen:NC18,Jellinggaard:PRB16}. QDs can also form unintentionally in a nanowire, e.g. by barriers at interfaces~\cite{Chang:PRL13,Lee:NN14,Chang:NN15,Deng:S16,Lee:PRB17}. (b) Top panel: charge stability diagram of a normal QD as a function of the bias voltage, $V$, and the gate voltage, $V_{g}$. The dot occupation (0, 1 or 2) is well-defined inside the Coulomb diamonds. Bottom panel: phase diagram of a hybrid QD as a function of $V_{g}$ and the QD-SC coupling, $\Gamma_S$, normalized to the charging energy, $U=e^2/2C$ ($e$ being the electron charge and $C$ the QD capacitance). In the weak coupling limit, $\Gamma_S/U \ll 1$, the ground state is a spin-doublet when the dot is occupied by an odd number of electrons. Conversely, for $\Gamma_S/U \gg 1$, the ground state is a spin-singlet irrespective of the dot occupancy. The precise boundary between both states can be obtained by experimentally tuning the ratio $\Gamma_S/U $ \cite{Lee:PRB17} in very good agreement with theoretical results obtained by a superconducting analog of the Anderson model \cite{Zitko:PRB15}. (c) \red{Transport spectroscopy of} ABSs formed within the SC gap by the Yu-Shiba-Rusinov mechanism, where the confined spin (impurity) is screened by itinerant quasiparticles. \red{Resonant $dI/dV$ peaks are observed when the chemical potential of the N probe matches} the bound state energy, $\pm E_{BS}$, which represents the excitation energy from the ground state of the QD-SC system to an excited state. \red{The transport cycle first involves the tunneling of an electron (hole) to the QD-SC system, changing its parity, followed by an Andreev reflection process whereby a Cooper pair is formed (broken) in SC and a hole (electron) is reflected to the probe.} (d)~Diagram of the possible transitions between ground and excited states of a hybrid QD. An external magnetic field, $B$, splits the doublet state by the Zeeman energy, $2V_Z$. Top panel: when the ground state is the doublet, the bound state energy increases with $B$ (green arrow). The transition between the two spin-polarized states is not visible by tunneling spectroscopy (red arrow). Bottom panel: when the ground state is the singlet, both transitions to the spin-polarized excited states are visible, $E_{BS}^{\uparrow}$ and $E_{BS}^{\downarrow}$. (e) Subgap spectrum of a QD with a single SC electrode as a function of $V_g$ at $B = 0$. Crossings of the \editP{bound state resonances at $V = 0$ signal} transitions between singlet and doublet ground states. Data reproduced from Ref. \cite{Jellinggaard:PRB16}. (f) The \editP{bound states} only split in the presence of an external $B$ when the ground state is the singlet. (g) Zeeman splitting of the bound states as a function of $B$. The dashed vertical line underscores a quantum phase transition (QPT) whereby the ground state of the system turns from the singlet to a spin-polarized state. Data reproduced from Ref. \cite{Lee:NN14}.}
\end{figure*}

For the junctions above, it was assumed that the channel connecting the SC leads allowed for coherent transport through a ballistic nanowire segment. By contrast, in QDs, charges localize in the channel and the effect of a finite electrostatic charging  energy $U$ must be taken into account. QDs can be 
formed in a nanowire by e.g. inducing barriers with electrostatic gates [Fig. \ref{fig:Shiba}(a)]. At low temperatures and for low bias voltages, transport is blocked by the large $U$ and the system is in the so-called Coulomb blockade regime with a well defined number of electrons $n$.  
Current flow is only possible at discrete degeneracy points where the energies of 
the $n$ and $n+1$ charge states become degenerate. 
Given the strong confinement in nanoscale QDs, $U$ can 
easily exceed $\Delta$ in the electrodes,
resulting in an interesting interplay between single-electron charge transport, localized spins and superconductivity \cite{De-Franceschi:NN10}.

The formation of ABSs can be understood by considering a single 
QD level coupled to a superconducting electrode. If the level is singly occupied, it holds an unpaired spin, i.e.\ a spin-doublet ground state [Fig. \ref{fig:Shiba} (b)]. Conceptually, this scenario is identical to having an isolated magnetic impurity in a superconducting host. 
As shown by Yu, Shiba and Rusinov (YSR) in the 1960s 
\cite{Yu:APS65,Shiba:PTP68,Rusinov:SPJ69}, 
the magnetic impurity induces localized bound states within the SC gap. At a critical exchange coupling the system undergoes a quantum phase transition to a magnetically screened, spin-singlet ground state. Conversely, at weaker coupling, 
the system maintains its original doublet state. 
While the above YSR picture applies for classical magnetic impurities, a full quantum treatment naturally leads to the physics of the Kondo effect \cite{Hewson:93} where, despite the absence of screening electrons within $\Delta$ of the electrodes,  
the localized spin can still be screened by the above-gap quasiparticles in the SC. As in normal metals, Kondo physics 
sets in below a characteristic temperature $T_K$,  
which results in singlet-doublet transitions 
occurring at $k_B T_K / \Delta\sim 0.3$. Early work on hybrid dots indicated the importance of Kondo-like correlations \cite{Buitelaar:PRL02,Sand-Jespersen:PRL07a}, while more recent experimental work has provided precise boundaries for the transition \cite{Lee:PRB17}. Figure \ref{fig:Shiba}(b) 
shows the generic phase diagram 
of a hybrid QD as a function of dot parameters \cite{Zitko:PRB15,Lee:NN14,Lee:PRB17}.

ABSs in QDs can be detected by transport spectroscopy \cite{Eichler:PRL07,Grove-Rasmussen:PRB09,Pillet:NP10,Deacon:PRL10,Chang:PRL13,Lee:NN14,Kumar:PRB14,Jellinggaard:PRB16,Li:PRB17,Island:PRL17}, 
whereby $dI/dV$ is measured as a function of bias voltage $V$. 
The sub-gap transport reflects resonant Andreev reflection processes 
at voltages matching the energy difference $E_{BS}$ between the ground and the excited state of the QD [Figs. \ref{fig:Shiba} (c) and (d)]. 
This results in $dI/dV$ peaks located symmetrically around $V=0$, corresponding to ABS resonances at energies $\pm E_{BS}$. Figure \ref{fig:Shiba} (e) shows a typical transport spectrum, 
where ABSs are visible as ridges below the gap. 
As the charge state, and thereby the parity, of the dot is tuned, the ground state switches between the singlet and doublet states, as reflected by the 
ABS crossings at zero bias. Remarkably, the ground state remains a singlet in some odd-occupancy regions 
due to the strong screening discussed above, which leads to avoided ABS crossings in the spectra. 
The experimental phase diagram of the QD-S system has been explored~\cite{Lee:PRB17,Jellinggaard:PRB16,Li:PRB17}, finding excellent quantitative agreement with theory \cite{Andersen:PRL11a,Zitko:PRB15}. In some cases, however, one needs to go beyond the bulk treatment of the SC above (to include soft gaps, finite-length effects, etc) in order to understand the complex ABS spectra of finite-length proximitized nanowires \cite{Su:PRL18,Junger:CP19}. Transport spectroscopy of ABSs can also be performed by replacing the N probe by a weakly coupled superconductor. Here, all spectroscopical features are shifted by $\Delta$ \cite{Lee:PRL12,Kumar:PRB14}. 
ABSs exist also in coupled hybrid dot systems \cite{Su:NC17,Grove-Rasmussen:NC18} where one can observe YSR screening of higher spin states and a more intricate phase diagram than Fig. \ref{fig:Shiba} (b) \cite{Grove-Rasmussen:NC18,Saldana:A18}. \editE{We note that YSR states have also been studied in STM experiments as reviewed e.g. in Ref. \cite{Heinrich:PSS18}.}

In an external magnetic field, the Zeeman effect lifts the spin degeneracy of the doublet state. This strongly impacts the transport spectra of the ABSs 
 [Fig. \ref{fig:Shiba}(d)]. 
In case of a singlet ground state, two (parity-changing) transitions are allowed, thanks to the splitting of the excited doublet state. In contrast, when the ground state is a doublet, 
only one transition remains accessible independent of $B$. As a result, the ABSs shift to higher energies but 
do not split. 
Figure \ref{fig:Shiba} (f) depicts these two distinct 
behaviors of the ABSs 
at finite $B$ \cite{Lee:NN14}. 
Interestingly, for high enough 
fields, the lowest-energy, spin-split ABSs can cross the Fermi level, denoting a quantum phase transition from the singlet ground state to a spin-polarized state \cite{Lee:NN14, Jellinggaard:PRB16}. This transition represents a parity crossing 
and appears as a zero-bias peak 
 at the critical field
[Fig. \ref{fig:Shiba} (g)]. 
While the transition is a true crossing, the peak can persist at $V=0$ for a wider range of $B$ owing to the broadening of ABS resonances
or to repulsion with other states or the 
gap edge \cite{Lee:NN14,Jellinggaard:PRB16,Chen:PRL19}.

In addition to the above ABS spectroscopy, the physics of a hybrid QD can also be captured by measurements of the Josephson supercurrent in a S-QD-S geometry \cite{Dam:N06,Delagrange:PRB15,Maurand:PRX12,Estrada-Saldana:SA19}. 
\editE{Notably, QDs have also been used to investigate MBSs in various device configurations \cite{Deng:NL12,Deng:SR14,Deng:S16,Deng:PRB18}.}

\section{Low-density nanowires and MBS\lowercase{s}}
\label{sec:mbs}

\begin{figure*}
\includegraphics[width=\textwidth]{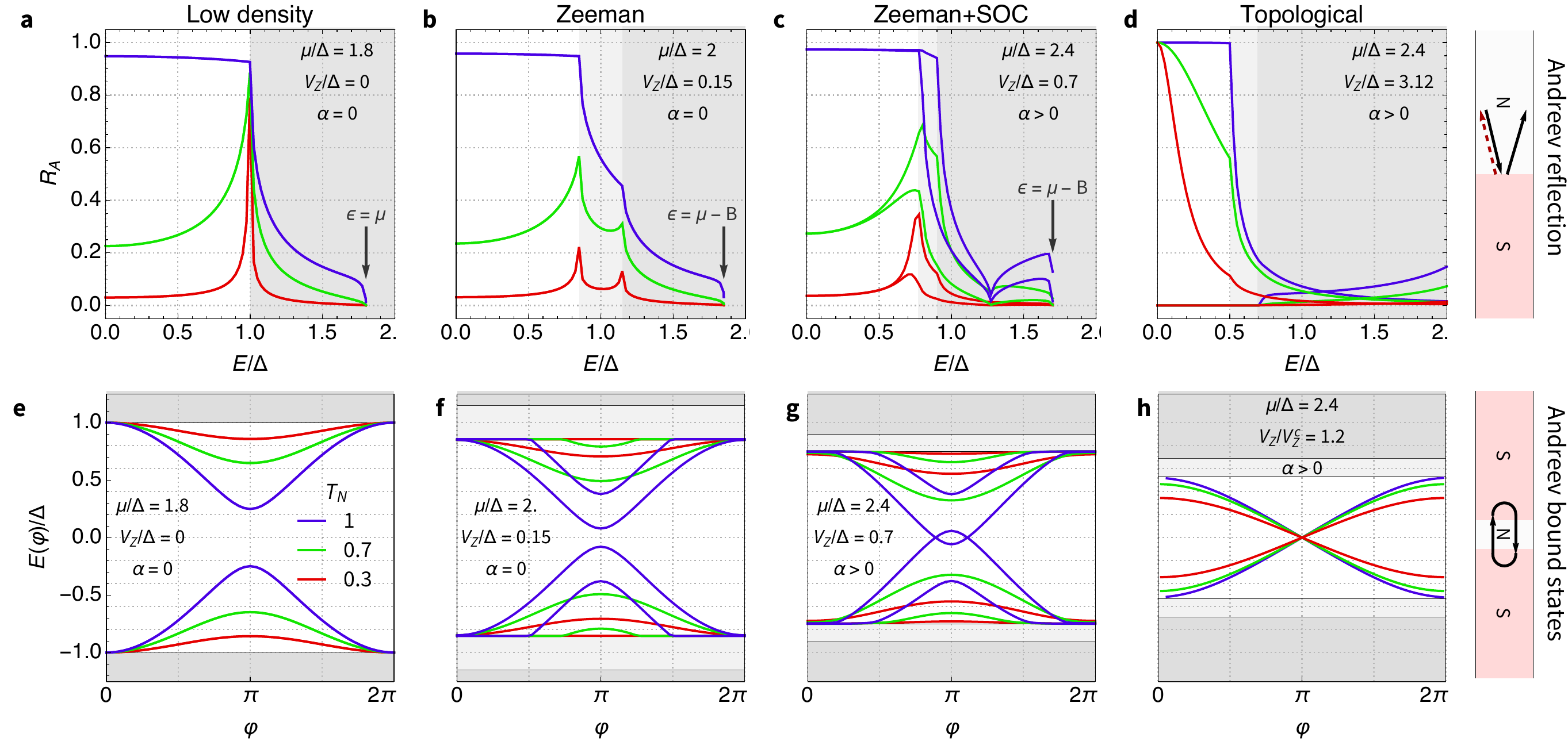}
\caption{\label{fig:Andreev2} \editP{\textbf{Theory of Andreev reflection and bound state formation in low-density NS and SNS nanowire junctions:}}
Andreev reflection probability $R_A$ (top row) and ABS energy $E(\varphi)$ (bottom row) computed within a tight-binding approach in \editE{constant} $\mu\sim\Delta$ low-density NS and short SNS junctions of varying normal-state transparency $T_N$, respectively. Energies are normalized to the SC gap $\Delta$ at Zeeman energy $V_Z=0$. Panels (a,e) correspond to zero Zeeman and spin-orbit (SO) coupling $\alpha$. \editP{Note that at low-densities the Andreev limit $\Delta\ll \mu$ is not satisfied. Hence, the $\{\varphi=\pi, T_N=1\}$ ABS crossing at zero energy expected from Eq. \eqref{Eq.1} becomes an anticrossing due to $\Delta$-induced normal reflection (in contrast to Fig. \ref{Fig1}).} In panels (b,f) we see that a small $V_Z$ splits the gap edge into two sectors $\Delta_\pm$, with their respective quasiparticle continuum colored in different shades of gray. All curves in (a,b,e) are double, one per incident spin channel (a,b) or level (e). The addition of $\alpha$ in (c,g) breaks the remaining spin-symmetry of $R_A$ around the Zeeman field direction (along the wire). This is visible as a splitting of same color curves. As one cranks up $V_Z$, $\Delta_{-}$ goes to zero and at $V_Z^c\equiv\sqrt{\Delta^2+\mu^2}$ undergoes a band inversion. Upon its reopening for $V_Z>V_Z^c$, a zero-energy scattering resonance arises in the Andreev reflection of the NS junction (panel d). \editE{The resonance manifests as a universal  $R_A=1$ at the Fermi level $E=0$, regardless of junction transparency $T_N$. It is the result of the emergence of a MBS at the NS junction.
In a short SNS geometry, two such MBSs emerge that hybridize into an ABS. Its energy $E(\varphi)$ is detached from the continuum at $\varphi=0,2\pi$ for $T_N<1$ and exhibits a protected zero-energy parity crossing at $\varphi=\pi$ (panel h), yielding a $4\pi$-periodic state at fixed parity.} Note that subgap curves in (d,h) are effectively spinless.}
\end{figure*}

\subsection{ABSs in trivial SNS junctions with SO coupling and Zeeman field}

As the Fermi energy $\mu$ of \editP{a nanowire SNS junction} is reduced (low density regime), it may become comparable to other energy scales in the problem, such as the SO energy \editP{$E_{SO}=m^*\alpha^2/2\hbar^2$ (where $\alpha$ is the SO coupling and $m^*$ the effective mass)}, the Zeeman energy $V_Z$ at the junction, or the gap $\Delta$ of the SCs at either side, see Box Fig. \ref{fig:box1}. 
The Andreev reflection at a \editP{low-density NS interface deviates} considerably from the standard picture described in Sec. \ref{sec:formabs}. Figure \ref{fig:Andreev2} (a-c) shows the typical dependence of $R_A$ with energy for a single channel contact when both N and SC sides have a common Fermi energy, SO coupling and Zeeman. Similarly, the Andreev spectrum of the corresponding low-density SNS nanowire junction is no longer well described by the conventional Eq. (\ref{Eq.1}), even in the short junction limit, see Fig. \ref{fig:Andreev2} (e-g). Note in particular that the parity crossing present at $\varphi=\pi$ in high-density transparent junctions becomes an anticrossing \editP{even at $T_N=1$ as soon as the Andreev limit $\Delta\ll \mu$ is not satisfied. This contrasts with the protected ($T_N$-independent) $\varphi=\pi$ crossing in topological SNS junctions, as we will see}.

\editP{To understand the main low-density corrections we consider first the case of an SNS junction in which $\mu$ becomes comparable to the SO energy $\mu\sim E_{SO}$, while still remaining in the Andreev limit $\Delta\ll\mu$. We further consider the realistic complication that the SO coupling $\alpha$ and the Zeeman field $V_Z$ are largely confined to the normal part of the nanowire. The Fermi energy $\mu_N$ in N is also assumed to differ} from that of the SC contacts $\mu_S$. The corresponding bandstructures will thus exhibit a Fermi momentum mismatch, which reduces Andreev reflection and affects the resulting ABS spectrum. In a nominally perfect, single mode SNS junction of nanowire length $L_N$ with $V_Z=0$, Eq. (\ref{Eq.1}) can be generalized to \cite{Cheng:PRB12} 
\begin{equation}
E(\varphi) = \Delta \sqrt{1-\frac{\sin^2(\varphi/2)}{1+\kappa \sin^2(k_0 L_N)}},
\label{ABS-SO1}
\end{equation}
where $\kappa=[({k^S_F})^2-k^2_0]/(2k^S_{F}k_0)$ captures the effect of momentum mismatch acting as an effective barrier at each interface, with a transmission $T_N=\frac{1}{1+\kappa \sin^2(k_0 L_N)}$ that is smaller than $1$, except at resonant values of the nanowire length $k_0L_N=n\pi$, $n\in\mathbb{Z}$. Here the SC and N Fermi wavevectors are $k^{S,N}_{F} = \sqrt{2m^*\mu_{S,N}}/\hbar$, and $k_0=\sqrt{(k^N_F)^2+4k^2_{SO}}$. This $k_0$ depends also on the SO momentum $k_{SO}=m^*\alpha/\hbar^2$, that captures the momentum band shift of the two spin sectors in the nanowire (see Box \ref{box:1}).
$E(\varphi)$ of Eq. (\ref{ABS-SO1}) remains doubly degenerate for all $\varphi$ despite the shift $k_{SO}$ of the two spin sectors; electron-hole pairs can still form in a similar manner as for a spin-degenerate single parabolic dispersion, see Fig. \ref{fig:box1}. \editP{While Eq. \eqref{ABS-SO1} still yields a zero energy crossing at $T_N=1$ and $\varphi=\pi$, it captures the fact that $T_N<1$ even with nominally perfect contacts due to the momentum mismatch.}

In the absence of a Zeeman field, spin splitting of the ABS spectrum can be achieved \editP{by a nonzero $\varphi$} in a two-subband model with intersubband coupling. Specifically, mixing between the two lowest transverse subbands \editP{in a low density regime may produce a strongly} spin-dependent Fermi velocity  $v^\uparrow_F\neq v^\downarrow_F$, and hence coherence lengths $\xi_{\uparrow/\downarrow}=\frac{\hbar v^{\uparrow/\downarrow}_F}{\Delta}$, which leads to spin-dependent quantization conditions  according to Eq. (\ref{ABSquantization}). \editR{For $T_N=1$, and assuming $\lambda_i=L_N/\xi_i\ll 1$ or $E_i\ll \Delta$}, the ABSs can be written as \cite{Park:PRB17} 
\begin{equation}
E_i(\varphi) = \pm \Delta \frac{\cos(\varphi/2)}{1+\lambda_i \sin(\varphi/2)}.
\label{ABS-SO2}
\end{equation}
The spin splitting between ABSs reads
\begin{equation}
E_\uparrow(\varphi)-E_\downarrow(\varphi) = \frac{\Delta(\lambda_\uparrow-\lambda_\downarrow)\sin(\varphi)}{2[1+\lambda_\uparrow \sin(\varphi/2)][1+\lambda_\downarrow \sin(\varphi/2)]}.
\label{ABS-SO3}
\end{equation}
This phase-dependent spin splitting is finite for $\varphi\neq 0,\pi$, and comes from the difference in coherence lengths and Fermi velocities. Spin-degeneracy at $\varphi=0$  and $\varphi=\pi$ is protected by time-reversal symmetry. The combined effect of Zeeman and SO coupling on the Andreev level spectra of single channel nanowires has been studied in Ref. \cite{Heck:PRB17,Dmytruk:PRB18}. Among others, an important consequence of the interplay of $V_Z$ and $\alpha$ is the strong suppression of the $g$-factor owing to SO coupling and/or high electron density. This $g$-factor renormalization drastically changes the spin splitting of Andreev levels for increasing magnetic fields.

\editE{The theory of spin-split ABS formation outlined above has been confirmed by recent experiments in a circuit quantum electrodynamics geometry using InAs nanowires \cite{Tosi:PRX19,Hays:A19}.}

\subsection{Emergence of MBSs}
\label{sec:formmbs} 

In Fig. \ref{fig:Andreev2} (e-g) we have illustrated the strong effect of SO coupling and Zeeman fields in the ABS spectrum of a low-density SNS nanowire junction.
When the SC contacts are taken as low-density proximitized nanowires, the Oreg-Lutchyn minimal model predicts that a sufficiently strong $V_Z>V_Z^c$ will make them undergo a topological phase transition, with MBSs at each interface. Their presence results in a topologically protected $R_A=1$ Andreev reflection amplitude at $E=0$, see Fig. \ref{fig:Andreev2} (d), and a protected $\varphi=\pi$ parity crossing  \editP{of SNS ABSs} for all transparencies, (h). The parity crossing is robust regardless of the microscopic channel configuration of the junction, and ideally gives rise to the topological Josephson effect, characterized by $4\pi$-periodic supercurrents as a function of $\varphi$ at fixed parity \cite{Kitaev:PU01,Kwon:LTP04,Lutchyn:PRL10, Oreg:PRL10}. \editP{The $2\pi$-periodic  $E(\varphi)$ solution in the trivial phase, Eq. \eqref{Eq.1}, transforms in the topological regime into $E(\varphi)\approx \pm\sqrt{T_N}\Delta\cos(\varphi/2)$, with different signs for opposite parities~\cite{Kwon:LTP04}}.

\begin{figure*}
\includegraphics[width=\textwidth]{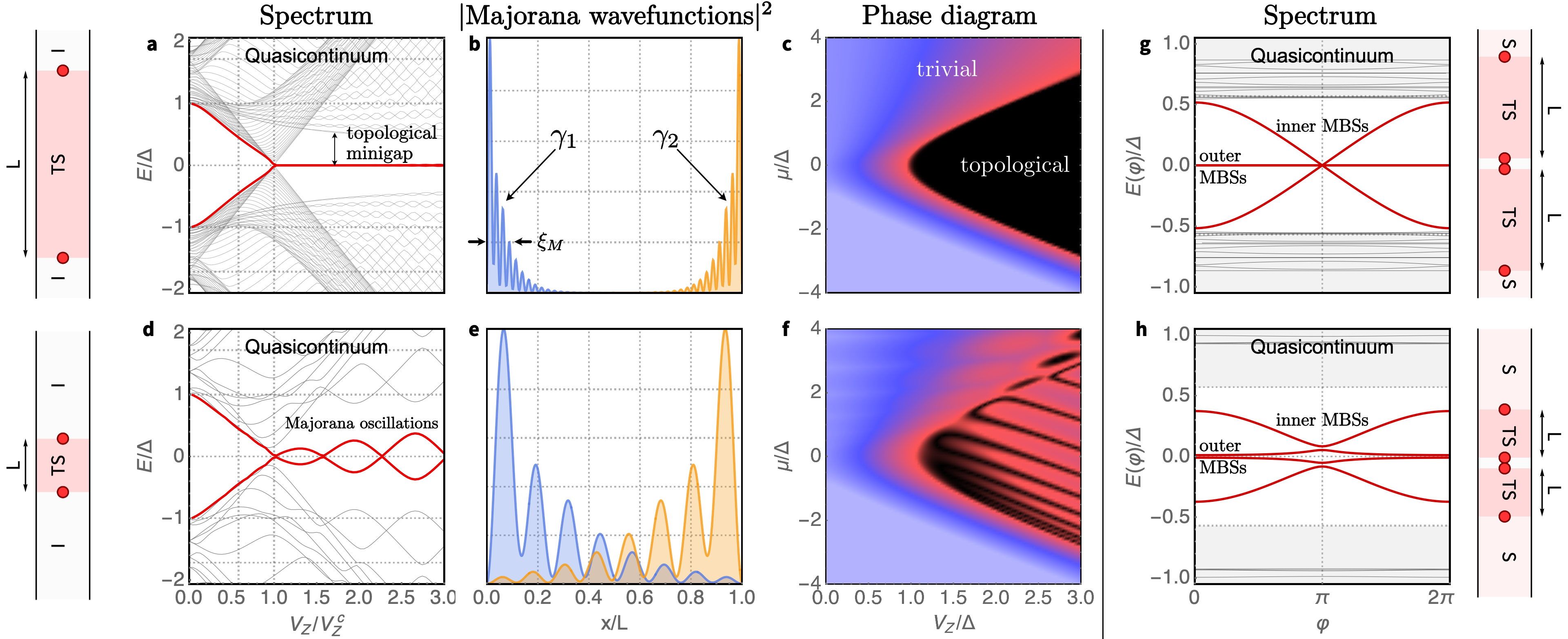}
\caption{\label{fig:NW} \textbf{
\editP{Theory of MBS formation and length-dependence in finite nanowires and Josephson junctions:}} Panels (a-f) correspond to  an ideal, uniform proximitized nanowire of \editP{length} $L=3\mathrm{\mu m}$ (a-c) and  $L=0.6\mathrm{\mu m}$ (d-f) distance between the two insulators \editP{(possibly vacuum)}. When the nanowire length $L$ is long as compared to the SC coherence length, the low energy spectrum as a function of Zeeman splitting $V_Z$ has the characteristic shape shown in (a). For $V_Z<V_Z^c$ the system is trivially gapped, but this gap decreases until it reaches zero at the topological phase transition at $V_Z^c$. From the latter, two Majorana zero modes around the ends of the nanowire emerge, whose exponentially decaying wave functions in the Majorana basis (see Box \ref{box:2}) are shown in (b). For shorter lengths, these MBSs \editP{hybridize} into fermions of oscillatory energy around zero (d), with overlapping wave functions (e). The energy of the lowest excitation clearly traces the topological transition at $V_Z^c\equiv\sqrt{\Delta^2+\mu^2}$, producing the phase diagrams of (c,f) as a function of Zeeman energy $V_Z$ and chemical potential $\mu$. The two Majoranas at either side of a $V_Z>V_Z^c$ short TS-N-TS junction combine into the characteristic low energy ABS spectrum $E(\varphi)\sim\cos(\varphi/2)$ of a topological Josephson junction that vanishes at $\varphi=\pi$ (g). This parity crossing leads to a $4\pi$-periodic ground state when fixing quasiparticle parity, and to the so-called topological Josephson effect.  \editP{The exact zero modes correspond to the two outer MBSs, decoupled from the two $\varphi$-dependent junction (inner) MBSs for long $L$. The parity crossing at $\varphi=\pi$ becomes lifted by four-Majorana overlaps for short TS nanowires, that couples inner and outer MBSs}, thereby destroying the $4\pi$ periodicity (h). In the schematics at the left and right edges of the figure, $I$ stands for insulator, S for trivial SC, TS for topological SC and the red circles symbolize the presence of MBSs at the junctions. \label{fig:Majoranabasics}}
\end{figure*}

Figure \ref{fig:Majoranabasics} shows a complementary picture of the topological transition in a low-density, isolated ISI uniform nanowire  of length $L$ (where $I$ stands for `insulator'), both in the long (a-c) and  short (d-f) nanowire regime. As $V_Z>V_Z^c$, a MBS appears localized at each end of the SC region, with zero energy in the large $L$ limit, or with characteristic Majorana oscillations around zero for shorter $L$, resulting from their hybridization into conventional fermions due to their finite overlap. The lowest energy level \editP{$E_\mathrm{min}$ [red in (a,d)]} clearly traces the topological phase diagram for large $L$, \editP{panel (c)}. It is interesting to note the role of finite $L$ in the topological Josephson effect \editP{[compare panels (g,h)]}. Due to the overlap of the \editP{`inner'} MBSs in the junction and the `outer' MBSs at the opposite ends of the nanowires, the $4\pi$ Josephson periodicity is destroyed under an adiabatic $\varphi(t)$, and a non-topological $2\pi$-periodic Josephson effect is restored~\cite{Pikulin:PRB12,SanJose:PRL12}. A similar effect is expected from quasiparticle poisoning (exchange of quasiparticles with the junction's environment which breaks parity conservation) and by higher-energy quasiparticle excitation \cite{SanJose:PRL12}.

The role of SO coupling is crucial for the physics of MBSs. For $\alpha=0$ and $V_Z$ larger than $\Delta$ the spectrum is gapless (the magnetic field just kills superconductivity), so that no localized MBSs emerge, while for $V_Z<\Delta$ the system has a gap.
The addition of SO coupling radically transforms this picture, and enables a topological \emph{minigap} to emerge at $V_Z>V_Z^c$. 
The minigap can be shown to be effectively $p$-wave, and hence topologically non-trivial. The Majorana zero modes at the ends of a $V_Z>V_Z^c$ nanowire are in fact a manifestation of the bulk-boundary correspondence of this topological gap. They are thus topologically protected states.
The extension of the Majorana wavefunction is the coherence length corresponding to the minigap (also known as the Majorana length $\xi_M$ \cite{Klinovaja:PRB12,Mishmash:PRB16}) and is hence smaller for stronger SO coupling. The Majorana oscillatory hybridization is thus exponentially suppressed by both a strong SO (minigap) and nanowire length. In both limits, an exact Majorana topological protected zero mode is recovered at each end of the nanowire.

\subsection{MBS spectroscopy}

\begin{figure*}
\includegraphics[width=\textwidth]{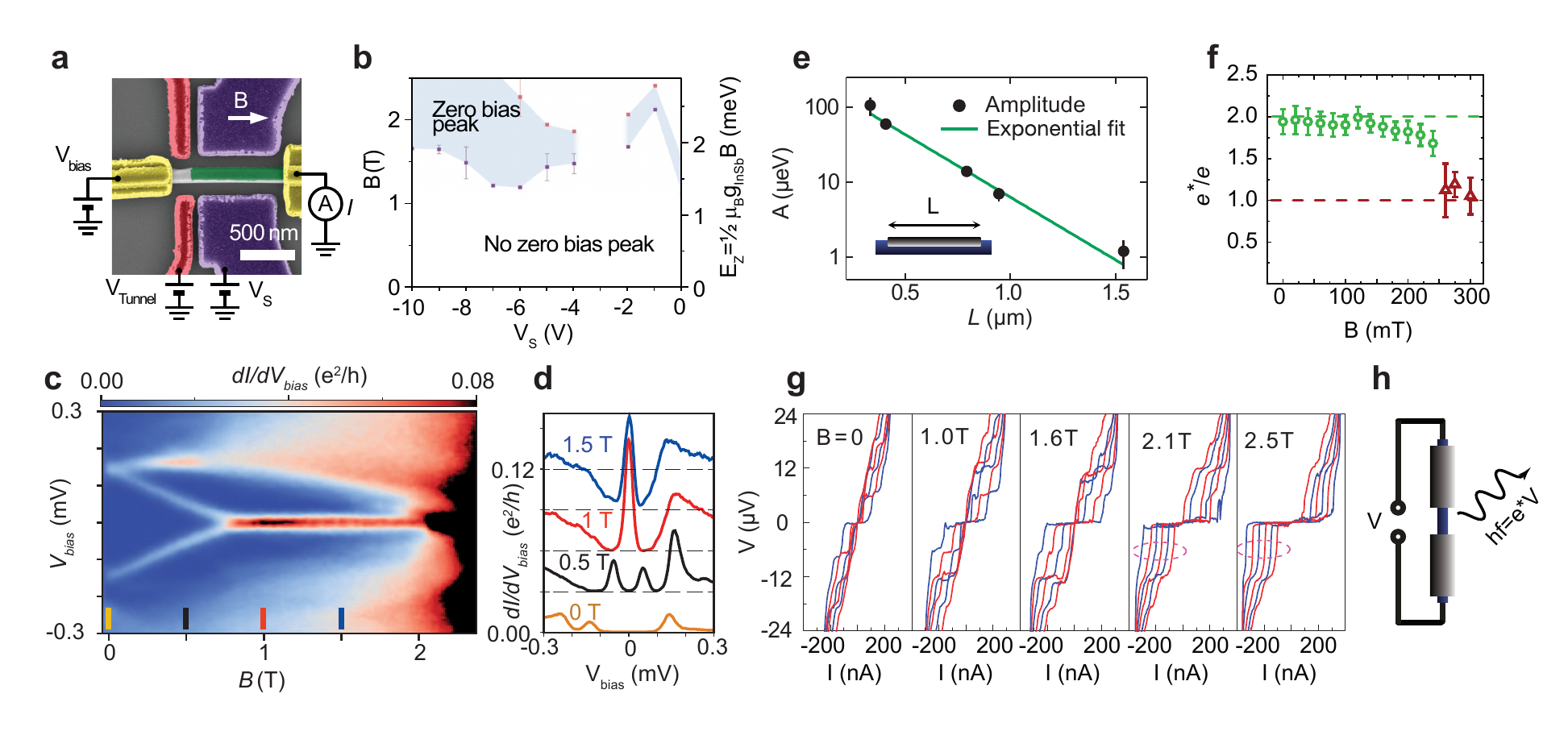}
\caption{\label{fig:tunneling_experiment}\editE{\textbf{Experimental signatures in the search for MBSs.} (a) A false color scanning electron micrograph of a device built for zero bias conductance peak (ZBP) measurements between the proximitized segment (nanowire covered by the SC, in green) and the normal metal ohmic contact (in yellow). The red electrostatic gates tune the transparency of the tunnel barrier and the purple gates change the electrochemical potential of the proximitized segment. The magnetic field $B$ points along the nanowire. (b) A ZBP phase diagram measured on an InSb nanowire covered by a NbTiN superconductor, reproduced from Ref. \cite{Gul:NN18}. (c) Experimental data of a robust ZBP taken on an InAs nanowire with epitaxial aluminum leads, reproduced from Ref.~\cite{Deng:S16}. (d) Linecuts taken at $B$-fields indicated by colored labels in panel (c). (e) The amplitude of lowest-energy-level oscillations in a finite-sized island device of length $L$, decreasing exponentially with increasing $L$ \cite{Albrecht:N16}. (f, g) AC Josephson effect experiments using the geometry sketched in (h). The characteristic frequency is proportional to the applied bias $f = e^\star V/h$ with $e^\star=2e$ for the conventional Josephson effect and $e^\star=e$ for a topological Josephson junction. The halving of $e^\star$ and the $4\pi$ periodicity of the topological Josephson effect was demonstrated as a function of $B$ by a frequency-sensitive measurement of the Josephson radiation of an InAs/Al nanowire junction [(f) taken from Ref. \cite{Laroche:NC19}] and by missing odd Shapiro steps in etched InSb/Nb junctions [(g) taken from Ref. \cite{Rokhinson:NP12}].}}
\end{figure*}

\editE{In this section, we outline the experimental techniques used to probe potential Majorana zero modes. First, tunneling spectroscopy as described in Sec. \ref{sec:ABSspectroscopy} has been performed extensively in nanowires in devices of the kind shown in Fig. \ref{fig:tunneling_experiment} (a), \editP{with the chemical potential in the nanowire controlled by the purple gate voltage $V_S$.}
The gate-defined tunnel barrier (red gates) allows the conductance through the junction to probe the local density of states at the left end of the hybrid nanowire (green), typically exhibiting a \editP{roughly} BCS-like gap
[see the orange linecut in Fig.~\ref{fig:tunneling_experiment} (d)]. In these experiments \cite{Mourik:S12,Das:NP12,Deng:S16,Chen:SA17,Gul:NN18,Chen:PRL19,Grivnin:NC19}, the expected signature of a Majorana zero mode is a zero bias conductance peak above a threshold magnetic field [see Fig.~\ref{fig:tunneling_experiment} (c)], resulting from the resonant Andreev reflection \editE{on} the MBS at the junction. \editR{This zero bias conductance peak is broadened due to both coupling to the normal lead (tunneling broadening $\Gamma$) and temperature (thermal broadening  $k_B T$). In the tunneling dominated regime $k_B T\ll \Gamma$, theory predicts that the peak should exhibit a universal quantized value of $2e^2/h$ \cite{Law:PRL09,Flensberg:PRB10,Wimmer:NJP11}. While most experiments yield much lower conductance values (see e.g. Fig.~\ref{fig:tunneling_experiment} (c,d) \cite{Deng:S16}), consistent with the thermally broadened regime $k_B T\gtrsim \Gamma$ \cite{Prada:PRB12,Setiawan:PRB17}, some experiments have reported scaling with the ratio $k_B T/\Gamma$ and saturation values close to the ideal $2e^2/h$ limit at low temperatures \cite{Nichele:PRL17}.}
Further comparison with theory [Fig.~\ref{fig:NW} (c)] can be performed by mapping the presence of the zero bias conductance peak as a function of the magnetic field $B$ and the gate voltage $V_S$ to create a phase diagram (see Fig.~\ref{fig:tunneling_experiment} (b) 
\cite{Chen:SA17,Gul:NN18}). }

\editP{Another class of experiments targets superconducting islands,
i.e., proximitized nanowires in a floating island geometry and characterized by combined superconducting and Coulomb blockade phenomenology. The Coulomb peak periodicity of the islands is found to transition from approximately $2e$- to $e$-periodic under a finite Zeeman field due to the appearance of subgap near-zero modes \cite{Albrecht:N16, Shen:NC18, Vaitiekenas:S20}. The peak positions as a function of gate voltage show deviations from perfect periodicity, which are interpreted as energy splittings of the subgap states. The splittings were shown to oscillate around zero energy with Zeeman field, with an overall oscillation amplitude that decreases exponentially with increasing island length $L$ (Fig.~\ref{fig:tunneling_experiment} (e) \cite{Albrecht:N16, Vaitiekenas:S20})}. The oscillations and their cutoff length $\xi_M\sim 260\,$nm have been interpreted as resulting from Majorana splittings, see Fig.~\ref{fig:NW} (b,e).

Apart from tunneling spectroscopy measurements, in order to detect MBSs one can also explore dynamical detection techniques in SNS junctions. In Sec. \ref{sec:formabs} we discussed the ABS spectrum and concluded that in a finite-transparency, high-density, short junction, they exhibit an avoided crossing at $\varphi=\pi$, while in Sec. \ref{sec:formmbs} we saw that in the topological phase, $E(\varphi)$ has a protected crossing at $\pi$, leading to the $4\pi$-periodic Josephson effect. 
At first glance, a tempting experimental detection of the topological Josephson
junction is to directly measure the gapless nature of the ABSs,
$E(\varphi)$, or the corresponding current-phase relation, $I_A(\varphi)
\sim dE(\varphi)/d\varphi$. However, as mentioned in the preceding section, in a finite length system of length $L$,
the overlap between the `inner' and `outer' Majorana wavefunctions restores the avoided crossing with an
energy scale \editP{$\sim \exp(-L/\xi_M)$}
\cite{vanHeck:PRB11,Pikulin:PRB12,SanJose:PRL12}. In addition, the
tunneling of unpaired non-equilibrium quasiparticles enables relaxation to the
parity ground state, resulting in a trivial, $2\pi$-periodic behavior on
timescales much longer than the parity poisoning time of the system
\cite{Lutchyn:PRL10,Houzet:PRL13}.

\editP{Due to these challenges, the experimental detection efforts of the $4\pi$-periodic Andreev
levels have typically focused on dynamical detection techniques based on the ac Josephson
effect \cite{Josephson:PL62}, \editE{Fig. \ref{fig:tunneling_experiment} (h)}. In conventional $2\pi$-periodic
Josephson junctions \editE{in the tunneling limit $T_N\ll 1$} a junction bias $V$ produces an oscillating
supercurrent $I(t)=I_C \sin{\left(2\pi f t\right)}$, with $f/V=2e/h\approx 486\,\textrm{MHz}/\mu $V \cite{Parker:PRL67}. In the topological Josephson
effect, the $4\pi$-periodicity of subgap states translate into a halving of the frequency $f/V=e/h$. This halving becomes visible} in Shapiro step measurements
\cite{Shapiro:PRL63}, where the junction is irradiated at a frequency $f$ in
the microwave domain. The dc component of $I(V)$ develops discrete
voltage steps with a spacing of $V_{2\pi}=hf/2e$ and $V_{4\pi}=hf/e$ for the
trivial and topological state, respectively
\cite{Dominguez:PRB12,Houzet:PRL13,Sau:PRB17}.
While the disappearance of the first voltage step was repeatedly observed, \editE{Fig. \ref{fig:tunneling_experiment} (g)},
higher odd steps typically persist in experiments \cite{Rokhinson:NP12,
Kamata:PRB18}. It has been argued that the interpretation of the measurements needs to
include the deviations from the tunnel junction behavior, such as non-sinusoidal supercurrents
\cite{Dominguez:PRB12}, overheating effects \cite{DeCecco:PRB16,
LeCalvez:CP19}, \editE{capacitive shunting \cite{Pico-Cortes:PRB17}}, Landau-Zener tunneling between the Andreev bands and to the
quasiparticle continuum \cite{Pikulin:PRB12,Sau:PRB17,Virtanen:PRB13}. Furthermore, the
addition of several non-topological ABSs has a non-trivial
effect on the observed Shapiro steps \cite{Dominguez:PRB12,Dominguez:PRB17}.

Another class of experiments rely on the direct spectroscopical detection of the
Josephson radiation of voltage-biased junctions, which is expected to be
centered at $f_{2\pi}=2eV/h$ or at $f_{4\pi}=eV/h$ \cite{SanJose:PRL12}. This
transition has been observed in InAs/Al nanowire Josephson
junctions integrated with an on-chip SIS microwave detector \editE{(Fig.~\ref{fig:tunneling_experiment} (f), \cite{Laroche:NC19})},
and by using a conventional microwave amplifier chain \cite{Kamata:PRB18}.

It should be noted that additional measurement schemes were proposed to
observe the $4\pi$-periodic Josephson effects as a probe for topological
superconductivity. These utilize Shapiro steps in the low-frequency regime
\cite{Sau:PRB17}, Andreev level pair excitations in long junctions
\cite{Vayrynen:PRB15},  critical current measurements \cite{San-Jose:PRL14,Tiira:NC17,Cayao:PRB17}, or the shape of switching current histograms
\cite{Peng:PRB16}.

\section{MBS\lowercase{s} beyond the minimal model}
\label{sec:beyond}

\subsection{Extensions of the minimal model}

The minimal Oreg-Lutchyn model has proven to be a first useful guide to investigate the physics of Majorana nanowires. However, discrepancies between its predictions and experimental observations have motivated extensions that provide a more complete understanding of the experimental system.
A natural extension of the 1D single band model is to allow for multiple subbands in the nanowire \cite{Potter:PRL10,Potter:PRB11,Lutchyn:PRL11,Lutchyn:PRB11}. This results in a more complicated phase diagram, depending on the number of occupied bands and their relative energies. Additionally, the orbital effects of the magnetic field (i.e. the magnetic flux across the nanowire section) may become relevant, especially when the number of occupied subbands is increased~\cite{Nijholt:PRB16}. They have been shown to dramatically modify the topological phase diagram ~\cite{Nijholt:PRB16,Winkler:PRB19} [see Fig. \ref{fig:beyond} (b)] and the dispersion of states in the nanowire, leading to large effective $g$-factors \cite{Nilsson:NL09,Winkler:PRL17} and suppressed topological gaps \cite{Nijholt:PRB16}. Although numerical simulations of multiband wires can shed additional light on the experimental results, they tend to depend strongly on details such as the geometry and effective parameter values which are not always experimentally accessible. 

While initial experiments generally suffered from unwanted quasiparticle states inside the superconducting gap [referred to as ``soft gap''~\cite{Mourik:S12,Takei:PRL13,Winkler:PRB19}, see Fig. \ref{fig:beyond} (c)], clean superconducting gaps comparable to the bulk gap of the parent SC have since been achieved~\cite{Chang:NN15,Zhang:NC17} by engineering epitaxial interfaces between the two material systems~\cite{Krogstrup:NM15,Gazibegovic:N17}. Both the ``soft gap'' issue \cite{Stanescu:PRB13} and the large gaps measured in later experiments ignited interest in a more complete description of the superconducting proximity effect in these systems. This includes pair breaking effects that suppress superconductivity beyond a critical value of the magnetic field, or a more accurate model for the induced pairing in the form of an energy-dependent anomalous self-energy. The latter extends the regime of weak coupling between the semiconductor and the SC, wherein the induced superconducting gap is simply proportional to the coupling strength between the two systems. It was found that in the opposite, strong coupling regime, the band structure of the nanowire is significantly altered, resulting in a strong renormalization of model parameters~\cite{Cole:PRB15}. It has also been demonstrated that the proximity effect can strongly depend on the thickness of the SC film~\cite{Reeg:PRB17a,Reeg:PRB18,Awoga:PRL19}. The SC-semiconductor coupling has furthermore been found to depend on the details of the electrostatic environment~\cite{Antipov:PRX18,Mikkelsen:PRX18}, resulting in gate voltage dependent effective parameters such as the $g$-factor~\cite{Vaitiekenas:PRL18,Pan:PRB19}, the SO coupling \cite{Escribano:A20} and the induced gap~\cite{Moor:NJP18}.

A notable disagreement between most experiments and the minimal model revolves around the Majorana oscillations. The oscillatory energy splittings are predicted to be regular and grow with Zeeman field \cite{Lim:PRB12,Prada:PRB12, Das-Sarma:PRB12, Rainis:PRB13,Sharma:A20}, while in most experiments robust zero-bias peaks appear without oscillations \cite{Deng:S16,Zhang:NC17}. Several model extensions have been explored that predict a reduction or suppression of oscillations, such as interactions with a dielectric environment or among carriers \cite{Das-Sarma:PRB12,Dominguez:NQM17,Escribano:BJN18}, orbital effects \cite{Dmytruk:PRB18a}, dissipation \cite{Liu:PRB17,Danon:PRB17,Avila:CP19} or non-uniform \editP{potentials \cite{Penaranda:PRB18,Sharma:A20}, pairing \cite{Fleckenstein:PRB18} or SO coupling \cite{Cao:PRL19}}. A further common disagreement is a lack of visible bandgap-closing and reopening in some experiments \cite{Mourik:S12,Deng:S16,Vaitiekenas:PRL18}, which is a key feature of the model's topological transition. This has been explained as the result of poor visibility resulting from tunnel probe smoothness \cite{Prada:PRB12,Stanescu:PRL12} and even by a lack of bulk transition altogether \cite{Huang:PRB18}, as will be discussed in Sec. \ref{sec:controversy}.

The topological phase transitions in these extended models are generally calculated using the chemical potential $\mu$ and the Zeeman energy $V_Z$. However, the control parameters used in experiments are gate voltages and magnetic fields. Calculating the phase diagram in terms of gate voltages requires a self-consistent treatment of the electrostatics~\cite{Vuik:NJP16}. While some progress has been made in self-consistent Schr\"{o}dinger-Poisson calculation for 3D device geometries \cite{Woods:PRB18,Antipov:PRX18,Escribano:BJN18,Winkler:PRB19} [see Fig. \ref{fig:beyond} (a)], this remains a difficult problem to solve reliably. In addition to electrostatic modifications of the phase diagram, interaction effects have been demonstrated to play a role in the low energy spectrum of Majorana nanowires \cite{Vuik:NJP16,Dominguez:NQM17,Escribano:BJN18}. 

An immediate effect of a self-consistent description of nanowire junctions, both for electrostatics and the proximity effect, is a smoothening of the pairing and Fermi energy profiles~\cite{Prada:PRB12}, which can no longer be assumed piecewise-constant as in the minimal model. Smooth $\Delta(\bm{r})$, $\mu(\bm{r})$ at a junction have been shown to give rise to near-zero modes without the need of a topological bulk. We devote the next subsections to these and other types of non-topological zero modes.

\begin{table*}
\includegraphics[width=\textwidth]{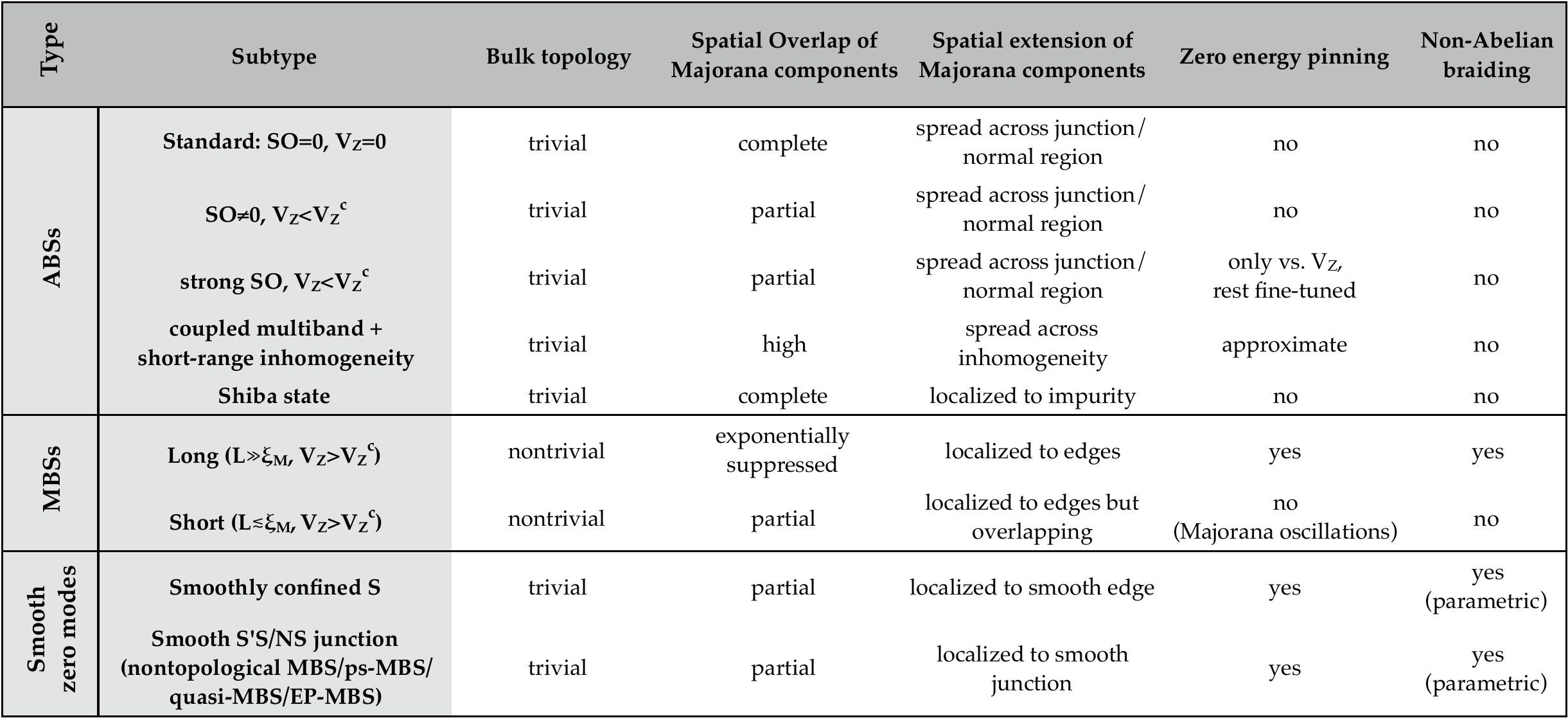}
\caption{\label{tab:mbs} \editP{\textbf{Classification of near-zero-energy subgap states in proximitized nanowire systems.} We divide the possible near-zero-energy subgap states in three main types: ABSs,
topological MBSs and zero modes produced by smooth inhomogeneities in a trivial nanowire. Each of these is distinguished by the band-topology of the nanowire bulk, the amount of spatial overlap between the Majorana components of the state, their spatial location and extension, whether the state's energy remains pinned to zero as system parameters are perturbed, and whether they are expected to exhibit spatial and/or parametric non-Abelian braiding statistics. 
See Sec. \ref{sec:beyond} for a discussion.}}
\end{table*}

\subsection{\editE{Zero energy pinning} with a topologically trivial bulk}
\label{sec:pinning}

The combination of multiband wires with disorder has been shown \cite{Bagrets:PRL12,Rainis:PRB13,Beenakker:RMP15} to produce topologically trivial zero energy states \editE{in class D Hamiltonians \cite{Altland:PRB97,Schnyder:PRB08}.}
Since the advent of cleaner experiments, it has become possible to distinguish disorder-based mechanisms from zero bias peaks of different origin, as the former are associated to specific observable features (e.g. soft gap, low transport peak heights) that have been optimized away. \editP{Strong interband coupling in multimode clean wires with a single short-range potential inhomogeneity have also been shown to conspire to produce approximate zero energy states of non-topological states \cite{Woods:PRB19,Chen:PRL19}}.

Near-zero bound states can also be generically present in a tunneling spectroscopy nanowire setup if there is a non-superconducting section between the tunnel barrier and the superconducting wire \cite{Prada:PRB12,Chevallier:PRB12,Cayao:PRB15,Liu:PRB17a,Ptok:PRB17,Moore:PRB18,Penaranda:PRB18,Reeg:PRB18,Reeg:PRB18b,Vuik:SP19,Avila:CP19,Stanescu:PRB19} [Fig.~\ref{fig:beyond} (d)]. Such an N region can host ABSs that become spin-polarized under a Zeeman field and may thus be tuned to zero energy, much like the Shiba states, possibly with a strongly renormalized $g$-factor due to SO coupling \cite{Dmytruk:PRB18}. In the simplest situation, these are readily distinguished because their zero energy \editP{results from} fine tuning parameters such as $B$ to specific values, unlike for topological MBSs. Under some circumstances, however, these modes can become \emph{pinned} to zero or near-zero energy for an extended range in magnetic field and other control parameters, resembling the behavior expected from MBSs, but with the SC in the topologically trivial phase \cite{Prada:PRB12,Kells:PRB12,Stanescu:PRB13,Liu:PRB17a,Moore:PRB18a,Penaranda:PRB18,Reeg:PRB18b,Liu:PRB18,Moore:PRB18,Vuik:SP19,Avila:CP19,Stanescu:PRB19}. 

In the case of isolated NS nanowire junctions, we can distinguish two main mechanisms for zero-energy pinning of a non-topological \editP{zero mode}: smooth confinement \cite{Moore:PRB18a,Penaranda:PRB18,Vuik:SP19,Avila:CP19,Stanescu:PRB19} and SO-induced pinning \cite{Reeg:PRB18b}. Both effects ultimately cause an enhanced Andreev reflection of a normal electron on the trivial SC. In the case of a junction with spatially smooth parameters, the momentum transfer required for normal reflection at the junction is suppressed, and hence Andreev reflection dominates. Under these conditions, states at the junctions decouple into two sectors around different Fermi wavevector and spin (due to the SO coupling and the Zeeman field) \cite{Kells:PRB12,Vuik:SP19,Penaranda:PRB18}, each of which behaves as an independent topological $p$-wave SC that gives rise to a zero-energy MBS decoupled from its partner. The Majorana wavefunction corresponding to the two wavevectors are centered at different positions along the wire and exhibit different spatial profiles (oscillatory exponential and smooth gaussian, respectively \cite{Penaranda:PRB18,Stanescu:PRB19}), see Fig. \ref{fig:beyond} (e).

A similar pinning effect can be caused by SO coupling. For large SO, the effective $g$-factor is strongly renormalized and ABSs can become largely insensitive to magnetic fields~\cite{Dmytruk:PRB18}. When the length of the N section is further tuned to an approximately odd-integer multiple of the SO length, an ABS will appear pinned near zero energy respect to $V_Z$~\cite{Reeg:PRB18b}. This SO-induced pinning does not require junction smoothness, but the above Fabry-Perot resonance condition on length must be satisfied.

A third route towards stabilising zero modes belonging to a nominally trivial bulk has been proposed in topologically trivial nanowires open to fermion reservoirs (which is a standard geometry in NS junctions used to perform transport spectroscopy). When such a nanowire becomes coupled to the reservoir, it can develop an `exceptional point' (EP) bifurcation in its complex (non-Hermitian) spectrum, see Fig. \ref{fig:beyond} (f), where the real part of the lowest quasibound Bogoliubov mode becomes robustly pinned to zero energy as the imaginary part bifurcates. This kind of non-Hermitian topological transition
stabilizes a couple of quasibound states at the contact with different decay rates. One of the two may become essentially non-decaying after the exceptional point bifurcation, thus becoming a stable Majorana zero mode without the need of a bulk topological transition. \editP{An EP requires a finite coupling asymmetry of the two Majorana components to the reservoir.}
Sources of asymmetry include finite length \cite{Pikulin:PRB13,Avila:CP19}, smooth potentials ~\cite{Avila:CP19}, spin-polarized leads \cite{San-Jose:SR16}, etc. Research into Majorana states in open systems for quantum computation purposes is still in its early stages. The field is advancing rapidly, however, with e.g. new non-Hermitian topological classification theories being developed recently \cite{Leykam:PRL17,Shen:PRL18,Gong:PRX18,McGinley:PRB19} that extend band-topological concepts to open systems where these do not strictly apply.

\subsection{The MBS vs. ABS controversy}
\label{sec:controversy}

\begin{figure*}
\includegraphics[width=\textwidth]{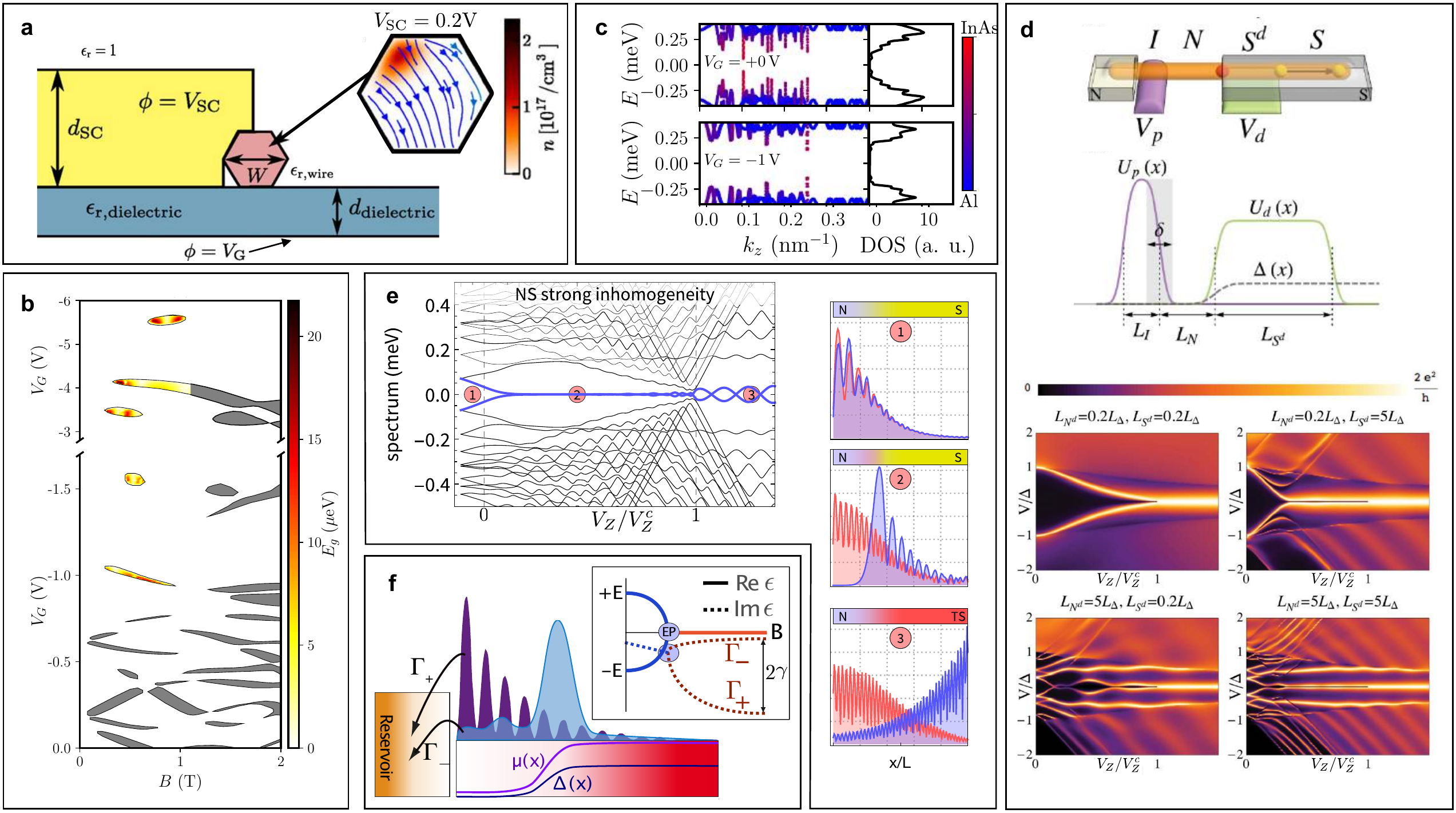}
\caption{\label{fig:beyond} \editE{\textbf{Illustrative examples of theoretical results beyond the minimal model}. (a) Schr\"odinger-Poisson computations in a realistic nanowire set-up including the electrostatic environment yield a non-homogeneous charge density distribution and electric field in the wire's cross section \cite{Vuik:NJP16}. (b) Topological phase diagram of a nanowire covered by an epitaxial SC shell on three of its facets \cite{Winkler:PRB19}. The presence of several filled subbands and the orbital effects due to the magnetic flux along the nanowire dramatically alter the phase diagram [compare to Fig. \ref{fig:Majoranabasics} (c)]. The topological minigap is overlaid in selected regions (it has not been calculated in  gray areas). (c) \editE{Effect of the backgate voltage on the induced gap. For more negative voltage (bottom panel), the wave function is pushed closer to the SC, resulting in an increased coupling and a hard gap. In contrast, a more positive voltage (bottom panel) results in a decreased coupling and a softer gap \cite{Winkler:PRB19}}. (d) A normal region in a proximitized nanowire, bounded by a smooth (screened) confinement potential and pairing can explain the emergence of non-topological zero modes in the trivial $V_Z<V_Z^c$ phase, a reduced visibility of the band inversion at the critical $V_Z=V_Z^c$ and the appearance of additional finite-energy subgap states, all apparent in $dI/dV$ simulations \cite{Prada:PRB12}. (e) Low energy spectrum vs Zeeman splitting $V_Z$ at an NS smooth junction in a finite length nanowire. At $V_Z=0$ there exist several ABSs below the induced gap (0.3 meV in this simulation) located in the normal region. As $V_Z$ increases, the lowest energy mode (in blue) approaches zero energy and remains pinned to zero for an extended $V_Z$ range before entering the  bulk topological phase at $V^c_Z$. After this point a topological minigap opens and MBSs oscillations ensue. The wave functions (in the Majorana basis) along the nanowire of the lowest energy mode are shown to the right, at three values of $V_Z$ (numbered circles). In 1 and 2 the entire wire is in the trivial phase, while in 3 the right part has become topological. The wavefunctions of the non-topological robust zero modes in 2 exhibit partial spatial overlap and different (fast and slow) wavevector components (red and blue, respectively) \cite{Penaranda:PRB18}. Note that the blue wavefunction in 2 is centered at the smooth NS junction, while it shifts to the right end after the topological transition. (f) Opening a (trivial or topological) nanowire to a fermionic reservoir may induce a decoupling of the two Majorana components of a low-energy fermionic state (blue and purple), which then acquire complex energies with distinct imaginary parts $\Gamma_\pm$, and real energies exactly pinned to zero. This happens at a so-called exceptional point (EP), which constitutes a non-Hermitian topological transition~\cite{Avila:CP19}. The evolution of $\Gamma_\pm$ and the real energies with magnetic field $B$ across the EP are shown in the inset (solid and dashed lines). The bifurcated $\Gamma_\pm$ represent different decay rates of the two Majorana components into the reservoir. The decoupling requires partial separation of the two Majorana wavefunctions in the isolated wire. After decoupling, the two Majorana components become zero energy quasibound states dubbed EP-MBSs. One of the two EP-MBSs (in blue) may develop a vanishing decay rate $\Gamma_- \to 0$, thus becoming a non-decaying and robust zero energy eigenstate \cite{San-Jose:SR16}.}}
\end{figure*}

As studies began to unveil the above phenomenology beyond the minimal model, it became clear that many experimental hints of Majoranas could easily be mistaking zero modes of non-topological origin with MBSs resulting from a non-trivial bulk topology. 
\editR{Notable examples include experiments showing zero bias peaks robust against magnetic field variations, or even conductance values close to the ideal quantized $2e^2/h$ value. Until recently, both cases were considered strong signatures of emergent Majoranas  after a bulk topological transition, but a growing body of literature shows that this is not necessarily the case  \cite{Moore:PRB18,Vuik:SP19,Avila:CP19,Stanescu:PRB19,Yu:A20}. A prominent reason is the possibility of robust but trivial zero modes arising at} smooth inhomogeneities~\cite{Prada:PRB12,Kells:PRB12,Stanescu:PRB13,Roy:PRB13,Stanescu:PRB14,Liu:PRB17a,Fleckenstein:PRB18,Penaranda:PRB18,Reeg:PRB18b,Liu:PRB18,Moore:PRB18a,Moore:PRB18,Vuik:SP19,Avila:CP19,Stanescu:PRB19}.  
Instead of emerging from a band inversion at a critical $V_Z^c$, these subgap states are predicted to emerge as a lone ABS that detaches from the continuum as $V_Z$ increases, and gradually becomes pinned to zero energy with no intervening bulk topological transition or band inversion \cite{Prada:PRB12,Szumniak:PRB17,Penaranda:PRB18,Moore:PRB18a,Vuik:SP19}, see Fig. \ref{fig:beyond} (d). This telltale feature is often observed in experiments, see e.g. Fig. \ref{fig:tunneling_experiment} (c), and should be taken as a strong hint that the zero mode might not be the result of an underlying bulk topological transition. This type of zero mode has been dubbed a quasi-MBS \cite{Vuik:SP19}, partially-separated MBS (ps-MBS) \cite{Moore:PRB18a}, or non-topological MBS \cite{Avila:CP19}. Notably, these states are not localized at opposite edges of the nanowire but are instead confined to the inhomogeneity neighborhood, whose location is often uncontrolled. As shown in Fig. 
\ref{fig:beyond} (e), subpanel 2, they typically exhibit a substantial spatial overlap.

\editP{Quasi-MBSs could also complicate the interpretation of superconducting Coulomb islands experiments~\cite{Albrecht:N16,Shen:NC18,Vaitiekenas:S20}. The Coulomb peak spacing technique used there allows to extract the energy splitting of nanowire zero-bias anomalies with high precision, see Sec. \ref{sec:MBSspec}. These measurements show oscillatory splittings that decay exponentially with nanowire length. The decay is compatible with MBSs spatially separated by a gapped topological bulk, see Fig. \ref{fig:tunneling_experiment} (e). However, the observations may also be compatible with pairs of quasi-MBSs, as the splittings of the latter can become exponentially suppressed due to reasons other than spatial separation \footnote{As nanowire screening lengths are often in the hundreds of nanometers, it is plausible to expect smoother confinement potentials, and hence suppressed splittings, as the nanowire length is increased}. 
The oscillations, moreover, are found to decay with magnetic field, contrary to the behaviour expected from topological MBSs \cite{Chiu:PRB17}, and more in line with that of quasi-MBSs.}

\editP{This kind of interpretation loophole is likely unavoidable in almost any type of experiment using purely local probes, which explains the longevity of the MBS vs ABS controversy. Other than finding a quantitative match with theoretical models across a large portion of parameter space, it seems that the only way out of these ambiguities will require truly non-local experimental schemes, see Sec. \ref{sec:overlaps}. In view of this, it becomes crucial to consider all possible types of zero modes when interpreting current and future experiments}. 

A summary of the main types \editP{of zero modes} and their defining properties is given in Table \ref{tab:mbs}. We distinguish broadly between (a) conventional, topologically trivial ABSs and variations thereof, (b) MBSs of topological origin, and (c) zero modes produced by some form of smooth inhomogeneity with a trivial bulk.
\editP{The first group includes the SNS ABSs of Figs. \ref{Fig1} and \ref{fig:Andreev2} (e-g), with or without SO coupling and Zeeman field. These states can be fine-tuned to zero energy by e.g a phase difference $\varphi\sim\pi$ across the junction or an adequate Zeeman field $V_Z$. The analogous ABSs in INS junctions, magnetic impurities or proximitized quantum dots are grouped under Shiba states, see Fig. \ref{fig:Shiba} (g). ABSs usually show no pinning to zero energy. However, ABSs INS junctions with strong SO coupling \cite{Reeg:PRB18b} or with short-range inhomogeneities and interband coupling \cite{Woods:PRB19,Chen:PRL19} may exhibit approximate zero-energy pinning as a function of some system parameters, as mentioned in Sec. \ref{sec:pinning}. In general all these ABSs have a high degree of spatial overlap of their Majorana components.} 
\editP{In the second group we consider the topological MBSs in nanowires with a topological bulk, both for short and long nanowires. The latter case corresponds to MBSs with exponentially small overlaps, the paradigmatic case highlighted by minimal models. In the last group we include all zero modes produced by sufficiently smooth inhomogeneities. We distinguish states in smoothly confined SC nanowires~\cite{Kells:PRB12} and the various forms of topologically trivial zero modes in smooth NS or S'S junctions \cite{Prada:PRB12,Stanescu:PRB13,Roy:PRB13,Stanescu:PRB14,Liu:PRB17a,Fleckenstein:PRB18,Penaranda:PRB18,Reeg:PRB18b,Liu:PRB18,Moore:PRB18a,Moore:PRB18,Vuik:SP19,Stanescu:PRB19}, including the exceptional point MBS (EP-MBS) generalization in open systems \cite{San-Jose:SR16,Avila:CP19}. The distinction between these subclasses is mostly historical, however, as the underlying mechanism for their formation is the same. All these states are characterized by a strong pinning as smoothness is increased, and partially overlapping wavefunctions.}

\editP{The debate on the interpretation of experimental signatures has often been framed in terms of ``true'' and ``fake'' MBSs, or actual MBSs (of the topological class above) and conventional zero energy ABSs. Such dichotomy has had the unfortunate side effect of establishing an imprecise terminology in some of the literature, whereby the term ``Majorana'' is used as a synonym of non-trivial band topology, instead of its original meaning of a self-conjugate zero-energy eigenstate. In truth, \emph{any} zero energy fermionic eigenstate  $c$ of a hybrid system, regardless of its origin, can be formally expressed as the sum of two self-conjugate Majorana eigenstates $c = \gamma_1+i\gamma_2$ \cite{Kitaev:PU01}, see Box. \ref{box:2}. This includes zero-energy trivial-bulk quasi- or ps-MBSs. 
}

\editP{As we expand upon in the next section, what makes MBSs of topological origin special is the exponential suppression of the spatial overlap between the $\gamma_1$ and $\gamma_2$ wavefunctions (i.e. the degree of non-locality of $c$).
This suppression, however, requires nanowires longer than the Majorana length, $L\gg\xi_M$, and of sufficiently uniformity so that MBSs are truly located at their ends (i.e. no additional zero modes appear in the bulk at uncontrolled inhomogeneities or defects). In real, finite-length nanowires these stringent conditions need not be satisfied. In a generic case, a useful formulation of the MBSs vs ABSs debate is based on whether a given robust zero-energy mode, regardless of its trivial or non-trivial bulk topology, has a sufficiently small Majorana overlap for a given application. Some requirements, such as resilience to arbitrary local noise or the possibility of spatial braiding, demand exponentially suppressed overlaps, while others, such as parametric non-Abelian braiding (i.e. relative phase manipulations without spatial displacements), merely need that a local probe may be selectively coupled to a single Majorana \cite{Vuik:SP19,San-Jose:SR16,Avila:CP19}. In such cases, ps-MBSs may be good enough.}

\subsection{Protection against errors and MBS overlaps}
\label{sec:overlaps}

Topological quantum computation was proposed as a way to achieve scalability through the hardware-level resilience of Majorana-based qubits \editP{to arbitrary local noise, in principle guaranteed by spatial non-locality}. The Majorana qubit is defined in terms of the occupation of non-local fermion states such as $c = \gamma_1+i\gamma_2$ \cite{Kitaev:PU01}, see Box \ref{box:2}. 
As efforts develop towards realising this promise, different error-inducing mechanisms have been identified and studied for topological MBS qubits, such as those created by a coupling to ungapped \cite{Budich:PRB12} or gapped \cite{Goldstein:PRB11,Rainis:PRB12} fermionic baths \editP{(quasiparticle poisoning)}, as well as to fluctuating bosonic fields \cite{Pedrocchi:PRL15} (e.g. phonons \cite{Knapp:PRB18,Aseev:PRB19}, 
photons \cite{Trif:PRL12,Schmidt:PRL13,Dmytruk:PRB15}, 
thermal fluctuations of a gate potential \cite{Aseev:PRB18,Schmidt:PRB12,Lai:PRB18}, or electromagnetic environments \cite{Knapp:PRB18}). One must also consider the errors induced by qubit manipulation, such as unwanted excitations created by nonadiabatic manipulation \cite{Scheurer:PRB13,Sekania:PRB17}, which are largely controlled by the superconducting minigap.

Given the likely ubiquity of non-topological \editP{zero modes} in realistic devices \cite{Chen:PRL19}, particularly when including QDs \editP{and screened barriers} as basic elements of many proposed schemes for topological quantum computing \cite{Hoffman:PRB16,Wakatsuki:PRB14,Karzig:PRB17,Plugge:NJP17}, it has become important to understand whether the protection of topological MBSs applies in some form also to non-topological \editP{zero modes}. \editP{A precise answer requires quantifying the Majorana overlap of a given zero mode in a sample.}
Traditional experimental schemes to measure the subgap spectrum of nanowires, such as tunneling spectroscopy, rely on local probes, so that they do not directly access the degree of non-locality of a given zero bias anomaly. An alternative, though still local scheme has been proposed to extract a quantitative estimate of the degree of MBS overlap \cite{Prada:PRB17,Clarke:PRB17,Schuray:PRB17,Penaranda:PRB18}. It consists of measuring tunneling spectroscopy into the end of the nanowire through a QD in series, which reveals the asymmetric coupling of spin-polarized states in the QD with the two spatially separated Majorana components of the zero mode. Such a scheme was implemented in a recent experiment \cite{Deng:PRB18} that demonstrated a varying degree of wavefunction overlap in otherwise similar zero modes.
\editP{Multiple tunnel probes have also been employed recently in search of clearer and less ambiguous evidence of zero-mode non-locality and non-trivial topology~\cite{Grivnin:NC19,Anselmetti:PRB19,Menard:PRL20,Puglia:A20}.}


\section{Summary and outlook}

We have reviewed the remarkable advances towards characterizing the detailed structure of ABSs in hybrid nanowires with strong SO coupling, and related systems. {In ideal nanowires, ABSs evolve with magnetic fields and gates, developing a spatial separation of their wavefunction components and ultimately transforming into robust, non-local, topological Majorana zero energy modes beyond a critical Zeeman field. Experimental observations, however, deviate from the predictions of minimal models, making evident the need to incorporate extensions, such as the electrostatic environment, multimode and orbital effects, spatial inhomogeneities and momentum mixing. These additions introduce a complex, non-universal phenomenology}. Theoretical work has identified alternative routes towards stabilizing zero modes, different from a bulk topological phase transition. Such routes include using smooth and/or spin-dependent confinement, or exceptional point bifurcations in open (non-Hermitian) systems. However, the wavefunction non-locality obtained by such means is generally poor. Given the importance of non-locality for the protection of MBSs, their resilience against noise and the possibility of carrying out braiding operations on them, it has become a major focus point in current experiments to detect and quantify the degree of Majorana overlap. First results in this area have been obtained using purely local probes.  These have intrinsic limitations, unfortunately, and can only suggest, not demonstrate, Majorana non-locality. Experiments are underway, nevertheless, to exploit truly non-local measurements in more complex nanowire setups without such limitations~\cite{Zhang:NC19,Frolov:NP20,Aguado-Kouwenhoven:PT20}. The ultimate demonstration, non-Abelian braiding, remains an open challenge. 

Nonlocality and braiding are the cornerstones behind the original promise of Majorana applications, namely, to harness the hardware-level resilience of Majorana qubits to solve the scalability problem of quantum computers. \editE{This is an ambitious long-term endeavour that will require solving important challenges to finally bring Majoranas out of the lab. Majorana nanowires will likely have an important role in this journey, as they represent the most readily accessible system of this kind and the most explored topological superconductor so far. However, they also have some disadvantages such as the unwanted presence of nontopological in-gap states or the necessity to subject the wire to strong magnetic fields. So far, partially-covered nanowires have been intensively explored, but other alternatives with some interesting advantages are beginning to be studied, such as full-shell nanowires \cite{Vaitiekenas:S20,Penaranda:PRR20} and ferromagnetic hybrid structures \cite{Vaitiekenas:A20}. Additionally, the recent demonstration of gate-tunable nanowire-based superconducting qubits, so-called gatemons \cite{Larsen:PRL15,Lange:PRL15}, which include full-shell nanowires \cite{Sabonis:A20} and junctions with quantum dots \cite{Bagerbos:PRL20,Kringhof:PRL20},  opens new possibilities for studying Majorana physics in hybrid architectures that use more mature technologies such as circuit QED and transmon qubits \cite{Ginossar:N14,Trif:PRB18,Kesselman:S19,Avila:A20a,Avila:A20}.} 

\editE{Nevertheless, it is entirely possible that radically different Majorana platforms will be discovered that exhibit cleaner and larger gaps, improved Majorana non-locality, easier braiding, and crucially, are more amenable to scalable fabrication and integration. Leading explorations include experiments on shallow quantum wells that can be proximitized by epitaxial superconductors, while being compatible with lithographic patterning techniques \cite{Suominen:PRL17}. On such platforms crossed-Andreev reflection has been proposed as an alternative way to stabilize Majoranas \cite{Finocchiaro:PRL18,Thakurathi:PRB18}. Also, the use of two-dimensional van der Waals crystals has received attention recently \cite{Young:N14,Lee:NP17}, where the SO coupling, required for generating MBSs, could be replaced by carrier chirality and intrinsic interactions \cite{San-Jose:PRX15}. Solving the material-science side of the challenge is crucial before tackling the more applied problems of engineering, operating and integrating coherent ensembles of topological quantum gates.} We expect these efforts to continue fertilizing condensed-matter research with novel and exciting possibilities.

\section{Author contributions}
L. P. K. initiated this review and E. P. coordinated the project. All authors discussed the general structure of the manuscript. M. W. A. M. and A. G. wrote ``ABS spectroscopy", ``MBS spectroscopy" and contributed to ``Extensions of the minimal model". E. J. H. L., J. N. and R. A. wrote ``ABSs in QDs".  J. K. and D. L. contributed to ``Zero-energy pinning with a topologically trivial bulk" and `Protection against errors and MBS overlaps". E. P., P. S.-J. and R. A. wrote everything else. All authors reviewed and polished the manuscript.

\acknowledgements
Research supported by the Spanish Ministry of Science, Innovation and Universities through grants FIS2015-65706-P, FIS2015-64654-P, FIS2016-80434-P, FIS2017-84860-R, PCI2018-093026 and PGC2018-097018-B-I00 (AEI/FEDER, EU), the Ram\'on y Cajal programme grant RYC-2011-09345 and RYC-2015-17973, the Mar\'ia de Maeztu Programme for Units of Excellence in R\&D (MDM-2014-0377), the European Union's Horizon 2020 research and innovation programme under grant agreements Nos 828948 (FETOPEN AndQC), 127900 (Quantera SuperTOP), the European Research Council (ERC) Starting Grant agreements 716559 (TOPOQDot), 757725 (ETOPEX) and 804988 (SiMS), the Netherlands Organization for Scientific Research (NWO), Microsoft, the Danish National Research Foundation, the Carlsberg Foundation, and the Swiss National Science Foundation and NCCR QSIT. We also acknowledge support from CSIC Research Platform on Quantum Technologies PTI-001.

\appendix

\section{Box A -- Proximitized nanowire model}
\label{box:1}
\begin{figure}
\includegraphics[width=\columnwidth]{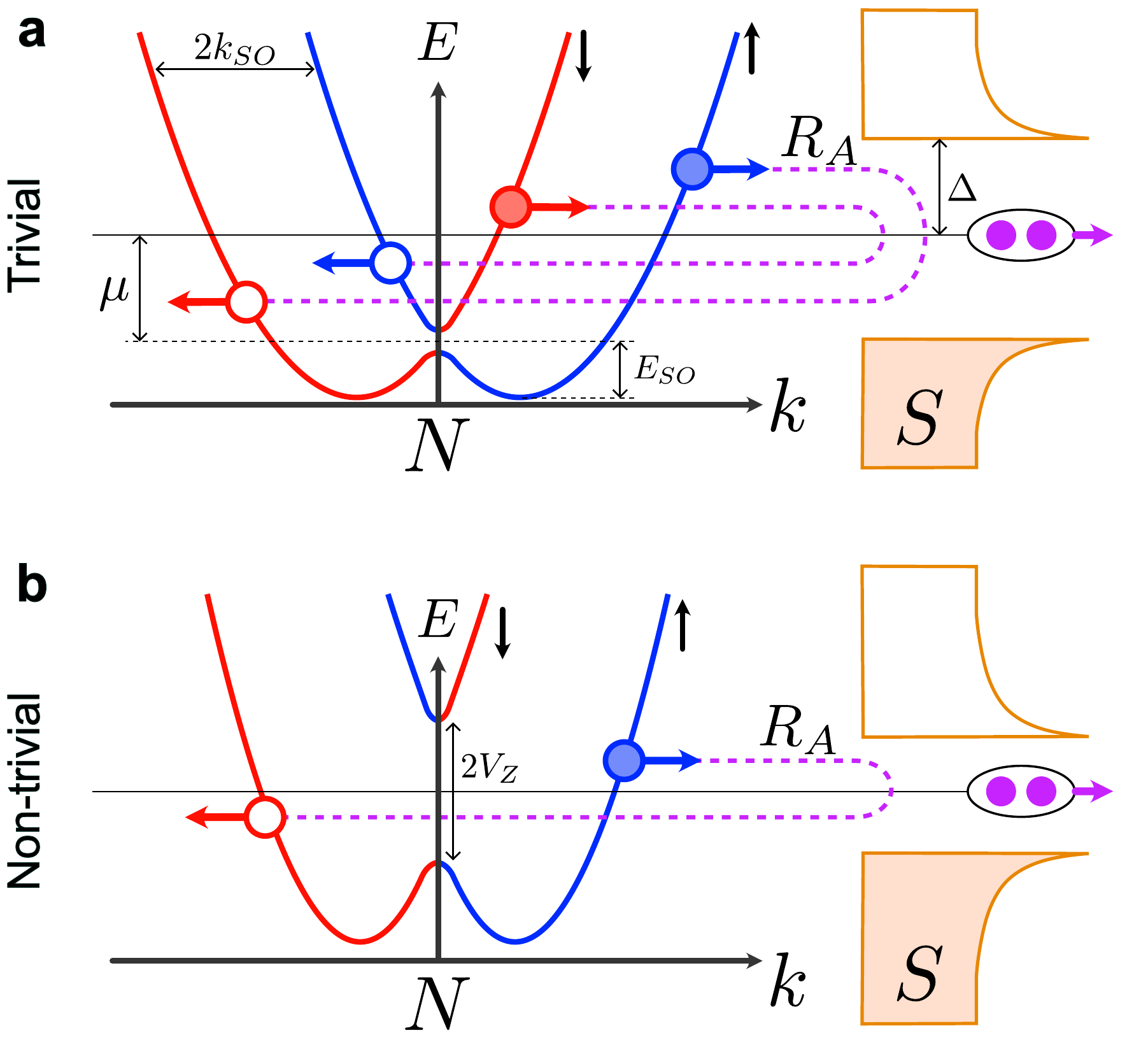}
\caption{\label{fig:box1} Band structure and subgap Andreev scattering processes of a semiconducting nanowire in contact with a SC. The nanowire SO coupling is responsible for the $2k_{SO}$ and $E_{SO}$ shifts, and a longitudinal Zeeman field is responsible for the $2V_Z$ splitting at $k=0$. Depending on whether $V_Z<V_Z^c$ (a) or $V_Z>V_Z^c$ (b), the proximitized nanowire is a topologically trivial ($s$-wave) or non-trivial SC (effectively spinless $p$-wave).}
\end{figure}

The starting point of the proximitized nanowire model (R. M. Lutchyn \emph{et al}~\cite{Lutchyn:PRL10} and Y. Oreg \emph{et al}~\cite{Oreg:PRL10}) is a Hamiltonian describing a 1D semiconducting nanowire with Rashba spin-orbit (SO) interaction and in the presence of an external magnetic field $B$ perpendicular to the Rashba field (here, we assume that $B$ is applied parallel to the nanowire axis $x$):
\[
H_w=\frac{1}{2}\int dx \Psi^\dagger(x)\mathcal{H}(x)\Psi(x),
\]
with 
\begin{equation}
\mathcal{H}_w(x)=\left(-\frac{\hbar^2\partial^2_x}{2m^*}-\mu-i\alpha\partial_x\sigma_y\right)\tau_z+V_Z\sigma_x,
\end{equation}
where $m^*$ is the effective mass of the semiconductor, $\mu$ its chemical potential, $\alpha$ the SO coupling and $V_Z=\frac{1}{2}g\mu_B B$ the Zeeman energy produced by $B$, given in terms of the nanowire's $g$-factor (with $\mu_B$ being the Bohr's magneton). $\Psi(x)=(\psi^\dagger_\uparrow,\psi^\dagger_\downarrow,\psi_\downarrow,-\psi_\uparrow)$ are Nambu spinors and $\sigma$ and $\tau$ Pauli matrices in spin and particle-hole space, respectively.

Solving the above Hamiltonian in reciprocal space yields a dispersion relation of the form $E_{k,\pm}=\frac{\hbar^{2}k^{2}}{2m^*}-\mu\pm\sqrt{V^2_Z+\alpha^{2}k^{2}}$. In the absence of Zeeman field, $V_Z=0$, the Rashba term removes the spin degeneracy of the 1D parabolic band and gives rise to two parabolas shifted relative to each other along the momentum axis (each by an amount $k_{SO}=m^*\alpha/\hbar^2$) and displaced down in energy by an amount $E_{SO}=m^*\alpha^2/2\hbar^2$, where $\hbar$ the reduced Planck's constant, see Fig. \ref{fig:box1} (a). These parabolas correspond to spin up and spin down projections along the spin-quantization axis fixed by the Rashba coupling (here $\sigma_y$). On the other hand, a finite Zeeman $V_Z\neq0$ mixes both spins and hence removes the spin degeneracy at $k=0$ by opening up a gap of size $2V_Z$, Fig. \ref{fig:box1} (b). These split bands define two sectors + and - of opposite helicity at large and small momenta, respectively. If $|\mu|<|V_Z|$, only the + sector is present around the Fermi energy (helical regime). Projecting a standard $s$-wave pairing term $H_s=\sum_k\Delta\psi^\dagger_{\uparrow,k}\psi^\dagger_{\downarrow,-k}+H. c.$ onto the helical basis one obtains intraband (spinless) pairing terms of the form $H_p=\sum_k\sum_{i=\pm}\Delta^i_k \psi^\dagger_{i,k}\psi^\dagger_{i,-k}+H. c.$, with $\Delta^{\pm}_k=\frac{\mp i\alpha k\Delta}{2\sqrt{V_Z^2+\alpha^2k^2}}$ having so-called $p$-wave symmetry $\Delta_k=-\Delta_{-k}$. In the helical regime, the minimal Hamiltonian $H=H_w+H_s$ is thus a realistic implementation of the Kitaev model for 1D  $p$-wave superconductivity \cite{Kitaev:PU01}. When the applied Zeeman field is larger than the critical value $V_Z^c = \sqrt{\Delta^2+\mu^2}$, the 1D SC becomes topological and hosts MBSs at its ends.

To implement this proposal in experimentally realizable systems, one needs semiconductors with large $g$-factors in order to achieve a large $V_Z$ under moderate external magnetic fields $B$ below the critical field of the SC. A good proximity effect with conventional SCs and a large Rashba energy are also necessary. Last but not least, one needs to be able to keep the chemical potential $\mu$ of the nanowire close to zero (in order to reach the helical regime with spin-momentum locking for moderate $B$), despite the proximity to the SC.

\section{Box B -- Majorana basis}
\label{box:2}
\begin{figure}
\includegraphics[width=\columnwidth]{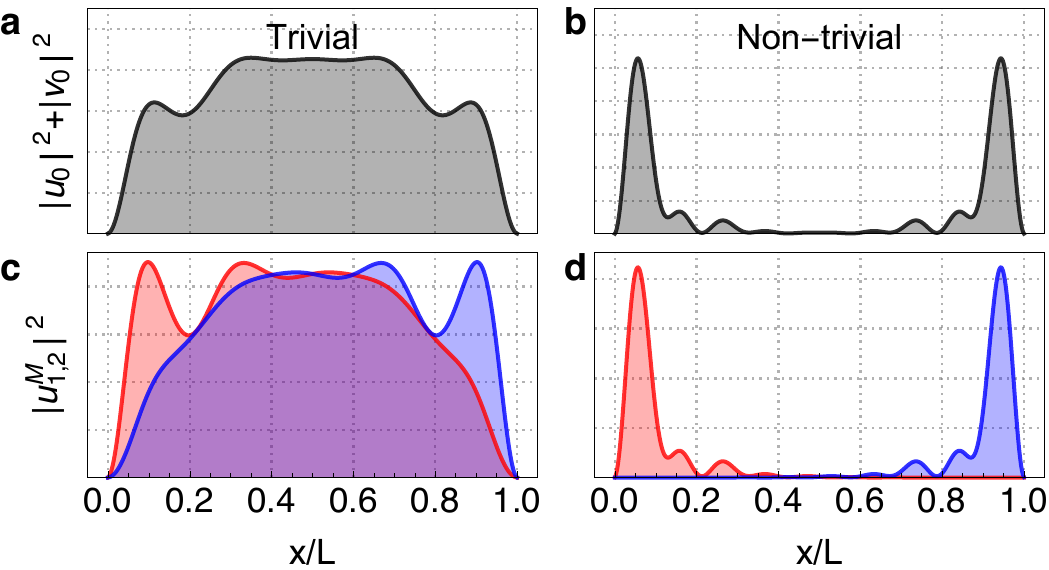}
\caption{\label{fig:majoranabasis} Wavefunction of the lowest energy state $\psi_0$ in a uniform $L=1\mathrm{\mu m}$ Majorana nanowire for a trivial $V_Z=0.5V_Z^c$ (a) and a non-trivial $V_Z=1.4V_Z^c$ (b) cases. (c,d) Wavefunctions of the corresponding Majorana components $u^M_{1,2}(x)$ in the Majorana basis. In the trivial regime the left (red) and right (blue) Majorana components strongly overlap (c), whereas in the topological regime they move apart and concentrate at the ends of the nanowire (d).}
\end{figure}

A Bogoliubov-de Gennes eigenstate in a superconducting system is an excitation $|\psi_n\rangle = \psi_n^\dagger|\mathrm{BCS}\rangle$ of energy $\epsilon_n$ over its ground state $|\mathrm{BCS}\rangle$ that consists of a superposition of one electron and one hole quasiparticles,
\[
\psi_n=\int dx\, \sum_\sigma\left[u_{n\sigma}(x)\psi_\sigma(x) + v_{n\sigma}(x)\psi^\dagger_\sigma(x)\right]. 
\]
Here $\psi_\sigma^\dagger(x)$ and $\psi_\sigma(x)$ create and destroy a quasiparticle of spin $\sigma$ perfectly localized at point $x$, respectively, and $u_{n\sigma}(x),v_{n\sigma}(x)$ are electron/hole wavefunctions.

If the energy of a given eigenstate $|\psi_0\rangle=\psi_0^\dagger|\mathrm{BCS}\rangle$ becomes negligibly small $\epsilon_0\approx 0$ as in the case of a topological Majorana nanowire, the eigenstates $|\mathrm{BCS}\rangle$ and $|\psi_0\rangle$ are both degenerate ground states, of even and odd fermionic parity, respectively. If we denote $|\psi_\textrm{even}\rangle\equiv |\mathrm{BCS}\rangle$ and $|\psi_\textrm{odd}\rangle\equiv|\psi_0\rangle$, we find that $\psi_0$ and $\psi_0^\dagger$ switch between the two
\[
\left(\begin{array}{c}
|\psi_\textrm{even}\rangle \\ |\psi_\textrm{odd}\rangle
\end{array}\right) = 
\left(\begin{array}{cc}
0 & \psi_0\\ \psi^\dagger_0 & 0
\end{array}\right)
\left(\begin{array}{c}
|\psi_\textrm{even}\rangle \\ |\psi_\textrm{odd}\rangle
\end{array}\right).
\]
The matrix elements of $\psi_0, \psi_0^\dagger$ in this subspace are therefore $\langle \psi_\textrm{even,odd}|\psi_0|\psi_\textrm{even,odd}\rangle = (\sigma_1+i\sigma_2)/2$ and $\langle \psi_\textrm{even,odd}|\psi^\dagger_0|\psi_\textrm{even,odd}\rangle = (\sigma_1-i\sigma_2)/2$, where $\sigma_i$ are Pauli matrices.

By performing a unitary rotation to the so-called Majorana basis, the eigenstate operators $\psi_0, \psi^\dagger_0$ can be decomposed into two Majorana operators that satisfy self-conjugation, $\gamma_1=\gamma_1^\dagger$ and $\gamma_2=\gamma_2^\dagger$, so that
\begin{equation}\label{majoranabasis}
\begin{array}{ll}
\psi_0 =(\gamma_1+i\gamma_2)/2,& \psi^\dagger_0=(\gamma_1-i\gamma_2)/2;\\
\gamma_1 =\psi^\dagger_0+\psi_0,& \gamma_2=i(\psi^\dagger_0-\psi_0).
\end{array}
\end{equation}
Each Majorana operator corresponds to a fermionic eigenstate, in the sense that $\{\gamma^\dagger_i,\gamma_j\}=2\delta_{ij}$, but the Majorana reality property also implies that, unlike a conventional fermion, $\gamma_i^2=1$. The matrix elements of $\gamma_i$ in the ground state subspace are $\langle \psi_\textrm{even,odd}|\gamma_i|\psi_\textrm{even,odd}\rangle = \sigma_i$. 

The Majorana states created by $\gamma_1$ and $\gamma_2$ are sometimes intuitively described as half-fermions, as they always come in pairs and any two in a system can be combined to create a conventional fermion as above. In a topological Majorana nanowire, the wavefunction $u^M_{i,\sigma}(x)$ of 
\[
\gamma_i=\int dx\, \sum_\sigma\left[u^M_{i\sigma}(x)\psi_\sigma(x) + {u^M_{i\sigma}}^*(x)\psi^\dagger_\sigma(x)\right] 
\]
is localized at either end of the nanowire, unlike $u_{0,\sigma}(x), v_{0,\sigma}(x)$ of $\psi_0$, that occupies both ends, see Fig. \ref{fig:majoranabasis}. The latter is hence called a non-local fermion.

The above transformations, Eqs. (\ref{majoranabasis}), can be applied to an arbitrary ABS $\psi_n$ of finite energy. In such case, the resulting Majorana states are not eigenstates. However, the decomposition still allows to examine the degree of Majorana non-locality of $\psi_n$ by computing the overlap between the corresponding $u^M_{1,\sigma}(x)$ and $u^M_{2,\sigma}(x)$.

\bibliography{biblioreview}

\begin{thebibliography}{276}%
\makeatletter
\providecommand \@ifxundefined [1]{%
 \@ifx{#1\undefined}
}%
\providecommand \@ifnum [1]{%
 \ifnum #1\expandafter \@firstoftwo
 \else \expandafter \@secondoftwo
 \fi
}%
\providecommand \@ifx [1]{%
 \ifx #1\expandafter \@firstoftwo
 \else \expandafter \@secondoftwo
 \fi
}%
\providecommand \natexlab [1]{#1}%
\providecommand \enquote  [1]{``#1''}%
\providecommand \bibnamefont  [1]{#1}%
\providecommand \bibfnamefont [1]{#1}%
\providecommand \citenamefont [1]{#1}%
\providecommand \href@noop [0]{\@secondoftwo}%
\providecommand \href [0]{\begingroup \@sanitize@url \@href}%
\providecommand \@href[1]{\@@startlink{#1}\@@href}%
\providecommand \@@href[1]{\endgroup#1\@@endlink}%
\providecommand \@sanitize@url [0]{\catcode `\\12\catcode `\$12\catcode
  `\&12\catcode `\#12\catcode `\^12\catcode `\_12\catcode `\%12\relax}%
\providecommand \@@startlink[1]{}%
\providecommand \@@endlink[0]{}%
\providecommand \url  [0]{\begingroup\@sanitize@url \@url }%
\providecommand \@url [1]{\endgroup\@href {#1}{\urlprefix }}%
\providecommand \urlprefix  [0]{URL }%
\providecommand \Eprint [0]{\href }%
\providecommand \doibase [0]{http://dx.doi.org/}%
\providecommand \selectlanguage [0]{\@gobble}%
\providecommand \bibinfo  [0]{\@secondoftwo}%
\providecommand \bibfield  [0]{\@secondoftwo}%
\providecommand \translation [1]{[#1]}%
\providecommand \BibitemOpen [0]{}%
\providecommand \bibitemStop [0]{}%
\providecommand \bibitemNoStop [0]{.\EOS\space}%
\providecommand \EOS [0]{\spacefactor3000\relax}%
\providecommand \BibitemShut  [1]{\csname bibitem#1\endcsname}%
\let\auto@bib@innerbib\@empty
\bibitem [{\citenamefont {Kamerlingh~Onnes}(1911)}]{Kamerlingh-Onnes:CPLUL11}%
  \BibitemOpen
  \bibfield  {author} {\bibinfo {author} {\bibfnamefont {H.}~\bibnamefont
  {Kamerlingh~Onnes}},\ }\href {https://ci.nii.ac.jp/naid/10027961803/en/}
  {\bibfield  {journal} {\bibinfo  {journal} {Comm. Phys. Lab. Univ. Leiden}\
  }\textbf {\bibinfo {volume} {122}},\ \bibinfo {pages} {122} (\bibinfo {year}
  {1911})}\BibitemShut {NoStop}%
\bibitem [{\citenamefont {van Delft}\ and\ \citenamefont
  {Kes}(2010)}]{vanDelft:PT2010}%
  \BibitemOpen
  \bibfield  {author} {\bibinfo {author} {\bibfnamefont {D.}~\bibnamefont {van
  Delft}}\ and\ \bibinfo {author} {\bibfnamefont {P.}~\bibnamefont {Kes}},\
  }\href {\doibase 10.1063/1.3490499} {\bibfield  {journal} {\bibinfo
  {journal} {Phys. Today}\ }\textbf {\bibinfo {volume} {63}},\ \bibinfo {pages}
  {38} (\bibinfo {year} {2010})}\BibitemShut {NoStop}%
\bibitem [{\citenamefont {De~Gennes}(2018)}]{De-Gennes:18}%
  \BibitemOpen
  \bibfield  {author} {\bibinfo {author} {\bibfnamefont {P.-G.}\ \bibnamefont
  {De~Gennes}},\ }\href@noop {} {\emph {\bibinfo {title} {Superconductivity of
  metals and alloys}}}\ (\bibinfo  {publisher} {CRC Press},\ \bibinfo {year}
  {2018})\BibitemShut {NoStop}%
\bibitem [{\citenamefont {Tinkham}(2004)}]{Tinkham:04}%
  \BibitemOpen
  \bibfield  {author} {\bibinfo {author} {\bibfnamefont {M.}~\bibnamefont
  {Tinkham}},\ }\href@noop {} {\emph {\bibinfo {title} {Introduction to
  superconductivity}}}\ (\bibinfo  {publisher} {Courier Corporation},\ \bibinfo
  {year} {2004})\BibitemShut {NoStop}%
\bibitem [{\citenamefont {Martin}(2019)}]{Martin:PT2019}%
  \BibitemOpen
  \bibfield  {author} {\bibinfo {author} {\bibfnamefont {J.~D.}\ \bibnamefont
  {Martin}},\ }\href {\doibase 10.1063/PT.3.4110} {\bibfield  {journal}
  {\bibinfo  {journal} {Phys. Today}\ }\textbf {\bibinfo {volume} {72}},\
  \bibinfo {pages} {30} (\bibinfo {year} {2019})}\BibitemShut {NoStop}%
\bibitem [{\citenamefont {Bardeen}\ \emph {et~al.}(1957)\citenamefont
  {Bardeen}, \citenamefont {Cooper},\ and\ \citenamefont
  {Schrieffer}}]{Bardeen:PR57}%
  \BibitemOpen
  \bibfield  {author} {\bibinfo {author} {\bibfnamefont {J.}~\bibnamefont
  {Bardeen}}, \bibinfo {author} {\bibfnamefont {L.~N.}\ \bibnamefont {Cooper}},
  \ and\ \bibinfo {author} {\bibfnamefont {J.~R.}\ \bibnamefont {Schrieffer}},\
  }\href {\doibase 10.1103/PhysRev.106.162} {\bibfield  {journal} {\bibinfo
  {journal} {Phys. Rev.}\ }\textbf {\bibinfo {volume} {106}},\ \bibinfo {pages}
  {162} (\bibinfo {year} {1957})}\BibitemShut {NoStop}%
\bibitem [{\citenamefont {Cooper}(1956)}]{Cooper:1956}%
  \BibitemOpen
  \bibfield  {author} {\bibinfo {author} {\bibfnamefont {L.~N.}\ \bibnamefont
  {Cooper}},\ }\href {\doibase 10.1103/PhysRev.104.1189} {\bibfield  {journal}
  {\bibinfo  {journal} {Phys. Rev.}\ }\textbf {\bibinfo {volume} {104}},\
  \bibinfo {pages} {1189} (\bibinfo {year} {1956})}\BibitemShut {NoStop}%
\bibitem [{\citenamefont {Ginzburg}\ and\ \citenamefont
  {Landau}(1950)}]{Ginzburg:ZETF50}%
  \BibitemOpen
  \bibfield  {author} {\bibinfo {author} {\bibfnamefont {V.~L.}\ \bibnamefont
  {Ginzburg}}\ and\ \bibinfo {author} {\bibfnamefont {L.~D.}\ \bibnamefont
  {Landau}},\ }\href@noop {} {\bibfield  {journal} {\bibinfo  {journal} {Zh.
  Eksp. Teor. Fiz.}\ }\textbf {\bibinfo {volume} {20}},\ \bibinfo {pages}
  {1064} (\bibinfo {year} {1950})}\BibitemShut {NoStop}%
\bibitem [{\citenamefont {Cyrot}(1973)}]{Cyrot:ROPIP73}%
  \BibitemOpen
  \bibfield  {author} {\bibinfo {author} {\bibfnamefont {M.}~\bibnamefont
  {Cyrot}},\ }\href {\doibase 10.1088/0034-4885/36/2/001} {\bibfield  {journal}
  {\bibinfo  {journal} {Rep. Prog. Phys.}\ }\textbf {\bibinfo {volume} {36}},\
  \bibinfo {pages} {103} (\bibinfo {year} {1973})}\BibitemShut {NoStop}%
\bibitem [{\citenamefont {Caroli}\ \emph {et~al.}(1964)\citenamefont {Caroli},
  \citenamefont {Gennes},\ and\ \citenamefont {Matricon}}]{Caroli:PL64}%
  \BibitemOpen
  \bibfield  {author} {\bibinfo {author} {\bibfnamefont {C.}~\bibnamefont
  {Caroli}}, \bibinfo {author} {\bibfnamefont {P.~D.}\ \bibnamefont {Gennes}},
  \ and\ \bibinfo {author} {\bibfnamefont {J.}~\bibnamefont {Matricon}},\
  }\href {\doibase https://doi.org/10.1016/0031-9163(64)90375-0} {\bibfield
  {journal} {\bibinfo  {journal} {Phys. Lett.}\ }\textbf {\bibinfo {volume}
  {9}},\ \bibinfo {pages} {307 } (\bibinfo {year} {1964})}\BibitemShut
  {NoStop}%
\bibitem [{\citenamefont {Yu}(1965)}]{Yu:APS65}%
  \BibitemOpen
  \bibfield  {author} {\bibinfo {author} {\bibfnamefont {L.}~\bibnamefont
  {Yu}},\ }\href {http://wulixb.iphy.ac.cn/CN/Y1965/V21/I1/75} {\bibfield
  {journal} {\bibinfo  {journal} {Acta. Phys. Sin.}\ }\textbf {\bibinfo
  {volume} {21}},\ \bibinfo {pages} {75} (\bibinfo {year} {1965})}\BibitemShut
  {NoStop}%
\bibitem [{\citenamefont {Shiba}(1968)}]{Shiba:PTP68}%
  \BibitemOpen
  \bibfield  {author} {\bibinfo {author} {\bibfnamefont {H.}~\bibnamefont
  {Shiba}},\ }\href {https://doi.org/10.1143/PTP.40.435} {\bibfield  {journal}
  {\bibinfo  {journal} {Prog. Theor. Phys.}\ }\textbf {\bibinfo {volume}
  {40}},\ \bibinfo {pages} {435} (\bibinfo {year} {1968})}\BibitemShut
  {NoStop}%
\bibitem [{\citenamefont {Rusinov}(1969)}]{Rusinov:SPJ69}%
  \BibitemOpen
  \bibfield  {author} {\bibinfo {author} {\bibfnamefont {A.}~\bibnamefont
  {Rusinov}},\ }\href
  {http://www.jetp.ac.ru/cgi-bin/e/index/e/29/6/p1101?a=list} {\bibfield
  {journal} {\bibinfo  {journal} {Sov. Phys. JETP}\ }\textbf {\bibinfo {volume}
  {9}},\ \bibinfo {pages} {85} (\bibinfo {year} {1969})}\BibitemShut {NoStop}%
\bibitem [{\citenamefont {Blonder}\ \emph {et~al.}(1982)\citenamefont
  {Blonder}, \citenamefont {Tinkham},\ and\ \citenamefont
  {Klapwijk}}]{Blonder:PRB82}%
  \BibitemOpen
  \bibfield  {author} {\bibinfo {author} {\bibfnamefont {G.~E.}\ \bibnamefont
  {Blonder}}, \bibinfo {author} {\bibfnamefont {M.}~\bibnamefont {Tinkham}}, \
  and\ \bibinfo {author} {\bibfnamefont {T.~M.}\ \bibnamefont {Klapwijk}},\
  }\href {\doibase 10.1103/PhysRevB.25.4515} {\bibfield  {journal} {\bibinfo
  {journal} {Phys. Rev. B}\ }\textbf {\bibinfo {volume} {25}},\ \bibinfo
  {pages} {4515} (\bibinfo {year} {1982})}\BibitemShut {NoStop}%
\bibitem [{\citenamefont {Andreev}(1964)}]{Andreev:SPJ64}%
  \BibitemOpen
  \bibfield  {author} {\bibinfo {author} {\bibfnamefont {A.~F.}\ \bibnamefont
  {Andreev}},\ }\href
  {http://www.jetp.ac.ru/cgi-bin/e/index/e/19/5/p1228?a=list} {\bibfield
  {journal} {\bibinfo  {journal} {Sov. Phys. JETP}\ }\textbf {\bibinfo {volume}
  {19}},\ \bibinfo {pages} {1228} (\bibinfo {year} {1964})}\BibitemShut
  {NoStop}%
\bibitem [{\citenamefont {Andreev}(1966)}]{Andreev:SPJ66}%
  \BibitemOpen
  \bibfield  {author} {\bibinfo {author} {\bibfnamefont {A.~F.}\ \bibnamefont
  {Andreev}},\ }\href
  {http://www.jetp.ac.ru/cgi-bin/e/index/e/22/2/p455?a=list} {\bibfield
  {journal} {\bibinfo  {journal} {Sov. Phys. JETP}\ }\textbf {\bibinfo {volume}
  {22}},\ \bibinfo {pages} {18} (\bibinfo {year} {1966})}\BibitemShut {NoStop}%
\bibitem [{\citenamefont {De~Gennes}\ and\ \citenamefont
  {Saint-James}(1963)}]{De-Gennes:PL63}%
  \BibitemOpen
  \bibfield  {author} {\bibinfo {author} {\bibfnamefont {P.}~\bibnamefont
  {De~Gennes}}\ and\ \bibinfo {author} {\bibfnamefont {D.}~\bibnamefont
  {Saint-James}},\ }\href {https://doi.org/10.1016/0031-9163(63)90148-3}
  {\bibfield  {journal} {\bibinfo  {journal} {Phys. Lett.}\ }\textbf {\bibinfo
  {volume} {4}} (\bibinfo {year} {1963})}\BibitemShut {NoStop}%
\bibitem [{\citenamefont {Kulik}(1970)}]{Kulik:SPJ70}%
  \BibitemOpen
  \bibfield  {author} {\bibinfo {author} {\bibfnamefont {I.~O.}\ \bibnamefont
  {Kulik}},\ }\href@noop {} {\bibfield  {journal} {\bibinfo  {journal} {Sov.
  Phys. JETP}\ }\textbf {\bibinfo {volume} {30}},\ \bibinfo {pages} {944}
  (\bibinfo {year} {1970})}\BibitemShut {NoStop}%
\bibitem [{\citenamefont {Sauls}(2018)}]{Sauls:PTRSA18}%
  \BibitemOpen
  \bibfield  {author} {\bibinfo {author} {\bibfnamefont {J.~A.}\ \bibnamefont
  {Sauls}},\ }\href
  {http://rsta.royalsocietypublishing.org/content/376/2125/20180140} {\bibfield
   {journal} {\bibinfo  {journal} {Philos. Trans. Royal Soc. A}\ }\textbf
  {\bibinfo {volume} {376}} (\bibinfo {year} {2018})}\BibitemShut {NoStop}%
\bibitem [{\citenamefont {Hasan}\ and\ \citenamefont
  {Kane}(2010)}]{Hasan:RMP10}%
  \BibitemOpen
  \bibfield  {author} {\bibinfo {author} {\bibfnamefont {M.~Z.}\ \bibnamefont
  {Hasan}}\ and\ \bibinfo {author} {\bibfnamefont {C.~L.}\ \bibnamefont
  {Kane}},\ }\href {\doibase 10.1103/RevModPhys.82.3045} {\bibfield  {journal}
  {\bibinfo  {journal} {Rev. Mod. Phys.}\ }\textbf {\bibinfo {volume} {82}},\
  \bibinfo {pages} {3045} (\bibinfo {year} {2010})}\BibitemShut {NoStop}%
\bibitem [{\citenamefont {Qi}\ and\ \citenamefont {Zhang}(2011)}]{Qi:RMP11}%
  \BibitemOpen
  \bibfield  {author} {\bibinfo {author} {\bibfnamefont {X.-L.}\ \bibnamefont
  {Qi}}\ and\ \bibinfo {author} {\bibfnamefont {S.-C.}\ \bibnamefont {Zhang}},\
  }\href {\doibase 10.1103/RevModPhys.83.1057} {\bibfield  {journal} {\bibinfo
  {journal} {Rev. Mod. Phys.}\ }\textbf {\bibinfo {volume} {83}},\ \bibinfo
  {pages} {1057} (\bibinfo {year} {2011})}\BibitemShut {NoStop}%
\bibitem [{\citenamefont {Thouless}\ \emph {et~al.}(1982)\citenamefont
  {Thouless}, \citenamefont {Kohmoto}, \citenamefont {Nightingale},\ and\
  \citenamefont {den Nijs}}]{Thouless:PRL82}%
  \BibitemOpen
  \bibfield  {author} {\bibinfo {author} {\bibfnamefont {D.~J.}\ \bibnamefont
  {Thouless}}, \bibinfo {author} {\bibfnamefont {M.}~\bibnamefont {Kohmoto}},
  \bibinfo {author} {\bibfnamefont {M.~P.}\ \bibnamefont {Nightingale}}, \ and\
  \bibinfo {author} {\bibfnamefont {M.}~\bibnamefont {den Nijs}},\ }\href
  {\doibase 10.1103/PhysRevLett.49.405} {\bibfield  {journal} {\bibinfo
  {journal} {Phys. Rev. Lett.}\ }\textbf {\bibinfo {volume} {49}},\ \bibinfo
  {pages} {405} (\bibinfo {year} {1982})}\BibitemShut {NoStop}%
\bibitem [{\citenamefont {Leijnse}\ and\ \citenamefont
  {Flensberg}(2012)}]{Leijnse:SSAT12}%
  \BibitemOpen
  \bibfield  {author} {\bibinfo {author} {\bibfnamefont {M.}~\bibnamefont
  {Leijnse}}\ and\ \bibinfo {author} {\bibfnamefont {K.}~\bibnamefont
  {Flensberg}},\ }\href {https://doi.org/10.1088/0268-1242/27/12/124003}
  {\bibfield  {journal} {\bibinfo  {journal} {Semicond. Sci. Technol.}\
  }\textbf {\bibinfo {volume} {27}},\ \bibinfo {pages} {124003} (\bibinfo
  {year} {2012})}\BibitemShut {NoStop}%
\bibitem [{\citenamefont {Alicea}(2012)}]{Alicea:RPP12}%
  \BibitemOpen
  \bibfield  {author} {\bibinfo {author} {\bibfnamefont {J.}~\bibnamefont
  {Alicea}},\ }\href {https://doi.org/10.1088/0034-4885/75/7/076501} {\bibfield
   {journal} {\bibinfo  {journal} {Rep. Prog. Phys.}\ }\textbf {\bibinfo
  {volume} {75}},\ \bibinfo {pages} {076501} (\bibinfo {year}
  {2012})}\BibitemShut {NoStop}%
\bibitem [{\citenamefont {Beenakker}(2013)}]{Beenakker:ARCMP13}%
  \BibitemOpen
  \bibfield  {author} {\bibinfo {author} {\bibfnamefont {C.}~\bibnamefont
  {Beenakker}},\ }\href {\doibase 10.1146/annurev-conmatphys-030212-184337}
  {\bibfield  {journal} {\bibinfo  {journal} {Annu. Rev. Cond. Mat. Phys.}\
  }\textbf {\bibinfo {volume} {4}},\ \bibinfo {pages} {113} (\bibinfo {year}
  {2013})}\BibitemShut {NoStop}%
\bibitem [{\citenamefont {Sato}\ and\ \citenamefont
  {Fujimoto}(2016)}]{Sato:JPSJ16}%
  \BibitemOpen
  \bibfield  {author} {\bibinfo {author} {\bibfnamefont {M.}~\bibnamefont
  {Sato}}\ and\ \bibinfo {author} {\bibfnamefont {S.}~\bibnamefont
  {Fujimoto}},\ }\href {\doibase 10.7566/JPSJ.85.072001} {\bibfield  {journal}
  {\bibinfo  {journal} {J. Phys. Soc. Jpn}\ }\textbf {\bibinfo {volume} {85}},\
  \bibinfo {pages} {072001} (\bibinfo {year} {2016})}\BibitemShut {NoStop}%
\bibitem [{\citenamefont {Aguado}(2017)}]{Aguado:RNC17}%
  \BibitemOpen
  \bibfield  {author} {\bibinfo {author} {\bibfnamefont {R.}~\bibnamefont
  {Aguado}},\ }\href {\doibase 10.1393/ncr/i2017-10141-9} {\bibfield  {journal}
  {\bibinfo  {journal} {Riv. Nuovo Cimento}\ }\textbf {\bibinfo {volume}
  {40}},\ \bibinfo {pages} {523} (\bibinfo {year} {2017})}\BibitemShut
  {NoStop}%
\bibitem [{\citenamefont {Sato}\ and\ \citenamefont
  {Ando}(2017)}]{Sato:ROPIP17}%
  \BibitemOpen
  \bibfield  {author} {\bibinfo {author} {\bibfnamefont {M.}~\bibnamefont
  {Sato}}\ and\ \bibinfo {author} {\bibfnamefont {Y.}~\bibnamefont {Ando}},\
  }\href {http://stacks.iop.org/0034-4885/80/i=7/a=076501} {\bibfield
  {journal} {\bibinfo  {journal} {Rep. Prog. Phys.}\ }\textbf {\bibinfo
  {volume} {80}},\ \bibinfo {pages} {076501} (\bibinfo {year}
  {2017})}\BibitemShut {NoStop}%
\bibitem [{\citenamefont {Salomaa}\ and\ \citenamefont
  {Volovik}(1988)}]{Salomaa:PRB88}%
  \BibitemOpen
  \bibfield  {author} {\bibinfo {author} {\bibfnamefont {M.}~\bibnamefont
  {Salomaa}}\ and\ \bibinfo {author} {\bibfnamefont {G.}~\bibnamefont
  {Volovik}},\ }\href {https://doi.org/10.1103/PhysRevB.37.9298} {\bibfield
  {journal} {\bibinfo  {journal} {Phys. Rev. B}\ }\textbf {\bibinfo {volume}
  {37}},\ \bibinfo {pages} {9298} (\bibinfo {year} {1988})}\BibitemShut
  {NoStop}%
\bibitem [{\citenamefont {Volovik}\ and\ \citenamefont
  {Volovik}(2009)}]{Volovik:09}%
  \BibitemOpen
  \bibfield  {author} {\bibinfo {author} {\bibfnamefont {G.~E.}\ \bibnamefont
  {Volovik}}\ and\ \bibinfo {author} {\bibfnamefont {G.}~\bibnamefont
  {Volovik}},\ }\href@noop {} {\emph {\bibinfo {title} {The universe in a
  helium droplet}}},\ Vol.\ \bibinfo {volume} {117}\ (\bibinfo  {publisher}
  {Oxford University Press New York},\ \bibinfo {year} {2009})\BibitemShut
  {NoStop}%
\bibitem [{\citenamefont {Read}\ and\ \citenamefont
  {Green}(2000)}]{Read:PRB00}%
  \BibitemOpen
  \bibfield  {author} {\bibinfo {author} {\bibfnamefont {N.}~\bibnamefont
  {Read}}\ and\ \bibinfo {author} {\bibfnamefont {D.}~\bibnamefont {Green}},\
  }\href {\doibase 10.1103/PhysRevB.61.10267} {\bibfield  {journal} {\bibinfo
  {journal} {Phys. Rev. B}\ }\textbf {\bibinfo {volume} {61}},\ \bibinfo
  {pages} {10267} (\bibinfo {year} {2000})}\BibitemShut {NoStop}%
\bibitem [{\citenamefont {Kitaev}(2001)}]{Kitaev:PU01}%
  \BibitemOpen
  \bibfield  {author} {\bibinfo {author} {\bibfnamefont {A.~Y.}\ \bibnamefont
  {Kitaev}},\ }\href {http://stacks.iop.org/1063-7869/44/i=10S/a=S29}
  {\bibfield  {journal} {\bibinfo  {journal} {Phys. Usp.}\ }\textbf {\bibinfo
  {volume} {44}},\ \bibinfo {pages} {131} (\bibinfo {year} {2001})}\BibitemShut
  {NoStop}%
\bibitem [{\citenamefont {Sato}\ and\ \citenamefont
  {Fujimoto}(2009)}]{Sato:PRB09}%
  \BibitemOpen
  \bibfield  {author} {\bibinfo {author} {\bibfnamefont {M.}~\bibnamefont
  {Sato}}\ and\ \bibinfo {author} {\bibfnamefont {S.}~\bibnamefont
  {Fujimoto}},\ }\href {\doibase 10.1103/PhysRevB.79.094504} {\bibfield
  {journal} {\bibinfo  {journal} {Phys. Rev. B}\ }\textbf {\bibinfo {volume}
  {79}},\ \bibinfo {pages} {094504} (\bibinfo {year} {2009})}\BibitemShut
  {NoStop}%
\bibitem [{\citenamefont {Majorana}(1937)}]{Majorana:INC137}%
  \BibitemOpen
  \bibfield  {author} {\bibinfo {author} {\bibfnamefont {E.}~\bibnamefont
  {Majorana}},\ }\href {https://doi.org/10.1007/BF02961314} {\bibfield
  {journal} {\bibinfo  {journal} {Il Nuovo Cimento}\ }\textbf {\bibinfo
  {volume} {14}},\ \bibinfo {pages} {171} (\bibinfo {year} {1937})}\BibitemShut
  {NoStop}%
\bibitem [{\citenamefont {Nishida}\ \emph {et~al.}(2010)\citenamefont
  {Nishida}, \citenamefont {Santos},\ and\ \citenamefont
  {Chamon}}]{Nishida:PRB10}%
  \BibitemOpen
  \bibfield  {author} {\bibinfo {author} {\bibfnamefont {Y.}~\bibnamefont
  {Nishida}}, \bibinfo {author} {\bibfnamefont {L.}~\bibnamefont {Santos}}, \
  and\ \bibinfo {author} {\bibfnamefont {C.}~\bibnamefont {Chamon}},\ }\href
  {\doibase 10.1103/PhysRevB.82.144513} {\bibfield  {journal} {\bibinfo
  {journal} {Phys. Rev. B}\ }\textbf {\bibinfo {volume} {82}},\ \bibinfo
  {pages} {144513} (\bibinfo {year} {2010})}\BibitemShut {NoStop}%
\bibitem [{\citenamefont {Jackiw}\ and\ \citenamefont
  {Rossi}(1981)}]{Jackiw:NPB81}%
  \BibitemOpen
  \bibfield  {author} {\bibinfo {author} {\bibfnamefont {R.}~\bibnamefont
  {Jackiw}}\ and\ \bibinfo {author} {\bibfnamefont {P.}~\bibnamefont {Rossi}},\
  }\href {\doibase 10.1016/0550-3213(81)90044-4} {\bibfield  {journal}
  {\bibinfo  {journal} {Nucl. Phys. B}\ }\textbf {\bibinfo {volume} {190}},\
  \bibinfo {pages} {681 } (\bibinfo {year} {1981})}\BibitemShut {NoStop}%
\bibitem [{\citenamefont {Fukui}\ \emph {et~al.}(2012)\citenamefont {Fukui},
  \citenamefont {Shiozaki}, \citenamefont {Fujiwara},\ and\ \citenamefont
  {Fujimoto}}]{Fukui:JOTPSOJ12}%
  \BibitemOpen
  \bibfield  {author} {\bibinfo {author} {\bibfnamefont {T.}~\bibnamefont
  {Fukui}}, \bibinfo {author} {\bibfnamefont {K.}~\bibnamefont {Shiozaki}},
  \bibinfo {author} {\bibfnamefont {T.}~\bibnamefont {Fujiwara}}, \ and\
  \bibinfo {author} {\bibfnamefont {S.}~\bibnamefont {Fujimoto}},\ }\href
  {\doibase 10.1143/JPSJ.81.114602} {\bibfield  {journal} {\bibinfo  {journal}
  {J. Phys. Soc. Jpn.}\ }\textbf {\bibinfo {volume} {81}},\ \bibinfo {pages}
  {114602} (\bibinfo {year} {2012})}\BibitemShut {NoStop}%
\bibitem [{\citenamefont {Kitaev}(2003)}]{Kitaev:AOP03}%
  \BibitemOpen
  \bibfield  {author} {\bibinfo {author} {\bibfnamefont {A.}~\bibnamefont
  {Kitaev}},\ }\href {\doibase https://doi.org/10.1016/S0003-4916(02)00018-0}
  {\bibfield  {journal} {\bibinfo  {journal} {Ann. Phys.}\ }\textbf {\bibinfo
  {volume} {303}},\ \bibinfo {pages} {2 } (\bibinfo {year} {2003})}\BibitemShut
  {NoStop}%
\bibitem [{\citenamefont {Nayak}\ \emph {et~al.}(2008)\citenamefont {Nayak},
  \citenamefont {Simon}, \citenamefont {Stern}, \citenamefont {Freedman},\ and\
  \citenamefont {Das~Sarma}}]{Nayak:RMP08}%
  \BibitemOpen
  \bibfield  {author} {\bibinfo {author} {\bibfnamefont {C.}~\bibnamefont
  {Nayak}}, \bibinfo {author} {\bibfnamefont {S.~H.}\ \bibnamefont {Simon}},
  \bibinfo {author} {\bibfnamefont {A.}~\bibnamefont {Stern}}, \bibinfo
  {author} {\bibfnamefont {M.}~\bibnamefont {Freedman}}, \ and\ \bibinfo
  {author} {\bibfnamefont {S.}~\bibnamefont {Das~Sarma}},\ }\href {\doibase
  10.1103/RevModPhys.80.1083} {\bibfield  {journal} {\bibinfo  {journal} {Rev.
  Mod. Phys.}\ }\textbf {\bibinfo {volume} {80}},\ \bibinfo {pages} {1083}
  (\bibinfo {year} {2008})}\BibitemShut {NoStop}%
\bibitem [{\citenamefont {Fu}\ and\ \citenamefont {Kane}(2008)}]{Fu:PRL08}%
  \BibitemOpen
  \bibfield  {author} {\bibinfo {author} {\bibfnamefont {L.}~\bibnamefont
  {Fu}}\ and\ \bibinfo {author} {\bibfnamefont {C.~L.}\ \bibnamefont {Kane}},\
  }\href {\doibase 10.1103/PhysRevLett.100.096407} {\bibfield  {journal}
  {\bibinfo  {journal} {Phys. Rev. Lett.}\ }\textbf {\bibinfo {volume} {100}},\
  \bibinfo {pages} {096407} (\bibinfo {year} {2008})}\BibitemShut {NoStop}%
\bibitem [{\citenamefont {Lutchyn}\ \emph {et~al.}(2010)\citenamefont
  {Lutchyn}, \citenamefont {Sau},\ and\ \citenamefont
  {Das~Sarma}}]{Lutchyn:PRL10}%
  \BibitemOpen
  \bibfield  {author} {\bibinfo {author} {\bibfnamefont {R.~M.}\ \bibnamefont
  {Lutchyn}}, \bibinfo {author} {\bibfnamefont {J.~D.}\ \bibnamefont {Sau}}, \
  and\ \bibinfo {author} {\bibfnamefont {S.}~\bibnamefont {Das~Sarma}},\ }\href
  {\doibase 10.1103/PhysRevLett.105.077001} {\bibfield  {journal} {\bibinfo
  {journal} {Phys. Rev. Lett.}\ }\textbf {\bibinfo {volume} {105}},\ \bibinfo
  {pages} {077001} (\bibinfo {year} {2010})}\BibitemShut {NoStop}%
\bibitem [{\citenamefont {Oreg}\ \emph {et~al.}(2010)\citenamefont {Oreg},
  \citenamefont {Refael},\ and\ \citenamefont {von Oppen}}]{Oreg:PRL10}%
  \BibitemOpen
  \bibfield  {author} {\bibinfo {author} {\bibfnamefont {Y.}~\bibnamefont
  {Oreg}}, \bibinfo {author} {\bibfnamefont {G.}~\bibnamefont {Refael}}, \ and\
  \bibinfo {author} {\bibfnamefont {F.}~\bibnamefont {von Oppen}},\ }\href
  {\doibase 10.1103/PhysRevLett.105.177002} {\bibfield  {journal} {\bibinfo
  {journal} {Phys. Rev. Lett.}\ }\textbf {\bibinfo {volume} {105}},\ \bibinfo
  {pages} {177002} (\bibinfo {year} {2010})}\BibitemShut {NoStop}%
\bibitem [{\citenamefont {St\ifmmode~\check{r}\else \v{r}\fi{}eda}\ and\
  \citenamefont {\ifmmode~\check{S}\else \v{S}\fi{}eba}(2003)}]{Streda:PRL03}%
  \BibitemOpen
  \bibfield  {author} {\bibinfo {author} {\bibfnamefont {P.}~\bibnamefont
  {St\ifmmode~\check{r}\else \v{r}\fi{}eda}}\ and\ \bibinfo {author}
  {\bibfnamefont {P.}~\bibnamefont {\ifmmode~\check{S}\else \v{S}\fi{}eba}},\
  }\href {https://doi.org/10.1103/PhysRevLett.90.256601} {\bibfield  {journal}
  {\bibinfo  {journal} {Phys. Rev. Lett.}\ }\textbf {\bibinfo {volume} {90}},\
  \bibinfo {pages} {256601} (\bibinfo {year} {2003})}\BibitemShut {NoStop}%
\bibitem [{\citenamefont {Stanescu}\ and\ \citenamefont
  {Tewari}(2013{\natexlab{a}})}]{Stanescu:JOPCM13}%
  \BibitemOpen
  \bibfield  {author} {\bibinfo {author} {\bibfnamefont {T.~D.}\ \bibnamefont
  {Stanescu}}\ and\ \bibinfo {author} {\bibfnamefont {S.}~\bibnamefont
  {Tewari}},\ }\href {http://stacks.iop.org/0953-8984/25/i=23/a=233201}
  {\bibfield  {journal} {\bibinfo  {journal} {J. Phys.: Condens. Matter}\
  }\textbf {\bibinfo {volume} {25}},\ \bibinfo {pages} {233201} (\bibinfo
  {year} {2013}{\natexlab{a}})}\BibitemShut {NoStop}%
\bibitem [{\citenamefont {Lutchyn}\ \emph {et~al.}(2018)\citenamefont
  {Lutchyn}, \citenamefont {Bakkers}, \citenamefont {Kouwenhoven},
  \citenamefont {Krogstrup}, \citenamefont {Marcus},\ and\ \citenamefont
  {Oreg}}]{Lutchyn:NRM18}%
  \BibitemOpen
  \bibfield  {author} {\bibinfo {author} {\bibfnamefont {R.~M.}\ \bibnamefont
  {Lutchyn}}, \bibinfo {author} {\bibfnamefont {E.~P. A.~M.}\ \bibnamefont
  {Bakkers}}, \bibinfo {author} {\bibfnamefont {L.~P.}\ \bibnamefont
  {Kouwenhoven}}, \bibinfo {author} {\bibfnamefont {P.}~\bibnamefont
  {Krogstrup}}, \bibinfo {author} {\bibfnamefont {C.~M.}\ \bibnamefont
  {Marcus}}, \ and\ \bibinfo {author} {\bibfnamefont {Y.}~\bibnamefont
  {Oreg}},\ }\href {\doibase 10.1038/s41578-018-0003-1} {\bibfield  {journal}
  {\bibinfo  {journal} {Nat. Rev. Mater.}\ }\textbf {\bibinfo {volume} {3}},\
  \bibinfo {pages} {52} (\bibinfo {year} {2018})}\BibitemShut {NoStop}%
\bibitem [{\citenamefont {Mourik}\ \emph {et~al.}(2012)\citenamefont {Mourik},
  \citenamefont {Zuo}, \citenamefont {Frolov}, \citenamefont {Plissard},
  \citenamefont {Bakkers},\ and\ \citenamefont {Kouwenhoven}}]{Mourik:S12}%
  \BibitemOpen
  \bibfield  {author} {\bibinfo {author} {\bibfnamefont {V.}~\bibnamefont
  {Mourik}}, \bibinfo {author} {\bibfnamefont {K.}~\bibnamefont {Zuo}},
  \bibinfo {author} {\bibfnamefont {S.~M.}\ \bibnamefont {Frolov}}, \bibinfo
  {author} {\bibfnamefont {S.~R.}\ \bibnamefont {Plissard}}, \bibinfo {author}
  {\bibfnamefont {E.~P. A.~M.}\ \bibnamefont {Bakkers}}, \ and\ \bibinfo
  {author} {\bibfnamefont {L.~P.}\ \bibnamefont {Kouwenhoven}},\ }\href
  {\doibase 10.1126/science.1222360} {\bibfield  {journal} {\bibinfo  {journal}
  {Science}\ }\textbf {\bibinfo {volume} {336}},\ \bibinfo {pages} {1003}
  (\bibinfo {year} {2012})}\BibitemShut {NoStop}%
\bibitem [{\citenamefont {van Woerkom}\ \emph {et~al.}(2017)\citenamefont {van
  Woerkom}, \citenamefont {Proutski}, \citenamefont {van Heck}, \citenamefont
  {Bouman}, \citenamefont {V{\"a}yrynen}, \citenamefont {Glazman},
  \citenamefont {Krogstrup}, \citenamefont {Nyg{\aa}rd}, \citenamefont
  {Kouwenhoven},\ and\ \citenamefont {Geresdi}}]{Woerkom:NP17}%
  \BibitemOpen
  \bibfield  {author} {\bibinfo {author} {\bibfnamefont {D.~J.}\ \bibnamefont
  {van Woerkom}}, \bibinfo {author} {\bibfnamefont {A.}~\bibnamefont
  {Proutski}}, \bibinfo {author} {\bibfnamefont {B.}~\bibnamefont {van Heck}},
  \bibinfo {author} {\bibfnamefont {D.}~\bibnamefont {Bouman}}, \bibinfo
  {author} {\bibfnamefont {J.~I.}\ \bibnamefont {V{\"a}yrynen}}, \bibinfo
  {author} {\bibfnamefont {L.~I.}\ \bibnamefont {Glazman}}, \bibinfo {author}
  {\bibfnamefont {P.}~\bibnamefont {Krogstrup}}, \bibinfo {author}
  {\bibfnamefont {J.}~\bibnamefont {Nyg{\aa}rd}}, \bibinfo {author}
  {\bibfnamefont {L.~P.}\ \bibnamefont {Kouwenhoven}}, \ and\ \bibinfo {author}
  {\bibfnamefont {A.}~\bibnamefont {Geresdi}},\ }\href
  {http://dx.doi.org/10.1038/nphys4150} {\bibfield  {journal} {\bibinfo
  {journal} {Nat. Phys.}\ }\textbf {\bibinfo {volume} {13}},\ \bibinfo {pages}
  {876 EP } (\bibinfo {year} {2017})}\BibitemShut {NoStop}%
\bibitem [{\citenamefont {Tosi}\ \emph {et~al.}(2019)\citenamefont {Tosi},
  \citenamefont {Metzger}, \citenamefont {Goffman}, \citenamefont {Urbina},
  \citenamefont {Pothier}, \citenamefont {Park}, \citenamefont {Levy~Yeyati},
  \citenamefont {Nyg\aa{}rd},\ and\ \citenamefont {Krogstrup}}]{Tosi:PRX19}%
  \BibitemOpen
  \bibfield  {author} {\bibinfo {author} {\bibfnamefont {L.}~\bibnamefont
  {Tosi}}, \bibinfo {author} {\bibfnamefont {C.}~\bibnamefont {Metzger}},
  \bibinfo {author} {\bibfnamefont {M.~F.}\ \bibnamefont {Goffman}}, \bibinfo
  {author} {\bibfnamefont {C.}~\bibnamefont {Urbina}}, \bibinfo {author}
  {\bibfnamefont {H.}~\bibnamefont {Pothier}}, \bibinfo {author} {\bibfnamefont
  {S.}~\bibnamefont {Park}}, \bibinfo {author} {\bibfnamefont {A.}~\bibnamefont
  {Levy~Yeyati}}, \bibinfo {author} {\bibfnamefont {J.}~\bibnamefont
  {Nyg\aa{}rd}}, \ and\ \bibinfo {author} {\bibfnamefont {P.}~\bibnamefont
  {Krogstrup}},\ }\href {\doibase 10.1103/PhysRevX.9.011010} {\bibfield
  {journal} {\bibinfo  {journal} {Phys. Rev. X}\ }\textbf {\bibinfo {volume}
  {9}},\ \bibinfo {pages} {011010} (\bibinfo {year} {2019})}\BibitemShut
  {NoStop}%
\bibitem [{\citenamefont {Lee}\ \emph {et~al.}(2012)\citenamefont {Lee},
  \citenamefont {Jiang}, \citenamefont {Aguado}, \citenamefont {Katsaros},
  \citenamefont {Lieber},\ and\ \citenamefont {De~Franceschi}}]{Lee:PRL12}%
  \BibitemOpen
  \bibfield  {author} {\bibinfo {author} {\bibfnamefont {E.~J.~H.}\
  \bibnamefont {Lee}}, \bibinfo {author} {\bibfnamefont {X.}~\bibnamefont
  {Jiang}}, \bibinfo {author} {\bibfnamefont {R.}~\bibnamefont {Aguado}},
  \bibinfo {author} {\bibfnamefont {G.}~\bibnamefont {Katsaros}}, \bibinfo
  {author} {\bibfnamefont {C.~M.}\ \bibnamefont {Lieber}}, \ and\ \bibinfo
  {author} {\bibfnamefont {S.}~\bibnamefont {De~Franceschi}},\ }\href {\doibase
  10.1103/PhysRevLett.109.186802} {\bibfield  {journal} {\bibinfo  {journal}
  {Phys. Rev. Lett.}\ }\textbf {\bibinfo {volume} {109}},\ \bibinfo {pages}
  {186802} (\bibinfo {year} {2012})}\BibitemShut {NoStop}%
\bibitem [{\citenamefont {Lee}\ \emph {et~al.}(2014)\citenamefont {Lee},
  \citenamefont {Jiang}, \citenamefont {Houzet}, \citenamefont {Aguado},
  \citenamefont {Lieber},\ and\ \citenamefont {De~Franceschi}}]{Lee:NN14}%
  \BibitemOpen
  \bibfield  {author} {\bibinfo {author} {\bibfnamefont {E.~J.~H.}\
  \bibnamefont {Lee}}, \bibinfo {author} {\bibfnamefont {X.}~\bibnamefont
  {Jiang}}, \bibinfo {author} {\bibfnamefont {M.}~\bibnamefont {Houzet}},
  \bibinfo {author} {\bibfnamefont {R.}~\bibnamefont {Aguado}}, \bibinfo
  {author} {\bibfnamefont {C.~M.}\ \bibnamefont {Lieber}}, \ and\ \bibinfo
  {author} {\bibfnamefont {S.}~\bibnamefont {De~Franceschi}},\ }\href
  {http://dx.doi.org/10.1038/nnano.2013.267} {\bibfield  {journal} {\bibinfo
  {journal} {Nat. Nanotechnol.}\ }\textbf {\bibinfo {volume} {9}},\ \bibinfo
  {pages} {79} (\bibinfo {year} {2014})}\BibitemShut {NoStop}%
\bibitem [{\citenamefont {Lee}\ \emph {et~al.}(2017{\natexlab{a}})\citenamefont
  {Lee}, \citenamefont {Jiang}, \citenamefont {\ifmmode~\check{Z}\else
  \v{Z}\fi{}itko}, \citenamefont {Aguado}, \citenamefont {Lieber},\ and\
  \citenamefont {De~Franceschi}}]{Lee:PRB17}%
  \BibitemOpen
  \bibfield  {author} {\bibinfo {author} {\bibfnamefont {E.~J.~H.}\
  \bibnamefont {Lee}}, \bibinfo {author} {\bibfnamefont {X.}~\bibnamefont
  {Jiang}}, \bibinfo {author} {\bibfnamefont {R.}~\bibnamefont
  {\ifmmode~\check{Z}\else \v{Z}\fi{}itko}}, \bibinfo {author} {\bibfnamefont
  {R.}~\bibnamefont {Aguado}}, \bibinfo {author} {\bibfnamefont {C.~M.}\
  \bibnamefont {Lieber}}, \ and\ \bibinfo {author} {\bibfnamefont
  {S.}~\bibnamefont {De~Franceschi}},\ }\href {\doibase
  10.1103/PhysRevB.95.180502} {\bibfield  {journal} {\bibinfo  {journal} {Phys.
  Rev. B}\ }\textbf {\bibinfo {volume} {95}},\ \bibinfo {pages} {180502}
  (\bibinfo {year} {2017}{\natexlab{a}})}\BibitemShut {NoStop}%
\bibitem [{\citenamefont {Grove-Rasmussen}\ \emph {et~al.}(2018)\citenamefont
  {Grove-Rasmussen}, \citenamefont {Steffensen}, \citenamefont {Jellinggaard},
  \citenamefont {Madsen}, \citenamefont {{\v Z}itko}, \citenamefont {Paaske},\
  and\ \citenamefont {Nyg{\aa}rd}}]{Grove-Rasmussen:NC18}%
  \BibitemOpen
  \bibfield  {author} {\bibinfo {author} {\bibfnamefont {K.}~\bibnamefont
  {Grove-Rasmussen}}, \bibinfo {author} {\bibfnamefont {G.}~\bibnamefont
  {Steffensen}}, \bibinfo {author} {\bibfnamefont {A.}~\bibnamefont
  {Jellinggaard}}, \bibinfo {author} {\bibfnamefont {M.~H.}\ \bibnamefont
  {Madsen}}, \bibinfo {author} {\bibfnamefont {R.}~\bibnamefont {{\v Z}itko}},
  \bibinfo {author} {\bibfnamefont {J.}~\bibnamefont {Paaske}}, \ and\ \bibinfo
  {author} {\bibfnamefont {J.}~\bibnamefont {Nyg{\aa}rd}},\ }\href {\doibase
  10.1038/s41467-018-04683-x} {\bibfield  {journal} {\bibinfo  {journal} {Nat.
  Commun.}\ }\textbf {\bibinfo {volume} {9}},\ \bibinfo {pages} {2376}
  (\bibinfo {year} {2018})}\BibitemShut {NoStop}%
\bibitem [{\citenamefont {Su}\ \emph {et~al.}(2018)\citenamefont {Su},
  \citenamefont {Zarassi}, \citenamefont {Hsu}, \citenamefont {San-Jose},
  \citenamefont {Prada}, \citenamefont {Aguado}, \citenamefont {Lee},
  \citenamefont {Gazibegovic}, \citenamefont {Op~het Veld}, \citenamefont
  {Car}, \citenamefont {Plissard}, \citenamefont {Hocevar}, \citenamefont
  {Pendharkar}, \citenamefont {Lee}, \citenamefont {Logan}, \citenamefont
  {Palmstr\o{}m}, \citenamefont {Bakkers},\ and\ \citenamefont
  {Frolov}}]{Su:PRL18}%
  \BibitemOpen
  \bibfield  {author} {\bibinfo {author} {\bibfnamefont {Z.}~\bibnamefont
  {Su}}, \bibinfo {author} {\bibfnamefont {A.}~\bibnamefont {Zarassi}},
  \bibinfo {author} {\bibfnamefont {J.-F.}\ \bibnamefont {Hsu}}, \bibinfo
  {author} {\bibfnamefont {P.}~\bibnamefont {San-Jose}}, \bibinfo {author}
  {\bibfnamefont {E.}~\bibnamefont {Prada}}, \bibinfo {author} {\bibfnamefont
  {R.}~\bibnamefont {Aguado}}, \bibinfo {author} {\bibfnamefont {E.~J.~H.}\
  \bibnamefont {Lee}}, \bibinfo {author} {\bibfnamefont {S.}~\bibnamefont
  {Gazibegovic}}, \bibinfo {author} {\bibfnamefont {R.~L.~M.}\ \bibnamefont
  {Op~het Veld}}, \bibinfo {author} {\bibfnamefont {D.}~\bibnamefont {Car}},
  \bibinfo {author} {\bibfnamefont {S.~R.}\ \bibnamefont {Plissard}}, \bibinfo
  {author} {\bibfnamefont {M.}~\bibnamefont {Hocevar}}, \bibinfo {author}
  {\bibfnamefont {M.}~\bibnamefont {Pendharkar}}, \bibinfo {author}
  {\bibfnamefont {J.~S.}\ \bibnamefont {Lee}}, \bibinfo {author} {\bibfnamefont
  {J.~A.}\ \bibnamefont {Logan}}, \bibinfo {author} {\bibfnamefont {C.~J.}\
  \bibnamefont {Palmstr\o{}m}}, \bibinfo {author} {\bibfnamefont {E.~P. A.~M.}\
  \bibnamefont {Bakkers}}, \ and\ \bibinfo {author} {\bibfnamefont {S.~M.}\
  \bibnamefont {Frolov}},\ }\href {\doibase 10.1103/PhysRevLett.121.127705}
  {\bibfield  {journal} {\bibinfo  {journal} {Phys. Rev. Lett.}\ }\textbf
  {\bibinfo {volume} {121}},\ \bibinfo {pages} {127705} (\bibinfo {year}
  {2018})}\BibitemShut {NoStop}%
\bibitem [{\citenamefont {J{\"u}nger}\ \emph {et~al.}(2019)\citenamefont
  {J{\"u}nger}, \citenamefont {Baumgartner}, \citenamefont {Delagrange},
  \citenamefont {Chevallier}, \citenamefont {Lehmann}, \citenamefont {Nilsson},
  \citenamefont {Dick}, \citenamefont {Thelander},\ and\ \citenamefont
  {Sch{\"o}nenberger}}]{Junger:CP19}%
  \BibitemOpen
  \bibfield  {author} {\bibinfo {author} {\bibfnamefont {C.}~\bibnamefont
  {J{\"u}nger}}, \bibinfo {author} {\bibfnamefont {A.}~\bibnamefont
  {Baumgartner}}, \bibinfo {author} {\bibfnamefont {R.}~\bibnamefont
  {Delagrange}}, \bibinfo {author} {\bibfnamefont {D.}~\bibnamefont
  {Chevallier}}, \bibinfo {author} {\bibfnamefont {S.}~\bibnamefont {Lehmann}},
  \bibinfo {author} {\bibfnamefont {M.}~\bibnamefont {Nilsson}}, \bibinfo
  {author} {\bibfnamefont {K.~A.}\ \bibnamefont {Dick}}, \bibinfo {author}
  {\bibfnamefont {C.}~\bibnamefont {Thelander}}, \ and\ \bibinfo {author}
  {\bibfnamefont {C.}~\bibnamefont {Sch{\"o}nenberger}},\ }\href
  {https://doi.org/10.1038/s42005-019-0162-4} {\bibfield  {journal} {\bibinfo
  {journal} {Commun. Phys.}\ }\textbf {\bibinfo {volume} {2}},\ \bibinfo
  {pages} {76} (\bibinfo {year} {2019})}\BibitemShut {NoStop}%
\bibitem [{\citenamefont {Potter}\ and\ \citenamefont
  {Lee}(2010)}]{Potter:PRL10}%
  \BibitemOpen
  \bibfield  {author} {\bibinfo {author} {\bibfnamefont {A.~C.}\ \bibnamefont
  {Potter}}\ and\ \bibinfo {author} {\bibfnamefont {P.~A.}\ \bibnamefont
  {Lee}},\ }\href {\doibase 10.1103/PhysRevLett.105.227003} {\bibfield
  {journal} {\bibinfo  {journal} {Phys. Rev. Lett.}\ }\textbf {\bibinfo
  {volume} {105}},\ \bibinfo {pages} {227003} (\bibinfo {year}
  {2010})}\BibitemShut {NoStop}%
\bibitem [{\citenamefont {Potter}\ and\ \citenamefont
  {Lee}(2011)}]{Potter:PRB11}%
  \BibitemOpen
  \bibfield  {author} {\bibinfo {author} {\bibfnamefont {A.~C.}\ \bibnamefont
  {Potter}}\ and\ \bibinfo {author} {\bibfnamefont {P.~A.}\ \bibnamefont
  {Lee}},\ }\href {\doibase 10.1103/PhysRevB.83.094525} {\bibfield  {journal}
  {\bibinfo  {journal} {Phys. Rev. B}\ }\textbf {\bibinfo {volume} {83}},\
  \bibinfo {pages} {094525} (\bibinfo {year} {2011})}\BibitemShut {NoStop}%
\bibitem [{\citenamefont {Lutchyn}\ \emph {et~al.}(2011)\citenamefont
  {Lutchyn}, \citenamefont {Stanescu},\ and\ \citenamefont
  {Das~Sarma}}]{Lutchyn:PRL11}%
  \BibitemOpen
  \bibfield  {author} {\bibinfo {author} {\bibfnamefont {R.~M.}\ \bibnamefont
  {Lutchyn}}, \bibinfo {author} {\bibfnamefont {T.~D.}\ \bibnamefont
  {Stanescu}}, \ and\ \bibinfo {author} {\bibfnamefont {S.}~\bibnamefont
  {Das~Sarma}},\ }\href {\doibase 10.1103/PhysRevLett.106.127001} {\bibfield
  {journal} {\bibinfo  {journal} {Phys. Rev. Lett.}\ }\textbf {\bibinfo
  {volume} {106}},\ \bibinfo {pages} {127001} (\bibinfo {year}
  {2011})}\BibitemShut {NoStop}%
\bibitem [{\citenamefont {Lutchyn}\ and\ \citenamefont
  {Fisher}(2011)}]{Lutchyn:PRB11}%
  \BibitemOpen
  \bibfield  {author} {\bibinfo {author} {\bibfnamefont {R.~M.}\ \bibnamefont
  {Lutchyn}}\ and\ \bibinfo {author} {\bibfnamefont {M.~P.~A.}\ \bibnamefont
  {Fisher}},\ }\href {\doibase 10.1103/PhysRevB.84.214528} {\bibfield
  {journal} {\bibinfo  {journal} {Phys. Rev. B}\ }\textbf {\bibinfo {volume}
  {84}},\ \bibinfo {pages} {214528} (\bibinfo {year} {2011})}\BibitemShut
  {NoStop}%
\bibitem [{\citenamefont {Cole}\ \emph {et~al.}(2015)\citenamefont {Cole},
  \citenamefont {Das~Sarma},\ and\ \citenamefont {Stanescu}}]{Cole:PRB15}%
  \BibitemOpen
  \bibfield  {author} {\bibinfo {author} {\bibfnamefont {W.~S.}\ \bibnamefont
  {Cole}}, \bibinfo {author} {\bibfnamefont {S.}~\bibnamefont {Das~Sarma}}, \
  and\ \bibinfo {author} {\bibfnamefont {T.~D.}\ \bibnamefont {Stanescu}},\
  }\href {\doibase 10.1103/PhysRevB.92.174511} {\bibfield  {journal} {\bibinfo
  {journal} {Phys. Rev. B}\ }\textbf {\bibinfo {volume} {92}},\ \bibinfo
  {pages} {174511} (\bibinfo {year} {2015})}\BibitemShut {NoStop}%
\bibitem [{\citenamefont {Reeg}\ \emph {et~al.}(2017)\citenamefont {Reeg},
  \citenamefont {Loss},\ and\ \citenamefont {Klinovaja}}]{Reeg:PRB17a}%
  \BibitemOpen
  \bibfield  {author} {\bibinfo {author} {\bibfnamefont {C.}~\bibnamefont
  {Reeg}}, \bibinfo {author} {\bibfnamefont {D.}~\bibnamefont {Loss}}, \ and\
  \bibinfo {author} {\bibfnamefont {J.}~\bibnamefont {Klinovaja}},\ }\href
  {\doibase 10.1103/PhysRevB.96.125426} {\bibfield  {journal} {\bibinfo
  {journal} {Phys. Rev. B}\ }\textbf {\bibinfo {volume} {96}},\ \bibinfo
  {pages} {125426} (\bibinfo {year} {2017})}\BibitemShut {NoStop}%
\bibitem [{\citenamefont {Reeg}\ \emph
  {et~al.}(2018{\natexlab{a}})\citenamefont {Reeg}, \citenamefont {Loss},\ and\
  \citenamefont {Klinovaja}}]{Reeg:PRB18}%
  \BibitemOpen
  \bibfield  {author} {\bibinfo {author} {\bibfnamefont {C.}~\bibnamefont
  {Reeg}}, \bibinfo {author} {\bibfnamefont {D.}~\bibnamefont {Loss}}, \ and\
  \bibinfo {author} {\bibfnamefont {J.}~\bibnamefont {Klinovaja}},\ }\href
  {\doibase 10.1103/PhysRevB.97.165425} {\bibfield  {journal} {\bibinfo
  {journal} {Phys. Rev. B}\ }\textbf {\bibinfo {volume} {97}},\ \bibinfo
  {pages} {165425} (\bibinfo {year} {2018}{\natexlab{a}})}\BibitemShut
  {NoStop}%
\bibitem [{\citenamefont {Vaitiekenas}\ \emph {et~al.}(2018)\citenamefont
  {Vaitiekenas}, \citenamefont {Deng}, \citenamefont {Nyg\aa{}rd},
  \citenamefont {Krogstrup},\ and\ \citenamefont {Marcus}}]{Vaitiekenas:PRL18}%
  \BibitemOpen
  \bibfield  {author} {\bibinfo {author} {\bibfnamefont {S.}~\bibnamefont
  {Vaitiekenas}}, \bibinfo {author} {\bibfnamefont {M.-T.}\ \bibnamefont
  {Deng}}, \bibinfo {author} {\bibfnamefont {J.}~\bibnamefont {Nyg\aa{}rd}},
  \bibinfo {author} {\bibfnamefont {P.}~\bibnamefont {Krogstrup}}, \ and\
  \bibinfo {author} {\bibfnamefont {C.~M.}\ \bibnamefont {Marcus}},\ }\href
  {\doibase 10.1103/PhysRevLett.121.037703} {\bibfield  {journal} {\bibinfo
  {journal} {Phys. Rev. Lett.}\ }\textbf {\bibinfo {volume} {121}},\ \bibinfo
  {pages} {037703} (\bibinfo {year} {2018})}\BibitemShut {NoStop}%
\bibitem [{\citenamefont {Antipov}\ \emph {et~al.}(2018)\citenamefont
  {Antipov}, \citenamefont {Bargerbos}, \citenamefont {Winkler}, \citenamefont
  {Bauer}, \citenamefont {Rossi},\ and\ \citenamefont
  {Lutchyn}}]{Antipov:PRX18}%
  \BibitemOpen
  \bibfield  {author} {\bibinfo {author} {\bibfnamefont {A.~E.}\ \bibnamefont
  {Antipov}}, \bibinfo {author} {\bibfnamefont {A.}~\bibnamefont {Bargerbos}},
  \bibinfo {author} {\bibfnamefont {G.~W.}\ \bibnamefont {Winkler}}, \bibinfo
  {author} {\bibfnamefont {B.}~\bibnamefont {Bauer}}, \bibinfo {author}
  {\bibfnamefont {E.}~\bibnamefont {Rossi}}, \ and\ \bibinfo {author}
  {\bibfnamefont {R.~M.}\ \bibnamefont {Lutchyn}},\ }\href {\doibase
  10.1103/PhysRevX.8.031041} {\bibfield  {journal} {\bibinfo  {journal} {Phys.
  Rev. X}\ }\textbf {\bibinfo {volume} {8}},\ \bibinfo {pages} {031041}
  (\bibinfo {year} {2018})}\BibitemShut {NoStop}%
\bibitem [{\citenamefont {Dmytruk}\ \emph {et~al.}(2018)\citenamefont
  {Dmytruk}, \citenamefont {Chevallier}, \citenamefont {Loss},\ and\
  \citenamefont {Klinovaja}}]{Dmytruk:PRB18}%
  \BibitemOpen
  \bibfield  {author} {\bibinfo {author} {\bibfnamefont {O.}~\bibnamefont
  {Dmytruk}}, \bibinfo {author} {\bibfnamefont {D.}~\bibnamefont {Chevallier}},
  \bibinfo {author} {\bibfnamefont {D.}~\bibnamefont {Loss}}, \ and\ \bibinfo
  {author} {\bibfnamefont {J.}~\bibnamefont {Klinovaja}},\ }\href {\doibase
  10.1103/PhysRevB.98.165403} {\bibfield  {journal} {\bibinfo  {journal} {Phys.
  Rev. B}\ }\textbf {\bibinfo {volume} {98}},\ \bibinfo {pages} {165403}
  (\bibinfo {year} {2018})}\BibitemShut {NoStop}%
\bibitem [{\citenamefont {Pan}\ \emph {et~al.}(2019)\citenamefont {Pan},
  \citenamefont {Sau}, \citenamefont {Stanescu},\ and\ \citenamefont
  {Das~Sarma}}]{Pan:PRB19}%
  \BibitemOpen
  \bibfield  {author} {\bibinfo {author} {\bibfnamefont {H.}~\bibnamefont
  {Pan}}, \bibinfo {author} {\bibfnamefont {J.~D.}\ \bibnamefont {Sau}},
  \bibinfo {author} {\bibfnamefont {T.~D.}\ \bibnamefont {Stanescu}}, \ and\
  \bibinfo {author} {\bibfnamefont {S.}~\bibnamefont {Das~Sarma}},\ }\href
  {\doibase 10.1103/PhysRevB.99.054507} {\bibfield  {journal} {\bibinfo
  {journal} {Phys. Rev. B}\ }\textbf {\bibinfo {volume} {99}},\ \bibinfo
  {pages} {054507} (\bibinfo {year} {2019})}\BibitemShut {NoStop}%
\bibitem [{\citenamefont {Vuik}\ \emph {et~al.}(2016)\citenamefont {Vuik},
  \citenamefont {Eeltink}, \citenamefont {Akhmerov},\ and\ \citenamefont
  {Wimmer}}]{Vuik:NJP16}%
  \BibitemOpen
  \bibfield  {author} {\bibinfo {author} {\bibfnamefont {A.}~\bibnamefont
  {Vuik}}, \bibinfo {author} {\bibfnamefont {D.}~\bibnamefont {Eeltink}},
  \bibinfo {author} {\bibfnamefont {A.~R.}\ \bibnamefont {Akhmerov}}, \ and\
  \bibinfo {author} {\bibfnamefont {M.}~\bibnamefont {Wimmer}},\ }\href
  {http://stacks.iop.org/1367-2630/18/i=3/a=033013} {\bibfield  {journal}
  {\bibinfo  {journal} {New J. Phys.}\ }\textbf {\bibinfo {volume} {18}},\
  \bibinfo {pages} {033013} (\bibinfo {year} {2016})}\BibitemShut {NoStop}%
\bibitem [{\citenamefont {Dom{\'\i}nguez}\ \emph
  {et~al.}(2017{\natexlab{a}})\citenamefont {Dom{\'\i}nguez}, \citenamefont
  {Cayao}, \citenamefont {San-Jose}, \citenamefont {Aguado}, \citenamefont
  {Levy~Yeyati},\ and\ \citenamefont {Prada}}]{Dominguez:NQM17}%
  \BibitemOpen
  \bibfield  {author} {\bibinfo {author} {\bibfnamefont {F.}~\bibnamefont
  {Dom{\'\i}nguez}}, \bibinfo {author} {\bibfnamefont {J.}~\bibnamefont
  {Cayao}}, \bibinfo {author} {\bibfnamefont {P.}~\bibnamefont {San-Jose}},
  \bibinfo {author} {\bibfnamefont {R.}~\bibnamefont {Aguado}}, \bibinfo
  {author} {\bibfnamefont {A.}~\bibnamefont {Levy~Yeyati}}, \ and\ \bibinfo
  {author} {\bibfnamefont {E.}~\bibnamefont {Prada}},\ }\href {\doibase
  10.1038/s41535-017-0012-0} {\bibfield  {journal} {\bibinfo  {journal} {npj
  Quant. Mater.}\ }\textbf {\bibinfo {volume} {2}},\ \bibinfo {pages} {13}
  (\bibinfo {year} {2017}{\natexlab{a}})}\BibitemShut {NoStop}%
\bibitem [{\citenamefont {Mikkelsen}\ \emph {et~al.}(2018)\citenamefont
  {Mikkelsen}, \citenamefont {Kotetes}, \citenamefont {Krogstrup},\ and\
  \citenamefont {Flensberg}}]{Mikkelsen:PRX18}%
  \BibitemOpen
  \bibfield  {author} {\bibinfo {author} {\bibfnamefont {A.~E.~G.}\
  \bibnamefont {Mikkelsen}}, \bibinfo {author} {\bibfnamefont {P.}~\bibnamefont
  {Kotetes}}, \bibinfo {author} {\bibfnamefont {P.}~\bibnamefont {Krogstrup}},
  \ and\ \bibinfo {author} {\bibfnamefont {K.}~\bibnamefont {Flensberg}},\
  }\href {\doibase 10.1103/PhysRevX.8.031040} {\bibfield  {journal} {\bibinfo
  {journal} {Phys. Rev. X}\ }\textbf {\bibinfo {volume} {8}},\ \bibinfo {pages}
  {031040} (\bibinfo {year} {2018})}\BibitemShut {NoStop}%
\bibitem [{\citenamefont {de~Moor}\ \emph {et~al.}(2018)\citenamefont
  {de~Moor}, \citenamefont {Bommer}, \citenamefont {Xu}, \citenamefont
  {Winkler}, \citenamefont {Antipov}, \citenamefont {Bargerbos}, \citenamefont
  {Wang}, \citenamefont {van Loo}, \citenamefont {het Veld}, \citenamefont
  {Gazibegovic}, \citenamefont {Car}, \citenamefont {Logan}, \citenamefont
  {Pendharkar}, \citenamefont {Lee}, \citenamefont {Bakkers}, \citenamefont
  {Palmstr{\o}m}, \citenamefont {Lutchyn}, \citenamefont {Kouwenhoven},\ and\
  \citenamefont {Zhang}}]{Moor:NJP18}%
  \BibitemOpen
  \bibfield  {author} {\bibinfo {author} {\bibfnamefont {M.~W.~A.}\
  \bibnamefont {de~Moor}}, \bibinfo {author} {\bibfnamefont {J.~D.~S.}\
  \bibnamefont {Bommer}}, \bibinfo {author} {\bibfnamefont {D.}~\bibnamefont
  {Xu}}, \bibinfo {author} {\bibfnamefont {G.~W.}\ \bibnamefont {Winkler}},
  \bibinfo {author} {\bibfnamefont {A.~E.}\ \bibnamefont {Antipov}}, \bibinfo
  {author} {\bibfnamefont {A.}~\bibnamefont {Bargerbos}}, \bibinfo {author}
  {\bibfnamefont {G.}~\bibnamefont {Wang}}, \bibinfo {author} {\bibfnamefont
  {N.}~\bibnamefont {van Loo}}, \bibinfo {author} {\bibfnamefont {R.~L. M.~O.}\
  \bibnamefont {het Veld}}, \bibinfo {author} {\bibfnamefont {S.}~\bibnamefont
  {Gazibegovic}}, \bibinfo {author} {\bibfnamefont {D.}~\bibnamefont {Car}},
  \bibinfo {author} {\bibfnamefont {J.~A.}\ \bibnamefont {Logan}}, \bibinfo
  {author} {\bibfnamefont {M.}~\bibnamefont {Pendharkar}}, \bibinfo {author}
  {\bibfnamefont {J.~S.}\ \bibnamefont {Lee}}, \bibinfo {author} {\bibfnamefont
  {E.~P. A.~M.}\ \bibnamefont {Bakkers}}, \bibinfo {author} {\bibfnamefont
  {C.~J.}\ \bibnamefont {Palmstr{\o}m}}, \bibinfo {author} {\bibfnamefont
  {R.~M.}\ \bibnamefont {Lutchyn}}, \bibinfo {author} {\bibfnamefont {L.~P.}\
  \bibnamefont {Kouwenhoven}}, \ and\ \bibinfo {author} {\bibfnamefont
  {H.}~\bibnamefont {Zhang}},\ }\href
  {http://stacks.iop.org/1367-2630/20/i=10/a=103049} {\bibfield  {journal}
  {\bibinfo  {journal} {New J. Phys.}\ }\textbf {\bibinfo {volume} {20}},\
  \bibinfo {pages} {103049} (\bibinfo {year} {2018})}\BibitemShut {NoStop}%
\bibitem [{\citenamefont {Woods}\ \emph {et~al.}(2018)\citenamefont {Woods},
  \citenamefont {Stanescu},\ and\ \citenamefont {Das~Sarma}}]{Woods:PRB18}%
  \BibitemOpen
  \bibfield  {author} {\bibinfo {author} {\bibfnamefont {B.~D.}\ \bibnamefont
  {Woods}}, \bibinfo {author} {\bibfnamefont {T.~D.}\ \bibnamefont {Stanescu}},
  \ and\ \bibinfo {author} {\bibfnamefont {S.}~\bibnamefont {Das~Sarma}},\
  }\href {\doibase 10.1103/PhysRevB.98.035428} {\bibfield  {journal} {\bibinfo
  {journal} {Phys. Rev. B}\ }\textbf {\bibinfo {volume} {98}},\ \bibinfo
  {pages} {035428} (\bibinfo {year} {2018})}\BibitemShut {NoStop}%
\bibitem [{\citenamefont {Escribano}\ \emph {et~al.}(2018)\citenamefont
  {Escribano}, \citenamefont {Levy~Yeyati},\ and\ \citenamefont
  {Prada}}]{Escribano:BJN18}%
  \BibitemOpen
  \bibfield  {author} {\bibinfo {author} {\bibfnamefont {S.~D.}\ \bibnamefont
  {Escribano}}, \bibinfo {author} {\bibfnamefont {A.}~\bibnamefont
  {Levy~Yeyati}}, \ and\ \bibinfo {author} {\bibfnamefont {E.}~\bibnamefont
  {Prada}},\ }\href {\doibase 10.3762/bjnano.9.203} {\bibfield  {journal}
  {\bibinfo  {journal} {Beilstein J. Nanotechnol.}\ }\textbf {\bibinfo {volume}
  {9}},\ \bibinfo {pages} {2171} (\bibinfo {year} {2018})}\BibitemShut
  {NoStop}%
\bibitem [{\citenamefont {Winkler}\ \emph {et~al.}(2019)\citenamefont
  {Winkler}, \citenamefont {Antipov}, \citenamefont {van Heck}, \citenamefont
  {Soluyanov}, \citenamefont {Glazman}, \citenamefont {Wimmer},\ and\
  \citenamefont {Lutchyn}}]{Winkler:PRB19}%
  \BibitemOpen
  \bibfield  {author} {\bibinfo {author} {\bibfnamefont {G.~W.}\ \bibnamefont
  {Winkler}}, \bibinfo {author} {\bibfnamefont {A.~E.}\ \bibnamefont
  {Antipov}}, \bibinfo {author} {\bibfnamefont {B.}~\bibnamefont {van Heck}},
  \bibinfo {author} {\bibfnamefont {A.~A.}\ \bibnamefont {Soluyanov}}, \bibinfo
  {author} {\bibfnamefont {L.~I.}\ \bibnamefont {Glazman}}, \bibinfo {author}
  {\bibfnamefont {M.}~\bibnamefont {Wimmer}}, \ and\ \bibinfo {author}
  {\bibfnamefont {R.~M.}\ \bibnamefont {Lutchyn}},\ }\href {\doibase
  10.1103/PhysRevB.99.245408} {\bibfield  {journal} {\bibinfo  {journal} {Phys.
  Rev. B}\ }\textbf {\bibinfo {volume} {99}},\ \bibinfo {pages} {245408}
  (\bibinfo {year} {2019})}\BibitemShut {NoStop}%
\bibitem [{\citenamefont {Prada}\ \emph {et~al.}(2012)\citenamefont {Prada},
  \citenamefont {San-Jose},\ and\ \citenamefont {Aguado}}]{Prada:PRB12}%
  \BibitemOpen
  \bibfield  {author} {\bibinfo {author} {\bibfnamefont {E.}~\bibnamefont
  {Prada}}, \bibinfo {author} {\bibfnamefont {P.}~\bibnamefont {San-Jose}}, \
  and\ \bibinfo {author} {\bibfnamefont {R.}~\bibnamefont {Aguado}},\ }\href
  {\doibase 10.1103/PhysRevB.86.180503} {\bibfield  {journal} {\bibinfo
  {journal} {Phys. Rev. B}\ }\textbf {\bibinfo {volume} {86}},\ \bibinfo
  {pages} {180503(R)} (\bibinfo {year} {2012})}\BibitemShut {NoStop}%
\bibitem [{\citenamefont {Kells}\ \emph {et~al.}(2012)\citenamefont {Kells},
  \citenamefont {Meidan},\ and\ \citenamefont {Brouwer}}]{Kells:PRB12}%
  \BibitemOpen
  \bibfield  {author} {\bibinfo {author} {\bibfnamefont {G.}~\bibnamefont
  {Kells}}, \bibinfo {author} {\bibfnamefont {D.}~\bibnamefont {Meidan}}, \
  and\ \bibinfo {author} {\bibfnamefont {P.~W.}\ \bibnamefont {Brouwer}},\
  }\href {\doibase 10.1103/PhysRevB.86.100503} {\bibfield  {journal} {\bibinfo
  {journal} {Phys. Rev. B}\ }\textbf {\bibinfo {volume} {86}},\ \bibinfo
  {pages} {100503} (\bibinfo {year} {2012})}\BibitemShut {NoStop}%
\bibitem [{\citenamefont {Stanescu}\ and\ \citenamefont
  {Tewari}(2013{\natexlab{b}})}]{Stanescu:PRB13}%
  \BibitemOpen
  \bibfield  {author} {\bibinfo {author} {\bibfnamefont {T.~D.}\ \bibnamefont
  {Stanescu}}\ and\ \bibinfo {author} {\bibfnamefont {S.}~\bibnamefont
  {Tewari}},\ }\href {https://doi.org/10.1103/PhysRevB.87.140504} {\bibfield
  {journal} {\bibinfo  {journal} {Phys. Rev. B}\ }\textbf {\bibinfo {volume}
  {87}},\ \bibinfo {pages} {140504(R)} (\bibinfo {year}
  {2013}{\natexlab{b}})}\BibitemShut {NoStop}%
\bibitem [{\citenamefont {Roy}\ \emph {et~al.}(2013)\citenamefont {Roy},
  \citenamefont {Bondyopadhaya},\ and\ \citenamefont {Tewari}}]{Roy:PRB13}%
  \BibitemOpen
  \bibfield  {author} {\bibinfo {author} {\bibfnamefont {D.}~\bibnamefont
  {Roy}}, \bibinfo {author} {\bibfnamefont {N.}~\bibnamefont {Bondyopadhaya}},
  \ and\ \bibinfo {author} {\bibfnamefont {S.}~\bibnamefont {Tewari}},\ }\href
  {\doibase 10.1103/PhysRevB.88.020502} {\bibfield  {journal} {\bibinfo
  {journal} {Phys. Rev. B}\ }\textbf {\bibinfo {volume} {88}},\ \bibinfo
  {pages} {020502} (\bibinfo {year} {2013})}\BibitemShut {NoStop}%
\bibitem [{\citenamefont {Fleckenstein}\ \emph {et~al.}(2018)\citenamefont
  {Fleckenstein}, \citenamefont {Dom\'{\i}nguez}, \citenamefont
  {Traverso~Ziani},\ and\ \citenamefont {Trauzettel}}]{Fleckenstein:PRB18}%
  \BibitemOpen
  \bibfield  {author} {\bibinfo {author} {\bibfnamefont {C.}~\bibnamefont
  {Fleckenstein}}, \bibinfo {author} {\bibfnamefont {F.}~\bibnamefont
  {Dom\'{\i}nguez}}, \bibinfo {author} {\bibfnamefont {N.}~\bibnamefont
  {Traverso~Ziani}}, \ and\ \bibinfo {author} {\bibfnamefont {B.}~\bibnamefont
  {Trauzettel}},\ }\href {\doibase 10.1103/PhysRevB.97.155425} {\bibfield
  {journal} {\bibinfo  {journal} {Phys. Rev. B}\ }\textbf {\bibinfo {volume}
  {97}},\ \bibinfo {pages} {155425} (\bibinfo {year} {2018})}\BibitemShut
  {NoStop}%
\bibitem [{\citenamefont {Stanescu}\ and\ \citenamefont
  {Tewari}(2014)}]{Stanescu:PRB14}%
  \BibitemOpen
  \bibfield  {author} {\bibinfo {author} {\bibfnamefont {T.~D.}\ \bibnamefont
  {Stanescu}}\ and\ \bibinfo {author} {\bibfnamefont {S.}~\bibnamefont
  {Tewari}},\ }\href {https://doi.org/10.1103/PhysRevB.89.220507} {\bibfield
  {journal} {\bibinfo  {journal} {Phys. Rev. B}\ }\textbf {\bibinfo {volume}
  {89}},\ \bibinfo {pages} {220507} (\bibinfo {year} {2014})}\BibitemShut
  {NoStop}%
\bibitem [{\citenamefont {Liu}\ \emph {et~al.}(2017{\natexlab{a}})\citenamefont
  {Liu}, \citenamefont {Sau}, \citenamefont {Stanescu},\ and\ \citenamefont
  {Das~Sarma}}]{Liu:PRB17a}%
  \BibitemOpen
  \bibfield  {author} {\bibinfo {author} {\bibfnamefont {C.-X.}\ \bibnamefont
  {Liu}}, \bibinfo {author} {\bibfnamefont {J.~D.}\ \bibnamefont {Sau}},
  \bibinfo {author} {\bibfnamefont {T.~D.}\ \bibnamefont {Stanescu}}, \ and\
  \bibinfo {author} {\bibfnamefont {S.}~\bibnamefont {Das~Sarma}},\ }\href
  {\doibase 10.1103/PhysRevB.96.075161} {\bibfield  {journal} {\bibinfo
  {journal} {Phys. Rev. B}\ }\textbf {\bibinfo {volume} {96}},\ \bibinfo
  {pages} {075161} (\bibinfo {year} {2017}{\natexlab{a}})}\BibitemShut
  {NoStop}%
\bibitem [{\citenamefont {Pe\~naranda}\ \emph {et~al.}(2018)\citenamefont
  {Pe\~naranda}, \citenamefont {Aguado}, \citenamefont {San-Jose},\ and\
  \citenamefont {Prada}}]{Penaranda:PRB18}%
  \BibitemOpen
  \bibfield  {author} {\bibinfo {author} {\bibfnamefont {F.}~\bibnamefont
  {Pe\~naranda}}, \bibinfo {author} {\bibfnamefont {R.}~\bibnamefont {Aguado}},
  \bibinfo {author} {\bibfnamefont {P.}~\bibnamefont {San-Jose}}, \ and\
  \bibinfo {author} {\bibfnamefont {E.}~\bibnamefont {Prada}},\ }\href
  {https://journals.aps.org/prb/abstract/10.1103/PhysRevB.98.235406} {\bibfield
   {journal} {\bibinfo  {journal} {Phys. Rev. B}\ }\textbf {\bibinfo {volume}
  {98}},\ \bibinfo {pages} {235406} (\bibinfo {year} {2018})}\BibitemShut
  {NoStop}%
\bibitem [{\citenamefont {Reeg}\ \emph
  {et~al.}(2018{\natexlab{b}})\citenamefont {Reeg}, \citenamefont {Dmytruk},
  \citenamefont {Chevallier}, \citenamefont {Loss},\ and\ \citenamefont
  {Klinovaja}}]{Reeg:PRB18b}%
  \BibitemOpen
  \bibfield  {author} {\bibinfo {author} {\bibfnamefont {C.}~\bibnamefont
  {Reeg}}, \bibinfo {author} {\bibfnamefont {O.}~\bibnamefont {Dmytruk}},
  \bibinfo {author} {\bibfnamefont {D.}~\bibnamefont {Chevallier}}, \bibinfo
  {author} {\bibfnamefont {D.}~\bibnamefont {Loss}}, \ and\ \bibinfo {author}
  {\bibfnamefont {J.}~\bibnamefont {Klinovaja}},\ }\href {\doibase
  10.1103/PhysRevB.98.245407} {\bibfield  {journal} {\bibinfo  {journal} {Phys.
  Rev. B}\ }\textbf {\bibinfo {volume} {98}},\ \bibinfo {pages} {245407}
  (\bibinfo {year} {2018}{\natexlab{b}})}\BibitemShut {NoStop}%
\bibitem [{\citenamefont {Liu}\ \emph {et~al.}(2018)\citenamefont {Liu},
  \citenamefont {Sau},\ and\ \citenamefont {Das~Sarma}}]{Liu:PRB18}%
  \BibitemOpen
  \bibfield  {author} {\bibinfo {author} {\bibfnamefont {C.-X.}\ \bibnamefont
  {Liu}}, \bibinfo {author} {\bibfnamefont {J.~D.}\ \bibnamefont {Sau}}, \ and\
  \bibinfo {author} {\bibfnamefont {S.}~\bibnamefont {Das~Sarma}},\ }\href
  {https://doi.org/10.1103/PhysRevB.97.214502} {\bibfield  {journal} {\bibinfo
  {journal} {Phys. Rev. B}\ }\textbf {\bibinfo {volume} {97}},\ \bibinfo
  {pages} {214502} (\bibinfo {year} {2018})}\BibitemShut {NoStop}%
\bibitem [{\citenamefont {Moore}\ \emph
  {et~al.}(2018{\natexlab{a}})\citenamefont {Moore}, \citenamefont {Stanescu},\
  and\ \citenamefont {Tewari}}]{Moore:PRB18a}%
  \BibitemOpen
  \bibfield  {author} {\bibinfo {author} {\bibfnamefont {C.}~\bibnamefont
  {Moore}}, \bibinfo {author} {\bibfnamefont {T.~D.}\ \bibnamefont {Stanescu}},
  \ and\ \bibinfo {author} {\bibfnamefont {S.}~\bibnamefont {Tewari}},\ }\href
  {\doibase 10.1103/PhysRevB.97.165302} {\bibfield  {journal} {\bibinfo
  {journal} {Phys. Rev. B}\ }\textbf {\bibinfo {volume} {97}},\ \bibinfo
  {pages} {165302} (\bibinfo {year} {2018}{\natexlab{a}})}\BibitemShut
  {NoStop}%
\bibitem [{\citenamefont {Moore}\ \emph
  {et~al.}(2018{\natexlab{b}})\citenamefont {Moore}, \citenamefont {Zeng},
  \citenamefont {Stanescu},\ and\ \citenamefont {Tewari}}]{Moore:PRB18}%
  \BibitemOpen
  \bibfield  {author} {\bibinfo {author} {\bibfnamefont {C.}~\bibnamefont
  {Moore}}, \bibinfo {author} {\bibfnamefont {C.}~\bibnamefont {Zeng}},
  \bibinfo {author} {\bibfnamefont {T.~D.}\ \bibnamefont {Stanescu}}, \ and\
  \bibinfo {author} {\bibfnamefont {S.}~\bibnamefont {Tewari}},\ }\href
  {\doibase 10.1103/PhysRevB.98.155314} {\bibfield  {journal} {\bibinfo
  {journal} {Phys. Rev. B}\ }\textbf {\bibinfo {volume} {98}},\ \bibinfo
  {pages} {155314} (\bibinfo {year} {2018}{\natexlab{b}})}\BibitemShut
  {NoStop}%
\bibitem [{\citenamefont {Vuik}\ \emph {et~al.}(2019)\citenamefont {Vuik},
  \citenamefont {Nijholt}, \citenamefont {Akhmerov},\ and\ \citenamefont
  {Wimmer}}]{Vuik:SP19}%
  \BibitemOpen
  \bibfield  {author} {\bibinfo {author} {\bibfnamefont {A.}~\bibnamefont
  {Vuik}}, \bibinfo {author} {\bibfnamefont {B.}~\bibnamefont {Nijholt}},
  \bibinfo {author} {\bibfnamefont {A.~R.}\ \bibnamefont {Akhmerov}}, \ and\
  \bibinfo {author} {\bibfnamefont {M.}~\bibnamefont {Wimmer}},\ }\href
  {\doibase 10.21468/SciPostPhys.7.5.061} {\bibfield  {journal} {\bibinfo
  {journal} {SciPost Phys.}\ }\textbf {\bibinfo {volume} {7}},\ \bibinfo
  {pages} {61} (\bibinfo {year} {2019})}\BibitemShut {NoStop}%
\bibitem [{\citenamefont {Awoga}\ \emph {et~al.}(2019)\citenamefont {Awoga},
  \citenamefont {Cayao},\ and\ \citenamefont {Black-Schaffer}}]{Awoga:PRL19}%
  \BibitemOpen
  \bibfield  {author} {\bibinfo {author} {\bibfnamefont {O.~A.}\ \bibnamefont
  {Awoga}}, \bibinfo {author} {\bibfnamefont {J.}~\bibnamefont {Cayao}}, \ and\
  \bibinfo {author} {\bibfnamefont {A.~M.}\ \bibnamefont {Black-Schaffer}},\
  }\href {\doibase 10.1103/PhysRevLett.123.117001} {\bibfield  {journal}
  {\bibinfo  {journal} {Phys. Rev. Lett.}\ }\textbf {\bibinfo {volume} {123}},\
  \bibinfo {pages} {117001} (\bibinfo {year} {2019})}\BibitemShut {NoStop}%
\bibitem [{\citenamefont {Avila}\ \emph {et~al.}(2019)\citenamefont {Avila},
  \citenamefont {Pe{\~n}aranda}, \citenamefont {Prada}, \citenamefont
  {San-Jose},\ and\ \citenamefont {Aguado}}]{Avila:CP19}%
  \BibitemOpen
  \bibfield  {author} {\bibinfo {author} {\bibfnamefont {J.}~\bibnamefont
  {Avila}}, \bibinfo {author} {\bibfnamefont {F.}~\bibnamefont
  {Pe{\~n}aranda}}, \bibinfo {author} {\bibfnamefont {E.}~\bibnamefont
  {Prada}}, \bibinfo {author} {\bibfnamefont {P.}~\bibnamefont {San-Jose}}, \
  and\ \bibinfo {author} {\bibfnamefont {R.}~\bibnamefont {Aguado}},\ }\href
  {https://www.nature.com/articles/s42005-019-0231-8} {\bibfield  {journal}
  {\bibinfo  {journal} {Commun. Phys.}\ }\textbf {\bibinfo {volume} {2}},\
  \bibinfo {pages} {133} (\bibinfo {year} {2019})}\BibitemShut {NoStop}%
\bibitem [{\citenamefont {Stanescu}\ and\ \citenamefont
  {Tewari}(2019)}]{Stanescu:PRB19}%
  \BibitemOpen
  \bibfield  {author} {\bibinfo {author} {\bibfnamefont {T.~D.}\ \bibnamefont
  {Stanescu}}\ and\ \bibinfo {author} {\bibfnamefont {S.}~\bibnamefont
  {Tewari}},\ }\href {\doibase 10.1103/PhysRevB.100.155429} {\bibfield
  {journal} {\bibinfo  {journal} {Phys. Rev. B}\ }\textbf {\bibinfo {volume}
  {100}},\ \bibinfo {pages} {155429} (\bibinfo {year} {2019})}\BibitemShut
  {NoStop}%
\bibitem [{\citenamefont {Woods}\ \emph {et~al.}(2019)\citenamefont {Woods},
  \citenamefont {Chen}, \citenamefont {Frolov},\ and\ \citenamefont
  {Stanescu}}]{Woods:PRB19}%
  \BibitemOpen
  \bibfield  {author} {\bibinfo {author} {\bibfnamefont {B.~D.}\ \bibnamefont
  {Woods}}, \bibinfo {author} {\bibfnamefont {J.}~\bibnamefont {Chen}},
  \bibinfo {author} {\bibfnamefont {S.~M.}\ \bibnamefont {Frolov}}, \ and\
  \bibinfo {author} {\bibfnamefont {T.~D.}\ \bibnamefont {Stanescu}},\ }\href
  {\doibase 10.1103/PhysRevB.100.125407} {\bibfield  {journal} {\bibinfo
  {journal} {Phys. Rev. B}\ }\textbf {\bibinfo {volume} {100}},\ \bibinfo
  {pages} {125407} (\bibinfo {year} {2019})}\BibitemShut {NoStop}%
\bibitem [{\citenamefont {Prada}\ \emph {et~al.}(2017)\citenamefont {Prada},
  \citenamefont {Aguado},\ and\ \citenamefont {San-Jose}}]{Prada:PRB17}%
  \BibitemOpen
  \bibfield  {author} {\bibinfo {author} {\bibfnamefont {E.}~\bibnamefont
  {Prada}}, \bibinfo {author} {\bibfnamefont {R.}~\bibnamefont {Aguado}}, \
  and\ \bibinfo {author} {\bibfnamefont {P.}~\bibnamefont {San-Jose}},\ }\href
  {\doibase 10.1103/PhysRevB.96.085418} {\bibfield  {journal} {\bibinfo
  {journal} {Phys. Rev. B}\ }\textbf {\bibinfo {volume} {96}},\ \bibinfo
  {pages} {085418} (\bibinfo {year} {2017})}\BibitemShut {NoStop}%
\bibitem [{\citenamefont {Deng}\ \emph {et~al.}(2018)\citenamefont {Deng},
  \citenamefont {Vaitiek\ifmmode~\dot{e}\else \.{e}\fi{}nas}, \citenamefont
  {Prada}, \citenamefont {San-Jose}, \citenamefont {Nyg\aa{}rd}, \citenamefont
  {Krogstrup}, \citenamefont {Aguado},\ and\ \citenamefont
  {Marcus}}]{Deng:PRB18}%
  \BibitemOpen
  \bibfield  {author} {\bibinfo {author} {\bibfnamefont {M.-T.}\ \bibnamefont
  {Deng}}, \bibinfo {author} {\bibfnamefont {S.}~\bibnamefont
  {Vaitiek\ifmmode~\dot{e}\else \.{e}\fi{}nas}}, \bibinfo {author}
  {\bibfnamefont {E.}~\bibnamefont {Prada}}, \bibinfo {author} {\bibfnamefont
  {P.}~\bibnamefont {San-Jose}}, \bibinfo {author} {\bibfnamefont
  {J.}~\bibnamefont {Nyg\aa{}rd}}, \bibinfo {author} {\bibfnamefont
  {P.}~\bibnamefont {Krogstrup}}, \bibinfo {author} {\bibfnamefont
  {R.}~\bibnamefont {Aguado}}, \ and\ \bibinfo {author} {\bibfnamefont {C.~M.}\
  \bibnamefont {Marcus}},\ }\href {\doibase 10.1103/PhysRevB.98.085125}
  {\bibfield  {journal} {\bibinfo  {journal} {Phys. Rev. B}\ }\textbf {\bibinfo
  {volume} {98}},\ \bibinfo {pages} {085125} (\bibinfo {year}
  {2018})}\BibitemShut {NoStop}%
\bibitem [{\citenamefont {San-Jose}\ \emph {et~al.}(2016)\citenamefont
  {San-Jose}, \citenamefont {Cayao}, \citenamefont {Prada},\ and\ \citenamefont
  {Aguado}}]{San-Jose:SR16}%
  \BibitemOpen
  \bibfield  {author} {\bibinfo {author} {\bibfnamefont {P.}~\bibnamefont
  {San-Jose}}, \bibinfo {author} {\bibfnamefont {J.}~\bibnamefont {Cayao}},
  \bibinfo {author} {\bibfnamefont {E.}~\bibnamefont {Prada}}, \ and\ \bibinfo
  {author} {\bibfnamefont {R.}~\bibnamefont {Aguado}},\ }\href
  {http://dx.doi.org/10.1038/srep21427} {\bibfield  {journal} {\bibinfo
  {journal} {Sci. Rep.}\ }\textbf {\bibinfo {volume} {6}},\ \bibinfo {pages}
  {21427} (\bibinfo {year} {2016})}\BibitemShut {NoStop}%
\bibitem [{\citenamefont {Goldstein}\ and\ \citenamefont
  {Chamon}(2011)}]{Goldstein:PRB11}%
  \BibitemOpen
  \bibfield  {author} {\bibinfo {author} {\bibfnamefont {G.}~\bibnamefont
  {Goldstein}}\ and\ \bibinfo {author} {\bibfnamefont {C.}~\bibnamefont
  {Chamon}},\ }\href {\doibase 10.1103/PhysRevB.84.205109} {\bibfield
  {journal} {\bibinfo  {journal} {Phys. Rev. B}\ }\textbf {\bibinfo {volume}
  {84}},\ \bibinfo {pages} {205109} (\bibinfo {year} {2011})}\BibitemShut
  {NoStop}%
\bibitem [{\citenamefont {Budich}\ \emph {et~al.}(2012)\citenamefont {Budich},
  \citenamefont {Walter},\ and\ \citenamefont {Trauzettel}}]{Budich:PRB12}%
  \BibitemOpen
  \bibfield  {author} {\bibinfo {author} {\bibfnamefont {J.~C.}\ \bibnamefont
  {Budich}}, \bibinfo {author} {\bibfnamefont {S.}~\bibnamefont {Walter}}, \
  and\ \bibinfo {author} {\bibfnamefont {B.}~\bibnamefont {Trauzettel}},\
  }\href {\doibase 10.1103/PhysRevB.85.121405} {\bibfield  {journal} {\bibinfo
  {journal} {Phys. Rev. B}\ }\textbf {\bibinfo {volume} {85}},\ \bibinfo
  {pages} {121405} (\bibinfo {year} {2012})}\BibitemShut {NoStop}%
\bibitem [{\citenamefont {Rainis}\ and\ \citenamefont
  {Loss}(2012)}]{Rainis:PRB12}%
  \BibitemOpen
  \bibfield  {author} {\bibinfo {author} {\bibfnamefont {D.}~\bibnamefont
  {Rainis}}\ and\ \bibinfo {author} {\bibfnamefont {D.}~\bibnamefont {Loss}},\
  }\href {\doibase 10.1103/PhysRevB.85.174533} {\bibfield  {journal} {\bibinfo
  {journal} {Phys. Rev. B}\ }\textbf {\bibinfo {volume} {85}},\ \bibinfo
  {pages} {174533} (\bibinfo {year} {2012})}\BibitemShut {NoStop}%
\bibitem [{\citenamefont {Trif}\ and\ \citenamefont
  {Tserkovnyak}(2012)}]{Trif:PRL12}%
  \BibitemOpen
  \bibfield  {author} {\bibinfo {author} {\bibfnamefont {M.}~\bibnamefont
  {Trif}}\ and\ \bibinfo {author} {\bibfnamefont {Y.}~\bibnamefont
  {Tserkovnyak}},\ }\href {\doibase 10.1103/PhysRevLett.109.257002} {\bibfield
  {journal} {\bibinfo  {journal} {Phys. Rev. Lett.}\ }\textbf {\bibinfo
  {volume} {109}},\ \bibinfo {pages} {257002} (\bibinfo {year}
  {2012})}\BibitemShut {NoStop}%
\bibitem [{\citenamefont {Schmidt}\ \emph {et~al.}(2012)\citenamefont
  {Schmidt}, \citenamefont {Rainis},\ and\ \citenamefont
  {Loss}}]{Schmidt:PRB12}%
  \BibitemOpen
  \bibfield  {author} {\bibinfo {author} {\bibfnamefont {M.~J.}\ \bibnamefont
  {Schmidt}}, \bibinfo {author} {\bibfnamefont {D.}~\bibnamefont {Rainis}}, \
  and\ \bibinfo {author} {\bibfnamefont {D.}~\bibnamefont {Loss}},\ }\href
  {\doibase 10.1103/PhysRevB.86.085414} {\bibfield  {journal} {\bibinfo
  {journal} {Phys. Rev. B}\ }\textbf {\bibinfo {volume} {86}},\ \bibinfo
  {pages} {085414} (\bibinfo {year} {2012})}\BibitemShut {NoStop}%
\bibitem [{\citenamefont {Schmidt}\ \emph {et~al.}(2013)\citenamefont
  {Schmidt}, \citenamefont {Nunnenkamp},\ and\ \citenamefont
  {Bruder}}]{Schmidt:PRL13}%
  \BibitemOpen
  \bibfield  {author} {\bibinfo {author} {\bibfnamefont {T.~L.}\ \bibnamefont
  {Schmidt}}, \bibinfo {author} {\bibfnamefont {A.}~\bibnamefont {Nunnenkamp}},
  \ and\ \bibinfo {author} {\bibfnamefont {C.}~\bibnamefont {Bruder}},\ }\href
  {\doibase 10.1103/PhysRevLett.110.107006} {\bibfield  {journal} {\bibinfo
  {journal} {Phys. Rev. Lett.}\ }\textbf {\bibinfo {volume} {110}},\ \bibinfo
  {pages} {107006} (\bibinfo {year} {2013})}\BibitemShut {NoStop}%
\bibitem [{\citenamefont {Scheurer}\ and\ \citenamefont
  {Shnirman}(2013)}]{Scheurer:PRB13}%
  \BibitemOpen
  \bibfield  {author} {\bibinfo {author} {\bibfnamefont {M.~S.}\ \bibnamefont
  {Scheurer}}\ and\ \bibinfo {author} {\bibfnamefont {A.}~\bibnamefont
  {Shnirman}},\ }\href {\doibase 10.1103/PhysRevB.88.064515} {\bibfield
  {journal} {\bibinfo  {journal} {Phys. Rev. B}\ }\textbf {\bibinfo {volume}
  {88}},\ \bibinfo {pages} {064515} (\bibinfo {year} {2013})}\BibitemShut
  {NoStop}%
\bibitem [{\citenamefont {Pedrocchi}\ and\ \citenamefont
  {DiVincenzo}(2015)}]{Pedrocchi:PRL15}%
  \BibitemOpen
  \bibfield  {author} {\bibinfo {author} {\bibfnamefont {F.~L.}\ \bibnamefont
  {Pedrocchi}}\ and\ \bibinfo {author} {\bibfnamefont {D.~P.}\ \bibnamefont
  {DiVincenzo}},\ }\href {\doibase 10.1103/PhysRevLett.115.120402} {\bibfield
  {journal} {\bibinfo  {journal} {Phys. Rev. Lett.}\ }\textbf {\bibinfo
  {volume} {115}},\ \bibinfo {pages} {120402} (\bibinfo {year}
  {2015})}\BibitemShut {NoStop}%
\bibitem [{\citenamefont {Dmytruk}\ \emph {et~al.}(2015)\citenamefont
  {Dmytruk}, \citenamefont {Trif},\ and\ \citenamefont
  {Simon}}]{Dmytruk:PRB15}%
  \BibitemOpen
  \bibfield  {author} {\bibinfo {author} {\bibfnamefont {O.}~\bibnamefont
  {Dmytruk}}, \bibinfo {author} {\bibfnamefont {M.}~\bibnamefont {Trif}}, \
  and\ \bibinfo {author} {\bibfnamefont {P.}~\bibnamefont {Simon}},\ }\href
  {\doibase 10.1103/PhysRevB.92.245432} {\bibfield  {journal} {\bibinfo
  {journal} {Phys. Rev. B}\ }\textbf {\bibinfo {volume} {92}},\ \bibinfo
  {pages} {245432} (\bibinfo {year} {2015})}\BibitemShut {NoStop}%
\bibitem [{\citenamefont {Sekania}\ \emph {et~al.}(2017)\citenamefont
  {Sekania}, \citenamefont {Plugge}, \citenamefont {Greiter}, \citenamefont
  {Thomale},\ and\ \citenamefont {Schmitteckert}}]{Sekania:PRB17}%
  \BibitemOpen
  \bibfield  {author} {\bibinfo {author} {\bibfnamefont {M.}~\bibnamefont
  {Sekania}}, \bibinfo {author} {\bibfnamefont {S.}~\bibnamefont {Plugge}},
  \bibinfo {author} {\bibfnamefont {M.}~\bibnamefont {Greiter}}, \bibinfo
  {author} {\bibfnamefont {R.}~\bibnamefont {Thomale}}, \ and\ \bibinfo
  {author} {\bibfnamefont {P.}~\bibnamefont {Schmitteckert}},\ }\href {\doibase
  10.1103/PhysRevB.96.094307} {\bibfield  {journal} {\bibinfo  {journal} {Phys.
  Rev. B}\ }\textbf {\bibinfo {volume} {96}},\ \bibinfo {pages} {094307}
  (\bibinfo {year} {2017})}\BibitemShut {NoStop}%
\bibitem [{\citenamefont {Knapp}\ \emph {et~al.}(2018)\citenamefont {Knapp},
  \citenamefont {Karzig}, \citenamefont {Lutchyn},\ and\ \citenamefont
  {Nayak}}]{Knapp:PRB18}%
  \BibitemOpen
  \bibfield  {author} {\bibinfo {author} {\bibfnamefont {C.}~\bibnamefont
  {Knapp}}, \bibinfo {author} {\bibfnamefont {T.}~\bibnamefont {Karzig}},
  \bibinfo {author} {\bibfnamefont {R.~M.}\ \bibnamefont {Lutchyn}}, \ and\
  \bibinfo {author} {\bibfnamefont {C.}~\bibnamefont {Nayak}},\ }\href
  {\doibase 10.1103/PhysRevB.97.125404} {\bibfield  {journal} {\bibinfo
  {journal} {Phys. Rev. B}\ }\textbf {\bibinfo {volume} {97}},\ \bibinfo
  {pages} {125404} (\bibinfo {year} {2018})}\BibitemShut {NoStop}%
\bibitem [{\citenamefont {Aseev}\ \emph {et~al.}(2018)\citenamefont {Aseev},
  \citenamefont {Klinovaja},\ and\ \citenamefont {Loss}}]{Aseev:PRB18}%
  \BibitemOpen
  \bibfield  {author} {\bibinfo {author} {\bibfnamefont {P.~P.}\ \bibnamefont
  {Aseev}}, \bibinfo {author} {\bibfnamefont {J.}~\bibnamefont {Klinovaja}}, \
  and\ \bibinfo {author} {\bibfnamefont {D.}~\bibnamefont {Loss}},\ }\href
  {\doibase 10.1103/PhysRevB.98.155414} {\bibfield  {journal} {\bibinfo
  {journal} {Phys. Rev. B}\ }\textbf {\bibinfo {volume} {98}},\ \bibinfo
  {pages} {155414} (\bibinfo {year} {2018})}\BibitemShut {NoStop}%
\bibitem [{\citenamefont {Lai}\ \emph {et~al.}(2018)\citenamefont {Lai},
  \citenamefont {Yang}, \citenamefont {Huang},\ and\ \citenamefont
  {Zhang}}]{Lai:PRB18}%
  \BibitemOpen
  \bibfield  {author} {\bibinfo {author} {\bibfnamefont {H.-L.}\ \bibnamefont
  {Lai}}, \bibinfo {author} {\bibfnamefont {P.-Y.}\ \bibnamefont {Yang}},
  \bibinfo {author} {\bibfnamefont {Y.-W.}\ \bibnamefont {Huang}}, \ and\
  \bibinfo {author} {\bibfnamefont {W.-M.}\ \bibnamefont {Zhang}},\ }\href
  {\doibase 10.1103/PhysRevB.97.054508} {\bibfield  {journal} {\bibinfo
  {journal} {Phys. Rev. B}\ }\textbf {\bibinfo {volume} {97}},\ \bibinfo
  {pages} {054508} (\bibinfo {year} {2018})}\BibitemShut {NoStop}%
\bibitem [{\citenamefont {Aseev}\ \emph {et~al.}(2019)\citenamefont {Aseev},
  \citenamefont {Marra}, \citenamefont {Stano}, \citenamefont {Klinovaja},\
  and\ \citenamefont {Loss}}]{Aseev:PRB19}%
  \BibitemOpen
  \bibfield  {author} {\bibinfo {author} {\bibfnamefont {P.~P.}\ \bibnamefont
  {Aseev}}, \bibinfo {author} {\bibfnamefont {P.}~\bibnamefont {Marra}},
  \bibinfo {author} {\bibfnamefont {P.}~\bibnamefont {Stano}}, \bibinfo
  {author} {\bibfnamefont {J.}~\bibnamefont {Klinovaja}}, \ and\ \bibinfo
  {author} {\bibfnamefont {D.}~\bibnamefont {Loss}},\ }\href {\doibase
  10.1103/PhysRevB.99.205435} {\bibfield  {journal} {\bibinfo  {journal} {Phys.
  Rev. B}\ }\textbf {\bibinfo {volume} {99}},\ \bibinfo {pages} {205435}
  (\bibinfo {year} {2019})}\BibitemShut {NoStop}%
\bibitem [{\citenamefont {Deng}\ \emph {et~al.}(2016)\citenamefont {Deng},
  \citenamefont {Vaitiekenas}, \citenamefont {Hansen}, \citenamefont {Danon},
  \citenamefont {Leijnse}, \citenamefont {Flensberg}, \citenamefont
  {Nyg{\aa}rd}, \citenamefont {Krogstrup},\ and\ \citenamefont
  {Marcus}}]{Deng:S16}%
  \BibitemOpen
  \bibfield  {author} {\bibinfo {author} {\bibfnamefont {M.~T.}\ \bibnamefont
  {Deng}}, \bibinfo {author} {\bibfnamefont {S.}~\bibnamefont {Vaitiekenas}},
  \bibinfo {author} {\bibfnamefont {E.~B.}\ \bibnamefont {Hansen}}, \bibinfo
  {author} {\bibfnamefont {J.}~\bibnamefont {Danon}}, \bibinfo {author}
  {\bibfnamefont {M.}~\bibnamefont {Leijnse}}, \bibinfo {author} {\bibfnamefont
  {K.}~\bibnamefont {Flensberg}}, \bibinfo {author} {\bibfnamefont
  {J.}~\bibnamefont {Nyg{\aa}rd}}, \bibinfo {author} {\bibfnamefont
  {P.}~\bibnamefont {Krogstrup}}, \ and\ \bibinfo {author} {\bibfnamefont
  {C.~M.}\ \bibnamefont {Marcus}},\ }\href {\doibase 10.1126/science.aaf3961}
  {\bibfield  {journal} {\bibinfo  {journal} {Science}\ }\textbf {\bibinfo
  {volume} {354}},\ \bibinfo {pages} {1557} (\bibinfo {year}
  {2016})}\BibitemShut {NoStop}%
\bibitem [{\citenamefont {Nichele}\ \emph {et~al.}(2017)\citenamefont
  {Nichele}, \citenamefont {Drachmann}, \citenamefont {Whiticar}, \citenamefont
  {O'Farrell}, \citenamefont {Suominen}, \citenamefont {Fornieri},
  \citenamefont {Wang}, \citenamefont {Gardner}, \citenamefont {Thomas},
  \citenamefont {Hatke}, \citenamefont {Krogstrup}, \citenamefont {Manfra},
  \citenamefont {Flensberg},\ and\ \citenamefont {Marcus}}]{Nichele:PRL17}%
  \BibitemOpen
  \bibfield  {author} {\bibinfo {author} {\bibfnamefont {F.}~\bibnamefont
  {Nichele}}, \bibinfo {author} {\bibfnamefont {A.~C.~C.}\ \bibnamefont
  {Drachmann}}, \bibinfo {author} {\bibfnamefont {A.~M.}\ \bibnamefont
  {Whiticar}}, \bibinfo {author} {\bibfnamefont {E.~C.~T.}\ \bibnamefont
  {O'Farrell}}, \bibinfo {author} {\bibfnamefont {H.~J.}\ \bibnamefont
  {Suominen}}, \bibinfo {author} {\bibfnamefont {A.}~\bibnamefont {Fornieri}},
  \bibinfo {author} {\bibfnamefont {T.}~\bibnamefont {Wang}}, \bibinfo {author}
  {\bibfnamefont {G.~C.}\ \bibnamefont {Gardner}}, \bibinfo {author}
  {\bibfnamefont {C.}~\bibnamefont {Thomas}}, \bibinfo {author} {\bibfnamefont
  {A.~T.}\ \bibnamefont {Hatke}}, \bibinfo {author} {\bibfnamefont
  {P.}~\bibnamefont {Krogstrup}}, \bibinfo {author} {\bibfnamefont {M.~J.}\
  \bibnamefont {Manfra}}, \bibinfo {author} {\bibfnamefont {K.}~\bibnamefont
  {Flensberg}}, \ and\ \bibinfo {author} {\bibfnamefont {C.~M.}\ \bibnamefont
  {Marcus}},\ }\href {\doibase 10.1103/PhysRevLett.119.136803} {\bibfield
  {journal} {\bibinfo  {journal} {Phys. Rev. Lett.}\ }\textbf {\bibinfo
  {volume} {119}},\ \bibinfo {pages} {136803} (\bibinfo {year}
  {2017})}\BibitemShut {NoStop}%
\bibitem [{\citenamefont {Zhang}\ \emph {et~al.}(2017)\citenamefont {Zhang},
  \citenamefont {G{\"u}l}, \citenamefont {Conesa-Boj}, \citenamefont {Nowak},
  \citenamefont {Wimmer}, \citenamefont {Zuo}, \citenamefont {Mourik},
  \citenamefont {de~Vries}, \citenamefont {van Veen}, \citenamefont {de~Moor},
  \citenamefont {Bommer}, \citenamefont {van Woerkom}, \citenamefont {Car},
  \citenamefont {Plissard}, \citenamefont {Bakkers}, \citenamefont
  {Quintero-P{\'e}rez}, \citenamefont {Cassidy}, \citenamefont {Koelling},
  \citenamefont {Goswami}, \citenamefont {Watanabe}, \citenamefont
  {Taniguchi},\ and\ \citenamefont {Kouwenhoven}}]{Zhang:NC17}%
  \BibitemOpen
  \bibfield  {author} {\bibinfo {author} {\bibfnamefont {H.}~\bibnamefont
  {Zhang}}, \bibinfo {author} {\bibfnamefont {{\"O}.}~\bibnamefont {G{\"u}l}},
  \bibinfo {author} {\bibfnamefont {S.}~\bibnamefont {Conesa-Boj}}, \bibinfo
  {author} {\bibfnamefont {M.}~\bibnamefont {Nowak}}, \bibinfo {author}
  {\bibfnamefont {M.}~\bibnamefont {Wimmer}}, \bibinfo {author} {\bibfnamefont
  {K.}~\bibnamefont {Zuo}}, \bibinfo {author} {\bibfnamefont {V.}~\bibnamefont
  {Mourik}}, \bibinfo {author} {\bibfnamefont {F.~K.}\ \bibnamefont
  {de~Vries}}, \bibinfo {author} {\bibfnamefont {J.}~\bibnamefont {van Veen}},
  \bibinfo {author} {\bibfnamefont {M.~W.~A.}\ \bibnamefont {de~Moor}},
  \bibinfo {author} {\bibfnamefont {J.~D.~S.}\ \bibnamefont {Bommer}}, \bibinfo
  {author} {\bibfnamefont {D.~J.}\ \bibnamefont {van Woerkom}}, \bibinfo
  {author} {\bibfnamefont {D.}~\bibnamefont {Car}}, \bibinfo {author}
  {\bibfnamefont {S.~R.}\ \bibnamefont {Plissard}}, \bibinfo {author}
  {\bibfnamefont {E.~P. A.~M.}\ \bibnamefont {Bakkers}}, \bibinfo {author}
  {\bibfnamefont {M.}~\bibnamefont {Quintero-P{\'e}rez}}, \bibinfo {author}
  {\bibfnamefont {M.~C.}\ \bibnamefont {Cassidy}}, \bibinfo {author}
  {\bibfnamefont {S.}~\bibnamefont {Koelling}}, \bibinfo {author}
  {\bibfnamefont {S.}~\bibnamefont {Goswami}}, \bibinfo {author} {\bibfnamefont
  {K.}~\bibnamefont {Watanabe}}, \bibinfo {author} {\bibfnamefont
  {T.}~\bibnamefont {Taniguchi}}, \ and\ \bibinfo {author} {\bibfnamefont
  {L.~P.}\ \bibnamefont {Kouwenhoven}},\ }\href
  {http://dx.doi.org/10.1038/ncomms16025} {\bibfield  {journal} {\bibinfo
  {journal} {Nat. Commun.}\ }\textbf {\bibinfo {volume} {8}},\ \bibinfo {pages}
  {16025 EP } (\bibinfo {year} {2017})}\BibitemShut {NoStop}%
\bibitem [{\citenamefont {Nadj-Perge}\ \emph {et~al.}(2013)\citenamefont
  {Nadj-Perge}, \citenamefont {Drozdov}, \citenamefont {Bernevig},\ and\
  \citenamefont {Yazdani}}]{Nadj-Perge:PRB13}%
  \BibitemOpen
  \bibfield  {author} {\bibinfo {author} {\bibfnamefont {S.}~\bibnamefont
  {Nadj-Perge}}, \bibinfo {author} {\bibfnamefont {I.~K.}\ \bibnamefont
  {Drozdov}}, \bibinfo {author} {\bibfnamefont {B.~A.}\ \bibnamefont
  {Bernevig}}, \ and\ \bibinfo {author} {\bibfnamefont {A.}~\bibnamefont
  {Yazdani}},\ }\href {\doibase 10.1103/PhysRevB.88.020407} {\bibfield
  {journal} {\bibinfo  {journal} {Phys. Rev. B}\ }\textbf {\bibinfo {volume}
  {88}},\ \bibinfo {pages} {020407} (\bibinfo {year} {2013})}\BibitemShut
  {NoStop}%
\bibitem [{\citenamefont {Nadj-Perge}\ \emph {et~al.}(2014)\citenamefont
  {Nadj-Perge}, \citenamefont {Drozdov}, \citenamefont {Li}, \citenamefont
  {Chen}, \citenamefont {Jeon}, \citenamefont {Seo}, \citenamefont {MacDonald},
  \citenamefont {Bernevig},\ and\ \citenamefont {Yazdani}}]{Nadj-Perge:S14}%
  \BibitemOpen
  \bibfield  {author} {\bibinfo {author} {\bibfnamefont {S.}~\bibnamefont
  {Nadj-Perge}}, \bibinfo {author} {\bibfnamefont {I.~K.}\ \bibnamefont
  {Drozdov}}, \bibinfo {author} {\bibfnamefont {J.}~\bibnamefont {Li}},
  \bibinfo {author} {\bibfnamefont {H.}~\bibnamefont {Chen}}, \bibinfo {author}
  {\bibfnamefont {S.}~\bibnamefont {Jeon}}, \bibinfo {author} {\bibfnamefont
  {J.}~\bibnamefont {Seo}}, \bibinfo {author} {\bibfnamefont {A.~H.}\
  \bibnamefont {MacDonald}}, \bibinfo {author} {\bibfnamefont {B.~A.}\
  \bibnamefont {Bernevig}}, \ and\ \bibinfo {author} {\bibfnamefont
  {A.}~\bibnamefont {Yazdani}},\ }\href {\doibase 10.1126/science.1259327}
  {\bibfield  {journal} {\bibinfo  {journal} {Science}\ }\textbf {\bibinfo
  {volume} {346}},\ \bibinfo {pages} {602} (\bibinfo {year}
  {2014})}\BibitemShut {NoStop}%
\bibitem [{\citenamefont {M{\'e}nard}\ \emph {et~al.}(2017)\citenamefont
  {M{\'e}nard}, \citenamefont {Guissart}, \citenamefont {Brun}, \citenamefont
  {Leriche}, \citenamefont {Trif}, \citenamefont {Debontridder}, \citenamefont
  {Demaille}, \citenamefont {Roditchev}, \citenamefont {Simon},\ and\
  \citenamefont {Cren}}]{Menard:NC17}%
  \BibitemOpen
  \bibfield  {author} {\bibinfo {author} {\bibfnamefont {G.~C.}\ \bibnamefont
  {M{\'e}nard}}, \bibinfo {author} {\bibfnamefont {S.}~\bibnamefont
  {Guissart}}, \bibinfo {author} {\bibfnamefont {C.}~\bibnamefont {Brun}},
  \bibinfo {author} {\bibfnamefont {R.~T.}\ \bibnamefont {Leriche}}, \bibinfo
  {author} {\bibfnamefont {M.}~\bibnamefont {Trif}}, \bibinfo {author}
  {\bibfnamefont {F.}~\bibnamefont {Debontridder}}, \bibinfo {author}
  {\bibfnamefont {D.}~\bibnamefont {Demaille}}, \bibinfo {author}
  {\bibfnamefont {D.}~\bibnamefont {Roditchev}}, \bibinfo {author}
  {\bibfnamefont {P.}~\bibnamefont {Simon}}, \ and\ \bibinfo {author}
  {\bibfnamefont {T.}~\bibnamefont {Cren}},\ }\href {\doibase
  10.1038/s41467-017-02192-x} {\bibfield  {journal} {\bibinfo  {journal} {Nat.
  Commun.}\ }\textbf {\bibinfo {volume} {8}},\ \bibinfo {pages} {2040}
  (\bibinfo {year} {2017})}\BibitemShut {NoStop}%
\bibitem [{\citenamefont {M{\'e}nard}\ \emph {et~al.}(2019)\citenamefont
  {M{\'e}nard}, \citenamefont {Mesaros}, \citenamefont {Brun}, \citenamefont
  {Debontridder}, \citenamefont {Roditchev}, \citenamefont {Simon},\ and\
  \citenamefont {Cren}}]{Menard:NC19}%
  \BibitemOpen
  \bibfield  {author} {\bibinfo {author} {\bibfnamefont {G.~C.}\ \bibnamefont
  {M{\'e}nard}}, \bibinfo {author} {\bibfnamefont {A.}~\bibnamefont {Mesaros}},
  \bibinfo {author} {\bibfnamefont {C.}~\bibnamefont {Brun}}, \bibinfo {author}
  {\bibfnamefont {F.}~\bibnamefont {Debontridder}}, \bibinfo {author}
  {\bibfnamefont {D.}~\bibnamefont {Roditchev}}, \bibinfo {author}
  {\bibfnamefont {P.}~\bibnamefont {Simon}}, \ and\ \bibinfo {author}
  {\bibfnamefont {T.}~\bibnamefont {Cren}},\ }\href {\doibase
  10.1038/s41467-019-10397-5} {\bibfield  {journal} {\bibinfo  {journal} {Nat.
  Commun.}\ }\textbf {\bibinfo {volume} {10}},\ \bibinfo {pages} {2587}
  (\bibinfo {year} {2019})}\BibitemShut {NoStop}%
\bibitem [{\citenamefont {Palacio-Morales}\ \emph {et~al.}(2019)\citenamefont
  {Palacio-Morales}, \citenamefont {Mascot}, \citenamefont {Cocklin},
  \citenamefont {Kim}, \citenamefont {Rachel}, \citenamefont {Morr},\ and\
  \citenamefont {Wiesendanger}}]{Palacio-Morales:SA19}%
  \BibitemOpen
  \bibfield  {author} {\bibinfo {author} {\bibfnamefont {A.}~\bibnamefont
  {Palacio-Morales}}, \bibinfo {author} {\bibfnamefont {E.}~\bibnamefont
  {Mascot}}, \bibinfo {author} {\bibfnamefont {S.}~\bibnamefont {Cocklin}},
  \bibinfo {author} {\bibfnamefont {H.}~\bibnamefont {Kim}}, \bibinfo {author}
  {\bibfnamefont {S.}~\bibnamefont {Rachel}}, \bibinfo {author} {\bibfnamefont
  {D.~K.}\ \bibnamefont {Morr}}, \ and\ \bibinfo {author} {\bibfnamefont
  {R.}~\bibnamefont {Wiesendanger}},\ }\href
  {https://advances.sciencemag.org/content/5/7/eaav6600} {\bibfield  {journal}
  {\bibinfo  {journal} {Sci. Adv.}\ }\textbf {\bibinfo {volume} {5}} (\bibinfo
  {year} {2019})}\BibitemShut {NoStop}%
\bibitem [{\citenamefont {Fu}\ and\ \citenamefont {Kane}(2009)}]{Fu:PRB09}%
  \BibitemOpen
  \bibfield  {author} {\bibinfo {author} {\bibfnamefont {L.}~\bibnamefont
  {Fu}}\ and\ \bibinfo {author} {\bibfnamefont {C.~L.}\ \bibnamefont {Kane}},\
  }\href {\doibase 10.1103/PhysRevB.79.161408} {\bibfield  {journal} {\bibinfo
  {journal} {Phys. Rev. B}\ }\textbf {\bibinfo {volume} {79}},\ \bibinfo
  {pages} {161408} (\bibinfo {year} {2009})}\BibitemShut {NoStop}%
\bibitem [{\citenamefont {Wiedenmann}\ \emph {et~al.}(2016)\citenamefont
  {Wiedenmann}, \citenamefont {Bocquillon}, \citenamefont {Deacon},
  \citenamefont {Hartinger}, \citenamefont {Herrmann}, \citenamefont
  {Klapwijk}, \citenamefont {Maier}, \citenamefont {Ames}, \citenamefont
  {Br{\"u}ne}, \citenamefont {Gould}, \citenamefont {Oiwa}, \citenamefont
  {Ishibashi}, \citenamefont {Tarucha}, \citenamefont {Buhmann},\ and\
  \citenamefont {Molenkamp}}]{Wiedenmann:NC16}%
  \BibitemOpen
  \bibfield  {author} {\bibinfo {author} {\bibfnamefont {J.}~\bibnamefont
  {Wiedenmann}}, \bibinfo {author} {\bibfnamefont {E.}~\bibnamefont
  {Bocquillon}}, \bibinfo {author} {\bibfnamefont {R.~S.}\ \bibnamefont
  {Deacon}}, \bibinfo {author} {\bibfnamefont {S.}~\bibnamefont {Hartinger}},
  \bibinfo {author} {\bibfnamefont {O.}~\bibnamefont {Herrmann}}, \bibinfo
  {author} {\bibfnamefont {T.~M.}\ \bibnamefont {Klapwijk}}, \bibinfo {author}
  {\bibfnamefont {L.}~\bibnamefont {Maier}}, \bibinfo {author} {\bibfnamefont
  {C.}~\bibnamefont {Ames}}, \bibinfo {author} {\bibfnamefont {C.}~\bibnamefont
  {Br{\"u}ne}}, \bibinfo {author} {\bibfnamefont {C.}~\bibnamefont {Gould}},
  \bibinfo {author} {\bibfnamefont {A.}~\bibnamefont {Oiwa}}, \bibinfo {author}
  {\bibfnamefont {K.}~\bibnamefont {Ishibashi}}, \bibinfo {author}
  {\bibfnamefont {S.}~\bibnamefont {Tarucha}}, \bibinfo {author} {\bibfnamefont
  {H.}~\bibnamefont {Buhmann}}, \ and\ \bibinfo {author} {\bibfnamefont
  {L.~W.}\ \bibnamefont {Molenkamp}},\ }\href {\doibase 10.1038/ncomms10303}
  {\bibfield  {journal} {\bibinfo  {journal} {Nat. Commun.}\ }\textbf {\bibinfo
  {volume} {7}},\ \bibinfo {pages} {10303} (\bibinfo {year}
  {2016})}\BibitemShut {NoStop}%
\bibitem [{\citenamefont {Suominen}\ \emph {et~al.}(2017)\citenamefont
  {Suominen}, \citenamefont {Kjaergaard}, \citenamefont {Hamilton},
  \citenamefont {Shabani}, \citenamefont {Palmstr\o{}m}, \citenamefont
  {Marcus},\ and\ \citenamefont {Nichele}}]{Suominen:PRL17}%
  \BibitemOpen
  \bibfield  {author} {\bibinfo {author} {\bibfnamefont {H.~J.}\ \bibnamefont
  {Suominen}}, \bibinfo {author} {\bibfnamefont {M.}~\bibnamefont
  {Kjaergaard}}, \bibinfo {author} {\bibfnamefont {A.~R.}\ \bibnamefont
  {Hamilton}}, \bibinfo {author} {\bibfnamefont {J.}~\bibnamefont {Shabani}},
  \bibinfo {author} {\bibfnamefont {C.~J.}\ \bibnamefont {Palmstr\o{}m}},
  \bibinfo {author} {\bibfnamefont {C.~M.}\ \bibnamefont {Marcus}}, \ and\
  \bibinfo {author} {\bibfnamefont {F.}~\bibnamefont {Nichele}},\ }\href
  {\doibase 10.1103/PhysRevLett.119.176805} {\bibfield  {journal} {\bibinfo
  {journal} {Phys. Rev. Lett.}\ }\textbf {\bibinfo {volume} {119}},\ \bibinfo
  {pages} {176805} (\bibinfo {year} {2017})}\BibitemShut {NoStop}%
\bibitem [{\citenamefont {Bretheau}\ \emph {et~al.}(2013)\citenamefont
  {Bretheau}, \citenamefont {Girit}, \citenamefont {Pothier}, \citenamefont
  {Esteve},\ and\ \citenamefont {Urbina}}]{Bretheau:N13}%
  \BibitemOpen
  \bibfield  {author} {\bibinfo {author} {\bibfnamefont {L.}~\bibnamefont
  {Bretheau}}, \bibinfo {author} {\bibfnamefont {C.~O.}\ \bibnamefont {Girit}},
  \bibinfo {author} {\bibfnamefont {H.}~\bibnamefont {Pothier}}, \bibinfo
  {author} {\bibfnamefont {D.}~\bibnamefont {Esteve}}, \ and\ \bibinfo {author}
  {\bibfnamefont {C.}~\bibnamefont {Urbina}},\ }\href
  {http://dx.doi.org/10.1038/nature12315} {\bibfield  {journal} {\bibinfo
  {journal} {Nature}\ }\textbf {\bibinfo {volume} {499}},\ \bibinfo {pages}
  {312} (\bibinfo {year} {2013})}\BibitemShut {NoStop}%
\bibitem [{\citenamefont {Janvier}\ \emph {et~al.}(2015)\citenamefont
  {Janvier}, \citenamefont {Tosi}, \citenamefont {Bretheau}, \citenamefont
  {Girit}, \citenamefont {Stern}, \citenamefont {Bertet}, \citenamefont
  {Joyez}, \citenamefont {Vion}, \citenamefont {Esteve}, \citenamefont
  {Goffman}, \citenamefont {Pothier},\ and\ \citenamefont
  {Urbina}}]{Janvier:S15}%
  \BibitemOpen
  \bibfield  {author} {\bibinfo {author} {\bibfnamefont {C.}~\bibnamefont
  {Janvier}}, \bibinfo {author} {\bibfnamefont {L.}~\bibnamefont {Tosi}},
  \bibinfo {author} {\bibfnamefont {L.}~\bibnamefont {Bretheau}}, \bibinfo
  {author} {\bibfnamefont {{\c C}.~{\"O}.}\ \bibnamefont {Girit}}, \bibinfo
  {author} {\bibfnamefont {M.}~\bibnamefont {Stern}}, \bibinfo {author}
  {\bibfnamefont {P.}~\bibnamefont {Bertet}}, \bibinfo {author} {\bibfnamefont
  {P.}~\bibnamefont {Joyez}}, \bibinfo {author} {\bibfnamefont
  {D.}~\bibnamefont {Vion}}, \bibinfo {author} {\bibfnamefont {D.}~\bibnamefont
  {Esteve}}, \bibinfo {author} {\bibfnamefont {M.~F.}\ \bibnamefont {Goffman}},
  \bibinfo {author} {\bibfnamefont {H.}~\bibnamefont {Pothier}}, \ and\
  \bibinfo {author} {\bibfnamefont {C.}~\bibnamefont {Urbina}},\ }\href
  {\doibase 10.1126/science.aab2179} {\bibfield  {journal} {\bibinfo  {journal}
  {Science}\ }\textbf {\bibinfo {volume} {349}},\ \bibinfo {pages} {1199}
  (\bibinfo {year} {2015})}\BibitemShut {NoStop}%
\bibitem [{\citenamefont {Pillet}\ \emph {et~al.}(2010)\citenamefont {Pillet},
  \citenamefont {Quay}, \citenamefont {Morfin}, \citenamefont {Bena},
  \citenamefont {Levy~Yeyati},\ and\ \citenamefont {Joyez}}]{Pillet:NP10}%
  \BibitemOpen
  \bibfield  {author} {\bibinfo {author} {\bibfnamefont {J.-D.}\ \bibnamefont
  {Pillet}}, \bibinfo {author} {\bibfnamefont {C.~H.~L.}\ \bibnamefont {Quay}},
  \bibinfo {author} {\bibfnamefont {P.}~\bibnamefont {Morfin}}, \bibinfo
  {author} {\bibfnamefont {C.}~\bibnamefont {Bena}}, \bibinfo {author}
  {\bibfnamefont {A.}~\bibnamefont {Levy~Yeyati}}, \ and\ \bibinfo {author}
  {\bibfnamefont {P.}~\bibnamefont {Joyez}},\ }\href {\doibase
  10.1038/nphys1811} {\bibfield  {journal} {\bibinfo  {journal} {Nat. Phys.}\
  }\textbf {\bibinfo {volume} {6}},\ \bibinfo {pages} {965} (\bibinfo {year}
  {2010})}\BibitemShut {NoStop}%
\bibitem [{\citenamefont {Eichler}\ \emph {et~al.}(2007)\citenamefont
  {Eichler}, \citenamefont {Weiss}, \citenamefont {Oberholzer}, \citenamefont
  {Sch{\"o}nenberger}, \citenamefont {Levy~Yeyati}, \citenamefont {Cuevas},\
  and\ \citenamefont {Mart{\'\i}n-Rodero}}]{Eichler:PRL07}%
  \BibitemOpen
  \bibfield  {author} {\bibinfo {author} {\bibfnamefont {A.}~\bibnamefont
  {Eichler}}, \bibinfo {author} {\bibfnamefont {M.}~\bibnamefont {Weiss}},
  \bibinfo {author} {\bibfnamefont {S.}~\bibnamefont {Oberholzer}}, \bibinfo
  {author} {\bibfnamefont {C.}~\bibnamefont {Sch{\"o}nenberger}}, \bibinfo
  {author} {\bibfnamefont {A.}~\bibnamefont {Levy~Yeyati}}, \bibinfo {author}
  {\bibfnamefont {J.~C.}\ \bibnamefont {Cuevas}}, \ and\ \bibinfo {author}
  {\bibfnamefont {A.}~\bibnamefont {Mart{\'\i}n-Rodero}},\ }\href
  {http://dx.doi.org/10.1103/PhysRevLett.99.126602} {\bibfield  {journal}
  {\bibinfo  {journal} {Phys. Rev. Lett.}\ }\textbf {\bibinfo {volume} {99}}
  (\bibinfo {year} {2007})}\BibitemShut {NoStop}%
\bibitem [{\citenamefont {Dirks}\ \emph {et~al.}(2011)\citenamefont {Dirks},
  \citenamefont {Hughes}, \citenamefont {Lal}, \citenamefont {Uchoa},
  \citenamefont {Chen}, \citenamefont {Chialvo}, \citenamefont {Goldbart},\
  and\ \citenamefont {Mason}}]{Dirks:NP11}%
  \BibitemOpen
  \bibfield  {author} {\bibinfo {author} {\bibfnamefont {T.}~\bibnamefont
  {Dirks}}, \bibinfo {author} {\bibfnamefont {T.~L.}\ \bibnamefont {Hughes}},
  \bibinfo {author} {\bibfnamefont {S.}~\bibnamefont {Lal}}, \bibinfo {author}
  {\bibfnamefont {B.}~\bibnamefont {Uchoa}}, \bibinfo {author} {\bibfnamefont
  {Y.-F.}\ \bibnamefont {Chen}}, \bibinfo {author} {\bibfnamefont
  {C.}~\bibnamefont {Chialvo}}, \bibinfo {author} {\bibfnamefont {P.~M.}\
  \bibnamefont {Goldbart}}, \ and\ \bibinfo {author} {\bibfnamefont
  {N.}~\bibnamefont {Mason}},\ }\href {http://dx.doi.org/10.1038/nphys1911}
  {\bibfield  {journal} {\bibinfo  {journal} {Nat. Phys.}\ }\textbf {\bibinfo
  {volume} {7}},\ \bibinfo {pages} {386} (\bibinfo {year} {2011})}\BibitemShut
  {NoStop}%
\bibitem [{\citenamefont {Deacon}\ \emph {et~al.}(2010)\citenamefont {Deacon},
  \citenamefont {Tanaka}, \citenamefont {Oiwa}, \citenamefont {Sakano},
  \citenamefont {Yoshida}, \citenamefont {Shibata}, \citenamefont {Hirakawa},\
  and\ \citenamefont {Tarucha}}]{Deacon:PRL10}%
  \BibitemOpen
  \bibfield  {author} {\bibinfo {author} {\bibfnamefont {R.~S.}\ \bibnamefont
  {Deacon}}, \bibinfo {author} {\bibfnamefont {Y.}~\bibnamefont {Tanaka}},
  \bibinfo {author} {\bibfnamefont {A.}~\bibnamefont {Oiwa}}, \bibinfo {author}
  {\bibfnamefont {R.}~\bibnamefont {Sakano}}, \bibinfo {author} {\bibfnamefont
  {K.}~\bibnamefont {Yoshida}}, \bibinfo {author} {\bibfnamefont
  {K.}~\bibnamefont {Shibata}}, \bibinfo {author} {\bibfnamefont
  {K.}~\bibnamefont {Hirakawa}}, \ and\ \bibinfo {author} {\bibfnamefont
  {S.}~\bibnamefont {Tarucha}},\ }\href {\doibase
  10.1103/PhysRevLett.104.076805} {\bibfield  {journal} {\bibinfo  {journal}
  {Phys. Rev. Lett.}\ }\textbf {\bibinfo {volume} {104}},\ \bibinfo {pages}
  {076805} (\bibinfo {year} {2010})}\BibitemShut {NoStop}%
\bibitem [{\citenamefont {Estrada Salda\~na}\ \emph {et~al.}(2018)\citenamefont
  {Estrada Salda\~na}, \citenamefont {Vekris}, \citenamefont {Steffensen},
  \citenamefont {\ifmmode~\check{Z}\else \v{Z}\fi{}itko}, \citenamefont
  {Krogstrup}, \citenamefont {Paaske}, \citenamefont {Grove-Rasmussen},\ and\
  \citenamefont {Nyg\aa{}rd}}]{Estrada-Saldana:PRL18}%
  \BibitemOpen
  \bibfield  {author} {\bibinfo {author} {\bibfnamefont {J.~C.}\ \bibnamefont
  {Estrada Salda\~na}}, \bibinfo {author} {\bibfnamefont {A.}~\bibnamefont
  {Vekris}}, \bibinfo {author} {\bibfnamefont {G.}~\bibnamefont {Steffensen}},
  \bibinfo {author} {\bibfnamefont {R.}~\bibnamefont {\ifmmode~\check{Z}\else
  \v{Z}\fi{}itko}}, \bibinfo {author} {\bibfnamefont {P.}~\bibnamefont
  {Krogstrup}}, \bibinfo {author} {\bibfnamefont {J.}~\bibnamefont {Paaske}},
  \bibinfo {author} {\bibfnamefont {K.}~\bibnamefont {Grove-Rasmussen}}, \ and\
  \bibinfo {author} {\bibfnamefont {J.}~\bibnamefont {Nyg\aa{}rd}},\ }\href
  {\doibase 10.1103/PhysRevLett.121.257701} {\bibfield  {journal} {\bibinfo
  {journal} {Phys. Rev. Lett.}\ }\textbf {\bibinfo {volume} {121}},\ \bibinfo
  {pages} {257701} (\bibinfo {year} {2018})}\BibitemShut {NoStop}%
\bibitem [{\citenamefont {Hays}\ \emph {et~al.}(2018)\citenamefont {Hays},
  \citenamefont {de~Lange}, \citenamefont {Serniak}, \citenamefont {van
  Woerkom}, \citenamefont {Bouman}, \citenamefont {Krogstrup}, \citenamefont
  {Nyg\aa{}rd}, \citenamefont {Geresdi},\ and\ \citenamefont
  {Devoret}}]{Hays:PRL18}%
  \BibitemOpen
  \bibfield  {author} {\bibinfo {author} {\bibfnamefont {M.}~\bibnamefont
  {Hays}}, \bibinfo {author} {\bibfnamefont {G.}~\bibnamefont {de~Lange}},
  \bibinfo {author} {\bibfnamefont {K.}~\bibnamefont {Serniak}}, \bibinfo
  {author} {\bibfnamefont {D.~J.}\ \bibnamefont {van Woerkom}}, \bibinfo
  {author} {\bibfnamefont {D.}~\bibnamefont {Bouman}}, \bibinfo {author}
  {\bibfnamefont {P.}~\bibnamefont {Krogstrup}}, \bibinfo {author}
  {\bibfnamefont {J.}~\bibnamefont {Nyg\aa{}rd}}, \bibinfo {author}
  {\bibfnamefont {A.}~\bibnamefont {Geresdi}}, \ and\ \bibinfo {author}
  {\bibfnamefont {M.~H.}\ \bibnamefont {Devoret}},\ }\href {\doibase
  10.1103/PhysRevLett.121.047001} {\bibfield  {journal} {\bibinfo  {journal}
  {Phys. Rev. Lett.}\ }\textbf {\bibinfo {volume} {121}},\ \bibinfo {pages}
  {047001} (\bibinfo {year} {2018})}\BibitemShut {NoStop}%
\bibitem [{\citenamefont {Plugge}\ \emph {et~al.}(2017)\citenamefont {Plugge},
  \citenamefont {Rasmussen}, \citenamefont {Egger},\ and\ \citenamefont
  {Flensberg}}]{Plugge:NJP17}%
  \BibitemOpen
  \bibfield  {author} {\bibinfo {author} {\bibfnamefont {S.}~\bibnamefont
  {Plugge}}, \bibinfo {author} {\bibfnamefont {A.}~\bibnamefont {Rasmussen}},
  \bibinfo {author} {\bibfnamefont {R.}~\bibnamefont {Egger}}, \ and\ \bibinfo
  {author} {\bibfnamefont {K.}~\bibnamefont {Flensberg}},\ }\href {\doibase
  https://doi.org/10.1088/1367-2630/aa54e1} {\bibfield  {journal} {\bibinfo
  {journal} {New J. Phys.}\ }\textbf {\bibinfo {volume} {19}},\ \bibinfo
  {pages} {012001} (\bibinfo {year} {2017})}\BibitemShut {NoStop}%
\bibitem [{\citenamefont {Fulga}\ \emph {et~al.}(2013)\citenamefont {Fulga},
  \citenamefont {Haim}, \citenamefont {Akhmerov},\ and\ \citenamefont
  {Oreg}}]{Fulga:NJOP13}%
  \BibitemOpen
  \bibfield  {author} {\bibinfo {author} {\bibfnamefont {I.~C.}\ \bibnamefont
  {Fulga}}, \bibinfo {author} {\bibfnamefont {A.}~\bibnamefont {Haim}},
  \bibinfo {author} {\bibfnamefont {A.~R.}\ \bibnamefont {Akhmerov}}, \ and\
  \bibinfo {author} {\bibfnamefont {Y.}~\bibnamefont {Oreg}},\ }\href {\doibase
  10.1088/1367-2630/15/4/045020} {\bibfield  {journal} {\bibinfo  {journal}
  {New J. Phys.}\ }\textbf {\bibinfo {volume} {15}},\ \bibinfo {pages} {045020}
  (\bibinfo {year} {2013})}\BibitemShut {NoStop}%
\bibitem [{\citenamefont {K{\"u}mmel}(1969)}]{Kummel:ZPA69}%
  \BibitemOpen
  \bibfield  {author} {\bibinfo {author} {\bibfnamefont {R.}~\bibnamefont
  {K{\"u}mmel}},\ }\href {\doibase 10.1007/BF01392426} {\bibfield  {journal}
  {\bibinfo  {journal} {Z. Phys. A}\ }\textbf {\bibinfo {volume} {218}},\
  \bibinfo {pages} {472} (\bibinfo {year} {1969})}\BibitemShut {NoStop}%
\bibitem [{\citenamefont {Xiang}\ \emph {et~al.}(2006)\citenamefont {Xiang},
  \citenamefont {Vidan}, \citenamefont {Tinkham}, \citenamefont {Westervelt},\
  and\ \citenamefont {Lieber}}]{Xiang:NN06}%
  \BibitemOpen
  \bibfield  {author} {\bibinfo {author} {\bibfnamefont {J.}~\bibnamefont
  {Xiang}}, \bibinfo {author} {\bibfnamefont {A.}~\bibnamefont {Vidan}},
  \bibinfo {author} {\bibfnamefont {M.}~\bibnamefont {Tinkham}}, \bibinfo
  {author} {\bibfnamefont {R.~M.}\ \bibnamefont {Westervelt}}, \ and\ \bibinfo
  {author} {\bibfnamefont {C.~M.}\ \bibnamefont {Lieber}},\ }\href {\doibase
  10.1038/nnano.2006.140} {\bibfield  {journal} {\bibinfo  {journal} {Nature
  Nanotechnol.}\ }\textbf {\bibinfo {volume} {1}},\ \bibinfo {pages} {208}
  (\bibinfo {year} {2006})}\BibitemShut {NoStop}%
\bibitem [{\citenamefont {Ridderbos}\ \emph {et~al.}(2019)\citenamefont
  {Ridderbos}, \citenamefont {Brauns}, \citenamefont {Li}, \citenamefont
  {Bakkers}, \citenamefont {Brinkman}, \citenamefont {van~der Wiel},\ and\
  \citenamefont {Zwanenburg}}]{Ridderbos:PRM19}%
  \BibitemOpen
  \bibfield  {author} {\bibinfo {author} {\bibfnamefont {J.}~\bibnamefont
  {Ridderbos}}, \bibinfo {author} {\bibfnamefont {M.}~\bibnamefont {Brauns}},
  \bibinfo {author} {\bibfnamefont {A.}~\bibnamefont {Li}}, \bibinfo {author}
  {\bibfnamefont {E.~P. A.~M.}\ \bibnamefont {Bakkers}}, \bibinfo {author}
  {\bibfnamefont {A.}~\bibnamefont {Brinkman}}, \bibinfo {author}
  {\bibfnamefont {W.~G.}\ \bibnamefont {van~der Wiel}}, \ and\ \bibinfo
  {author} {\bibfnamefont {F.~A.}\ \bibnamefont {Zwanenburg}},\ }\href
  {\doibase 10.1103/PhysRevMaterials.3.084803} {\bibfield  {journal} {\bibinfo
  {journal} {Phys. Rev. Mater.}\ }\textbf {\bibinfo {volume} {3}},\ \bibinfo
  {pages} {084803} (\bibinfo {year} {2019})}\BibitemShut {NoStop}%
\bibitem [{\citenamefont {Jespersen}\ \emph {et~al.}(2009)\citenamefont
  {Jespersen}, \citenamefont {Polianski}, \citenamefont {S{\o}rensen},
  \citenamefont {Flensberg},\ and\ \citenamefont
  {Nyg{\aa}rd}}]{Jespersen:NJP09}%
  \BibitemOpen
  \bibfield  {author} {\bibinfo {author} {\bibfnamefont {T.~S.}\ \bibnamefont
  {Jespersen}}, \bibinfo {author} {\bibfnamefont {M.~L.}\ \bibnamefont
  {Polianski}}, \bibinfo {author} {\bibfnamefont {C.~B.}\ \bibnamefont
  {S{\o}rensen}}, \bibinfo {author} {\bibfnamefont {K.}~\bibnamefont
  {Flensberg}}, \ and\ \bibinfo {author} {\bibfnamefont {J.}~\bibnamefont
  {Nyg{\aa}rd}},\ }\href {\doibase 10.1088/1367-2630/11/11/113025} {\bibfield
  {journal} {\bibinfo  {journal} {New J. Phys.}\ }\textbf {\bibinfo {volume}
  {11}},\ \bibinfo {pages} {113025} (\bibinfo {year} {2009})}\BibitemShut
  {NoStop}%
\bibitem [{\citenamefont {Doh}\ \emph {et~al.}(2005)\citenamefont {Doh},
  \citenamefont {van Dam}, \citenamefont {Roest}, \citenamefont {Bakkers},
  \citenamefont {Kouwenhoven},\ and\ \citenamefont {De~Franceschi}}]{Doh:S05}%
  \BibitemOpen
  \bibfield  {author} {\bibinfo {author} {\bibfnamefont {Y.-J.}\ \bibnamefont
  {Doh}}, \bibinfo {author} {\bibfnamefont {J.~A.}\ \bibnamefont {van Dam}},
  \bibinfo {author} {\bibfnamefont {A.~L.}\ \bibnamefont {Roest}}, \bibinfo
  {author} {\bibfnamefont {E.~P. A.~M.}\ \bibnamefont {Bakkers}}, \bibinfo
  {author} {\bibfnamefont {L.~P.}\ \bibnamefont {Kouwenhoven}}, \ and\ \bibinfo
  {author} {\bibfnamefont {S.}~\bibnamefont {De~Franceschi}},\ }\href {\doibase
  10.1126/science.1113523} {\bibfield  {journal} {\bibinfo  {journal}
  {Science}\ }\textbf {\bibinfo {volume} {309}},\ \bibinfo {pages} {272}
  (\bibinfo {year} {2005})}\BibitemShut {NoStop}%
\bibitem [{\citenamefont {G{\"u}nel}\ \emph {et~al.}(2012)\citenamefont
  {G{\"u}nel}, \citenamefont {Batov}, \citenamefont {Hardtdegen}, \citenamefont
  {Sladek}, \citenamefont {Winden}, \citenamefont {Weis}, \citenamefont
  {Panaitov}, \citenamefont {Gr{\"u}tzmacher},\ and\ \citenamefont
  {Sch{\"a}pers}}]{Gunel:JAP12}%
  \BibitemOpen
  \bibfield  {author} {\bibinfo {author} {\bibfnamefont {H.~Y.}\ \bibnamefont
  {G{\"u}nel}}, \bibinfo {author} {\bibfnamefont {I.~E.}\ \bibnamefont
  {Batov}}, \bibinfo {author} {\bibfnamefont {H.}~\bibnamefont {Hardtdegen}},
  \bibinfo {author} {\bibfnamefont {K.}~\bibnamefont {Sladek}}, \bibinfo
  {author} {\bibfnamefont {A.}~\bibnamefont {Winden}}, \bibinfo {author}
  {\bibfnamefont {K.}~\bibnamefont {Weis}}, \bibinfo {author} {\bibfnamefont
  {G.}~\bibnamefont {Panaitov}}, \bibinfo {author} {\bibfnamefont
  {D.}~\bibnamefont {Gr{\"u}tzmacher}}, \ and\ \bibinfo {author} {\bibfnamefont
  {T.}~\bibnamefont {Sch{\"a}pers}},\ }\href {\doibase 10.1063/1.4745024}
  {\bibfield  {journal} {\bibinfo  {journal} {J. App. Phys.}\ }\textbf
  {\bibinfo {volume} {112}},\ \bibinfo {pages} {034316} (\bibinfo {year}
  {2012})}\BibitemShut {NoStop}%
\bibitem [{\citenamefont {Goffman}\ \emph {et~al.}(2017)\citenamefont
  {Goffman}, \citenamefont {Urbina}, \citenamefont {Pothier}, \citenamefont
  {Nyg\aa{}rd}, \citenamefont {Marcus},\ and\ \citenamefont
  {Krogstrup}}]{Goffman:NJP17}%
  \BibitemOpen
  \bibfield  {author} {\bibinfo {author} {\bibfnamefont {M.~F.}\ \bibnamefont
  {Goffman}}, \bibinfo {author} {\bibfnamefont {C.}~\bibnamefont {Urbina}},
  \bibinfo {author} {\bibfnamefont {H.}~\bibnamefont {Pothier}}, \bibinfo
  {author} {\bibfnamefont {J.}~\bibnamefont {Nyg\aa{}rd}}, \bibinfo {author}
  {\bibfnamefont {C.~M.}\ \bibnamefont {Marcus}}, \ and\ \bibinfo {author}
  {\bibfnamefont {P.}~\bibnamefont {Krogstrup}},\ }\href {\doibase
  10.1088/1367-2630/aa7641} {\bibfield  {journal} {\bibinfo  {journal} {New J.
  Phys.}\ }\textbf {\bibinfo {volume} {19}},\ \bibinfo {pages} {092002}
  (\bibinfo {year} {2017})}\BibitemShut {NoStop}%
\bibitem [{\citenamefont {Nilsson}\ \emph {et~al.}(2012)\citenamefont
  {Nilsson}, \citenamefont {Samuelsson}, \citenamefont {Caroff},\ and\
  \citenamefont {Xu}}]{Nilsson:NL12}%
  \BibitemOpen
  \bibfield  {author} {\bibinfo {author} {\bibfnamefont {H.~A.}\ \bibnamefont
  {Nilsson}}, \bibinfo {author} {\bibfnamefont {P.}~\bibnamefont {Samuelsson}},
  \bibinfo {author} {\bibfnamefont {P.}~\bibnamefont {Caroff}}, \ and\ \bibinfo
  {author} {\bibfnamefont {H.~Q.}\ \bibnamefont {Xu}},\ }\href {\doibase
  10.1021/nl203380w} {\bibfield  {journal} {\bibinfo  {journal} {Nano Lett.}\
  }\textbf {\bibinfo {volume} {12}},\ \bibinfo {pages} {228} (\bibinfo {year}
  {2012})}\BibitemShut {NoStop}%
\bibitem [{\citenamefont {Deng}\ \emph {et~al.}(2012)\citenamefont {Deng},
  \citenamefont {Yu}, \citenamefont {Huang}, \citenamefont {Larsson},
  \citenamefont {Caroff},\ and\ \citenamefont {Xu}}]{Deng:NL12}%
  \BibitemOpen
  \bibfield  {author} {\bibinfo {author} {\bibfnamefont {M.~T.}\ \bibnamefont
  {Deng}}, \bibinfo {author} {\bibfnamefont {C.~L.}\ \bibnamefont {Yu}},
  \bibinfo {author} {\bibfnamefont {G.~Y.}\ \bibnamefont {Huang}}, \bibinfo
  {author} {\bibfnamefont {M.}~\bibnamefont {Larsson}}, \bibinfo {author}
  {\bibfnamefont {P.}~\bibnamefont {Caroff}}, \ and\ \bibinfo {author}
  {\bibfnamefont {H.~Q.}\ \bibnamefont {Xu}},\ }\href {\doibase
  10.1021/nl303758w} {\bibfield  {journal} {\bibinfo  {journal} {Nano Lett.}\
  }\textbf {\bibinfo {volume} {12}},\ \bibinfo {pages} {6414} (\bibinfo {year}
  {2012})}\BibitemShut {NoStop}%
\bibitem [{\citenamefont {Beenakker}(1992{\natexlab{a}})}]{Beenakker:92}%
  \BibitemOpen
  \bibfield  {author} {\bibinfo {author} {\bibfnamefont {C.}~\bibnamefont
  {Beenakker}},\ }in\ \href@noop {} {\emph {\bibinfo {booktitle} {Transport
  phenomena in mesoscopic systems: proceedings of the 14th Taniguchi symposium,
  Shima, Japan, November 10-14, 1991}}}\ (\bibinfo {organization}
  {Springer-Verlag},\ \bibinfo {year} {1992})\BibitemShut {NoStop}%
\bibitem [{\citenamefont {Beenakker}(1992{\natexlab{b}})}]{Beenakker:PRB92}%
  \BibitemOpen
  \bibfield  {author} {\bibinfo {author} {\bibfnamefont {C.~W.~J.}\
  \bibnamefont {Beenakker}},\ }\href {\doibase 10.1103/PhysRevB.46.12841}
  {\bibfield  {journal} {\bibinfo  {journal} {Phys. Rev. B}\ }\textbf {\bibinfo
  {volume} {46}},\ \bibinfo {pages} {12841} (\bibinfo {year}
  {1992}{\natexlab{b}})}\BibitemShut {NoStop}%
\bibitem [{\citenamefont {Likharev}(1979)}]{Likharev:RMP79}%
  \BibitemOpen
  \bibfield  {author} {\bibinfo {author} {\bibfnamefont {K.~K.}\ \bibnamefont
  {Likharev}},\ }\href {\doibase 10.1103/RevModPhys.51.101} {\bibfield
  {journal} {\bibinfo  {journal} {Rev. Mod. Phys.}\ }\textbf {\bibinfo {volume}
  {51}},\ \bibinfo {pages} {101} (\bibinfo {year} {1979})}\BibitemShut
  {NoStop}%
\bibitem [{\citenamefont {Furusaki}\ and\ \citenamefont
  {Tsukada}(1991)}]{Furusaki:PRB91}%
  \BibitemOpen
  \bibfield  {author} {\bibinfo {author} {\bibfnamefont {A.}~\bibnamefont
  {Furusaki}}\ and\ \bibinfo {author} {\bibfnamefont {M.}~\bibnamefont
  {Tsukada}},\ }\href {\doibase 10.1103/PhysRevB.43.10164} {\bibfield
  {journal} {\bibinfo  {journal} {Phys. Rev. B}\ }\textbf {\bibinfo {volume}
  {43}},\ \bibinfo {pages} {10164} (\bibinfo {year} {1991})}\BibitemShut
  {NoStop}%
\bibitem [{\citenamefont {Beenakker}\ and\ \citenamefont {van
  Houten}(1991)}]{Beenakker:PRL91}%
  \BibitemOpen
  \bibfield  {author} {\bibinfo {author} {\bibfnamefont {C.~W.~J.}\
  \bibnamefont {Beenakker}}\ and\ \bibinfo {author} {\bibfnamefont
  {H.}~\bibnamefont {van Houten}},\ }\href {\doibase
  10.1103/PhysRevLett.66.3056} {\bibfield  {journal} {\bibinfo  {journal}
  {Phys. Rev. Lett.}\ }\textbf {\bibinfo {volume} {66}},\ \bibinfo {pages}
  {3056} (\bibinfo {year} {1991})}\BibitemShut {NoStop}%
\bibitem [{\citenamefont {Bagwell}(1992)}]{Bagwell:PRB92}%
  \BibitemOpen
  \bibfield  {author} {\bibinfo {author} {\bibfnamefont {P.~F.}\ \bibnamefont
  {Bagwell}},\ }\href {\doibase 10.1103/PhysRevB.46.12573} {\bibfield
  {journal} {\bibinfo  {journal} {Phys. Rev. B}\ }\textbf {\bibinfo {volume}
  {46}},\ \bibinfo {pages} {12573} (\bibinfo {year} {1992})}\BibitemShut
  {NoStop}%
\bibitem [{\citenamefont {Furusaki}(1999)}]{Furusaki:SM99}%
  \BibitemOpen
  \bibfield  {author} {\bibinfo {author} {\bibfnamefont {A.}~\bibnamefont
  {Furusaki}},\ }\href {\doibase https://doi.org/10.1006/spmi.1999.0730}
  {\bibfield  {journal} {\bibinfo  {journal} {Superlattices Microstruct.}\
  }\textbf {\bibinfo {volume} {25}},\ \bibinfo {pages} {809 } (\bibinfo {year}
  {1999})}\BibitemShut {NoStop}%
\bibitem [{\citenamefont {Landauer}(1981)}]{Landauer:PLA81}%
  \BibitemOpen
  \bibfield  {author} {\bibinfo {author} {\bibfnamefont {R.}~\bibnamefont
  {Landauer}},\ }\href {\doibase
  http://dx.doi.org/10.1016/0375-9601(81)90230-9} {\bibfield  {journal}
  {\bibinfo  {journal} {Phys. Lett. A}\ }\textbf {\bibinfo {volume} {85}},\
  \bibinfo {pages} {91 } (\bibinfo {year} {1981})}\BibitemShut {NoStop}%
\bibitem [{\citenamefont {Josephson}(1962)}]{Josephson:PL62}%
  \BibitemOpen
  \bibfield  {author} {\bibinfo {author} {\bibfnamefont {B.~D.}\ \bibnamefont
  {Josephson}},\ }\href {\doibase 10.1016/0031-9163(62)91369-0} {\bibfield
  {journal} {\bibinfo  {journal} {Phys. Lett.}\ }\textbf {\bibinfo {volume}
  {1}},\ \bibinfo {pages} {251} (\bibinfo {year} {1962})}\BibitemShut {NoStop}%
\bibitem [{\citenamefont {Josephson}(1965)}]{Josephson:AP65}%
  \BibitemOpen
  \bibfield  {author} {\bibinfo {author} {\bibfnamefont {B.~D.}\ \bibnamefont
  {Josephson}},\ }\href {https://doi.org/10.1080/00018736500101091} {\bibfield
  {journal} {\bibinfo  {journal} {Adv. Phys.}\ }\textbf {\bibinfo {volume}
  {14}},\ \bibinfo {pages} {419} (\bibinfo {year} {1965})}\BibitemShut
  {NoStop}%
\bibitem [{\citenamefont {Kos}\ \emph {et~al.}(2013)\citenamefont {Kos},
  \citenamefont {Nigg},\ and\ \citenamefont {Glazman}}]{Kos:PRB13}%
  \BibitemOpen
  \bibfield  {author} {\bibinfo {author} {\bibfnamefont {F.}~\bibnamefont
  {Kos}}, \bibinfo {author} {\bibfnamefont {S.~E.}\ \bibnamefont {Nigg}}, \
  and\ \bibinfo {author} {\bibfnamefont {L.~I.}\ \bibnamefont {Glazman}},\
  }\href {\doibase 10.1103/PhysRevB.87.174521} {\bibfield  {journal} {\bibinfo
  {journal} {Phys. Rev. B}\ }\textbf {\bibinfo {volume} {87}},\ \bibinfo
  {pages} {174521} (\bibinfo {year} {2013})}\BibitemShut {NoStop}%
\bibitem [{\citenamefont {Hofheinz}\ \emph {et~al.}(2011)\citenamefont
  {Hofheinz}, \citenamefont {Portier}, \citenamefont {Baudouin}, \citenamefont
  {Joyez}, \citenamefont {Vion}, \citenamefont {Bertet}, \citenamefont
  {Roche},\ and\ \citenamefont {Esteve}}]{Hofheinz:PRL11}%
  \BibitemOpen
  \bibfield  {author} {\bibinfo {author} {\bibfnamefont {M.}~\bibnamefont
  {Hofheinz}}, \bibinfo {author} {\bibfnamefont {F.}~\bibnamefont {Portier}},
  \bibinfo {author} {\bibfnamefont {Q.}~\bibnamefont {Baudouin}}, \bibinfo
  {author} {\bibfnamefont {P.}~\bibnamefont {Joyez}}, \bibinfo {author}
  {\bibfnamefont {D.}~\bibnamefont {Vion}}, \bibinfo {author} {\bibfnamefont
  {P.}~\bibnamefont {Bertet}}, \bibinfo {author} {\bibfnamefont
  {P.}~\bibnamefont {Roche}}, \ and\ \bibinfo {author} {\bibfnamefont
  {D.}~\bibnamefont {Esteve}},\ }\href {\doibase
  10.1103/PhysRevLett.106.217005} {\bibfield  {journal} {\bibinfo  {journal}
  {Phys. Rev. Lett.}\ }\textbf {\bibinfo {volume} {106}},\ \bibinfo {pages}
  {217005} (\bibinfo {year} {2011})}\BibitemShut {NoStop}%
\bibitem [{\citenamefont {Holst}\ \emph {et~al.}(1994)\citenamefont {Holst},
  \citenamefont {Esteve}, \citenamefont {Urbina},\ and\ \citenamefont
  {Devoret}}]{Holst:PRL94}%
  \BibitemOpen
  \bibfield  {author} {\bibinfo {author} {\bibfnamefont {T.}~\bibnamefont
  {Holst}}, \bibinfo {author} {\bibfnamefont {D.}~\bibnamefont {Esteve}},
  \bibinfo {author} {\bibfnamefont {C.}~\bibnamefont {Urbina}}, \ and\ \bibinfo
  {author} {\bibfnamefont {M.~H.}\ \bibnamefont {Devoret}},\ }\href {\doibase
  10.1103/PhysRevLett.73.3455} {\bibfield  {journal} {\bibinfo  {journal}
  {Phys. Rev. Lett.}\ }\textbf {\bibinfo {volume} {73}},\ \bibinfo {pages}
  {3455} (\bibinfo {year} {1994})}\BibitemShut {NoStop}%
\bibitem [{\citenamefont {Blais}\ \emph {et~al.}(2004)\citenamefont {Blais},
  \citenamefont {Huang}, \citenamefont {Wallraff}, \citenamefont {Girvin},\
  and\ \citenamefont {Schoelkopf}}]{Blais:PRA04}%
  \BibitemOpen
  \bibfield  {author} {\bibinfo {author} {\bibfnamefont {A.}~\bibnamefont
  {Blais}}, \bibinfo {author} {\bibfnamefont {R.-S.}\ \bibnamefont {Huang}},
  \bibinfo {author} {\bibfnamefont {A.}~\bibnamefont {Wallraff}}, \bibinfo
  {author} {\bibfnamefont {S.~M.}\ \bibnamefont {Girvin}}, \ and\ \bibinfo
  {author} {\bibfnamefont {R.~J.}\ \bibnamefont {Schoelkopf}},\ }\href
  {\doibase 10.1103/PhysRevA.69.062320} {\bibfield  {journal} {\bibinfo
  {journal} {Phys. Rev. A}\ }\textbf {\bibinfo {volume} {69}},\ \bibinfo
  {pages} {062320} (\bibinfo {year} {2004})}\BibitemShut {NoStop}%
\bibitem [{\citenamefont {Car}\ \emph {et~al.}(2017)\citenamefont {Car},
  \citenamefont {Conesa-Boj}, \citenamefont {Zhang}, \citenamefont {Op~het
  Veld}, \citenamefont {de~Moor}, \citenamefont {Fadaly}, \citenamefont
  {G{\"u}l}, \citenamefont {K{\"o}lling}, \citenamefont {Plissard},
  \citenamefont {Toresen}, \citenamefont {Wimmer}, \citenamefont {Watanabe},
  \citenamefont {Taniguchi}, \citenamefont {Kouwenhoven},\ and\ \citenamefont
  {Bakkers}}]{Car:NL17}%
  \BibitemOpen
  \bibfield  {author} {\bibinfo {author} {\bibfnamefont {D.}~\bibnamefont
  {Car}}, \bibinfo {author} {\bibfnamefont {S.}~\bibnamefont {Conesa-Boj}},
  \bibinfo {author} {\bibfnamefont {H.}~\bibnamefont {Zhang}}, \bibinfo
  {author} {\bibfnamefont {R.~L.~M.}\ \bibnamefont {Op~het Veld}}, \bibinfo
  {author} {\bibfnamefont {M.~W.~A.}\ \bibnamefont {de~Moor}}, \bibinfo
  {author} {\bibfnamefont {E.~M.~T.}\ \bibnamefont {Fadaly}}, \bibinfo {author}
  {\bibfnamefont {{\"O}.}~\bibnamefont {G{\"u}l}}, \bibinfo {author}
  {\bibfnamefont {S.}~\bibnamefont {K{\"o}lling}}, \bibinfo {author}
  {\bibfnamefont {S.~R.}\ \bibnamefont {Plissard}}, \bibinfo {author}
  {\bibfnamefont {V.}~\bibnamefont {Toresen}}, \bibinfo {author} {\bibfnamefont
  {M.~T.}\ \bibnamefont {Wimmer}}, \bibinfo {author} {\bibfnamefont
  {K.}~\bibnamefont {Watanabe}}, \bibinfo {author} {\bibfnamefont
  {T.}~\bibnamefont {Taniguchi}}, \bibinfo {author} {\bibfnamefont {L.~P.}\
  \bibnamefont {Kouwenhoven}}, \ and\ \bibinfo {author} {\bibfnamefont {E.~P.
  A.~M.}\ \bibnamefont {Bakkers}},\ }\href {\doibase
  10.1021/acs.nanolett.6b03835} {\bibfield  {journal} {\bibinfo  {journal}
  {Nano Lett.}\ }\textbf {\bibinfo {volume} {17}},\ \bibinfo {pages} {721}
  (\bibinfo {year} {2017})}\BibitemShut {NoStop}%
\bibitem [{\citenamefont {Chang}\ \emph {et~al.}(2015)\citenamefont {Chang},
  \citenamefont {Albrecht}, \citenamefont {Jespersen}, \citenamefont
  {Kuemmeth}, \citenamefont {Krogstrup}, \citenamefont {Nyg{\aa}rd},\ and\
  \citenamefont {Marcus}}]{Chang:NN15}%
  \BibitemOpen
  \bibfield  {author} {\bibinfo {author} {\bibfnamefont {W.}~\bibnamefont
  {Chang}}, \bibinfo {author} {\bibfnamefont {S.~M.}\ \bibnamefont {Albrecht}},
  \bibinfo {author} {\bibfnamefont {T.~S.}\ \bibnamefont {Jespersen}}, \bibinfo
  {author} {\bibfnamefont {F.}~\bibnamefont {Kuemmeth}}, \bibinfo {author}
  {\bibfnamefont {P.}~\bibnamefont {Krogstrup}}, \bibinfo {author}
  {\bibfnamefont {J.}~\bibnamefont {Nyg{\aa}rd}}, \ and\ \bibinfo {author}
  {\bibfnamefont {C.~M.}\ \bibnamefont {Marcus}},\ }\href
  {http://dx.doi.org/10.1038/nnano.2014.306} {\bibfield  {journal} {\bibinfo
  {journal} {Nat. Nanotechnol.}\ }\textbf {\bibinfo {volume} {10}},\ \bibinfo
  {pages} {232} (\bibinfo {year} {2015})}\BibitemShut {NoStop}%
\bibitem [{\citenamefont {Anselmetti}\ \emph {et~al.}(2019)\citenamefont
  {Anselmetti}, \citenamefont {Martinez}, \citenamefont {M\'enard},
  \citenamefont {Puglia}, \citenamefont {Malinowski}, \citenamefont {Lee},
  \citenamefont {Choi}, \citenamefont {Pendharkar}, \citenamefont
  {Palmstr\o{}m}, \citenamefont {Marcus}, \citenamefont {Casparis},\ and\
  \citenamefont {Higginbotham}}]{Anselmetti:PRB19}%
  \BibitemOpen
  \bibfield  {author} {\bibinfo {author} {\bibfnamefont {G.~L.~R.}\
  \bibnamefont {Anselmetti}}, \bibinfo {author} {\bibfnamefont {E.~A.}\
  \bibnamefont {Martinez}}, \bibinfo {author} {\bibfnamefont {G.~C.}\
  \bibnamefont {M\'enard}}, \bibinfo {author} {\bibfnamefont {D.}~\bibnamefont
  {Puglia}}, \bibinfo {author} {\bibfnamefont {F.~K.}\ \bibnamefont
  {Malinowski}}, \bibinfo {author} {\bibfnamefont {J.~S.}\ \bibnamefont {Lee}},
  \bibinfo {author} {\bibfnamefont {S.}~\bibnamefont {Choi}}, \bibinfo {author}
  {\bibfnamefont {M.}~\bibnamefont {Pendharkar}}, \bibinfo {author}
  {\bibfnamefont {C.~J.}\ \bibnamefont {Palmstr\o{}m}}, \bibinfo {author}
  {\bibfnamefont {C.~M.}\ \bibnamefont {Marcus}}, \bibinfo {author}
  {\bibfnamefont {L.}~\bibnamefont {Casparis}}, \ and\ \bibinfo {author}
  {\bibfnamefont {A.~P.}\ \bibnamefont {Higginbotham}},\ }\href {\doibase
  10.1103/PhysRevB.100.205412} {\bibfield  {journal} {\bibinfo  {journal}
  {Phys. Rev. B}\ }\textbf {\bibinfo {volume} {100}},\ \bibinfo {pages}
  {205412} (\bibinfo {year} {2019})}\BibitemShut {NoStop}%
\bibitem [{\citenamefont {Spanton}\ \emph {et~al.}(2017)\citenamefont
  {Spanton}, \citenamefont {Deng}, \citenamefont {Vaitiek{\.e}nas},
  \citenamefont {Krogstrup}, \citenamefont {Nyg\aa{}rd}, \citenamefont
  {Marcus},\ and\ \citenamefont {Moler}}]{Spanton:NP17}%
  \BibitemOpen
  \bibfield  {author} {\bibinfo {author} {\bibfnamefont {E.~M.}\ \bibnamefont
  {Spanton}}, \bibinfo {author} {\bibfnamefont {M.}~\bibnamefont {Deng}},
  \bibinfo {author} {\bibfnamefont {S.}~\bibnamefont {Vaitiek{\.e}nas}},
  \bibinfo {author} {\bibfnamefont {P.}~\bibnamefont {Krogstrup}}, \bibinfo
  {author} {\bibfnamefont {J.}~\bibnamefont {Nyg\aa{}rd}}, \bibinfo {author}
  {\bibfnamefont {C.~M.}\ \bibnamefont {Marcus}}, \ and\ \bibinfo {author}
  {\bibfnamefont {K.~A.}\ \bibnamefont {Moler}},\ }\href
  {https://doi.org/10.1038/nphys4224} {\bibfield  {journal} {\bibinfo
  {journal} {Nat. Phys.}\ }\textbf {\bibinfo {volume} {13}},\ \bibinfo {pages}
  {1177} (\bibinfo {year} {2017})}\BibitemShut {NoStop}%
\bibitem [{\citenamefont {Hart}\ \emph {et~al.}(2019)\citenamefont {Hart},
  \citenamefont {Cui}, \citenamefont {M\'enard}, \citenamefont {Deng},
  \citenamefont {Antipov}, \citenamefont {Lutchyn}, \citenamefont {Krogstrup},
  \citenamefont {Marcus},\ and\ \citenamefont {Moler}}]{Hart:PRB19}%
  \BibitemOpen
  \bibfield  {author} {\bibinfo {author} {\bibfnamefont {S.}~\bibnamefont
  {Hart}}, \bibinfo {author} {\bibfnamefont {Z.}~\bibnamefont {Cui}}, \bibinfo
  {author} {\bibfnamefont {G.}~\bibnamefont {M\'enard}}, \bibinfo {author}
  {\bibfnamefont {M.}~\bibnamefont {Deng}}, \bibinfo {author} {\bibfnamefont
  {A.~E.}\ \bibnamefont {Antipov}}, \bibinfo {author} {\bibfnamefont {R.~M.}\
  \bibnamefont {Lutchyn}}, \bibinfo {author} {\bibfnamefont {P.}~\bibnamefont
  {Krogstrup}}, \bibinfo {author} {\bibfnamefont {C.~M.}\ \bibnamefont
  {Marcus}}, \ and\ \bibinfo {author} {\bibfnamefont {K.~A.}\ \bibnamefont
  {Moler}},\ }\href {\doibase 10.1103/PhysRevB.100.064523} {\bibfield
  {journal} {\bibinfo  {journal} {Phys. Rev. B}\ }\textbf {\bibinfo {volume}
  {100}},\ \bibinfo {pages} {064523} (\bibinfo {year} {2019})}\BibitemShut
  {NoStop}%
\bibitem [{\citenamefont {Rifkin}\ and\ \citenamefont
  {Deaver}(1976)}]{Rifkin:PRB76}%
  \BibitemOpen
  \bibfield  {author} {\bibinfo {author} {\bibfnamefont {R.}~\bibnamefont
  {Rifkin}}\ and\ \bibinfo {author} {\bibfnamefont {B.~S.}\ \bibnamefont
  {Deaver}},\ }\href {\doibase 10.1103/PhysRevB.13.3894} {\bibfield  {journal}
  {\bibinfo  {journal} {Phys. Rev. B}\ }\textbf {\bibinfo {volume} {13}},\
  \bibinfo {pages} {3894} (\bibinfo {year} {1976})}\BibitemShut {NoStop}%
\bibitem [{\citenamefont {Jellinggaard}\ \emph {et~al.}(2016)\citenamefont
  {Jellinggaard}, \citenamefont {Grove-Rasmussen}, \citenamefont {Madsen},\
  and\ \citenamefont {Nyg\aa{}rd}}]{Jellinggaard:PRB16}%
  \BibitemOpen
  \bibfield  {author} {\bibinfo {author} {\bibfnamefont {A.}~\bibnamefont
  {Jellinggaard}}, \bibinfo {author} {\bibfnamefont {K.}~\bibnamefont
  {Grove-Rasmussen}}, \bibinfo {author} {\bibfnamefont {M.~H.}\ \bibnamefont
  {Madsen}}, \ and\ \bibinfo {author} {\bibfnamefont {J.}~\bibnamefont
  {Nyg\aa{}rd}},\ }\href {\doibase 10.1103/PhysRevB.94.064520} {\bibfield
  {journal} {\bibinfo  {journal} {Phys. Rev. B}\ }\textbf {\bibinfo {volume}
  {94}},\ \bibinfo {pages} {064520} (\bibinfo {year} {2016})}\BibitemShut
  {NoStop}%
\bibitem [{\citenamefont {Chang}\ \emph {et~al.}(2013)\citenamefont {Chang},
  \citenamefont {Manucharyan}, \citenamefont {Jespersen}, \citenamefont
  {Nyg\aa{}rd},\ and\ \citenamefont {Marcus}}]{Chang:PRL13}%
  \BibitemOpen
  \bibfield  {author} {\bibinfo {author} {\bibfnamefont {W.}~\bibnamefont
  {Chang}}, \bibinfo {author} {\bibfnamefont {V.~E.}\ \bibnamefont
  {Manucharyan}}, \bibinfo {author} {\bibfnamefont {T.~S.}\ \bibnamefont
  {Jespersen}}, \bibinfo {author} {\bibfnamefont {J.}~\bibnamefont
  {Nyg\aa{}rd}}, \ and\ \bibinfo {author} {\bibfnamefont {C.~M.}\ \bibnamefont
  {Marcus}},\ }\href {\doibase 10.1103/PhysRevLett.110.217005} {\bibfield
  {journal} {\bibinfo  {journal} {Phys. Rev. Lett.}\ }\textbf {\bibinfo
  {volume} {110}},\ \bibinfo {pages} {217005} (\bibinfo {year}
  {2013})}\BibitemShut {NoStop}%
\bibitem [{\citenamefont {Zitko}\ \emph {et~al.}(2015)\citenamefont {Zitko},
  \citenamefont {Lim}, \citenamefont {L\'opez},\ and\ \citenamefont
  {Aguado}}]{Zitko:PRB15}%
  \BibitemOpen
  \bibfield  {author} {\bibinfo {author} {\bibfnamefont {R.}~\bibnamefont
  {Zitko}}, \bibinfo {author} {\bibfnamefont {J.~S.}\ \bibnamefont {Lim}},
  \bibinfo {author} {\bibfnamefont {R.}~\bibnamefont {L\'opez}}, \ and\
  \bibinfo {author} {\bibfnamefont {R.}~\bibnamefont {Aguado}},\ }\href
  {\doibase 10.1103/PhysRevB.91.045441} {\bibfield  {journal} {\bibinfo
  {journal} {Phys. Rev. B}\ }\textbf {\bibinfo {volume} {91}},\ \bibinfo
  {pages} {045441} (\bibinfo {year} {2015})}\BibitemShut {NoStop}%
\bibitem [{\citenamefont {De~Franceschi}\ \emph {et~al.}(2010)\citenamefont
  {De~Franceschi}, \citenamefont {Kouwenhoven}, \citenamefont
  {Sch{\"o}nenberger},\ and\ \citenamefont {Wernsdorfer}}]{De-Franceschi:NN10}%
  \BibitemOpen
  \bibfield  {author} {\bibinfo {author} {\bibfnamefont {S.}~\bibnamefont
  {De~Franceschi}}, \bibinfo {author} {\bibfnamefont {L.}~\bibnamefont
  {Kouwenhoven}}, \bibinfo {author} {\bibfnamefont {C.}~\bibnamefont
  {Sch{\"o}nenberger}}, \ and\ \bibinfo {author} {\bibfnamefont
  {W.}~\bibnamefont {Wernsdorfer}},\ }\href
  {https://doi.org/10.1038/nnano.2010.173} {\bibfield  {journal} {\bibinfo
  {journal} {Nat. Nanotechnol.}\ }\textbf {\bibinfo {volume} {5}},\ \bibinfo
  {pages} {703} (\bibinfo {year} {2010})}\BibitemShut {NoStop}%
\bibitem [{\citenamefont {Hewson}(1993)}]{Hewson:93}%
  \BibitemOpen
  \bibfield  {author} {\bibinfo {author} {\bibfnamefont {A.~C.}\ \bibnamefont
  {Hewson}},\ }\href {\doibase 10.1017/CBO9780511470752} {\emph {\bibinfo
  {title} {The Kondo Problem to Heavy Fermions}}},\ Cambridge Studies in
  Magnetism\ (\bibinfo  {publisher} {Cambridge University Press},\ \bibinfo
  {year} {1993})\BibitemShut {NoStop}%
\bibitem [{\citenamefont {Buitelaar}\ \emph {et~al.}(2002)\citenamefont
  {Buitelaar}, \citenamefont {Nussbaumer},\ and\ \citenamefont
  {Sch\"onenberger}}]{Buitelaar:PRL02}%
  \BibitemOpen
  \bibfield  {author} {\bibinfo {author} {\bibfnamefont {M.~R.}\ \bibnamefont
  {Buitelaar}}, \bibinfo {author} {\bibfnamefont {T.}~\bibnamefont
  {Nussbaumer}}, \ and\ \bibinfo {author} {\bibfnamefont {C.}~\bibnamefont
  {Sch\"onenberger}},\ }\href {\doibase 10.1103/PhysRevLett.89.256801}
  {\bibfield  {journal} {\bibinfo  {journal} {Phys. Rev. Lett.}\ }\textbf
  {\bibinfo {volume} {89}},\ \bibinfo {pages} {256801} (\bibinfo {year}
  {2002})}\BibitemShut {NoStop}%
\bibitem [{\citenamefont {Sand-Jespersen}\ \emph {et~al.}(2007)\citenamefont
  {Sand-Jespersen}, \citenamefont {Paaske}, \citenamefont {Andersen},
  \citenamefont {Grove-Rasmussen}, \citenamefont {J\o{}rgensen}, \citenamefont
  {Aagesen}, \citenamefont {S\o{}rensen}, \citenamefont {Lindelof},
  \citenamefont {Flensberg},\ and\ \citenamefont
  {Nyg\aa{}rd}}]{Sand-Jespersen:PRL07a}%
  \BibitemOpen
  \bibfield  {author} {\bibinfo {author} {\bibfnamefont {T.}~\bibnamefont
  {Sand-Jespersen}}, \bibinfo {author} {\bibfnamefont {J.}~\bibnamefont
  {Paaske}}, \bibinfo {author} {\bibfnamefont {B.~M.}\ \bibnamefont
  {Andersen}}, \bibinfo {author} {\bibfnamefont {K.}~\bibnamefont
  {Grove-Rasmussen}}, \bibinfo {author} {\bibfnamefont {H.~I.}\ \bibnamefont
  {J\o{}rgensen}}, \bibinfo {author} {\bibfnamefont {M.}~\bibnamefont
  {Aagesen}}, \bibinfo {author} {\bibfnamefont {C.~B.}\ \bibnamefont
  {S\o{}rensen}}, \bibinfo {author} {\bibfnamefont {P.~E.}\ \bibnamefont
  {Lindelof}}, \bibinfo {author} {\bibfnamefont {K.}~\bibnamefont {Flensberg}},
  \ and\ \bibinfo {author} {\bibfnamefont {J.}~\bibnamefont {Nyg\aa{}rd}},\
  }\href {\doibase 10.1103/PhysRevLett.99.126603} {\bibfield  {journal}
  {\bibinfo  {journal} {Phys. Rev. Lett.}\ }\textbf {\bibinfo {volume} {99}},\
  \bibinfo {pages} {126603} (\bibinfo {year} {2007})}\BibitemShut {NoStop}%
\bibitem [{\citenamefont {Grove-Rasmussen}\ \emph {et~al.}(2009)\citenamefont
  {Grove-Rasmussen}, \citenamefont {J\o{}rgensen}, \citenamefont {Andersen},
  \citenamefont {Paaske}, \citenamefont {Jespersen}, \citenamefont
  {Nyg\aa{}rd}, \citenamefont {Flensberg},\ and\ \citenamefont
  {Lindelof}}]{Grove-Rasmussen:PRB09}%
  \BibitemOpen
  \bibfield  {author} {\bibinfo {author} {\bibfnamefont {K.}~\bibnamefont
  {Grove-Rasmussen}}, \bibinfo {author} {\bibfnamefont {H.~I.}\ \bibnamefont
  {J\o{}rgensen}}, \bibinfo {author} {\bibfnamefont {B.~M.}\ \bibnamefont
  {Andersen}}, \bibinfo {author} {\bibfnamefont {J.}~\bibnamefont {Paaske}},
  \bibinfo {author} {\bibfnamefont {T.~S.}\ \bibnamefont {Jespersen}}, \bibinfo
  {author} {\bibfnamefont {J.}~\bibnamefont {Nyg\aa{}rd}}, \bibinfo {author}
  {\bibfnamefont {K.}~\bibnamefont {Flensberg}}, \ and\ \bibinfo {author}
  {\bibfnamefont {P.~E.}\ \bibnamefont {Lindelof}},\ }\href {\doibase
  10.1103/PhysRevB.79.134518} {\bibfield  {journal} {\bibinfo  {journal} {Phys.
  Rev. B}\ }\textbf {\bibinfo {volume} {79}},\ \bibinfo {pages} {134518}
  (\bibinfo {year} {2009})}\BibitemShut {NoStop}%
\bibitem [{\citenamefont {Kumar}\ \emph {et~al.}(2014)\citenamefont {Kumar},
  \citenamefont {Gaim}, \citenamefont {Steininger}, \citenamefont {Yeyati},
  \citenamefont {Mart\'{\i}n-Rodero}, \citenamefont {H\"uttel},\ and\
  \citenamefont {Strunk}}]{Kumar:PRB14}%
  \BibitemOpen
  \bibfield  {author} {\bibinfo {author} {\bibfnamefont {A.}~\bibnamefont
  {Kumar}}, \bibinfo {author} {\bibfnamefont {M.}~\bibnamefont {Gaim}},
  \bibinfo {author} {\bibfnamefont {D.}~\bibnamefont {Steininger}}, \bibinfo
  {author} {\bibfnamefont {A.~L.}\ \bibnamefont {Yeyati}}, \bibinfo {author}
  {\bibfnamefont {A.}~\bibnamefont {Mart\'{\i}n-Rodero}}, \bibinfo {author}
  {\bibfnamefont {A.~K.}\ \bibnamefont {H\"uttel}}, \ and\ \bibinfo {author}
  {\bibfnamefont {C.}~\bibnamefont {Strunk}},\ }\href {\doibase
  10.1103/PhysRevB.89.075428} {\bibfield  {journal} {\bibinfo  {journal} {Phys.
  Rev. B}\ }\textbf {\bibinfo {volume} {89}},\ \bibinfo {pages} {075428}
  (\bibinfo {year} {2014})}\BibitemShut {NoStop}%
\bibitem [{\citenamefont {Li}\ \emph {et~al.}(2017)\citenamefont {Li},
  \citenamefont {Kang}, \citenamefont {Caroff},\ and\ \citenamefont
  {Xu}}]{Li:PRB17}%
  \BibitemOpen
  \bibfield  {author} {\bibinfo {author} {\bibfnamefont {S.}~\bibnamefont
  {Li}}, \bibinfo {author} {\bibfnamefont {N.}~\bibnamefont {Kang}}, \bibinfo
  {author} {\bibfnamefont {P.}~\bibnamefont {Caroff}}, \ and\ \bibinfo {author}
  {\bibfnamefont {H.~Q.}\ \bibnamefont {Xu}},\ }\href {\doibase
  10.1103/PhysRevB.95.014515} {\bibfield  {journal} {\bibinfo  {journal} {Phys.
  Rev. B}\ }\textbf {\bibinfo {volume} {95}},\ \bibinfo {pages} {014515}
  (\bibinfo {year} {2017})}\BibitemShut {NoStop}%
\bibitem [{\citenamefont {Island}\ \emph {et~al.}(2017)\citenamefont {Island},
  \citenamefont {Gaudenzi}, \citenamefont {de~Bruijckere}, \citenamefont
  {Burzur\'{\i}}, \citenamefont {Franco}, \citenamefont {Mas-Torrent},
  \citenamefont {Rovira}, \citenamefont {Veciana}, \citenamefont {Klapwijk},
  \citenamefont {Aguado},\ and\ \citenamefont {van~der Zant}}]{Island:PRL17}%
  \BibitemOpen
  \bibfield  {author} {\bibinfo {author} {\bibfnamefont {J.~O.}\ \bibnamefont
  {Island}}, \bibinfo {author} {\bibfnamefont {R.}~\bibnamefont {Gaudenzi}},
  \bibinfo {author} {\bibfnamefont {J.}~\bibnamefont {de~Bruijckere}}, \bibinfo
  {author} {\bibfnamefont {E.}~\bibnamefont {Burzur\'{\i}}}, \bibinfo {author}
  {\bibfnamefont {C.}~\bibnamefont {Franco}}, \bibinfo {author} {\bibfnamefont
  {M.}~\bibnamefont {Mas-Torrent}}, \bibinfo {author} {\bibfnamefont
  {C.}~\bibnamefont {Rovira}}, \bibinfo {author} {\bibfnamefont
  {J.}~\bibnamefont {Veciana}}, \bibinfo {author} {\bibfnamefont {T.~M.}\
  \bibnamefont {Klapwijk}}, \bibinfo {author} {\bibfnamefont {R.}~\bibnamefont
  {Aguado}}, \ and\ \bibinfo {author} {\bibfnamefont {H.~S.~J.}\ \bibnamefont
  {van~der Zant}},\ }\href {\doibase 10.1103/PhysRevLett.118.117001} {\bibfield
   {journal} {\bibinfo  {journal} {Phys. Rev. Lett.}\ }\textbf {\bibinfo
  {volume} {118}},\ \bibinfo {pages} {117001} (\bibinfo {year}
  {2017})}\BibitemShut {NoStop}%
\bibitem [{\citenamefont {Andersen}\ \emph {et~al.}(2011)\citenamefont
  {Andersen}, \citenamefont {Flensberg}, \citenamefont {Koerting},\ and\
  \citenamefont {Paaske}}]{Andersen:PRL11a}%
  \BibitemOpen
  \bibfield  {author} {\bibinfo {author} {\bibfnamefont {B.~M.}\ \bibnamefont
  {Andersen}}, \bibinfo {author} {\bibfnamefont {K.}~\bibnamefont {Flensberg}},
  \bibinfo {author} {\bibfnamefont {V.}~\bibnamefont {Koerting}}, \ and\
  \bibinfo {author} {\bibfnamefont {J.}~\bibnamefont {Paaske}},\ }\href
  {\doibase 10.1103/PhysRevLett.107.256802} {\bibfield  {journal} {\bibinfo
  {journal} {Phys. Rev. Lett.}\ }\textbf {\bibinfo {volume} {107}},\ \bibinfo
  {pages} {256802} (\bibinfo {year} {2011})}\BibitemShut {NoStop}%
\bibitem [{\citenamefont {Su}\ \emph {et~al.}(2017)\citenamefont {Su},
  \citenamefont {Tacla}, \citenamefont {Hocevar}, \citenamefont {Car},
  \citenamefont {Plissard}, \citenamefont {Bakkers}, \citenamefont {Daley},
  \citenamefont {Pekker},\ and\ \citenamefont {Frolov}}]{Su:NC17}%
  \BibitemOpen
  \bibfield  {author} {\bibinfo {author} {\bibfnamefont {Z.}~\bibnamefont
  {Su}}, \bibinfo {author} {\bibfnamefont {A.~B.}\ \bibnamefont {Tacla}},
  \bibinfo {author} {\bibfnamefont {M.}~\bibnamefont {Hocevar}}, \bibinfo
  {author} {\bibfnamefont {D.}~\bibnamefont {Car}}, \bibinfo {author}
  {\bibfnamefont {S.~R.}\ \bibnamefont {Plissard}}, \bibinfo {author}
  {\bibfnamefont {E.~P. A.~M.}\ \bibnamefont {Bakkers}}, \bibinfo {author}
  {\bibfnamefont {A.~J.}\ \bibnamefont {Daley}}, \bibinfo {author}
  {\bibfnamefont {D.}~\bibnamefont {Pekker}}, \ and\ \bibinfo {author}
  {\bibfnamefont {S.~M.}\ \bibnamefont {Frolov}},\ }\href {\doibase
  10.1038/s41467-017-00665-7} {\bibfield  {journal} {\bibinfo  {journal} {Nat.
  Commun.}\ }\textbf {\bibinfo {volume} {8}},\ \bibinfo {pages} {585} (\bibinfo
  {year} {2017})}\BibitemShut {NoStop}%
\bibitem [{\citenamefont {Salda{\~n}a}\ \emph {et~al.}(2018)\citenamefont
  {Salda{\~n}a}, \citenamefont {Vekris}, \citenamefont {{\v Z}itko},
  \citenamefont {Steffensen}, \citenamefont {Krogstrup}, \citenamefont
  {Paaske}, \citenamefont {Grove-Rasmussen},\ and\ \citenamefont
  {Nyg{\aa}rd}}]{Saldana:A18}%
  \BibitemOpen
  \bibfield  {author} {\bibinfo {author} {\bibfnamefont {J.~C.~E.}\
  \bibnamefont {Salda{\~n}a}}, \bibinfo {author} {\bibfnamefont
  {A.}~\bibnamefont {Vekris}}, \bibinfo {author} {\bibfnamefont
  {R.}~\bibnamefont {{\v Z}itko}}, \bibinfo {author} {\bibfnamefont
  {G.}~\bibnamefont {Steffensen}}, \bibinfo {author} {\bibfnamefont
  {P.}~\bibnamefont {Krogstrup}}, \bibinfo {author} {\bibfnamefont
  {J.}~\bibnamefont {Paaske}}, \bibinfo {author} {\bibfnamefont
  {K.}~\bibnamefont {Grove-Rasmussen}}, \ and\ \bibinfo {author} {\bibfnamefont
  {J.}~\bibnamefont {Nyg{\aa}rd}},\ }\href {https://arxiv.org/abs/1812.09303}
  {\bibfield  {journal} {\bibinfo  {journal} {arXiv:1812.09303}\ } (\bibinfo
  {year} {2018})}\BibitemShut {NoStop}%
\bibitem [{\citenamefont {Heinrich}\ \emph {et~al.}(2018)\citenamefont
  {Heinrich}, \citenamefont {Pascual},\ and\ \citenamefont
  {Franke}}]{Heinrich:PSS18}%
  \BibitemOpen
  \bibfield  {author} {\bibinfo {author} {\bibfnamefont {B.~W.}\ \bibnamefont
  {Heinrich}}, \bibinfo {author} {\bibfnamefont {J.~I.}\ \bibnamefont
  {Pascual}}, \ and\ \bibinfo {author} {\bibfnamefont {K.~J.}\ \bibnamefont
  {Franke}},\ }\href {\doibase https://doi.org/10.1016/j.progsurf.2018.01.001}
  {\bibfield  {journal} {\bibinfo  {journal} {Prog. Surf. Sci.}\ }\textbf
  {\bibinfo {volume} {93}},\ \bibinfo {pages} {1 } (\bibinfo {year}
  {2018})}\BibitemShut {NoStop}%
\bibitem [{\citenamefont {Chen}\ \emph {et~al.}(2019)\citenamefont {Chen},
  \citenamefont {Woods}, \citenamefont {Yu}, \citenamefont {Hocevar},
  \citenamefont {Car}, \citenamefont {Plissard}, \citenamefont {Bakkers},
  \citenamefont {Stanescu},\ and\ \citenamefont {Frolov}}]{Chen:PRL19}%
  \BibitemOpen
  \bibfield  {author} {\bibinfo {author} {\bibfnamefont {J.}~\bibnamefont
  {Chen}}, \bibinfo {author} {\bibfnamefont {B.~D.}\ \bibnamefont {Woods}},
  \bibinfo {author} {\bibfnamefont {P.}~\bibnamefont {Yu}}, \bibinfo {author}
  {\bibfnamefont {M.}~\bibnamefont {Hocevar}}, \bibinfo {author} {\bibfnamefont
  {D.}~\bibnamefont {Car}}, \bibinfo {author} {\bibfnamefont {S.~R.}\
  \bibnamefont {Plissard}}, \bibinfo {author} {\bibfnamefont {E.~P. A.~M.}\
  \bibnamefont {Bakkers}}, \bibinfo {author} {\bibfnamefont {T.~D.}\
  \bibnamefont {Stanescu}}, \ and\ \bibinfo {author} {\bibfnamefont {S.~M.}\
  \bibnamefont {Frolov}},\ }\href {\doibase 10.1103/PhysRevLett.123.107703}
  {\bibfield  {journal} {\bibinfo  {journal} {Phys. Rev. Lett.}\ }\textbf
  {\bibinfo {volume} {123}},\ \bibinfo {pages} {107703} (\bibinfo {year}
  {2019})}\BibitemShut {NoStop}%
\bibitem [{\citenamefont {van Dam}\ \emph {et~al.}(2006)\citenamefont {van
  Dam}, \citenamefont {Nazarov}, \citenamefont {Bakkers}, \citenamefont
  {De~Franceschi},\ and\ \citenamefont {Kouwenhoven}}]{Dam:N06}%
  \BibitemOpen
  \bibfield  {author} {\bibinfo {author} {\bibfnamefont {J.~A.}\ \bibnamefont
  {van Dam}}, \bibinfo {author} {\bibfnamefont {Y.~V.}\ \bibnamefont
  {Nazarov}}, \bibinfo {author} {\bibfnamefont {E.~P. A.~M.}\ \bibnamefont
  {Bakkers}}, \bibinfo {author} {\bibfnamefont {S.}~\bibnamefont
  {De~Franceschi}}, \ and\ \bibinfo {author} {\bibfnamefont {L.~P.}\
  \bibnamefont {Kouwenhoven}},\ }\href {https://doi.org/10.1038/nature05018}
  {\bibfield  {journal} {\bibinfo  {journal} {Nature}\ }\textbf {\bibinfo
  {volume} {442}},\ \bibinfo {pages} {667} (\bibinfo {year}
  {2006})}\BibitemShut {NoStop}%
\bibitem [{\citenamefont {Delagrange}\ \emph {et~al.}(2015)\citenamefont
  {Delagrange}, \citenamefont {Luitz}, \citenamefont {Weil}, \citenamefont
  {Kasumov}, \citenamefont {Meden}, \citenamefont {Bouchiat},\ and\
  \citenamefont {Deblock}}]{Delagrange:PRB15}%
  \BibitemOpen
  \bibfield  {author} {\bibinfo {author} {\bibfnamefont {R.}~\bibnamefont
  {Delagrange}}, \bibinfo {author} {\bibfnamefont {D.~J.}\ \bibnamefont
  {Luitz}}, \bibinfo {author} {\bibfnamefont {R.}~\bibnamefont {Weil}},
  \bibinfo {author} {\bibfnamefont {A.}~\bibnamefont {Kasumov}}, \bibinfo
  {author} {\bibfnamefont {V.}~\bibnamefont {Meden}}, \bibinfo {author}
  {\bibfnamefont {H.}~\bibnamefont {Bouchiat}}, \ and\ \bibinfo {author}
  {\bibfnamefont {R.}~\bibnamefont {Deblock}},\ }\href {\doibase
  10.1103/PhysRevB.91.241401} {\bibfield  {journal} {\bibinfo  {journal} {Phys.
  Rev. B}\ }\textbf {\bibinfo {volume} {91}},\ \bibinfo {pages} {241401}
  (\bibinfo {year} {2015})}\BibitemShut {NoStop}%
\bibitem [{\citenamefont {Maurand}\ \emph {et~al.}(2012)\citenamefont
  {Maurand}, \citenamefont {Meng}, \citenamefont {Bonet}, \citenamefont
  {Florens}, \citenamefont {Marty},\ and\ \citenamefont
  {Wernsdorfer}}]{Maurand:PRX12}%
  \BibitemOpen
  \bibfield  {author} {\bibinfo {author} {\bibfnamefont {R.}~\bibnamefont
  {Maurand}}, \bibinfo {author} {\bibfnamefont {T.}~\bibnamefont {Meng}},
  \bibinfo {author} {\bibfnamefont {E.}~\bibnamefont {Bonet}}, \bibinfo
  {author} {\bibfnamefont {S.}~\bibnamefont {Florens}}, \bibinfo {author}
  {\bibfnamefont {L.}~\bibnamefont {Marty}}, \ and\ \bibinfo {author}
  {\bibfnamefont {W.}~\bibnamefont {Wernsdorfer}},\ }\href {\doibase
  10.1103/PhysRevX.2.011009} {\bibfield  {journal} {\bibinfo  {journal} {Phys.
  Rev. X}\ }\textbf {\bibinfo {volume} {2}},\ \bibinfo {pages} {011009}
  (\bibinfo {year} {2012})}\BibitemShut {NoStop}%
\bibitem [{\citenamefont {Estrada~Salda{\~n}a}\ \emph
  {et~al.}(2019)\citenamefont {Estrada~Salda{\~n}a}, \citenamefont {{\v
  Z}itko}, \citenamefont {Cleuziou}, \citenamefont {Lee}, \citenamefont
  {Zannier}, \citenamefont {Ercolani}, \citenamefont {Sorba}, \citenamefont
  {Aguado},\ and\ \citenamefont {De~Franceschi}}]{Estrada-Saldana:SA19}%
  \BibitemOpen
  \bibfield  {author} {\bibinfo {author} {\bibfnamefont {J.~C.}\ \bibnamefont
  {Estrada~Salda{\~n}a}}, \bibinfo {author} {\bibfnamefont {R.}~\bibnamefont
  {{\v Z}itko}}, \bibinfo {author} {\bibfnamefont {J.~P.}\ \bibnamefont
  {Cleuziou}}, \bibinfo {author} {\bibfnamefont {E.~J.~H.}\ \bibnamefont
  {Lee}}, \bibinfo {author} {\bibfnamefont {V.}~\bibnamefont {Zannier}},
  \bibinfo {author} {\bibfnamefont {D.}~\bibnamefont {Ercolani}}, \bibinfo
  {author} {\bibfnamefont {L.}~\bibnamefont {Sorba}}, \bibinfo {author}
  {\bibfnamefont {R.}~\bibnamefont {Aguado}}, \ and\ \bibinfo {author}
  {\bibfnamefont {S.}~\bibnamefont {De~Franceschi}},\ }\href
  {https://doi.org/10.1126/sciadv.aav1235} {\bibfield  {journal} {\bibinfo
  {journal} {Sci. Adv.}\ }\textbf {\bibinfo {volume} {5}} (\bibinfo {year}
  {2019})}\BibitemShut {NoStop}%
\bibitem [{\citenamefont {Deng}\ \emph {et~al.}(2014)\citenamefont {Deng},
  \citenamefont {Yu}, \citenamefont {Huang}, \citenamefont {Larsson},
  \citenamefont {Caroff},\ and\ \citenamefont {Xu}}]{Deng:SR14}%
  \BibitemOpen
  \bibfield  {author} {\bibinfo {author} {\bibfnamefont {M.~T.}\ \bibnamefont
  {Deng}}, \bibinfo {author} {\bibfnamefont {C.~L.}\ \bibnamefont {Yu}},
  \bibinfo {author} {\bibfnamefont {G.~Y.}\ \bibnamefont {Huang}}, \bibinfo
  {author} {\bibfnamefont {M.}~\bibnamefont {Larsson}}, \bibinfo {author}
  {\bibfnamefont {P.}~\bibnamefont {Caroff}}, \ and\ \bibinfo {author}
  {\bibfnamefont {H.~Q.}\ \bibnamefont {Xu}},\ }\href {\doibase
  10.1038/srep07261} {\bibfield  {journal} {\bibinfo  {journal} {Sci. Rep.}\
  }\textbf {\bibinfo {volume} {4}},\ \bibinfo {pages} {7261} (\bibinfo {year}
  {2014})}\BibitemShut {NoStop}%
\bibitem [{\citenamefont {Cheng}\ and\ \citenamefont
  {Lutchyn}(2012)}]{Cheng:PRB12}%
  \BibitemOpen
  \bibfield  {author} {\bibinfo {author} {\bibfnamefont {M.}~\bibnamefont
  {Cheng}}\ and\ \bibinfo {author} {\bibfnamefont {R.~M.}\ \bibnamefont
  {Lutchyn}},\ }\href {\doibase 10.1103/PhysRevB.86.134522} {\bibfield
  {journal} {\bibinfo  {journal} {Phys. Rev. B}\ }\textbf {\bibinfo {volume}
  {86}},\ \bibinfo {pages} {134522} (\bibinfo {year} {2012})}\BibitemShut
  {NoStop}%
\bibitem [{\citenamefont {Park}\ and\ \citenamefont
  {Levy~Yeyati}(2017)}]{Park:PRB17}%
  \BibitemOpen
  \bibfield  {author} {\bibinfo {author} {\bibfnamefont {S.}~\bibnamefont
  {Park}}\ and\ \bibinfo {author} {\bibfnamefont {A.}~\bibnamefont
  {Levy~Yeyati}},\ }\href {\doibase 10.1103/PhysRevB.96.125416} {\bibfield
  {journal} {\bibinfo  {journal} {Phys. Rev. B}\ }\textbf {\bibinfo {volume}
  {96}},\ \bibinfo {pages} {125416} (\bibinfo {year} {2017})}\BibitemShut
  {NoStop}%
\bibitem [{\citenamefont {van Heck}\ \emph {et~al.}(2017)\citenamefont {van
  Heck}, \citenamefont {V\"ayrynen},\ and\ \citenamefont
  {Glazman}}]{Heck:PRB17}%
  \BibitemOpen
  \bibfield  {author} {\bibinfo {author} {\bibfnamefont {B.}~\bibnamefont {van
  Heck}}, \bibinfo {author} {\bibfnamefont {J.~I.}\ \bibnamefont {V\"ayrynen}},
  \ and\ \bibinfo {author} {\bibfnamefont {L.~I.}\ \bibnamefont {Glazman}},\
  }\href {\doibase 10.1103/PhysRevB.96.075404} {\bibfield  {journal} {\bibinfo
  {journal} {Phys. Rev. B}\ }\textbf {\bibinfo {volume} {96}},\ \bibinfo
  {pages} {075404} (\bibinfo {year} {2017})}\BibitemShut {NoStop}%
\bibitem [{\citenamefont {Hays}\ \emph {et~al.}(2019)\citenamefont {Hays},
  \citenamefont {Fatemi}, \citenamefont {Serniak}, \citenamefont {Bouman},
  \citenamefont {Diamond}, \citenamefont {de~Lange}, \citenamefont {Krogstrup},
  \citenamefont {Nyg{\aa}rd}, \citenamefont {Geresdi},\ and\ \citenamefont
  {Devoret}}]{Hays:A19}%
  \BibitemOpen
  \bibfield  {author} {\bibinfo {author} {\bibfnamefont {M.}~\bibnamefont
  {Hays}}, \bibinfo {author} {\bibfnamefont {V.}~\bibnamefont {Fatemi}},
  \bibinfo {author} {\bibfnamefont {K.}~\bibnamefont {Serniak}}, \bibinfo
  {author} {\bibfnamefont {D.}~\bibnamefont {Bouman}}, \bibinfo {author}
  {\bibfnamefont {S.}~\bibnamefont {Diamond}}, \bibinfo {author} {\bibfnamefont
  {G.}~\bibnamefont {de~Lange}}, \bibinfo {author} {\bibfnamefont
  {P.}~\bibnamefont {Krogstrup}}, \bibinfo {author} {\bibfnamefont
  {J.}~\bibnamefont {Nyg{\aa}rd}}, \bibinfo {author} {\bibfnamefont
  {A.}~\bibnamefont {Geresdi}}, \ and\ \bibinfo {author} {\bibfnamefont
  {M.~H.}\ \bibnamefont {Devoret}},\ }\href
  {https://arxiv.org/pdf/1908.02800.pdf} {\bibfield  {journal} {\bibinfo
  {journal} {arXiv:1908.02800}\ } (\bibinfo {year} {2019})}\BibitemShut
  {NoStop}%
\bibitem [{\citenamefont {Kwon}\ \emph {et~al.}(2004)\citenamefont {Kwon},
  \citenamefont {Yakovenko},\ and\ \citenamefont {Sengupta}}]{Kwon:LTP04}%
  \BibitemOpen
  \bibfield  {author} {\bibinfo {author} {\bibfnamefont {H.-J.}\ \bibnamefont
  {Kwon}}, \bibinfo {author} {\bibfnamefont {V.~M.}\ \bibnamefont {Yakovenko}},
  \ and\ \bibinfo {author} {\bibfnamefont {K.}~\bibnamefont {Sengupta}},\
  }\href {https://doi.org/10.1063/1.1789931} {\bibfield  {journal} {\bibinfo
  {journal} {Low Temp. Phys.}\ }\textbf {\bibinfo {volume} {30}},\ \bibinfo
  {pages} {613} (\bibinfo {year} {2004})}\BibitemShut {NoStop}%
\bibitem [{\citenamefont {Pikulin}\ and\ \citenamefont
  {Nazarov}(2012)}]{Pikulin:PRB12}%
  \BibitemOpen
  \bibfield  {author} {\bibinfo {author} {\bibfnamefont {D.~I.}\ \bibnamefont
  {Pikulin}}\ and\ \bibinfo {author} {\bibfnamefont {Y.~V.}\ \bibnamefont
  {Nazarov}},\ }\href {\doibase 10.1103/PhysRevB.86.140504} {\bibfield
  {journal} {\bibinfo  {journal} {Phys. Rev. B}\ }\textbf {\bibinfo {volume}
  {86}},\ \bibinfo {pages} {140504} (\bibinfo {year} {2012})}\BibitemShut
  {NoStop}%
\bibitem [{\citenamefont {San-Jose}\ \emph {et~al.}(2012)\citenamefont
  {San-Jose}, \citenamefont {Prada},\ and\ \citenamefont
  {Aguado}}]{SanJose:PRL12}%
  \BibitemOpen
  \bibfield  {author} {\bibinfo {author} {\bibfnamefont {P.}~\bibnamefont
  {San-Jose}}, \bibinfo {author} {\bibfnamefont {E.}~\bibnamefont {Prada}}, \
  and\ \bibinfo {author} {\bibfnamefont {R.}~\bibnamefont {Aguado}},\ }\href
  {\doibase 10.1103/PhysRevLett.108.257001} {\bibfield  {journal} {\bibinfo
  {journal} {Phys. Rev. Lett.}\ }\textbf {\bibinfo {volume} {108}},\ \bibinfo
  {pages} {257001} (\bibinfo {year} {2012})}\BibitemShut {NoStop}%
\bibitem [{\citenamefont {Klinovaja}\ and\ \citenamefont
  {Loss}(2012)}]{Klinovaja:PRB12}%
  \BibitemOpen
  \bibfield  {author} {\bibinfo {author} {\bibfnamefont {J.}~\bibnamefont
  {Klinovaja}}\ and\ \bibinfo {author} {\bibfnamefont {D.}~\bibnamefont
  {Loss}},\ }\href {\doibase 10.1103/PhysRevB.86.085408} {\bibfield  {journal}
  {\bibinfo  {journal} {Phys. Rev. B}\ }\textbf {\bibinfo {volume} {86}},\
  \bibinfo {pages} {085408} (\bibinfo {year} {2012})}\BibitemShut {NoStop}%
\bibitem [{\citenamefont {Mishmash}\ \emph {et~al.}(2016)\citenamefont
  {Mishmash}, \citenamefont {Aasen}, \citenamefont {Higginbotham},\ and\
  \citenamefont {Alicea}}]{Mishmash:PRB16}%
  \BibitemOpen
  \bibfield  {author} {\bibinfo {author} {\bibfnamefont {R.~V.}\ \bibnamefont
  {Mishmash}}, \bibinfo {author} {\bibfnamefont {D.}~\bibnamefont {Aasen}},
  \bibinfo {author} {\bibfnamefont {A.~P.}\ \bibnamefont {Higginbotham}}, \
  and\ \bibinfo {author} {\bibfnamefont {J.}~\bibnamefont {Alicea}},\ }\href
  {\doibase 10.1103/PhysRevB.93.245404} {\bibfield  {journal} {\bibinfo
  {journal} {Phys. Rev. B}\ }\textbf {\bibinfo {volume} {93}},\ \bibinfo
  {pages} {245404} (\bibinfo {year} {2016})}\BibitemShut {NoStop}%
\bibitem [{\citenamefont {Das}\ \emph {et~al.}(2012)\citenamefont {Das},
  \citenamefont {Ronen}, \citenamefont {Most}, \citenamefont {Oreg},
  \citenamefont {Heiblum},\ and\ \citenamefont {Shtrikman}}]{Das:NP12}%
  \BibitemOpen
  \bibfield  {author} {\bibinfo {author} {\bibfnamefont {A.}~\bibnamefont
  {Das}}, \bibinfo {author} {\bibfnamefont {Y.}~\bibnamefont {Ronen}}, \bibinfo
  {author} {\bibfnamefont {Y.}~\bibnamefont {Most}}, \bibinfo {author}
  {\bibfnamefont {Y.}~\bibnamefont {Oreg}}, \bibinfo {author} {\bibfnamefont
  {M.}~\bibnamefont {Heiblum}}, \ and\ \bibinfo {author} {\bibfnamefont
  {H.}~\bibnamefont {Shtrikman}},\ }\href {https://doi.org/10.1038/nphys2479}
  {\bibfield  {journal} {\bibinfo  {journal} {Nat. Phys.}\ }\textbf {\bibinfo
  {volume} {8}},\ \bibinfo {pages} {887} (\bibinfo {year} {2012})}\BibitemShut
  {NoStop}%
\bibitem [{\citenamefont {Chen}\ \emph {et~al.}(2017)\citenamefont {Chen},
  \citenamefont {Yu}, \citenamefont {Stenger}, \citenamefont {Hocevar},
  \citenamefont {Car}, \citenamefont {Plissard}, \citenamefont {Bakkers},
  \citenamefont {Stanescu},\ and\ \citenamefont {Frolov}}]{Chen:SA17}%
  \BibitemOpen
  \bibfield  {author} {\bibinfo {author} {\bibfnamefont {J.}~\bibnamefont
  {Chen}}, \bibinfo {author} {\bibfnamefont {P.}~\bibnamefont {Yu}}, \bibinfo
  {author} {\bibfnamefont {J.}~\bibnamefont {Stenger}}, \bibinfo {author}
  {\bibfnamefont {M.}~\bibnamefont {Hocevar}}, \bibinfo {author} {\bibfnamefont
  {D.}~\bibnamefont {Car}}, \bibinfo {author} {\bibfnamefont {S.~R.}\
  \bibnamefont {Plissard}}, \bibinfo {author} {\bibfnamefont {E.~P. A.~M.}\
  \bibnamefont {Bakkers}}, \bibinfo {author} {\bibfnamefont {T.~D.}\
  \bibnamefont {Stanescu}}, \ and\ \bibinfo {author} {\bibfnamefont {S.~M.}\
  \bibnamefont {Frolov}},\ }\href
  {https://advances.sciencemag.org/content/3/9/e1701476} {\bibfield  {journal}
  {\bibinfo  {journal} {Sci. Adv.}\ }\textbf {\bibinfo {volume} {3}} (\bibinfo
  {year} {2017})}\BibitemShut {NoStop}%
\bibitem [{\citenamefont {G{\"u}l}\ \emph {et~al.}(2018)\citenamefont
  {G{\"u}l}, \citenamefont {Zhang}, \citenamefont {Bommer}, \citenamefont
  {de~Moor}, \citenamefont {Car}, \citenamefont {Plissard}, \citenamefont
  {Bakkers}, \citenamefont {Geresdi}, \citenamefont {Watanabe}, \citenamefont
  {Taniguchi},\ and\ \citenamefont {Kouwenhoven}}]{Gul:NN18}%
  \BibitemOpen
  \bibfield  {author} {\bibinfo {author} {\bibfnamefont {{\"O}.}~\bibnamefont
  {G{\"u}l}}, \bibinfo {author} {\bibfnamefont {H.}~\bibnamefont {Zhang}},
  \bibinfo {author} {\bibfnamefont {J.~D.~S.}\ \bibnamefont {Bommer}}, \bibinfo
  {author} {\bibfnamefont {M.~W.~A.}\ \bibnamefont {de~Moor}}, \bibinfo
  {author} {\bibfnamefont {D.}~\bibnamefont {Car}}, \bibinfo {author}
  {\bibfnamefont {S.~R.}\ \bibnamefont {Plissard}}, \bibinfo {author}
  {\bibfnamefont {E.~P. A.~M.}\ \bibnamefont {Bakkers}}, \bibinfo {author}
  {\bibfnamefont {A.}~\bibnamefont {Geresdi}}, \bibinfo {author} {\bibfnamefont
  {K.}~\bibnamefont {Watanabe}}, \bibinfo {author} {\bibfnamefont
  {T.}~\bibnamefont {Taniguchi}}, \ and\ \bibinfo {author} {\bibfnamefont
  {L.~P.}\ \bibnamefont {Kouwenhoven}},\ }\href {\doibase
  10.1038/s41565-017-0032-8} {\bibfield  {journal} {\bibinfo  {journal} {Nat.
  Nanotechnol.}\ }\textbf {\bibinfo {volume} {13}},\ \bibinfo {pages} {192}
  (\bibinfo {year} {2018})}\BibitemShut {NoStop}%
\bibitem [{\citenamefont {Grivnin}\ \emph {et~al.}(2019)\citenamefont
  {Grivnin}, \citenamefont {Bor}, \citenamefont {Heiblum}, \citenamefont
  {Oreg},\ and\ \citenamefont {Shtrikman}}]{Grivnin:NC19}%
  \BibitemOpen
  \bibfield  {author} {\bibinfo {author} {\bibfnamefont {A.}~\bibnamefont
  {Grivnin}}, \bibinfo {author} {\bibfnamefont {E.}~\bibnamefont {Bor}},
  \bibinfo {author} {\bibfnamefont {M.}~\bibnamefont {Heiblum}}, \bibinfo
  {author} {\bibfnamefont {Y.}~\bibnamefont {Oreg}}, \ and\ \bibinfo {author}
  {\bibfnamefont {H.}~\bibnamefont {Shtrikman}},\ }\href {\doibase
  10.1038/s41467-019-09771-0} {\bibfield  {journal} {\bibinfo  {journal} {Nat.
  Commun.}\ }\textbf {\bibinfo {volume} {10}},\ \bibinfo {pages} {1940}
  (\bibinfo {year} {2019})}\BibitemShut {NoStop}%
\bibitem [{\citenamefont {Law}\ \emph {et~al.}(2009)\citenamefont {Law},
  \citenamefont {Lee},\ and\ \citenamefont {Ng}}]{Law:PRL09}%
  \BibitemOpen
  \bibfield  {author} {\bibinfo {author} {\bibfnamefont {K.~T.}\ \bibnamefont
  {Law}}, \bibinfo {author} {\bibfnamefont {P.~A.}\ \bibnamefont {Lee}}, \ and\
  \bibinfo {author} {\bibfnamefont {T.~K.}\ \bibnamefont {Ng}},\ }\href
  {\doibase 10.1103/PhysRevLett.103.237001} {\bibfield  {journal} {\bibinfo
  {journal} {Phys. Rev. Lett.}\ }\textbf {\bibinfo {volume} {103}},\ \bibinfo
  {pages} {237001} (\bibinfo {year} {2009})}\BibitemShut {NoStop}%
\bibitem [{\citenamefont {Flensberg}(2010)}]{Flensberg:PRB10}%
  \BibitemOpen
  \bibfield  {author} {\bibinfo {author} {\bibfnamefont {K.}~\bibnamefont
  {Flensberg}},\ }\href {\doibase 10.1103/PhysRevB.82.180516} {\bibfield
  {journal} {\bibinfo  {journal} {Phys. Rev. B}\ }\textbf {\bibinfo {volume}
  {82}},\ \bibinfo {pages} {180516} (\bibinfo {year} {2010})}\BibitemShut
  {NoStop}%
\bibitem [{\citenamefont {Wimmer}\ \emph {et~al.}(2011)\citenamefont {Wimmer},
  \citenamefont {Akhmerov}, \citenamefont {Dahlhaus},\ and\ \citenamefont
  {Beenakker}}]{Wimmer:NJP11}%
  \BibitemOpen
  \bibfield  {author} {\bibinfo {author} {\bibfnamefont {M.}~\bibnamefont
  {Wimmer}}, \bibinfo {author} {\bibfnamefont {A.~R.}\ \bibnamefont
  {Akhmerov}}, \bibinfo {author} {\bibfnamefont {J.~P.}\ \bibnamefont
  {Dahlhaus}}, \ and\ \bibinfo {author} {\bibfnamefont {C.~W.~J.}\ \bibnamefont
  {Beenakker}},\ }\href {\doibase 10.1088/1367-2630/13/5/053016} {\bibfield
  {journal} {\bibinfo  {journal} {New J. Phys.}\ }\textbf {\bibinfo {volume}
  {13}},\ \bibinfo {pages} {053016} (\bibinfo {year} {2011})}\BibitemShut
  {NoStop}%
\bibitem [{\citenamefont {Setiawan}\ \emph {et~al.}(2017)\citenamefont
  {Setiawan}, \citenamefont {Liu}, \citenamefont {Sau},\ and\ \citenamefont
  {Das~Sarma}}]{Setiawan:PRB17}%
  \BibitemOpen
  \bibfield  {author} {\bibinfo {author} {\bibfnamefont {F.}~\bibnamefont
  {Setiawan}}, \bibinfo {author} {\bibfnamefont {C.-X.}\ \bibnamefont {Liu}},
  \bibinfo {author} {\bibfnamefont {J.~D.}\ \bibnamefont {Sau}}, \ and\
  \bibinfo {author} {\bibfnamefont {S.}~\bibnamefont {Das~Sarma}},\ }\href
  {\doibase 10.1103/PhysRevB.96.184520} {\bibfield  {journal} {\bibinfo
  {journal} {Phys. Rev. B}\ }\textbf {\bibinfo {volume} {96}},\ \bibinfo
  {pages} {184520} (\bibinfo {year} {2017})}\BibitemShut {NoStop}%
\bibitem [{\citenamefont {Albrecht}\ \emph {et~al.}(2016)\citenamefont
  {Albrecht}, \citenamefont {Higginbotham}, \citenamefont {Madsen},
  \citenamefont {Kuemmeth}, \citenamefont {Jespersen}, \citenamefont
  {Nyg{\aa}rd}, \citenamefont {Krogstrup},\ and\ \citenamefont
  {Marcus}}]{Albrecht:N16}%
  \BibitemOpen
  \bibfield  {author} {\bibinfo {author} {\bibfnamefont {S.~M.}\ \bibnamefont
  {Albrecht}}, \bibinfo {author} {\bibfnamefont {A.~P.}\ \bibnamefont
  {Higginbotham}}, \bibinfo {author} {\bibfnamefont {M.}~\bibnamefont
  {Madsen}}, \bibinfo {author} {\bibfnamefont {F.}~\bibnamefont {Kuemmeth}},
  \bibinfo {author} {\bibfnamefont {T.~S.}\ \bibnamefont {Jespersen}}, \bibinfo
  {author} {\bibfnamefont {J.}~\bibnamefont {Nyg{\aa}rd}}, \bibinfo {author}
  {\bibfnamefont {P.}~\bibnamefont {Krogstrup}}, \ and\ \bibinfo {author}
  {\bibfnamefont {C.~M.}\ \bibnamefont {Marcus}},\ }\href {\doibase
  10.1038/nature17162} {\bibfield  {journal} {\bibinfo  {journal} {Nature}\
  }\textbf {\bibinfo {volume} {531}},\ \bibinfo {pages} {206} (\bibinfo {year}
  {2016})}\BibitemShut {NoStop}%
\bibitem [{\citenamefont {Shen}\ \emph
  {et~al.}(2018{\natexlab{a}})\citenamefont {Shen}, \citenamefont {Heedt},
  \citenamefont {Borsoi}, \citenamefont {van Heck}, \citenamefont
  {Gazibegovic}, \citenamefont {Op~het Veld}, \citenamefont {Car},
  \citenamefont {Logan}, \citenamefont {Pendharkar}, \citenamefont {Ramakers},
  \citenamefont {Wang}, \citenamefont {Xu}, \citenamefont {Bouman},
  \citenamefont {Geresdi}, \citenamefont {Palmstr{\o}m}, \citenamefont
  {Bakkers},\ and\ \citenamefont {Kouwenhoven}}]{Shen:NC18}%
  \BibitemOpen
  \bibfield  {author} {\bibinfo {author} {\bibfnamefont {J.}~\bibnamefont
  {Shen}}, \bibinfo {author} {\bibfnamefont {S.}~\bibnamefont {Heedt}},
  \bibinfo {author} {\bibfnamefont {F.}~\bibnamefont {Borsoi}}, \bibinfo
  {author} {\bibfnamefont {B.}~\bibnamefont {van Heck}}, \bibinfo {author}
  {\bibfnamefont {S.}~\bibnamefont {Gazibegovic}}, \bibinfo {author}
  {\bibfnamefont {R.~L.~M.}\ \bibnamefont {Op~het Veld}}, \bibinfo {author}
  {\bibfnamefont {D.}~\bibnamefont {Car}}, \bibinfo {author} {\bibfnamefont
  {J.~A.}\ \bibnamefont {Logan}}, \bibinfo {author} {\bibfnamefont
  {M.}~\bibnamefont {Pendharkar}}, \bibinfo {author} {\bibfnamefont {S.~J.~J.}\
  \bibnamefont {Ramakers}}, \bibinfo {author} {\bibfnamefont {G.}~\bibnamefont
  {Wang}}, \bibinfo {author} {\bibfnamefont {D.}~\bibnamefont {Xu}}, \bibinfo
  {author} {\bibfnamefont {D.}~\bibnamefont {Bouman}}, \bibinfo {author}
  {\bibfnamefont {A.}~\bibnamefont {Geresdi}}, \bibinfo {author} {\bibfnamefont
  {C.~J.}\ \bibnamefont {Palmstr{\o}m}}, \bibinfo {author} {\bibfnamefont
  {E.~P. A.~M.}\ \bibnamefont {Bakkers}}, \ and\ \bibinfo {author}
  {\bibfnamefont {L.~P.}\ \bibnamefont {Kouwenhoven}},\ }\href {\doibase
  10.1038/s41467-018-07279-7} {\bibfield  {journal} {\bibinfo  {journal} {Nat.
  Commun.}\ }\textbf {\bibinfo {volume} {9}},\ \bibinfo {pages} {4801}
  (\bibinfo {year} {2018}{\natexlab{a}})}\BibitemShut {NoStop}%
\bibitem [{\citenamefont {Vaitiekenas}\ \emph {et~al.}(2020)\citenamefont
  {Vaitiekenas}, \citenamefont {Winkler}, \citenamefont {van Heck},
  \citenamefont {Karzig}, \citenamefont {Deng}, \citenamefont {Flensberg},
  \citenamefont {Glazman}, \citenamefont {Nayak}, \citenamefont {Krogstrup},
  \citenamefont {Lutchyn},\ and\ \citenamefont {Marcus}}]{Vaitiekenas:S20}%
  \BibitemOpen
  \bibfield  {author} {\bibinfo {author} {\bibfnamefont {S.}~\bibnamefont
  {Vaitiekenas}}, \bibinfo {author} {\bibfnamefont {G.~W.}\ \bibnamefont
  {Winkler}}, \bibinfo {author} {\bibfnamefont {B.}~\bibnamefont {van Heck}},
  \bibinfo {author} {\bibfnamefont {T.}~\bibnamefont {Karzig}}, \bibinfo
  {author} {\bibfnamefont {M.-T.}\ \bibnamefont {Deng}}, \bibinfo {author}
  {\bibfnamefont {K.}~\bibnamefont {Flensberg}}, \bibinfo {author}
  {\bibfnamefont {L.~I.}\ \bibnamefont {Glazman}}, \bibinfo {author}
  {\bibfnamefont {C.}~\bibnamefont {Nayak}}, \bibinfo {author} {\bibfnamefont
  {P.}~\bibnamefont {Krogstrup}}, \bibinfo {author} {\bibfnamefont {R.~M.}\
  \bibnamefont {Lutchyn}}, \ and\ \bibinfo {author} {\bibfnamefont {C.~M.}\
  \bibnamefont {Marcus}},\ }\href
  {https://science.sciencemag.org/content/367/6485/eaav3392} {\bibfield
  {journal} {\bibinfo  {journal} {Science}\ }\textbf {\bibinfo {volume} {367}}
  (\bibinfo {year} {2020})}\BibitemShut {NoStop}%
\bibitem [{\citenamefont {Laroche}\ \emph {et~al.}(2019)\citenamefont
  {Laroche}, \citenamefont {Bouman}, \citenamefont {van Woerkom}, \citenamefont
  {Proutski}, \citenamefont {Murthy}, \citenamefont {Pikulin}, \citenamefont
  {Nayak}, \citenamefont {van Gulik}, \citenamefont {Nyg\aa{}rd}, \citenamefont
  {Krogstrup}, \citenamefont {Kouwenhoven},\ and\ \citenamefont
  {Geresdi}}]{Laroche:NC19}%
  \BibitemOpen
  \bibfield  {author} {\bibinfo {author} {\bibfnamefont {D.}~\bibnamefont
  {Laroche}}, \bibinfo {author} {\bibfnamefont {D.}~\bibnamefont {Bouman}},
  \bibinfo {author} {\bibfnamefont {D.~J.}\ \bibnamefont {van Woerkom}},
  \bibinfo {author} {\bibfnamefont {A.}~\bibnamefont {Proutski}}, \bibinfo
  {author} {\bibfnamefont {C.}~\bibnamefont {Murthy}}, \bibinfo {author}
  {\bibfnamefont {D.~I.}\ \bibnamefont {Pikulin}}, \bibinfo {author}
  {\bibfnamefont {C.}~\bibnamefont {Nayak}}, \bibinfo {author} {\bibfnamefont
  {R.~J.~J.}\ \bibnamefont {van Gulik}}, \bibinfo {author} {\bibfnamefont
  {J.}~\bibnamefont {Nyg\aa{}rd}}, \bibinfo {author} {\bibfnamefont
  {P.}~\bibnamefont {Krogstrup}}, \bibinfo {author} {\bibfnamefont {L.~P.}\
  \bibnamefont {Kouwenhoven}}, \ and\ \bibinfo {author} {\bibfnamefont
  {A.}~\bibnamefont {Geresdi}},\ }\href
  {https://doi.org/10.1038/s41467-018-08161-2} {\bibfield  {journal} {\bibinfo
  {journal} {Nat. Commun.}\ }\textbf {\bibinfo {volume} {10}},\ \bibinfo
  {pages} {245} (\bibinfo {year} {2019})}\BibitemShut {NoStop}%
\bibitem [{\citenamefont {Rokhinson}\ \emph {et~al.}(2012)\citenamefont
  {Rokhinson}, \citenamefont {Liu},\ and\ \citenamefont
  {Furdyna}}]{Rokhinson:NP12}%
  \BibitemOpen
  \bibfield  {author} {\bibinfo {author} {\bibfnamefont {L.~P.}\ \bibnamefont
  {Rokhinson}}, \bibinfo {author} {\bibfnamefont {X.}~\bibnamefont {Liu}}, \
  and\ \bibinfo {author} {\bibfnamefont {J.~K.}\ \bibnamefont {Furdyna}},\
  }\href {http://dx.doi.org/10.1038/nphys2429} {\bibfield  {journal} {\bibinfo
  {journal} {Nat. Phys.}\ }\textbf {\bibinfo {volume} {8}},\ \bibinfo {pages}
  {795} (\bibinfo {year} {2012})}\BibitemShut {NoStop}%
\bibitem [{\citenamefont {van Heck}\ \emph {et~al.}(2011)\citenamefont {van
  Heck}, \citenamefont {Hassler}, \citenamefont {Akhmerov},\ and\ \citenamefont
  {Beenakker}}]{vanHeck:PRB11}%
  \BibitemOpen
  \bibfield  {author} {\bibinfo {author} {\bibfnamefont {B.}~\bibnamefont {van
  Heck}}, \bibinfo {author} {\bibfnamefont {F.}~\bibnamefont {Hassler}},
  \bibinfo {author} {\bibfnamefont {A.~R.}\ \bibnamefont {Akhmerov}}, \ and\
  \bibinfo {author} {\bibfnamefont {C.~W.~J.}\ \bibnamefont {Beenakker}},\
  }\href {\doibase 10.1103/PhysRevB.84.180502} {\bibfield  {journal} {\bibinfo
  {journal} {Phys. Rev. B}\ }\textbf {\bibinfo {volume} {84}},\ \bibinfo
  {pages} {180502} (\bibinfo {year} {2011})}\BibitemShut {NoStop}%
\bibitem [{\citenamefont {Houzet}\ \emph {et~al.}(2013)\citenamefont {Houzet},
  \citenamefont {Meyer}, \citenamefont {Badiane},\ and\ \citenamefont
  {Glazman}}]{Houzet:PRL13}%
  \BibitemOpen
  \bibfield  {author} {\bibinfo {author} {\bibfnamefont {M.}~\bibnamefont
  {Houzet}}, \bibinfo {author} {\bibfnamefont {J.~S.}\ \bibnamefont {Meyer}},
  \bibinfo {author} {\bibfnamefont {D.~M.}\ \bibnamefont {Badiane}}, \ and\
  \bibinfo {author} {\bibfnamefont {L.~I.}\ \bibnamefont {Glazman}},\ }\href
  {\doibase 10.1103/PhysRevLett.111.046401} {\bibfield  {journal} {\bibinfo
  {journal} {Phys. Rev. Lett.}\ }\textbf {\bibinfo {volume} {111}},\ \bibinfo
  {pages} {046401} (\bibinfo {year} {2013})}\BibitemShut {NoStop}%
\bibitem [{\citenamefont {Parker}\ \emph {et~al.}(1967)\citenamefont {Parker},
  \citenamefont {Taylor},\ and\ \citenamefont {Langenberg}}]{Parker:PRL67}%
  \BibitemOpen
  \bibfield  {author} {\bibinfo {author} {\bibfnamefont {W.~H.}\ \bibnamefont
  {Parker}}, \bibinfo {author} {\bibfnamefont {B.~N.}\ \bibnamefont {Taylor}},
  \ and\ \bibinfo {author} {\bibfnamefont {D.~N.}\ \bibnamefont {Langenberg}},\
  }\href {\doibase 10.1103/PhysRevLett.18.287} {\bibfield  {journal} {\bibinfo
  {journal} {Phys. Rev. Lett.}\ }\textbf {\bibinfo {volume} {18}},\ \bibinfo
  {pages} {287} (\bibinfo {year} {1967})}\BibitemShut {NoStop}%
\bibitem [{\citenamefont {Shapiro}(1963)}]{Shapiro:PRL63}%
  \BibitemOpen
  \bibfield  {author} {\bibinfo {author} {\bibfnamefont {S.}~\bibnamefont
  {Shapiro}},\ }\href {\doibase 10.1103/PhysRevLett.11.80} {\bibfield
  {journal} {\bibinfo  {journal} {Phys. Rev. Lett.}\ }\textbf {\bibinfo
  {volume} {11}},\ \bibinfo {pages} {80} (\bibinfo {year} {1963})}\BibitemShut
  {NoStop}%
\bibitem [{\citenamefont {Dom\'{\i}nguez}\ \emph {et~al.}(2012)\citenamefont
  {Dom\'{\i}nguez}, \citenamefont {Hassler},\ and\ \citenamefont
  {Platero}}]{Dominguez:PRB12}%
  \BibitemOpen
  \bibfield  {author} {\bibinfo {author} {\bibfnamefont {F.}~\bibnamefont
  {Dom\'{\i}nguez}}, \bibinfo {author} {\bibfnamefont {F.}~\bibnamefont
  {Hassler}}, \ and\ \bibinfo {author} {\bibfnamefont {G.}~\bibnamefont
  {Platero}},\ }\href {\doibase 10.1103/PhysRevB.86.140503} {\bibfield
  {journal} {\bibinfo  {journal} {Phys. Rev. B}\ }\textbf {\bibinfo {volume}
  {86}},\ \bibinfo {pages} {140503} (\bibinfo {year} {2012})}\BibitemShut
  {NoStop}%
\bibitem [{\citenamefont {Sau}\ and\ \citenamefont
  {Setiawan}(2017)}]{Sau:PRB17}%
  \BibitemOpen
  \bibfield  {author} {\bibinfo {author} {\bibfnamefont {J.~D.}\ \bibnamefont
  {Sau}}\ and\ \bibinfo {author} {\bibfnamefont {F.}~\bibnamefont {Setiawan}},\
  }\href {\doibase 10.1103/PhysRevB.95.060501} {\bibfield  {journal} {\bibinfo
  {journal} {Phys. Rev. B}\ }\textbf {\bibinfo {volume} {95}},\ \bibinfo
  {pages} {060501} (\bibinfo {year} {2017})}\BibitemShut {NoStop}%
\bibitem [{\citenamefont {Kamata}\ \emph {et~al.}(2018)\citenamefont {Kamata},
  \citenamefont {Deacon}, \citenamefont {Matsuo}, \citenamefont {Li},
  \citenamefont {Jeppesen}, \citenamefont {Samuelson}, \citenamefont {Xu},
  \citenamefont {Ishibashi},\ and\ \citenamefont {Tarucha}}]{Kamata:PRB18}%
  \BibitemOpen
  \bibfield  {author} {\bibinfo {author} {\bibfnamefont {H.}~\bibnamefont
  {Kamata}}, \bibinfo {author} {\bibfnamefont {R.~S.}\ \bibnamefont {Deacon}},
  \bibinfo {author} {\bibfnamefont {S.}~\bibnamefont {Matsuo}}, \bibinfo
  {author} {\bibfnamefont {K.}~\bibnamefont {Li}}, \bibinfo {author}
  {\bibfnamefont {S.}~\bibnamefont {Jeppesen}}, \bibinfo {author}
  {\bibfnamefont {L.}~\bibnamefont {Samuelson}}, \bibinfo {author}
  {\bibfnamefont {H.~Q.}\ \bibnamefont {Xu}}, \bibinfo {author} {\bibfnamefont
  {K.}~\bibnamefont {Ishibashi}}, \ and\ \bibinfo {author} {\bibfnamefont
  {S.}~\bibnamefont {Tarucha}},\ }\href {\doibase 10.1103/PhysRevB.98.041302}
  {\bibfield  {journal} {\bibinfo  {journal} {Phys. Rev. B}\ }\textbf {\bibinfo
  {volume} {98}},\ \bibinfo {pages} {041302} (\bibinfo {year}
  {2018})}\BibitemShut {NoStop}%
\bibitem [{\citenamefont {De~Cecco}\ \emph {et~al.}(2016)\citenamefont
  {De~Cecco}, \citenamefont {Le~Calvez}, \citenamefont {Sac\'ep\'e},
  \citenamefont {Winkelmann},\ and\ \citenamefont {Courtois}}]{DeCecco:PRB16}%
  \BibitemOpen
  \bibfield  {author} {\bibinfo {author} {\bibfnamefont {A.}~\bibnamefont
  {De~Cecco}}, \bibinfo {author} {\bibfnamefont {K.}~\bibnamefont {Le~Calvez}},
  \bibinfo {author} {\bibfnamefont {B.}~\bibnamefont {Sac\'ep\'e}}, \bibinfo
  {author} {\bibfnamefont {C.~B.}\ \bibnamefont {Winkelmann}}, \ and\ \bibinfo
  {author} {\bibfnamefont {H.}~\bibnamefont {Courtois}},\ }\href {\doibase
  10.1103/PhysRevB.93.180505} {\bibfield  {journal} {\bibinfo  {journal} {Phys.
  Rev. B}\ }\textbf {\bibinfo {volume} {93}},\ \bibinfo {pages} {180505}
  (\bibinfo {year} {2016})}\BibitemShut {NoStop}%
\bibitem [{\citenamefont {Le~Calvez}\ \emph {et~al.}(2019)\citenamefont
  {Le~Calvez}, \citenamefont {Veyrat}, \citenamefont {Gay}, \citenamefont
  {Plaindoux}, \citenamefont {Winkelmann}, \citenamefont {Courtois},\ and\
  \citenamefont {Sacepe}}]{LeCalvez:CP19}%
  \BibitemOpen
  \bibfield  {author} {\bibinfo {author} {\bibfnamefont {K.}~\bibnamefont
  {Le~Calvez}}, \bibinfo {author} {\bibfnamefont {L.}~\bibnamefont {Veyrat}},
  \bibinfo {author} {\bibfnamefont {F.}~\bibnamefont {Gay}}, \bibinfo {author}
  {\bibfnamefont {P.}~\bibnamefont {Plaindoux}}, \bibinfo {author}
  {\bibfnamefont {C.~B.}\ \bibnamefont {Winkelmann}}, \bibinfo {author}
  {\bibfnamefont {H.}~\bibnamefont {Courtois}}, \ and\ \bibinfo {author}
  {\bibfnamefont {B.}~\bibnamefont {Sacepe}},\ }\href
  {https://doi.org/10.1038/s42005-018-0100-x} {\bibfield  {journal} {\bibinfo
  {journal} {Commun. Phys.}\ }\textbf {\bibinfo {volume} {2}},\ \bibinfo
  {pages} {4} (\bibinfo {year} {2019})}\BibitemShut {NoStop}%
\bibitem [{\citenamefont {Pic\'o-Cort\'es}\ \emph {et~al.}(2017)\citenamefont
  {Pic\'o-Cort\'es}, \citenamefont {Dom\'{\i}nguez},\ and\ \citenamefont
  {Platero}}]{Pico-Cortes:PRB17}%
  \BibitemOpen
  \bibfield  {author} {\bibinfo {author} {\bibfnamefont {J.}~\bibnamefont
  {Pic\'o-Cort\'es}}, \bibinfo {author} {\bibfnamefont {F.}~\bibnamefont
  {Dom\'{\i}nguez}}, \ and\ \bibinfo {author} {\bibfnamefont {G.}~\bibnamefont
  {Platero}},\ }\href {\doibase 10.1103/PhysRevB.96.125438} {\bibfield
  {journal} {\bibinfo  {journal} {Phys. Rev. B}\ }\textbf {\bibinfo {volume}
  {96}},\ \bibinfo {pages} {125438} (\bibinfo {year} {2017})}\BibitemShut
  {NoStop}%
\bibitem [{\citenamefont {Virtanen}\ and\ \citenamefont
  {Recher}(2013)}]{Virtanen:PRB13}%
  \BibitemOpen
  \bibfield  {author} {\bibinfo {author} {\bibfnamefont {P.}~\bibnamefont
  {Virtanen}}\ and\ \bibinfo {author} {\bibfnamefont {P.}~\bibnamefont
  {Recher}},\ }\href {\doibase 10.1103/PhysRevB.88.144507} {\bibfield
  {journal} {\bibinfo  {journal} {Phys. Rev. B}\ }\textbf {\bibinfo {volume}
  {88}},\ \bibinfo {pages} {144507} (\bibinfo {year} {2013})}\BibitemShut
  {NoStop}%
\bibitem [{\citenamefont {Dom{\'\i}nguez}\ \emph
  {et~al.}(2017{\natexlab{b}})\citenamefont {Dom{\'\i}nguez}, \citenamefont
  {Kashuba}, \citenamefont {Bocquillon}, \citenamefont {Wiedenmann},
  \citenamefont {Deacon}, \citenamefont {Klapwijk}, \citenamefont {Platero},
  \citenamefont {Molenkamp}, \citenamefont {Trauzettel},\ and\ \citenamefont
  {Hankiewicz}}]{Dominguez:PRB17}%
  \BibitemOpen
  \bibfield  {author} {\bibinfo {author} {\bibfnamefont {F.}~\bibnamefont
  {Dom{\'\i}nguez}}, \bibinfo {author} {\bibfnamefont {O.}~\bibnamefont
  {Kashuba}}, \bibinfo {author} {\bibfnamefont {E.}~\bibnamefont {Bocquillon}},
  \bibinfo {author} {\bibfnamefont {J.}~\bibnamefont {Wiedenmann}}, \bibinfo
  {author} {\bibfnamefont {R.}~\bibnamefont {Deacon}}, \bibinfo {author}
  {\bibfnamefont {T.}~\bibnamefont {Klapwijk}}, \bibinfo {author}
  {\bibfnamefont {G.}~\bibnamefont {Platero}}, \bibinfo {author} {\bibfnamefont
  {L.}~\bibnamefont {Molenkamp}}, \bibinfo {author} {\bibfnamefont
  {B.}~\bibnamefont {Trauzettel}}, \ and\ \bibinfo {author} {\bibfnamefont
  {E.}~\bibnamefont {Hankiewicz}},\ }\href
  {https://journals.aps.org/prb/abstract/10.1103/PhysRevB.95.195430} {\bibfield
   {journal} {\bibinfo  {journal} {Phys. Rev. B}\ }\textbf {\bibinfo {volume}
  {95}},\ \bibinfo {pages} {195430} (\bibinfo {year}
  {2017}{\natexlab{b}})}\BibitemShut {NoStop}%
\bibitem [{\citenamefont {V\"ayrynen}\ \emph {et~al.}(2015)\citenamefont
  {V\"ayrynen}, \citenamefont {Rastelli}, \citenamefont {Belzig},\ and\
  \citenamefont {Glazman}}]{Vayrynen:PRB15}%
  \BibitemOpen
  \bibfield  {author} {\bibinfo {author} {\bibfnamefont {J.~I.}\ \bibnamefont
  {V\"ayrynen}}, \bibinfo {author} {\bibfnamefont {G.}~\bibnamefont
  {Rastelli}}, \bibinfo {author} {\bibfnamefont {W.}~\bibnamefont {Belzig}}, \
  and\ \bibinfo {author} {\bibfnamefont {L.~I.}\ \bibnamefont {Glazman}},\
  }\href {\doibase 10.1103/PhysRevB.92.134508} {\bibfield  {journal} {\bibinfo
  {journal} {Phys. Rev. B}\ }\textbf {\bibinfo {volume} {92}},\ \bibinfo
  {pages} {134508} (\bibinfo {year} {2015})}\BibitemShut {NoStop}%
\bibitem [{\citenamefont {San-Jose}\ \emph {et~al.}(2014)\citenamefont
  {San-Jose}, \citenamefont {Prada},\ and\ \citenamefont
  {Aguado}}]{San-Jose:PRL14}%
  \BibitemOpen
  \bibfield  {author} {\bibinfo {author} {\bibfnamefont {P.}~\bibnamefont
  {San-Jose}}, \bibinfo {author} {\bibfnamefont {E.}~\bibnamefont {Prada}}, \
  and\ \bibinfo {author} {\bibfnamefont {R.}~\bibnamefont {Aguado}},\ }\href
  {\doibase 10.1103/PhysRevLett.112.137001} {\bibfield  {journal} {\bibinfo
  {journal} {Phys. Rev. Lett.}\ }\textbf {\bibinfo {volume} {112}},\ \bibinfo
  {pages} {137001} (\bibinfo {year} {2014})}\BibitemShut {NoStop}%
\bibitem [{\citenamefont {Tiira}\ \emph {et~al.}(2017)\citenamefont {Tiira},
  \citenamefont {Strambini}, \citenamefont {Amado}, \citenamefont {Roddaro},
  \citenamefont {San-Jose}, \citenamefont {Aguado}, \citenamefont {Bergeret},
  \citenamefont {Ercolani}, \citenamefont {Sorba},\ and\ \citenamefont
  {Giazotto}}]{Tiira:NC17}%
  \BibitemOpen
  \bibfield  {author} {\bibinfo {author} {\bibfnamefont {J.}~\bibnamefont
  {Tiira}}, \bibinfo {author} {\bibfnamefont {E.}~\bibnamefont {Strambini}},
  \bibinfo {author} {\bibfnamefont {M.}~\bibnamefont {Amado}}, \bibinfo
  {author} {\bibfnamefont {S.}~\bibnamefont {Roddaro}}, \bibinfo {author}
  {\bibfnamefont {P.}~\bibnamefont {San-Jose}}, \bibinfo {author}
  {\bibfnamefont {R.}~\bibnamefont {Aguado}}, \bibinfo {author} {\bibfnamefont
  {F.~S.}\ \bibnamefont {Bergeret}}, \bibinfo {author} {\bibfnamefont
  {D.}~\bibnamefont {Ercolani}}, \bibinfo {author} {\bibfnamefont
  {L.}~\bibnamefont {Sorba}}, \ and\ \bibinfo {author} {\bibfnamefont
  {F.}~\bibnamefont {Giazotto}},\ }\href {\doibase 10.1038/ncomms14984}
  {\bibfield  {journal} {\bibinfo  {journal} {Nat. Commun.}\ }\textbf {\bibinfo
  {volume} {8}},\ \bibinfo {pages} {14984} (\bibinfo {year}
  {2017})}\BibitemShut {NoStop}%
\bibitem [{\citenamefont {Cayao}\ \emph {et~al.}(2017)\citenamefont {Cayao},
  \citenamefont {San-Jose}, \citenamefont {Black-Schaffer}, \citenamefont
  {Aguado},\ and\ \citenamefont {Prada}}]{Cayao:PRB17}%
  \BibitemOpen
  \bibfield  {author} {\bibinfo {author} {\bibfnamefont {J.}~\bibnamefont
  {Cayao}}, \bibinfo {author} {\bibfnamefont {P.}~\bibnamefont {San-Jose}},
  \bibinfo {author} {\bibfnamefont {A.~M.}\ \bibnamefont {Black-Schaffer}},
  \bibinfo {author} {\bibfnamefont {R.}~\bibnamefont {Aguado}}, \ and\ \bibinfo
  {author} {\bibfnamefont {E.}~\bibnamefont {Prada}},\ }\href {\doibase
  10.1103/PhysRevB.96.205425} {\bibfield  {journal} {\bibinfo  {journal} {Phys.
  Rev. B}\ }\textbf {\bibinfo {volume} {96}},\ \bibinfo {pages} {205425}
  (\bibinfo {year} {2017})}\BibitemShut {NoStop}%
\bibitem [{\citenamefont {Peng}\ \emph {et~al.}(2016)\citenamefont {Peng},
  \citenamefont {Pientka}, \citenamefont {Berg}, \citenamefont {Oreg},\ and\
  \citenamefont {von Oppen}}]{Peng:PRB16}%
  \BibitemOpen
  \bibfield  {author} {\bibinfo {author} {\bibfnamefont {Y.}~\bibnamefont
  {Peng}}, \bibinfo {author} {\bibfnamefont {F.}~\bibnamefont {Pientka}},
  \bibinfo {author} {\bibfnamefont {E.}~\bibnamefont {Berg}}, \bibinfo {author}
  {\bibfnamefont {Y.}~\bibnamefont {Oreg}}, \ and\ \bibinfo {author}
  {\bibfnamefont {F.}~\bibnamefont {von Oppen}},\ }\href {\doibase
  10.1103/PhysRevB.94.085409} {\bibfield  {journal} {\bibinfo  {journal} {Phys.
  Rev. B}\ }\textbf {\bibinfo {volume} {94}},\ \bibinfo {pages} {085409}
  (\bibinfo {year} {2016})}\BibitemShut {NoStop}%
\bibitem [{\citenamefont {Nijholt}\ and\ \citenamefont
  {Akhmerov}(2016)}]{Nijholt:PRB16}%
  \BibitemOpen
  \bibfield  {author} {\bibinfo {author} {\bibfnamefont {B.}~\bibnamefont
  {Nijholt}}\ and\ \bibinfo {author} {\bibfnamefont {A.~R.}\ \bibnamefont
  {Akhmerov}},\ }\href {http://dx.doi.org/10.1103/PhysRevB.93.235434}
  {\bibfield  {journal} {\bibinfo  {journal} {Phys. Rev. B}\ }\textbf {\bibinfo
  {volume} {93}} (\bibinfo {year} {2016})}\BibitemShut {NoStop}%
\bibitem [{\citenamefont {Nilsson}\ \emph {et~al.}(2009)\citenamefont
  {Nilsson}, \citenamefont {Caroff}, \citenamefont {Thelander}, \citenamefont
  {Larsson}, \citenamefont {Wagner}, \citenamefont {Wernersson}, \citenamefont
  {Samuelson},\ and\ \citenamefont {Xu}}]{Nilsson:NL09}%
  \BibitemOpen
  \bibfield  {author} {\bibinfo {author} {\bibfnamefont {H.~A.}\ \bibnamefont
  {Nilsson}}, \bibinfo {author} {\bibfnamefont {P.}~\bibnamefont {Caroff}},
  \bibinfo {author} {\bibfnamefont {C.}~\bibnamefont {Thelander}}, \bibinfo
  {author} {\bibfnamefont {M.}~\bibnamefont {Larsson}}, \bibinfo {author}
  {\bibfnamefont {J.~B.}\ \bibnamefont {Wagner}}, \bibinfo {author}
  {\bibfnamefont {L.-E.}\ \bibnamefont {Wernersson}}, \bibinfo {author}
  {\bibfnamefont {L.}~\bibnamefont {Samuelson}}, \ and\ \bibinfo {author}
  {\bibfnamefont {H.~Q.}\ \bibnamefont {Xu}},\ }\href {\doibase
  10.1021/nl901333a} {\bibfield  {journal} {\bibinfo  {journal} {Nano Lett.}\
  }\textbf {\bibinfo {volume} {9}},\ \bibinfo {pages} {3151} (\bibinfo {year}
  {2009})}\BibitemShut {NoStop}%
\bibitem [{\citenamefont {Winkler}\ \emph {et~al.}(2017)\citenamefont
  {Winkler}, \citenamefont {Varjas}, \citenamefont {Skolasinski}, \citenamefont
  {Soluyanov}, \citenamefont {Troyer},\ and\ \citenamefont
  {Wimmer}}]{Winkler:PRL17}%
  \BibitemOpen
  \bibfield  {author} {\bibinfo {author} {\bibfnamefont {G.~W.}\ \bibnamefont
  {Winkler}}, \bibinfo {author} {\bibfnamefont {D.}~\bibnamefont {Varjas}},
  \bibinfo {author} {\bibfnamefont {R.}~\bibnamefont {Skolasinski}}, \bibinfo
  {author} {\bibfnamefont {A.~A.}\ \bibnamefont {Soluyanov}}, \bibinfo {author}
  {\bibfnamefont {M.}~\bibnamefont {Troyer}}, \ and\ \bibinfo {author}
  {\bibfnamefont {M.}~\bibnamefont {Wimmer}},\ }\href {\doibase
  10.1103/PhysRevLett.119.037701} {\bibfield  {journal} {\bibinfo  {journal}
  {Phys. Rev. Lett.}\ }\textbf {\bibinfo {volume} {119}},\ \bibinfo {pages}
  {037701} (\bibinfo {year} {2017})}\BibitemShut {NoStop}%
\bibitem [{\citenamefont {Takei}\ \emph {et~al.}(2013)\citenamefont {Takei},
  \citenamefont {Fregoso}, \citenamefont {Hui}, \citenamefont {Lobos},\ and\
  \citenamefont {Das~Sarma}}]{Takei:PRL13}%
  \BibitemOpen
  \bibfield  {author} {\bibinfo {author} {\bibfnamefont {S.}~\bibnamefont
  {Takei}}, \bibinfo {author} {\bibfnamefont {B.~M.}\ \bibnamefont {Fregoso}},
  \bibinfo {author} {\bibfnamefont {H.-Y.}\ \bibnamefont {Hui}}, \bibinfo
  {author} {\bibfnamefont {A.~M.}\ \bibnamefont {Lobos}}, \ and\ \bibinfo
  {author} {\bibfnamefont {S.}~\bibnamefont {Das~Sarma}},\ }\href {\doibase
  10.1103/PhysRevLett.110.186803} {\bibfield  {journal} {\bibinfo  {journal}
  {Phys. Rev. Lett.}\ }\textbf {\bibinfo {volume} {110}},\ \bibinfo {pages}
  {186803} (\bibinfo {year} {2013})}\BibitemShut {NoStop}%
\bibitem [{\citenamefont {Krogstrup}\ \emph {et~al.}(2015)\citenamefont
  {Krogstrup}, \citenamefont {Ziino}, \citenamefont {Chang}, \citenamefont
  {Albrecht}, \citenamefont {Madsen}, \citenamefont {Johnson}, \citenamefont
  {Nyg{\aa}rd}, \citenamefont {Marcus},\ and\ \citenamefont
  {Jespersen}}]{Krogstrup:NM15}%
  \BibitemOpen
  \bibfield  {author} {\bibinfo {author} {\bibfnamefont {P.}~\bibnamefont
  {Krogstrup}}, \bibinfo {author} {\bibfnamefont {N.~L.~B.}\ \bibnamefont
  {Ziino}}, \bibinfo {author} {\bibfnamefont {W.}~\bibnamefont {Chang}},
  \bibinfo {author} {\bibfnamefont {S.~M.}\ \bibnamefont {Albrecht}}, \bibinfo
  {author} {\bibfnamefont {M.~H.}\ \bibnamefont {Madsen}}, \bibinfo {author}
  {\bibfnamefont {E.}~\bibnamefont {Johnson}}, \bibinfo {author} {\bibfnamefont
  {J.}~\bibnamefont {Nyg{\aa}rd}}, \bibinfo {author} {\bibfnamefont {C.~M.}\
  \bibnamefont {Marcus}}, \ and\ \bibinfo {author} {\bibfnamefont {T.~S.}\
  \bibnamefont {Jespersen}},\ }\href {http://dx.doi.org/10.1038/nmat4176}
  {\bibfield  {journal} {\bibinfo  {journal} {Nat. Mater.}\ }\textbf {\bibinfo
  {volume} {14}},\ \bibinfo {pages} {400} (\bibinfo {year} {2015})}\BibitemShut
  {NoStop}%
\bibitem [{\citenamefont {Gazibegovic}\ \emph {et~al.}(2017)\citenamefont
  {Gazibegovic}, \citenamefont {Car}, \citenamefont {Zhang}, \citenamefont
  {Balk}, \citenamefont {Logan}, \citenamefont {de~Moor}, \citenamefont
  {Cassidy}, \citenamefont {Schmits}, \citenamefont {Xu}, \citenamefont {Wang},
  \citenamefont {Krogstrup}, \citenamefont {Op~het Veld}, \citenamefont {Zuo},
  \citenamefont {Vos}, \citenamefont {Shen}, \citenamefont {Bouman},
  \citenamefont {Shojaei}, \citenamefont {Pennachio}, \citenamefont {Lee},
  \citenamefont {van Veldhoven}, \citenamefont {Koelling}, \citenamefont
  {Verheijen}, \citenamefont {Kouwenhoven}, \citenamefont {Palmstr{\o}m},\ and\
  \citenamefont {Bakkers}}]{Gazibegovic:N17}%
  \BibitemOpen
  \bibfield  {author} {\bibinfo {author} {\bibfnamefont {S.}~\bibnamefont
  {Gazibegovic}}, \bibinfo {author} {\bibfnamefont {D.}~\bibnamefont {Car}},
  \bibinfo {author} {\bibfnamefont {H.}~\bibnamefont {Zhang}}, \bibinfo
  {author} {\bibfnamefont {S.~C.}\ \bibnamefont {Balk}}, \bibinfo {author}
  {\bibfnamefont {J.~A.}\ \bibnamefont {Logan}}, \bibinfo {author}
  {\bibfnamefont {M.~W.~A.}\ \bibnamefont {de~Moor}}, \bibinfo {author}
  {\bibfnamefont {M.~C.}\ \bibnamefont {Cassidy}}, \bibinfo {author}
  {\bibfnamefont {R.}~\bibnamefont {Schmits}}, \bibinfo {author} {\bibfnamefont
  {D.}~\bibnamefont {Xu}}, \bibinfo {author} {\bibfnamefont {G.}~\bibnamefont
  {Wang}}, \bibinfo {author} {\bibfnamefont {P.}~\bibnamefont {Krogstrup}},
  \bibinfo {author} {\bibfnamefont {R.~L.~M.}\ \bibnamefont {Op~het Veld}},
  \bibinfo {author} {\bibfnamefont {K.}~\bibnamefont {Zuo}}, \bibinfo {author}
  {\bibfnamefont {Y.}~\bibnamefont {Vos}}, \bibinfo {author} {\bibfnamefont
  {J.}~\bibnamefont {Shen}}, \bibinfo {author} {\bibfnamefont {D.}~\bibnamefont
  {Bouman}}, \bibinfo {author} {\bibfnamefont {B.}~\bibnamefont {Shojaei}},
  \bibinfo {author} {\bibfnamefont {D.}~\bibnamefont {Pennachio}}, \bibinfo
  {author} {\bibfnamefont {J.~S.}\ \bibnamefont {Lee}}, \bibinfo {author}
  {\bibfnamefont {P.~J.}\ \bibnamefont {van Veldhoven}}, \bibinfo {author}
  {\bibfnamefont {S.}~\bibnamefont {Koelling}}, \bibinfo {author}
  {\bibfnamefont {M.~A.}\ \bibnamefont {Verheijen}}, \bibinfo {author}
  {\bibfnamefont {L.~P.}\ \bibnamefont {Kouwenhoven}}, \bibinfo {author}
  {\bibfnamefont {C.~J.}\ \bibnamefont {Palmstr{\o}m}}, \ and\ \bibinfo
  {author} {\bibfnamefont {E.~P. A.~M.}\ \bibnamefont {Bakkers}},\ }\href
  {http://dx.doi.org/10.1038/nature23468} {\bibfield  {journal} {\bibinfo
  {journal} {Nature}\ }\textbf {\bibinfo {volume} {548}},\ \bibinfo {pages}
  {434 EP } (\bibinfo {year} {2017})}\BibitemShut {NoStop}%
\bibitem [{\citenamefont {Escribano}\ \emph {et~al.}(2020)\citenamefont
  {Escribano}, \citenamefont {Yeyati},\ and\ \citenamefont
  {Prada}}]{Escribano:A20}%
  \BibitemOpen
  \bibfield  {author} {\bibinfo {author} {\bibfnamefont {S.~D.}\ \bibnamefont
  {Escribano}}, \bibinfo {author} {\bibfnamefont {A.~L.}\ \bibnamefont
  {Yeyati}}, \ and\ \bibinfo {author} {\bibfnamefont {E.}~\bibnamefont
  {Prada}},\ }\href {https://arxiv.org/abs/2001.04375} {\bibfield  {journal}
  {\bibinfo  {journal} {arXiv:2001.04375}\ } (\bibinfo {year}
  {2020})}\BibitemShut {NoStop}%
\bibitem [{\citenamefont {Lim}\ \emph {et~al.}(2012)\citenamefont {Lim},
  \citenamefont {Serra}, \citenamefont {L\'opez},\ and\ \citenamefont
  {Aguado}}]{Lim:PRB12}%
  \BibitemOpen
  \bibfield  {author} {\bibinfo {author} {\bibfnamefont {J.~S.}\ \bibnamefont
  {Lim}}, \bibinfo {author} {\bibfnamefont {L.~m.~c.}\ \bibnamefont {Serra}},
  \bibinfo {author} {\bibfnamefont {R.}~\bibnamefont {L\'opez}}, \ and\
  \bibinfo {author} {\bibfnamefont {R.}~\bibnamefont {Aguado}},\ }\href
  {\doibase 10.1103/PhysRevB.86.121103} {\bibfield  {journal} {\bibinfo
  {journal} {Phys. Rev. B}\ }\textbf {\bibinfo {volume} {86}},\ \bibinfo
  {pages} {121103} (\bibinfo {year} {2012})}\BibitemShut {NoStop}%
\bibitem [{\citenamefont {Das~Sarma}\ \emph {et~al.}(2012)\citenamefont
  {Das~Sarma}, \citenamefont {Sau},\ and\ \citenamefont
  {Stanescu}}]{Das-Sarma:PRB12}%
  \BibitemOpen
  \bibfield  {author} {\bibinfo {author} {\bibfnamefont {S.}~\bibnamefont
  {Das~Sarma}}, \bibinfo {author} {\bibfnamefont {J.~D.}\ \bibnamefont {Sau}},
  \ and\ \bibinfo {author} {\bibfnamefont {T.~D.}\ \bibnamefont {Stanescu}},\
  }\href {\doibase 10.1103/PhysRevB.86.220506} {\bibfield  {journal} {\bibinfo
  {journal} {Phys. Rev. B}\ }\textbf {\bibinfo {volume} {86}},\ \bibinfo
  {pages} {220506} (\bibinfo {year} {2012})}\BibitemShut {NoStop}%
\bibitem [{\citenamefont {Rainis}\ \emph {et~al.}(2013)\citenamefont {Rainis},
  \citenamefont {Trifunovic}, \citenamefont {Klinovaja},\ and\ \citenamefont
  {Loss}}]{Rainis:PRB13}%
  \BibitemOpen
  \bibfield  {author} {\bibinfo {author} {\bibfnamefont {D.}~\bibnamefont
  {Rainis}}, \bibinfo {author} {\bibfnamefont {L.}~\bibnamefont {Trifunovic}},
  \bibinfo {author} {\bibfnamefont {J.}~\bibnamefont {Klinovaja}}, \ and\
  \bibinfo {author} {\bibfnamefont {D.}~\bibnamefont {Loss}},\ }\href {\doibase
  10.1103/PhysRevB.87.024515} {\bibfield  {journal} {\bibinfo  {journal} {Phys.
  Rev. B}\ }\textbf {\bibinfo {volume} {87}},\ \bibinfo {pages} {024515}
  (\bibinfo {year} {2013})}\BibitemShut {NoStop}%
\bibitem [{\citenamefont {Sharma}\ \emph {et~al.}(2020)\citenamefont {Sharma},
  \citenamefont {Zeng}, \citenamefont {Stanescu},\ and\ \citenamefont
  {Tewari}}]{Sharma:A20}%
  \BibitemOpen
  \bibfield  {author} {\bibinfo {author} {\bibfnamefont {G.}~\bibnamefont
  {Sharma}}, \bibinfo {author} {\bibfnamefont {C.}~\bibnamefont {Zeng}},
  \bibinfo {author} {\bibfnamefont {T.~D.}\ \bibnamefont {Stanescu}}, \ and\
  \bibinfo {author} {\bibfnamefont {S.}~\bibnamefont {Tewari}},\ }\href
  {https://arxiv.org/abs/2001.10551} {\bibfield  {journal} {\bibinfo  {journal}
  {arXiv:2001.10551}\ } (\bibinfo {year} {2020})}\BibitemShut {NoStop}%
\bibitem [{\citenamefont {Dmytruk}\ and\ \citenamefont
  {Klinovaja}(2018)}]{Dmytruk:PRB18a}%
  \BibitemOpen
  \bibfield  {author} {\bibinfo {author} {\bibfnamefont {O.}~\bibnamefont
  {Dmytruk}}\ and\ \bibinfo {author} {\bibfnamefont {J.}~\bibnamefont
  {Klinovaja}},\ }\href {\doibase 10.1103/PhysRevB.97.155409} {\bibfield
  {journal} {\bibinfo  {journal} {Phys. Rev. B}\ }\textbf {\bibinfo {volume}
  {97}},\ \bibinfo {pages} {155409} (\bibinfo {year} {2018})}\BibitemShut
  {NoStop}%
\bibitem [{\citenamefont {Liu}\ \emph {et~al.}(2017{\natexlab{b}})\citenamefont
  {Liu}, \citenamefont {Sau},\ and\ \citenamefont {Das~Sarma}}]{Liu:PRB17}%
  \BibitemOpen
  \bibfield  {author} {\bibinfo {author} {\bibfnamefont {C.-X.}\ \bibnamefont
  {Liu}}, \bibinfo {author} {\bibfnamefont {J.~D.}\ \bibnamefont {Sau}}, \ and\
  \bibinfo {author} {\bibfnamefont {S.}~\bibnamefont {Das~Sarma}},\ }\href
  {\doibase 10.1103/PhysRevB.95.054502} {\bibfield  {journal} {\bibinfo
  {journal} {Phys. Rev. B}\ }\textbf {\bibinfo {volume} {95}},\ \bibinfo
  {pages} {054502} (\bibinfo {year} {2017}{\natexlab{b}})}\BibitemShut
  {NoStop}%
\bibitem [{\citenamefont {Danon}\ \emph {et~al.}(2017)\citenamefont {Danon},
  \citenamefont {Hansen},\ and\ \citenamefont {Flensberg}}]{Danon:PRB17}%
  \BibitemOpen
  \bibfield  {author} {\bibinfo {author} {\bibfnamefont {J.}~\bibnamefont
  {Danon}}, \bibinfo {author} {\bibfnamefont {E.~B.}\ \bibnamefont {Hansen}}, \
  and\ \bibinfo {author} {\bibfnamefont {K.}~\bibnamefont {Flensberg}},\ }\href
  {\doibase 10.1103/PhysRevB.96.125420} {\bibfield  {journal} {\bibinfo
  {journal} {Phys. Rev. B}\ }\textbf {\bibinfo {volume} {96}},\ \bibinfo
  {pages} {125420} (\bibinfo {year} {2017})}\BibitemShut {NoStop}%
\bibitem [{\citenamefont {Cao}\ \emph {et~al.}(2019)\citenamefont {Cao},
  \citenamefont {Zhang}, \citenamefont {L\"u}, \citenamefont {He},
  \citenamefont {Lu},\ and\ \citenamefont {Xie}}]{Cao:PRL19}%
  \BibitemOpen
  \bibfield  {author} {\bibinfo {author} {\bibfnamefont {Z.}~\bibnamefont
  {Cao}}, \bibinfo {author} {\bibfnamefont {H.}~\bibnamefont {Zhang}}, \bibinfo
  {author} {\bibfnamefont {H.-F.}\ \bibnamefont {L\"u}}, \bibinfo {author}
  {\bibfnamefont {W.-X.}\ \bibnamefont {He}}, \bibinfo {author} {\bibfnamefont
  {H.-Z.}\ \bibnamefont {Lu}}, \ and\ \bibinfo {author} {\bibfnamefont {X.~C.}\
  \bibnamefont {Xie}},\ }\href {\doibase 10.1103/PhysRevLett.122.147701}
  {\bibfield  {journal} {\bibinfo  {journal} {Phys. Rev. Lett.}\ }\textbf
  {\bibinfo {volume} {122}},\ \bibinfo {pages} {147701} (\bibinfo {year}
  {2019})}\BibitemShut {NoStop}%
\bibitem [{\citenamefont {Stanescu}\ \emph {et~al.}(2012)\citenamefont
  {Stanescu}, \citenamefont {Tewari}, \citenamefont {Sau},\ and\ \citenamefont
  {Das~Sarma}}]{Stanescu:PRL12}%
  \BibitemOpen
  \bibfield  {author} {\bibinfo {author} {\bibfnamefont {T.~D.}\ \bibnamefont
  {Stanescu}}, \bibinfo {author} {\bibfnamefont {S.}~\bibnamefont {Tewari}},
  \bibinfo {author} {\bibfnamefont {J.~D.}\ \bibnamefont {Sau}}, \ and\
  \bibinfo {author} {\bibfnamefont {S.}~\bibnamefont {Das~Sarma}},\ }\href
  {\doibase 10.1103/PhysRevLett.109.266402} {\bibfield  {journal} {\bibinfo
  {journal} {Phys. Rev. Lett.}\ }\textbf {\bibinfo {volume} {109}},\ \bibinfo
  {pages} {266402} (\bibinfo {year} {2012})}\BibitemShut {NoStop}%
\bibitem [{\citenamefont {Huang}\ \emph {et~al.}(2018)\citenamefont {Huang},
  \citenamefont {Pan}, \citenamefont {Liu}, \citenamefont {Sau}, \citenamefont
  {Stanescu},\ and\ \citenamefont {Das~Sarma}}]{Huang:PRB18}%
  \BibitemOpen
  \bibfield  {author} {\bibinfo {author} {\bibfnamefont {Y.}~\bibnamefont
  {Huang}}, \bibinfo {author} {\bibfnamefont {H.}~\bibnamefont {Pan}}, \bibinfo
  {author} {\bibfnamefont {C.-X.}\ \bibnamefont {Liu}}, \bibinfo {author}
  {\bibfnamefont {J.~D.}\ \bibnamefont {Sau}}, \bibinfo {author} {\bibfnamefont
  {T.~D.}\ \bibnamefont {Stanescu}}, \ and\ \bibinfo {author} {\bibfnamefont
  {S.}~\bibnamefont {Das~Sarma}},\ }\href {\doibase 10.1103/PhysRevB.98.144511}
  {\bibfield  {journal} {\bibinfo  {journal} {Phys. Rev. B}\ }\textbf {\bibinfo
  {volume} {98}},\ \bibinfo {pages} {144511} (\bibinfo {year}
  {2018})}\BibitemShut {NoStop}%
\bibitem [{\citenamefont {Bagrets}\ and\ \citenamefont
  {Altland}(2012)}]{Bagrets:PRL12}%
  \BibitemOpen
  \bibfield  {author} {\bibinfo {author} {\bibfnamefont {D.}~\bibnamefont
  {Bagrets}}\ and\ \bibinfo {author} {\bibfnamefont {A.}~\bibnamefont
  {Altland}},\ }\href {\doibase 10.1103/PhysRevLett.109.227005} {\bibfield
  {journal} {\bibinfo  {journal} {Phys. Rev. Lett.}\ }\textbf {\bibinfo
  {volume} {109}},\ \bibinfo {pages} {227005} (\bibinfo {year}
  {2012})}\BibitemShut {NoStop}%
\bibitem [{\citenamefont {Beenakker}(2015)}]{Beenakker:RMP15}%
  \BibitemOpen
  \bibfield  {author} {\bibinfo {author} {\bibfnamefont {C.~W.~J.}\
  \bibnamefont {Beenakker}},\ }\href {\doibase 10.1103/RevModPhys.87.1037}
  {\bibfield  {journal} {\bibinfo  {journal} {Rev. Mod. Phys.}\ }\textbf
  {\bibinfo {volume} {87}},\ \bibinfo {pages} {1037} (\bibinfo {year}
  {2015})}\BibitemShut {NoStop}%
\bibitem [{\citenamefont {Altland}\ and\ \citenamefont
  {Zirnbauer}(1997)}]{Altland:PRB97}%
  \BibitemOpen
  \bibfield  {author} {\bibinfo {author} {\bibfnamefont {A.}~\bibnamefont
  {Altland}}\ and\ \bibinfo {author} {\bibfnamefont {M.~R.}\ \bibnamefont
  {Zirnbauer}},\ }\href {\doibase 10.1103/PhysRevB.55.1142} {\bibfield
  {journal} {\bibinfo  {journal} {Phys. Rev. B}\ }\textbf {\bibinfo {volume}
  {55}},\ \bibinfo {pages} {1142} (\bibinfo {year} {1997})}\BibitemShut
  {NoStop}%
\bibitem [{\citenamefont {Schnyder}\ \emph {et~al.}(2008)\citenamefont
  {Schnyder}, \citenamefont {Ryu}, \citenamefont {Furusaki},\ and\
  \citenamefont {Ludwig}}]{Schnyder:PRB08}%
  \BibitemOpen
  \bibfield  {author} {\bibinfo {author} {\bibfnamefont {A.~P.}\ \bibnamefont
  {Schnyder}}, \bibinfo {author} {\bibfnamefont {S.}~\bibnamefont {Ryu}},
  \bibinfo {author} {\bibfnamefont {A.}~\bibnamefont {Furusaki}}, \ and\
  \bibinfo {author} {\bibfnamefont {A.~W.~W.}\ \bibnamefont {Ludwig}},\ }\href
  {\doibase 10.1103/PhysRevB.78.195125} {\bibfield  {journal} {\bibinfo
  {journal} {Phys. Rev. B}\ }\textbf {\bibinfo {volume} {78}},\ \bibinfo
  {pages} {195125} (\bibinfo {year} {2008})}\BibitemShut {NoStop}%
\bibitem [{\citenamefont {Chevallier}\ \emph {et~al.}(2012)\citenamefont
  {Chevallier}, \citenamefont {Sticlet}, \citenamefont {Simon},\ and\
  \citenamefont {Bena}}]{Chevallier:PRB12}%
  \BibitemOpen
  \bibfield  {author} {\bibinfo {author} {\bibfnamefont {D.}~\bibnamefont
  {Chevallier}}, \bibinfo {author} {\bibfnamefont {D.}~\bibnamefont {Sticlet}},
  \bibinfo {author} {\bibfnamefont {P.}~\bibnamefont {Simon}}, \ and\ \bibinfo
  {author} {\bibfnamefont {C.}~\bibnamefont {Bena}},\ }\href {\doibase
  10.1103/PhysRevB.85.235307} {\bibfield  {journal} {\bibinfo  {journal} {Phys.
  Rev. B}\ }\textbf {\bibinfo {volume} {85}},\ \bibinfo {pages} {235307}
  (\bibinfo {year} {2012})}\BibitemShut {NoStop}%
\bibitem [{\citenamefont {Cayao}\ \emph {et~al.}(2015)\citenamefont {Cayao},
  \citenamefont {Prada}, \citenamefont {San-Jose},\ and\ \citenamefont
  {Aguado}}]{Cayao:PRB15}%
  \BibitemOpen
  \bibfield  {author} {\bibinfo {author} {\bibfnamefont {J.}~\bibnamefont
  {Cayao}}, \bibinfo {author} {\bibfnamefont {E.}~\bibnamefont {Prada}},
  \bibinfo {author} {\bibfnamefont {P.}~\bibnamefont {San-Jose}}, \ and\
  \bibinfo {author} {\bibfnamefont {R.}~\bibnamefont {Aguado}},\ }\href
  {\doibase 10.1103/PhysRevB.91.024514} {\bibfield  {journal} {\bibinfo
  {journal} {Phys. Rev. B}\ }\textbf {\bibinfo {volume} {91}},\ \bibinfo
  {pages} {024514} (\bibinfo {year} {2015})}\BibitemShut {NoStop}%
\bibitem [{\citenamefont {Ptok}\ \emph {et~al.}(2017)\citenamefont {Ptok},
  \citenamefont {Kobia\l{}ka},\ and\ \citenamefont {Doma\ifmmode~\acute{n}\else
  \'{n}\fi{}ski}}]{Ptok:PRB17}%
  \BibitemOpen
  \bibfield  {author} {\bibinfo {author} {\bibfnamefont {A.}~\bibnamefont
  {Ptok}}, \bibinfo {author} {\bibfnamefont {A.}~\bibnamefont {Kobia\l{}ka}}, \
  and\ \bibinfo {author} {\bibfnamefont {T.}~\bibnamefont
  {Doma\ifmmode~\acute{n}\else \'{n}\fi{}ski}},\ }\href {\doibase
  10.1103/PhysRevB.96.195430} {\bibfield  {journal} {\bibinfo  {journal} {Phys.
  Rev. B}\ }\textbf {\bibinfo {volume} {96}},\ \bibinfo {pages} {195430}
  (\bibinfo {year} {2017})}\BibitemShut {NoStop}%
\bibitem [{\citenamefont {Pikulin}\ and\ \citenamefont
  {Nazarov}(2013)}]{Pikulin:PRB13}%
  \BibitemOpen
  \bibfield  {author} {\bibinfo {author} {\bibfnamefont {D.~I.}\ \bibnamefont
  {Pikulin}}\ and\ \bibinfo {author} {\bibfnamefont {Y.~V.}\ \bibnamefont
  {Nazarov}},\ }\href {\doibase 10.1103/PhysRevB.87.235421} {\bibfield
  {journal} {\bibinfo  {journal} {Phys. Rev. B}\ }\textbf {\bibinfo {volume}
  {87}},\ \bibinfo {pages} {235421} (\bibinfo {year} {2013})}\BibitemShut
  {NoStop}%
\bibitem [{\citenamefont {Leykam}\ \emph {et~al.}(2017)\citenamefont {Leykam},
  \citenamefont {Bliokh}, \citenamefont {Huang}, \citenamefont {Chong},\ and\
  \citenamefont {Nori}}]{Leykam:PRL17}%
  \BibitemOpen
  \bibfield  {author} {\bibinfo {author} {\bibfnamefont {D.}~\bibnamefont
  {Leykam}}, \bibinfo {author} {\bibfnamefont {K.~Y.}\ \bibnamefont {Bliokh}},
  \bibinfo {author} {\bibfnamefont {C.}~\bibnamefont {Huang}}, \bibinfo
  {author} {\bibfnamefont {Y.~D.}\ \bibnamefont {Chong}}, \ and\ \bibinfo
  {author} {\bibfnamefont {F.}~\bibnamefont {Nori}},\ }\href {\doibase
  10.1103/PhysRevLett.118.040401} {\bibfield  {journal} {\bibinfo  {journal}
  {Phys. Rev. Lett.}\ }\textbf {\bibinfo {volume} {118}},\ \bibinfo {pages}
  {040401} (\bibinfo {year} {2017})}\BibitemShut {NoStop}%
\bibitem [{\citenamefont {Shen}\ \emph
  {et~al.}(2018{\natexlab{b}})\citenamefont {Shen}, \citenamefont {Zhen},\ and\
  \citenamefont {Fu}}]{Shen:PRL18}%
  \BibitemOpen
  \bibfield  {author} {\bibinfo {author} {\bibfnamefont {H.}~\bibnamefont
  {Shen}}, \bibinfo {author} {\bibfnamefont {B.}~\bibnamefont {Zhen}}, \ and\
  \bibinfo {author} {\bibfnamefont {L.}~\bibnamefont {Fu}},\ }\href {\doibase
  10.1103/PhysRevLett.120.146402} {\bibfield  {journal} {\bibinfo  {journal}
  {Phys. Rev. Lett.}\ }\textbf {\bibinfo {volume} {120}},\ \bibinfo {pages}
  {146402} (\bibinfo {year} {2018}{\natexlab{b}})}\BibitemShut {NoStop}%
\bibitem [{\citenamefont {Gong}\ \emph {et~al.}(2018)\citenamefont {Gong},
  \citenamefont {Ashida}, \citenamefont {Kawabata}, \citenamefont {Takasan},
  \citenamefont {Higashikawa},\ and\ \citenamefont {Ueda}}]{Gong:PRX18}%
  \BibitemOpen
  \bibfield  {author} {\bibinfo {author} {\bibfnamefont {Z.}~\bibnamefont
  {Gong}}, \bibinfo {author} {\bibfnamefont {Y.}~\bibnamefont {Ashida}},
  \bibinfo {author} {\bibfnamefont {K.}~\bibnamefont {Kawabata}}, \bibinfo
  {author} {\bibfnamefont {K.}~\bibnamefont {Takasan}}, \bibinfo {author}
  {\bibfnamefont {S.}~\bibnamefont {Higashikawa}}, \ and\ \bibinfo {author}
  {\bibfnamefont {M.}~\bibnamefont {Ueda}},\ }\href {\doibase
  10.1103/PhysRevX.8.031079} {\bibfield  {journal} {\bibinfo  {journal} {Phys.
  Rev. X}\ }\textbf {\bibinfo {volume} {8}},\ \bibinfo {pages} {031079}
  (\bibinfo {year} {2018})}\BibitemShut {NoStop}%
\bibitem [{\citenamefont {McGinley}\ and\ \citenamefont
  {Cooper}(2019)}]{McGinley:PRB19}%
  \BibitemOpen
  \bibfield  {author} {\bibinfo {author} {\bibfnamefont {M.}~\bibnamefont
  {McGinley}}\ and\ \bibinfo {author} {\bibfnamefont {N.~R.}\ \bibnamefont
  {Cooper}},\ }\href {\doibase 10.1103/PhysRevB.99.075148} {\bibfield
  {journal} {\bibinfo  {journal} {Phys. Rev. B}\ }\textbf {\bibinfo {volume}
  {99}},\ \bibinfo {pages} {075148} (\bibinfo {year} {2019})}\BibitemShut
  {NoStop}%
\bibitem [{\citenamefont {Yu}\ \emph {et~al.}(2020)\citenamefont {Yu},
  \citenamefont {Chen}, \citenamefont {Gomanko}, \citenamefont {Badawy},
  \citenamefont {Bakkers}, \citenamefont {Zuo}, \citenamefont {Mourik},\ and\
  \citenamefont {Frolov}}]{Yu:A20}%
  \BibitemOpen
  \bibfield  {author} {\bibinfo {author} {\bibfnamefont {P.}~\bibnamefont
  {Yu}}, \bibinfo {author} {\bibfnamefont {J.}~\bibnamefont {Chen}}, \bibinfo
  {author} {\bibfnamefont {M.}~\bibnamefont {Gomanko}}, \bibinfo {author}
  {\bibfnamefont {G.}~\bibnamefont {Badawy}}, \bibinfo {author} {\bibfnamefont
  {E.~P. A.~M.}\ \bibnamefont {Bakkers}}, \bibinfo {author} {\bibfnamefont
  {K.}~\bibnamefont {Zuo}}, \bibinfo {author} {\bibfnamefont {V.}~\bibnamefont
  {Mourik}}, \ and\ \bibinfo {author} {\bibfnamefont {S.~M.}\ \bibnamefont
  {Frolov}},\ }\href {https://arxiv.org/abs/2004.08583} {\bibfield  {journal}
  {\bibinfo  {journal} {arXiv:2004.08583}\ } (\bibinfo {year}
  {2020})}\BibitemShut {NoStop}%
\bibitem [{\citenamefont {Szumniak}\ \emph {et~al.}(2017)\citenamefont
  {Szumniak}, \citenamefont {Chevallier}, \citenamefont {Loss},\ and\
  \citenamefont {Klinovaja}}]{Szumniak:PRB17}%
  \BibitemOpen
  \bibfield  {author} {\bibinfo {author} {\bibfnamefont {P.}~\bibnamefont
  {Szumniak}}, \bibinfo {author} {\bibfnamefont {D.}~\bibnamefont
  {Chevallier}}, \bibinfo {author} {\bibfnamefont {D.}~\bibnamefont {Loss}}, \
  and\ \bibinfo {author} {\bibfnamefont {J.}~\bibnamefont {Klinovaja}},\ }\href
  {\doibase 10.1103/PhysRevB.96.041401} {\bibfield  {journal} {\bibinfo
  {journal} {Phys. Rev. B}\ }\textbf {\bibinfo {volume} {96}},\ \bibinfo
  {pages} {041401} (\bibinfo {year} {2017})}\BibitemShut {NoStop}%
\bibitem [{Note1()}]{Note1}%
  \BibitemOpen
  \bibinfo {note} {As nanowire screening lengths are often in the hundreds of
  nanometers, it is plausible to expect smoother confinement potentials, and
  hence suppressed splittings, as the nanowire length is increased}\BibitemShut
  {NoStop}%
\bibitem [{\citenamefont {Chiu}\ \emph {et~al.}(2017)\citenamefont {Chiu},
  \citenamefont {Sau},\ and\ \citenamefont {Das~Sarma}}]{Chiu:PRB17}%
  \BibitemOpen
  \bibfield  {author} {\bibinfo {author} {\bibfnamefont {C.-K.}\ \bibnamefont
  {Chiu}}, \bibinfo {author} {\bibfnamefont {J.~D.}\ \bibnamefont {Sau}}, \
  and\ \bibinfo {author} {\bibfnamefont {S.}~\bibnamefont {Das~Sarma}},\ }\href
  {\doibase 10.1103/PhysRevB.96.054504} {\bibfield  {journal} {\bibinfo
  {journal} {Phys. Rev. B}\ }\textbf {\bibinfo {volume} {96}},\ \bibinfo
  {pages} {054504} (\bibinfo {year} {2017})}\BibitemShut {NoStop}%
\bibitem [{\citenamefont {Hoffman}\ \emph {et~al.}(2016)\citenamefont
  {Hoffman}, \citenamefont {Schrade}, \citenamefont {Klinovaja},\ and\
  \citenamefont {Loss}}]{Hoffman:PRB16}%
  \BibitemOpen
  \bibfield  {author} {\bibinfo {author} {\bibfnamefont {S.}~\bibnamefont
  {Hoffman}}, \bibinfo {author} {\bibfnamefont {C.}~\bibnamefont {Schrade}},
  \bibinfo {author} {\bibfnamefont {J.}~\bibnamefont {Klinovaja}}, \ and\
  \bibinfo {author} {\bibfnamefont {D.}~\bibnamefont {Loss}},\ }\href {\doibase
  10.1103/PhysRevB.94.045316} {\bibfield  {journal} {\bibinfo  {journal} {Phys.
  Rev. B}\ }\textbf {\bibinfo {volume} {94}},\ \bibinfo {pages} {045316}
  (\bibinfo {year} {2016})}\BibitemShut {NoStop}%
\bibitem [{\citenamefont {Wakatsuki}\ \emph {et~al.}(2014)\citenamefont
  {Wakatsuki}, \citenamefont {Ezawa},\ and\ \citenamefont
  {Nagaosa}}]{Wakatsuki:PRB14}%
  \BibitemOpen
  \bibfield  {author} {\bibinfo {author} {\bibfnamefont {R.}~\bibnamefont
  {Wakatsuki}}, \bibinfo {author} {\bibfnamefont {M.}~\bibnamefont {Ezawa}}, \
  and\ \bibinfo {author} {\bibfnamefont {N.}~\bibnamefont {Nagaosa}},\ }\href
  {\doibase 10.1103/PhysRevB.89.174514} {\bibfield  {journal} {\bibinfo
  {journal} {Phys. Rev. B}\ }\textbf {\bibinfo {volume} {89}},\ \bibinfo
  {pages} {174514} (\bibinfo {year} {2014})}\BibitemShut {NoStop}%
\bibitem [{\citenamefont {Karzig}\ \emph {et~al.}(2017)\citenamefont {Karzig},
  \citenamefont {Knapp}, \citenamefont {Lutchyn}, \citenamefont {Bonderson},
  \citenamefont {Hastings}, \citenamefont {Nayak}, \citenamefont {Alicea},
  \citenamefont {Flensberg}, \citenamefont {Plugge}, \citenamefont {Oreg},
  \citenamefont {Marcus},\ and\ \citenamefont {Freedman}}]{Karzig:PRB17}%
  \BibitemOpen
  \bibfield  {author} {\bibinfo {author} {\bibfnamefont {T.}~\bibnamefont
  {Karzig}}, \bibinfo {author} {\bibfnamefont {C.}~\bibnamefont {Knapp}},
  \bibinfo {author} {\bibfnamefont {R.~M.}\ \bibnamefont {Lutchyn}}, \bibinfo
  {author} {\bibfnamefont {P.}~\bibnamefont {Bonderson}}, \bibinfo {author}
  {\bibfnamefont {M.~B.}\ \bibnamefont {Hastings}}, \bibinfo {author}
  {\bibfnamefont {C.}~\bibnamefont {Nayak}}, \bibinfo {author} {\bibfnamefont
  {J.}~\bibnamefont {Alicea}}, \bibinfo {author} {\bibfnamefont
  {K.}~\bibnamefont {Flensberg}}, \bibinfo {author} {\bibfnamefont
  {S.}~\bibnamefont {Plugge}}, \bibinfo {author} {\bibfnamefont
  {Y.}~\bibnamefont {Oreg}}, \bibinfo {author} {\bibfnamefont {C.~M.}\
  \bibnamefont {Marcus}}, \ and\ \bibinfo {author} {\bibfnamefont {M.~H.}\
  \bibnamefont {Freedman}},\ }\href {\doibase 10.1103/PhysRevB.95.235305}
  {\bibfield  {journal} {\bibinfo  {journal} {Phys. Rev. B}\ }\textbf {\bibinfo
  {volume} {95}},\ \bibinfo {pages} {235305} (\bibinfo {year}
  {2017})}\BibitemShut {NoStop}%
\bibitem [{\citenamefont {Clarke}(2017)}]{Clarke:PRB17}%
  \BibitemOpen
  \bibfield  {author} {\bibinfo {author} {\bibfnamefont {D.~J.}\ \bibnamefont
  {Clarke}},\ }\href {\doibase 10.1103/PhysRevB.96.201109} {\bibfield
  {journal} {\bibinfo  {journal} {Phys. Rev. B}\ }\textbf {\bibinfo {volume}
  {96}},\ \bibinfo {pages} {201109} (\bibinfo {year} {2017})}\BibitemShut
  {NoStop}%
\bibitem [{\citenamefont {Schuray}\ \emph {et~al.}(2017)\citenamefont
  {Schuray}, \citenamefont {Weithofer},\ and\ \citenamefont
  {Recher}}]{Schuray:PRB17}%
  \BibitemOpen
  \bibfield  {author} {\bibinfo {author} {\bibfnamefont {A.}~\bibnamefont
  {Schuray}}, \bibinfo {author} {\bibfnamefont {L.}~\bibnamefont {Weithofer}},
  \ and\ \bibinfo {author} {\bibfnamefont {P.}~\bibnamefont {Recher}},\ }\href
  {\doibase 10.1103/PhysRevB.96.085417} {\bibfield  {journal} {\bibinfo
  {journal} {Phys. Rev. B}\ }\textbf {\bibinfo {volume} {96}},\ \bibinfo
  {pages} {085417} (\bibinfo {year} {2017})}\BibitemShut {NoStop}%
\bibitem [{\citenamefont {M\'enard}\ \emph {et~al.}(2020)\citenamefont
  {M\'enard}, \citenamefont {Anselmetti}, \citenamefont {Martinez},
  \citenamefont {Puglia}, \citenamefont {Malinowski}, \citenamefont {Lee},
  \citenamefont {Choi}, \citenamefont {Pendharkar}, \citenamefont
  {Palmstr\o{}m}, \citenamefont {Flensberg}, \citenamefont {Marcus},
  \citenamefont {Casparis},\ and\ \citenamefont {Higginbotham}}]{Menard:PRL20}%
  \BibitemOpen
  \bibfield  {author} {\bibinfo {author} {\bibfnamefont {G.~C.}\ \bibnamefont
  {M\'enard}}, \bibinfo {author} {\bibfnamefont {G.~L.~R.}\ \bibnamefont
  {Anselmetti}}, \bibinfo {author} {\bibfnamefont {E.~A.}\ \bibnamefont
  {Martinez}}, \bibinfo {author} {\bibfnamefont {D.}~\bibnamefont {Puglia}},
  \bibinfo {author} {\bibfnamefont {F.~K.}\ \bibnamefont {Malinowski}},
  \bibinfo {author} {\bibfnamefont {J.~S.}\ \bibnamefont {Lee}}, \bibinfo
  {author} {\bibfnamefont {S.}~\bibnamefont {Choi}}, \bibinfo {author}
  {\bibfnamefont {M.}~\bibnamefont {Pendharkar}}, \bibinfo {author}
  {\bibfnamefont {C.~J.}\ \bibnamefont {Palmstr\o{}m}}, \bibinfo {author}
  {\bibfnamefont {K.}~\bibnamefont {Flensberg}}, \bibinfo {author}
  {\bibfnamefont {C.~M.}\ \bibnamefont {Marcus}}, \bibinfo {author}
  {\bibfnamefont {L.}~\bibnamefont {Casparis}}, \ and\ \bibinfo {author}
  {\bibfnamefont {A.~P.}\ \bibnamefont {Higginbotham}},\ }\href {\doibase
  10.1103/PhysRevLett.124.036802} {\bibfield  {journal} {\bibinfo  {journal}
  {Phys. Rev. Lett.}\ }\textbf {\bibinfo {volume} {124}},\ \bibinfo {pages}
  {036802} (\bibinfo {year} {2020})}\BibitemShut {NoStop}%
\bibitem [{\citenamefont {Puglia}\ \emph {et~al.}(2020)\citenamefont {Puglia},
  \citenamefont {Martinez}, \citenamefont {M{\'e}nard}, \citenamefont
  {P{\"o}schl}, \citenamefont {Gronin}, \citenamefont {Gardner}, \citenamefont
  {Kallaher}, \citenamefont {Manfra}, \citenamefont {Marcus}, \citenamefont
  {Higginbotham},\ and\ \citenamefont {Casparis}}]{Puglia:A20}%
  \BibitemOpen
  \bibfield  {author} {\bibinfo {author} {\bibfnamefont {D.}~\bibnamefont
  {Puglia}}, \bibinfo {author} {\bibfnamefont {E.~A.}\ \bibnamefont
  {Martinez}}, \bibinfo {author} {\bibfnamefont {G.~C.}\ \bibnamefont
  {M{\'e}nard}}, \bibinfo {author} {\bibfnamefont {A.}~\bibnamefont
  {P{\"o}schl}}, \bibinfo {author} {\bibfnamefont {S.}~\bibnamefont {Gronin}},
  \bibinfo {author} {\bibfnamefont {G.~C.}\ \bibnamefont {Gardner}}, \bibinfo
  {author} {\bibfnamefont {R.}~\bibnamefont {Kallaher}}, \bibinfo {author}
  {\bibfnamefont {M.~J.}\ \bibnamefont {Manfra}}, \bibinfo {author}
  {\bibfnamefont {C.~M.}\ \bibnamefont {Marcus}}, \bibinfo {author}
  {\bibfnamefont {A.~P.}\ \bibnamefont {Higginbotham}}, \ and\ \bibinfo
  {author} {\bibfnamefont {L.}~\bibnamefont {Casparis}},\ }\href
  {https://arxiv.org/abs/2006.01275} {\bibfield  {journal} {\bibinfo  {journal}
  {arXiv:2006.01275}\ } (\bibinfo {year} {2020})}\BibitemShut {NoStop}%
\bibitem [{\citenamefont {Zhang}\ \emph {et~al.}(2019)\citenamefont {Zhang},
  \citenamefont {Liu}, \citenamefont {Wimmer},\ and\ \citenamefont
  {Kouwenhoven}}]{Zhang:NC19}%
  \BibitemOpen
  \bibfield  {author} {\bibinfo {author} {\bibfnamefont {H.}~\bibnamefont
  {Zhang}}, \bibinfo {author} {\bibfnamefont {D.~E.}\ \bibnamefont {Liu}},
  \bibinfo {author} {\bibfnamefont {M.}~\bibnamefont {Wimmer}}, \ and\ \bibinfo
  {author} {\bibfnamefont {L.~P.}\ \bibnamefont {Kouwenhoven}},\ }\href
  {\doibase 10.1038/s41467-019-13133-1} {\bibfield  {journal} {\bibinfo
  {journal} {Nat. Commun.}\ }\textbf {\bibinfo {volume} {10}},\ \bibinfo
  {pages} {5128} (\bibinfo {year} {2019})}\BibitemShut {NoStop}%
\bibitem [{\citenamefont {Frolov}\ \emph {et~al.}(2020)\citenamefont {Frolov},
  \citenamefont {Manfra},\ and\ \citenamefont {Sau}}]{Frolov:NP20}%
  \BibitemOpen
  \bibfield  {author} {\bibinfo {author} {\bibfnamefont {S.~M.}\ \bibnamefont
  {Frolov}}, \bibinfo {author} {\bibfnamefont {M.~J.}\ \bibnamefont {Manfra}},
  \ and\ \bibinfo {author} {\bibfnamefont {J.~D.}\ \bibnamefont {Sau}},\ }\href
  {\doibase 10.1038/s41567-020-0925-6} {\bibfield  {journal} {\bibinfo
  {journal} {Nature Physics}\ }\textbf {\bibinfo {volume} {16}},\ \bibinfo
  {pages} {718} (\bibinfo {year} {2020})}\BibitemShut {NoStop}%
\bibitem [{\citenamefont {Aguado}\ and\ \citenamefont
  {Kouwenhoven}(2020)}]{Aguado-Kouwenhoven:PT20}%
  \BibitemOpen
  \bibfield  {author} {\bibinfo {author} {\bibfnamefont {R.}~\bibnamefont
  {Aguado}}\ and\ \bibinfo {author} {\bibfnamefont {L.~P.}\ \bibnamefont
  {Kouwenhoven}},\ }\href {https://doi.org/10.1063/PT.3.4499} {\bibfield
  {journal} {\bibinfo  {journal} {Phys. Today}\ }\textbf {\bibinfo {volume}
  {73}},\ \bibinfo {pages} {44} (\bibinfo {year} {2020})}\BibitemShut {NoStop}%
\bibitem [{\citenamefont {Pe\~naranda}\ \emph {et~al.}(2020)\citenamefont
  {Pe\~naranda}, \citenamefont {Aguado}, \citenamefont {San-Jose},\ and\
  \citenamefont {Prada}}]{Penaranda:PRR20}%
  \BibitemOpen
  \bibfield  {author} {\bibinfo {author} {\bibfnamefont {F.}~\bibnamefont
  {Pe\~naranda}}, \bibinfo {author} {\bibfnamefont {R.}~\bibnamefont {Aguado}},
  \bibinfo {author} {\bibfnamefont {P.}~\bibnamefont {San-Jose}}, \ and\
  \bibinfo {author} {\bibfnamefont {E.}~\bibnamefont {Prada}},\ }\href@noop {}
  {\bibfield  {journal} {\bibinfo  {journal} {Phys. Rev. Research}\ }\textbf
  {\bibinfo {volume} {2}},\ \bibinfo {pages} {023171} (\bibinfo {year}
  {2020})}\BibitemShut {NoStop}%
\bibitem [{\citenamefont {Vaitiek{\.e}nas}\ \emph {et~al.}(2020)\citenamefont
  {Vaitiek{\.e}nas}, \citenamefont {Liu}, \citenamefont {Krogstrup},\ and\
  \citenamefont {Marcus}}]{Vaitiekenas:A20}%
  \BibitemOpen
  \bibfield  {author} {\bibinfo {author} {\bibfnamefont {S.}~\bibnamefont
  {Vaitiek{\.e}nas}}, \bibinfo {author} {\bibfnamefont {Y.}~\bibnamefont
  {Liu}}, \bibinfo {author} {\bibfnamefont {P.}~\bibnamefont {Krogstrup}}, \
  and\ \bibinfo {author} {\bibfnamefont {C.~M.}\ \bibnamefont {Marcus}},\
  }\href@noop {} {\bibfield  {journal} {\bibinfo  {journal} {arXiv:2004.02226}\
  } (\bibinfo {year} {2020})}\BibitemShut {NoStop}%
\bibitem [{\citenamefont {Larsen}\ \emph {et~al.}(2015)\citenamefont {Larsen},
  \citenamefont {Petersson}, \citenamefont {Kuemmeth}, \citenamefont
  {Jespersen}, \citenamefont {Krogstrup}, \citenamefont {Nyg\aa{}rd},\ and\
  \citenamefont {Marcus}}]{Larsen:PRL15}%
  \BibitemOpen
  \bibfield  {author} {\bibinfo {author} {\bibfnamefont {T.~W.}\ \bibnamefont
  {Larsen}}, \bibinfo {author} {\bibfnamefont {K.~D.}\ \bibnamefont
  {Petersson}}, \bibinfo {author} {\bibfnamefont {F.}~\bibnamefont {Kuemmeth}},
  \bibinfo {author} {\bibfnamefont {T.~S.}\ \bibnamefont {Jespersen}}, \bibinfo
  {author} {\bibfnamefont {P.}~\bibnamefont {Krogstrup}}, \bibinfo {author}
  {\bibfnamefont {J.}~\bibnamefont {Nyg\aa{}rd}}, \ and\ \bibinfo {author}
  {\bibfnamefont {C.~M.}\ \bibnamefont {Marcus}},\ }\href@noop {} {\bibfield
  {journal} {\bibinfo  {journal} {Phys. Rev. Lett.}\ }\textbf {\bibinfo
  {volume} {115}},\ \bibinfo {pages} {127001} (\bibinfo {year}
  {2015})}\BibitemShut {NoStop}%
\bibitem [{\citenamefont {de~Lange}\ \emph {et~al.}(2015)\citenamefont
  {de~Lange}, \citenamefont {van Heck}, \citenamefont {Bruno}, \citenamefont
  {van Woerkom}, \citenamefont {Geresdi}, \citenamefont {Plissard},
  \citenamefont {Bakkers}, \citenamefont {Akhmerov},\ and\ \citenamefont
  {DiCarlo}}]{Lange:PRL15}%
  \BibitemOpen
  \bibfield  {author} {\bibinfo {author} {\bibfnamefont {G.}~\bibnamefont
  {de~Lange}}, \bibinfo {author} {\bibfnamefont {B.}~\bibnamefont {van Heck}},
  \bibinfo {author} {\bibfnamefont {A.}~\bibnamefont {Bruno}}, \bibinfo
  {author} {\bibfnamefont {D.~J.}\ \bibnamefont {van Woerkom}}, \bibinfo
  {author} {\bibfnamefont {A.}~\bibnamefont {Geresdi}}, \bibinfo {author}
  {\bibfnamefont {S.~R.}\ \bibnamefont {Plissard}}, \bibinfo {author}
  {\bibfnamefont {E.~P. A.~M.}\ \bibnamefont {Bakkers}}, \bibinfo {author}
  {\bibfnamefont {A.~R.}\ \bibnamefont {Akhmerov}}, \ and\ \bibinfo {author}
  {\bibfnamefont {L.}~\bibnamefont {DiCarlo}},\ }\href@noop {} {\bibfield
  {journal} {\bibinfo  {journal} {Phys. Rev. Lett.}\ }\textbf {\bibinfo
  {volume} {115}},\ \bibinfo {pages} {127002} (\bibinfo {year}
  {2015})}\BibitemShut {NoStop}%
\bibitem [{\citenamefont {Sabonis}\ \emph {et~al.}(2020)\citenamefont
  {Sabonis}, \citenamefont {Erlandsson}, \citenamefont {Kringh{\o}j},
  \citenamefont {van Heck}, \citenamefont {Larsen}, \citenamefont {Petkovic},
  \citenamefont {Krogstrup}, \citenamefont {Petersson},\ and\ \citenamefont
  {Marcus}}]{Sabonis:A20}%
  \BibitemOpen
  \bibfield  {author} {\bibinfo {author} {\bibfnamefont {D.}~\bibnamefont
  {Sabonis}}, \bibinfo {author} {\bibfnamefont {O.}~\bibnamefont {Erlandsson}},
  \bibinfo {author} {\bibfnamefont {A.}~\bibnamefont {Kringh{\o}j}}, \bibinfo
  {author} {\bibfnamefont {B.}~\bibnamefont {van Heck}}, \bibinfo {author}
  {\bibfnamefont {T.~W.}\ \bibnamefont {Larsen}}, \bibinfo {author}
  {\bibfnamefont {I.}~\bibnamefont {Petkovic}}, \bibinfo {author}
  {\bibfnamefont {P.}~\bibnamefont {Krogstrup}}, \bibinfo {author}
  {\bibfnamefont {K.~D.}\ \bibnamefont {Petersson}}, \ and\ \bibinfo {author}
  {\bibfnamefont {C.~M.}\ \bibnamefont {Marcus}},\ }\href@noop {} {\bibfield
  {journal} {\bibinfo  {journal} {arXiv:2005.01748}\ } (\bibinfo {year}
  {2020})}\BibitemShut {NoStop}%
\bibitem [{\citenamefont {Bargerbos}\ \emph {et~al.}(2020)\citenamefont
  {Bargerbos}, \citenamefont {Uilhoorn}, \citenamefont {Yang}, \citenamefont
  {Krogstrup}, \citenamefont {Kouwenhoven}, \citenamefont {de~Lange},
  \citenamefont {van Heck},\ and\ \citenamefont {Kou}}]{Bagerbos:PRL20}%
  \BibitemOpen
  \bibfield  {author} {\bibinfo {author} {\bibfnamefont {A.}~\bibnamefont
  {Bargerbos}}, \bibinfo {author} {\bibfnamefont {W.}~\bibnamefont {Uilhoorn}},
  \bibinfo {author} {\bibfnamefont {C.-K.}\ \bibnamefont {Yang}}, \bibinfo
  {author} {\bibfnamefont {P.}~\bibnamefont {Krogstrup}}, \bibinfo {author}
  {\bibfnamefont {L.~P.}\ \bibnamefont {Kouwenhoven}}, \bibinfo {author}
  {\bibfnamefont {G.}~\bibnamefont {de~Lange}}, \bibinfo {author}
  {\bibfnamefont {B.}~\bibnamefont {van Heck}}, \ and\ \bibinfo {author}
  {\bibfnamefont {A.}~\bibnamefont {Kou}},\ }\href {\doibase
  10.1103/PhysRevLett.124.246802} {\bibfield  {journal} {\bibinfo  {journal}
  {Phys. Rev. Lett.}\ }\textbf {\bibinfo {volume} {124}},\ \bibinfo {pages}
  {246802} (\bibinfo {year} {2020})}\BibitemShut {NoStop}%
\bibitem [{\citenamefont {Kringh\o{}j}\ \emph {et~al.}(2020)\citenamefont
  {Kringh\o{}j}, \citenamefont {van Heck}, \citenamefont {Larsen},
  \citenamefont {Erlandsson}, \citenamefont {Sabonis}, \citenamefont
  {Krogstrup}, \citenamefont {Casparis}, \citenamefont {Petersson},\ and\
  \citenamefont {Marcus}}]{Kringhof:PRL20}%
  \BibitemOpen
  \bibfield  {author} {\bibinfo {author} {\bibfnamefont {A.}~\bibnamefont
  {Kringh\o{}j}}, \bibinfo {author} {\bibfnamefont {B.}~\bibnamefont {van
  Heck}}, \bibinfo {author} {\bibfnamefont {T.~W.}\ \bibnamefont {Larsen}},
  \bibinfo {author} {\bibfnamefont {O.}~\bibnamefont {Erlandsson}}, \bibinfo
  {author} {\bibfnamefont {D.}~\bibnamefont {Sabonis}}, \bibinfo {author}
  {\bibfnamefont {P.}~\bibnamefont {Krogstrup}}, \bibinfo {author}
  {\bibfnamefont {L.}~\bibnamefont {Casparis}}, \bibinfo {author}
  {\bibfnamefont {K.~D.}\ \bibnamefont {Petersson}}, \ and\ \bibinfo {author}
  {\bibfnamefont {C.~M.}\ \bibnamefont {Marcus}},\ }\href {\doibase
  10.1103/PhysRevLett.124.246803} {\bibfield  {journal} {\bibinfo  {journal}
  {Phys. Rev. Lett.}\ }\textbf {\bibinfo {volume} {124}},\ \bibinfo {pages}
  {246803} (\bibinfo {year} {2020})}\BibitemShut {NoStop}%
\bibitem [{\citenamefont {Ginossar}\ and\ \citenamefont
  {Grosfeld}(2014)}]{Ginossar:N14}%
  \BibitemOpen
  \bibfield  {author} {\bibinfo {author} {\bibfnamefont {E.}~\bibnamefont
  {Ginossar}}\ and\ \bibinfo {author} {\bibfnamefont {E.}~\bibnamefont
  {Grosfeld}},\ }\href@noop {} {\bibfield  {journal} {\bibinfo  {journal}
  {Nature communications}\ }\textbf {\bibinfo {volume} {5}},\ \bibinfo {pages}
  {4772} (\bibinfo {year} {2014})}\BibitemShut {NoStop}%
\bibitem [{\citenamefont {Trif}\ \emph {et~al.}(2018)\citenamefont {Trif},
  \citenamefont {Dmytruk}, \citenamefont {Bouchiat}, \citenamefont {Aguado},\
  and\ \citenamefont {Simon}}]{Trif:PRB18}%
  \BibitemOpen
  \bibfield  {author} {\bibinfo {author} {\bibfnamefont {M.}~\bibnamefont
  {Trif}}, \bibinfo {author} {\bibfnamefont {O.}~\bibnamefont {Dmytruk}},
  \bibinfo {author} {\bibfnamefont {H.}~\bibnamefont {Bouchiat}}, \bibinfo
  {author} {\bibfnamefont {R.}~\bibnamefont {Aguado}}, \ and\ \bibinfo {author}
  {\bibfnamefont {P.}~\bibnamefont {Simon}},\ }\href {\doibase
  10.1103/PhysRevB.97.041415} {\bibfield  {journal} {\bibinfo  {journal} {Phys.
  Rev. B}\ }\textbf {\bibinfo {volume} {97}},\ \bibinfo {pages} {041415}
  (\bibinfo {year} {2018})}\BibitemShut {NoStop}%
\bibitem [{\citenamefont {Keselman}\ \emph {et~al.}(2019)\citenamefont
  {Keselman}, \citenamefont {Murthy}, \citenamefont {van Heck},\ and\
  \citenamefont {Bauer}}]{Kesselman:S19}%
  \BibitemOpen
  \bibfield  {author} {\bibinfo {author} {\bibfnamefont {A.}~\bibnamefont
  {Keselman}}, \bibinfo {author} {\bibfnamefont {C.}~\bibnamefont {Murthy}},
  \bibinfo {author} {\bibfnamefont {B.}~\bibnamefont {van Heck}}, \ and\
  \bibinfo {author} {\bibfnamefont {B.}~\bibnamefont {Bauer}},\ }\href
  {\doibase 10.21468/SciPostPhys.7.4.050} {\bibfield  {journal} {\bibinfo
  {journal} {SciPost Phys.}\ }\textbf {\bibinfo {volume} {7}},\ \bibinfo
  {pages} {50} (\bibinfo {year} {2019})}\BibitemShut {NoStop}%
\bibitem [{\citenamefont {Avila}\ \emph
  {et~al.}(2020{\natexlab{a}})\citenamefont {Avila}, \citenamefont {Prada},
  \citenamefont {San-Jose},\ and\ \citenamefont {Aguado}}]{Avila:A20a}%
  \BibitemOpen
  \bibfield  {author} {\bibinfo {author} {\bibfnamefont {J.}~\bibnamefont
  {Avila}}, \bibinfo {author} {\bibfnamefont {E.}~\bibnamefont {Prada}},
  \bibinfo {author} {\bibfnamefont {P.}~\bibnamefont {San-Jose}}, \ and\
  \bibinfo {author} {\bibfnamefont {R.}~\bibnamefont {Aguado}},\ }\href
  {https://arxiv.org/abs/2003.02852} {\bibfield  {journal} {\bibinfo  {journal}
  {arXiv:2003.02852}\ } (\bibinfo {year} {2020}{\natexlab{a}})}\BibitemShut
  {NoStop}%
\bibitem [{\citenamefont {Avila}\ \emph
  {et~al.}(2020{\natexlab{b}})\citenamefont {Avila}, \citenamefont {Prada},
  \citenamefont {San-Jose},\ and\ \citenamefont {Aguado}}]{Avila:A20}%
  \BibitemOpen
  \bibfield  {author} {\bibinfo {author} {\bibfnamefont {J.}~\bibnamefont
  {Avila}}, \bibinfo {author} {\bibfnamefont {E.}~\bibnamefont {Prada}},
  \bibinfo {author} {\bibfnamefont {P.}~\bibnamefont {San-Jose}}, \ and\
  \bibinfo {author} {\bibfnamefont {R.}~\bibnamefont {Aguado}},\ }\href
  {https://arxiv.org/abs/2003.02858} {\bibfield  {journal} {\bibinfo  {journal}
  {arXiv:2003.02858}\ } (\bibinfo {year} {2020}{\natexlab{b}})}\BibitemShut
  {NoStop}%
\bibitem [{\citenamefont {Finocchiaro}\ \emph {et~al.}(2018)\citenamefont
  {Finocchiaro}, \citenamefont {Guinea},\ and\ \citenamefont
  {San-Jose}}]{Finocchiaro:PRL18}%
  \BibitemOpen
  \bibfield  {author} {\bibinfo {author} {\bibfnamefont {F.}~\bibnamefont
  {Finocchiaro}}, \bibinfo {author} {\bibfnamefont {F.}~\bibnamefont {Guinea}},
  \ and\ \bibinfo {author} {\bibfnamefont {P.}~\bibnamefont {San-Jose}},\
  }\href@noop {} {\bibfield  {journal} {\bibinfo  {journal} {Phys. Rev. Lett.}\
  }\textbf {\bibinfo {volume} {120}},\ \bibinfo {pages} {116801} (\bibinfo
  {year} {2018})}\BibitemShut {NoStop}%
\bibitem [{\citenamefont {Thakurathi}\ \emph {et~al.}(2018)\citenamefont
  {Thakurathi}, \citenamefont {Simon}, \citenamefont {Mandal}, \citenamefont
  {Klinovaja},\ and\ \citenamefont {Loss}}]{Thakurathi:PRB18}%
  \BibitemOpen
  \bibfield  {author} {\bibinfo {author} {\bibfnamefont {M.}~\bibnamefont
  {Thakurathi}}, \bibinfo {author} {\bibfnamefont {P.}~\bibnamefont {Simon}},
  \bibinfo {author} {\bibfnamefont {I.}~\bibnamefont {Mandal}}, \bibinfo
  {author} {\bibfnamefont {J.}~\bibnamefont {Klinovaja}}, \ and\ \bibinfo
  {author} {\bibfnamefont {D.}~\bibnamefont {Loss}},\ }\href@noop {} {\bibfield
   {journal} {\bibinfo  {journal} {Phys. Rev. B}\ }\textbf {\bibinfo {volume}
  {97}},\ \bibinfo {pages} {045415} (\bibinfo {year} {2018})}\BibitemShut
  {NoStop}%
\bibitem [{\citenamefont {Young}\ \emph {et~al.}(2014)\citenamefont {Young},
  \citenamefont {Sanchez-Yamagishi}, \citenamefont {Hunt}, \citenamefont
  {Choi}, \citenamefont {Watanabe}, \citenamefont {Taniguchi}, \citenamefont
  {Ashoori},\ and\ \citenamefont {Jarillo-Herrero}}]{Young:N14}%
  \BibitemOpen
  \bibfield  {author} {\bibinfo {author} {\bibfnamefont {A.~F.}\ \bibnamefont
  {Young}}, \bibinfo {author} {\bibfnamefont {J.~D.}\ \bibnamefont
  {Sanchez-Yamagishi}}, \bibinfo {author} {\bibfnamefont {B.}~\bibnamefont
  {Hunt}}, \bibinfo {author} {\bibfnamefont {S.~H.}\ \bibnamefont {Choi}},
  \bibinfo {author} {\bibfnamefont {K.}~\bibnamefont {Watanabe}}, \bibinfo
  {author} {\bibfnamefont {T.}~\bibnamefont {Taniguchi}}, \bibinfo {author}
  {\bibfnamefont {R.~C.}\ \bibnamefont {Ashoori}}, \ and\ \bibinfo {author}
  {\bibfnamefont {P.}~\bibnamefont {Jarillo-Herrero}},\ }\href@noop {}
  {\bibfield  {journal} {\bibinfo  {journal} {Nature}\ }\textbf {\bibinfo
  {volume} {505}},\ \bibinfo {pages} {528} (\bibinfo {year}
  {2014})}\BibitemShut {NoStop}%
\bibitem [{\citenamefont {Lee}\ \emph {et~al.}(2017{\natexlab{b}})\citenamefont
  {Lee}, \citenamefont {Huang}, \citenamefont {Efetov}, \citenamefont {Wei},
  \citenamefont {Hart}, \citenamefont {Taniguchi}, \citenamefont {Watanabe},
  \citenamefont {Yacoby},\ and\ \citenamefont {Kim}}]{Lee:NP17}%
  \BibitemOpen
  \bibfield  {author} {\bibinfo {author} {\bibfnamefont {G.-H.}\ \bibnamefont
  {Lee}}, \bibinfo {author} {\bibfnamefont {K.-F.}\ \bibnamefont {Huang}},
  \bibinfo {author} {\bibfnamefont {D.~K.}\ \bibnamefont {Efetov}}, \bibinfo
  {author} {\bibfnamefont {D.~S.}\ \bibnamefont {Wei}}, \bibinfo {author}
  {\bibfnamefont {S.}~\bibnamefont {Hart}}, \bibinfo {author} {\bibfnamefont
  {T.}~\bibnamefont {Taniguchi}}, \bibinfo {author} {\bibfnamefont
  {K.}~\bibnamefont {Watanabe}}, \bibinfo {author} {\bibfnamefont
  {A.}~\bibnamefont {Yacoby}}, \ and\ \bibinfo {author} {\bibfnamefont
  {P.}~\bibnamefont {Kim}},\ }\href@noop {} {\bibfield  {journal} {\bibinfo
  {journal} {Nat Phys}\ }\textbf {\bibinfo {volume} {13}},\ \bibinfo {pages}
  {693} (\bibinfo {year} {2017}{\natexlab{b}})}\BibitemShut {NoStop}%
\bibitem [{\citenamefont {San-Jose}\ \emph {et~al.}(2015)\citenamefont
  {San-Jose}, \citenamefont {Lado}, \citenamefont {Aguado}, \citenamefont
  {Guinea},\ and\ \citenamefont {Fern\'andez-Rossier}}]{San-Jose:PRX15}%
  \BibitemOpen
  \bibfield  {author} {\bibinfo {author} {\bibfnamefont {P.}~\bibnamefont
  {San-Jose}}, \bibinfo {author} {\bibfnamefont {J.~L.}\ \bibnamefont {Lado}},
  \bibinfo {author} {\bibfnamefont {R.}~\bibnamefont {Aguado}}, \bibinfo
  {author} {\bibfnamefont {F.}~\bibnamefont {Guinea}}, \ and\ \bibinfo {author}
  {\bibfnamefont {J.}~\bibnamefont {Fern\'andez-Rossier}},\ }\href@noop {}
  {\bibfield  {journal} {\bibinfo  {journal} {Phys. Rev. X}\ }\textbf {\bibinfo
  {volume} {5}},\ \bibinfo {pages} {041042} (\bibinfo {year}
  {2015})}\BibitemShut {NoStop}%
\end{thebibliography}%

\end{document}